\documentclass{aa}  

\pdfoutput=1

\usepackage[english]{babel}  
\usepackage{graphicx}  
\usepackage[utf8]{inputenc}  
\usepackage{microtype}  
\usepackage{booktabs}  
\usepackage{lscape}
\usepackage[squaren]{SIunits}
\usepackage{natbib}
\usepackage{amsmath}
\usepackage{amssymb}
\usepackage{txfonts}
\usepackage[caption=false]{subfig}
\usepackage{tikz}
\usepackage{hyperref}

\newcommand{\tk}{T_\mathrm{{K}}}
\newcommand{\trot}{T_\mathrm{{rot}}}

\newcommand{\td}{T_\mathrm{{d}}}

\newcommand{\nhtwo}{n(\mathrm{H_2})}
\newcommand{\coldhtwo}{N(\mathrm{H_2})}
\newcommand{\co}{C$^{18}$O}

\newcommand{\pot}[1]{10^{#1}}

\newcommand{\vlsr}{V_\mathrm{{LSR}}}

\newcommand{\cm}{\usk\centi \metre}

\newcommand{\pc}{\usk\mathrm{pc}}
\newcommand{\pctab}{\mathrm{pc}}
\newcommand{\hii}{H\textsc{ii}}
\newcommand{\msun}{\usk\mathrm{M_\odot}} 
\newcommand{\msuntab}{\mathrm{M_\odot}} 
\newcommand{\jy}{\usk\mathrm{Jy}}
\newcommand{\jytab}{\mathrm{Jy}}
\newcommand{\yr}{\usk\mathrm{yr}}

\newcommand{\lsun}{\usk\mathrm{L_\odot}}
\newcommand{\lsuntab}{\mathrm{L_\odot}}
\newcommand{\lbol}{L_\mathrm{bol}}

\newcommand{\kms}{\usk\kilo\metre\usk\second^{-1}}
\newcommand{\kmstab}{\kilo\metre\usk\second^{-1}}
\newcommand{\kel}{\usk\kelvin}
\newcommand{\mum}{\usk\micro\metre}
\newcommand{\mm}{\usk\milli\metre}
\newcommand{\asec}{^{\prime\prime}}
\newcommand{\unc}[2]{_{#1}^{#2}}
\newcommand{\mvir}{M_\mathrm{vir}}

\newcommand{\msest}{M}

\bibpunct{(}{)}{;}{a}{}{,}

\title{Physical properties of high-mass clumps in different stages of evolution\thanks{Partly based on observations with the Herschel satellite. {\it Herschel} is an ESA space observatory with science instruments provided by European-led Principal Investigator consortia and with important participation from NASA.}}
\author{A. Giannetti \inst{\ref{ira}, \ref{unibo}}
\and J. Brand \inst{\ref{ira}, \ref{arc}}
\and \'A. S\'anchez-Monge \inst{\ref{arcetri}}
\and F. Fontani \inst{\ref{arcetri}}
\and R. Cesaroni \inst{\ref{arcetri}}
\and M. T. Beltr\'{a}n\inst{\ref{arcetri}}
\and S. Molinari \inst{\ref{iaps}}
\and R. Dodson \inst{\ref{austr}}
\and M. J. Rioja \inst{\ref{austr}, \ref{rioja}}}
\institute{INAF-Istituto di Radioastronomia, Via Gobetti 101, 40129, Bologna, Italy \label{ira}
\and Dipartimento di Astronomia, Universit\`{a} di Bologna, Via Ranzani 1, 40127, Bologna, Italy \label{unibo}
\and Italian ALMA Regional Centre, Via Gobetti 101, 40129, Bologna, Italy \label{arc}
\and INAF-Osservatorio Astrofisico di Arcetri, Largo E. Fermi 5, 50125, Firenze, Italy \label{arcetri}
\and INAF-Istituto di Astrofisica e Planetologia Spaziali, Via Fosso del Cavaliere 100, 00133, Roma, Italy \label{iaps}
\and ICRAR, University of Western Australia, Perth, Australia \label{austr}
\and Observatorio Astron\'omico Nacional (OAN), Apartado 112, E-28803, Alcala de Henares, Spain \label{rioja}}

\abstract
{
The details of the process of massive star formation are still elusive. A complete characterization of the first stages of the process from an observational point of view is needed to constrain theories on the subject. In the past 20 years we have made a thorough investigation of colour-selected IRAS sources over the whole sky. The sources in the northern hemisphere were studied in detail and used to derive an evolutionary sequence based on their spectral energy distribution.%
}
{
To investigate the first stages of the process of high-mass star formation, we selected a sample of massive clumps previously observed with the Swedish-ESO Submillimetre Telescope at $1.2\mm$ and with the ATNF Australia Telescope Compact Array at $1.3\cm$. We want to characterize the physical conditions in such sources, and test whether their properties depend on the evolutionary stage of the clump.%
}
{
With ATCA we observed the selected sources in the NH$_3$(1,1) and (2,2) transitions and in the H$_2$O($6_{16}-5_{23}$) maser line. Ammonia lines are a very good temperature probe that allow us to accurately determine the mass and the column, volume, and surface densities of the clumps. We also collected all data available to construct the spectral energy distribution of the individual clumps and to determine if star formation is already occurring through observations of its most common signposts, thus putting constraints on the evolutionary stage of the source. We fitted the spectral energy distribution between $1.2\mm$ and $70\usk \mu \metre$ with a modified black body to derive the dust temperature and independently determine the mass.%
}
{
We find that the clumps are cold ($T\sim10-30\kel$), massive ($M\sim\pot{2}-\pot{3}\msun$), and dense ($\nhtwo\gtrsim\pot{5}\cm^{-3}$) and that they have high column densities ($\coldhtwo\sim\pot{23}\cm^{-2}$). All clumps appear to be potentially able to form high-mass stars. The most massive clumps appear to be gravitationally unstable, if the only sources of support against collapse are turbulence and thermal pressure, which possibly indicates that the magnetic field is important in stabilizing them.%
}
{
After investigating how the average properties depend on the evolutionary phase of the source, we find that the temperature and central density progressively increase with time. Sources likely hosting a ZAMS star show a steeper radial dependence of the volume density and tend to be more compact than starless clumps.%
}

\offprints{A. Giannetti (agianne@ira.inaf.it)}                                                                                              

\begin{document}  
\maketitle

\section{Introduction}

Massive stars spend a significant part ($\gtrsim10\%$) of their lives embedded in their parental molecular cloud, making it difficult to investigate their early evolutionary stages.
The discovery of IR-dark clouds \citep[IRDCs; e.g., ][]{Perault+96, Egan+98} seen in absorption against the mid-IR Galactic background made it possible to identify the most likely birthplaces of high-mass stars. These clouds are usually filamentary, hosting complexes of cold ($T\lesssim25\usk\kelvin$) and dense ($n\gtrsim\pot{5}\cm^{-3}$) clumps, with high H$_2$ column densities ($N\gtrsim\pot{23}\cm^{-2}$) and masses that usually exceed $100\msun$ (though not all of them will form massive stars; e.g., \citealt{KauffmannPillai10}).
Clouds with such low temperatures have a spectral energy distribution (SED) that peaks at far-IR (FIR) wavelengths and are optically thin in the millimetre/submillimetre regime. The emission at these wavelengths usually matches the IR absorption very well \citep[e.g., ][]{Rathborne+06, Pillai+06}, and makes it easy to identify the cold and dense gas concentrations. Some clumps within IRDCs show signs of active star formation, such as $24\mum$ emission, presence of extended excess emission at $4.5\mum$\footnote{The excess at $4.5\mum$, typically named Extended Green Object or ``green fuzzy'' is commonly interpreted as arising from $\mathrm{H_2}$ and CO lines, likely tracing shocks \citep[e.g., ][]{Noriega-Crespo+04, Marston+04}.}, masers and SiO emission from outflows (e.g., \citealt{Beuther+05, Rathborne+05, Chambers+09}).

The pre-/proto-stellar phase for high-mass stars is very short ($\lesssim3\times\pot{4}\yr$), according to statistical studies \citep{Motte+07}, when compared to the low-mass regime \citep[$\sim 3\times\pot{5}\yr$,][]{Kirk+05}, because the accretion timescale is longer than the Kelvin-Helmoltz timescale and nuclear fusion starts before the star has reached its final mass. Therefore, the entire life of the protostar and part of the main sequence is spent inside the parental clump. Objects in these early phases of evolution and their influence on the surrounding material, can be investigated at frequencies that can penetrate the cocoon in which the objects are enshrouded.

In the last 2 decades we have made a thorough investigation of a sample of luminous IRAS sources distributed over the whole sky, selected on the basis of FIR colours typical of YSOs \citep[e.g, ][]{Palla+91}. Our expectation that this sample contains high-mass YSOs in different evolutionary stages has been supported by a large number of observations at both low- and high-angular resolution \citep[e.g., ][]{Molinari+96, Molinari+98a, Molinari+98b, Brand+01, Fontani+05, Beltran+06}. 
Those with $\delta<30\degree$ have been observed with the SEST in the continuum at 1.2-mm (SIMBA) and in CS \citep{Fontani+05, Beltran+06}. The mm-continuum maps often show the presence of several clumps around a single IRAS source. A comparison with MSX (and later Spitzer) images revealed that some of these clumps are associated with mid-IR emission, while others appear IR-dark \citep{Beltran+06}.

\begin{table}[tbp]
\centering
\caption{Central coordinates of the observed fields, names of the clumps in the field and their coordinates.}
\label{tab:positions}
\tiny
\begin{tabular}{ccccc}
\toprule
 \multicolumn{2}{c}{Phase Centre (J2000)}                           & Clump        & \multicolumn{2}{c}{Clump Coordinates (J2000)} \\
\midrule
 RA                           & DEC                         &              & RA              & DEC                 \\
\midrule
 08:49:35.13                  &$-$44:11:59.0                & 08477-4359c1 & 08:49:35.13     &$-$44:11:59.0        \\ 
 09:00:40.50                  &$-$47:25:55.0                & 08589-4714c1 & 09:00:39.71     &$-$47:26:11.0        \\ 
 10:10:41.70                  &$-$57:44:36.0                & 10088-5730c2 & 10:10:41.70     &$-$57:44:36.0        \\
 12:32:52.10                  &$-$61:35:42.0                & 12300-6119c1 & 12:32:49.86     &$-$61:35:34.0        \\ 
 13:07:09.40                  &$-$63:47:12.0                & 13039-6331c1 & 13:07:08.19     &$-$63:47:12.0        \\
                {13:59:33.04} &$                -$61:49:13.0& 13560-6133c1 & 13:59:31.91     &$-$61:48:41.0        \\ 
                              &$                 $          & 13560-6133c2 & 13:59:33.04     &$-$61:49:13.0        \\ 
 13:59:55.50                  &$-$61:24:25.0                & 13563-6109c1 & 13:59:57.73     &$-$61:24:33.0        \\ 
                {14:20:21.74} &$                -$61:31:13.0& 14166-6118c1 & 14:20:19.50     &$-$61:31:53.0        \\
                              &$                 $          & 14166-6118c2 & 14:20:21.74     &$-$61:31:13.0        \\ 
 14:22:21.54                  &$-$61:06:42.0                & 14183-6050c3 & 14:22:21.54     &$-$61:06:42.0        \\ 
 15:07:32.52                  &$-$58:40:33.0                & 15038-5828c1 & 15:07:32.52     &$-$58:40:33.0        \\ 
 15:11:07.90                  &$-$59:06:30.0                & 15072-5855c1 & 15:11:08.94     &$-$59:06:46.0        \\ 
                {15:31:44.17} &$                -$56:32:08.0& 15278-5620c1 & 15:31:45.13     &$-$56:30:48.0        \\ 
                              &$                 $          & 15278-5620c2 & 15:31:44.17     &$-$56:32:08.0        \\ 
 15:48:40.82                  &$-$53:40:35.0                & 15454-5335c2 & 15:48:40.82     &$-$53:40:35.0        \\
 15:51:28.24                  &$-$54:31:42.0                & 15470-5419c1 & 15:51:28.24     &$-$54:31:42.0        \\ 
 15:51:01.62                  &$-$54:26:46.0                & 15470-5419c3 & 15:51:01.62     &$-$54:26:46.0        \\ 
 15:50:56.12                  &$-$54:30:38.0                & 15470-5419c4 & 15:50:56.12     &$-$54:30:38.0        \\ 
                {15:59:36.20} &$                -$52:22:58.0& 15557-5215c1 & 15:59:40.57     &$-$52:23:30.0        \\ 
                              &$                 $          & 15557-5215c2 & 15:59:36.20     &$-$52:22:58.0        \\ 
 15:59:39.70                  &$-$52:25:14.0                & 15557-5215c3 & 15:59:39.70     &$-$52:25:14.0        \\ 
 16:01:52.83                  &$-$53:11:57.0                & 15579-5303c1 & 16:01:46.60     &$-$53:11:41.0        \\ 
 16:02:08.86                  &$-$53:08:53.0                & 15579-5303c3 & 16:02:08.86     &$-$53:08:53.0        \\ 
 16:10:06.61                  &$-$50:50:29.0                & 16061-5048c1 & 16:10:06.61     &$-$50:50:29.0        \\ 
 16:09:57.30                  &$-$50:57:09.0                & 16061-5048c2 & 16:10:02.38     &$-$50:49:33.0        \\ 
 16:10:06.61                  &$-$50:57:09.0                & 16061-5048c4 & 16:10:06.61     &$-$50:57:09.0        \\ 
 16:13:05.20                  &$-$50:23:05.0                & 16093-5015c1 & 16:13:01.85     &$-$50:22:41.0        \\ 
                {16:12:55.46} &$                -$51:43:22.0& 16093-5128c1 & 16:12:49.45     &$-$51:43:30.0        \\ 
                              &$                 $          & 16093-5128c2 & 16:12:55.46     &$-$51:43:22.0        \\ 
 16:12:49.45                  &$-$51:36:34.0                & 16093-5128c8 & 16:12:49.45     &$-$51:36:34.0        \\ 
                {16:20:24.51} &$                -$49:35:34.0& 16164-4929c2 & 16:20:18.75     &$-$49:34:54.0        \\
                              &$                 $          & 16164-4929c3 & 16:20:24.51     &$-$49:35:34.0        \\ 
 16:20:31.92                  &$-$49:35:26.0                & 16164-4929c6 & 16:20:31.92     &$-$49:35:26.0        \\ 
 16:20:24.33                  &$-$48:44:58.0                & 16164-4837c2 & 16:20:24.33     &$-$48:44:58.0        \\ 
 16:29:00.89                  &$-$48:50:31.0                & 16254-4844c1 & 16:29:00.89     &$-$48:50:31.0        \\ 
 16:47:01.70                  &$-$41:15:18.0                & 16428-4109c1 & 16:47:01.70     &$-$41:15:18.0        \\ 
 16:46:46.81                  &$-$41:14:22.0                & 16428-4109c2 & 16:46:46.81     &$-$41:14:22.0        \\ 
 16:47:33.13                  &$-$45:22:51.0                & 16435-4515c3 & 16:47:33.13     &$-$45:22:51.0        \\ 
 16:51:44.59                  &$-$44:46:50.0                & 16482-4443c2 & 16:51:44.59     &$-$44:46:50.0        \\ 
 17:00:33.38                  &$-$42:25:18.0                & 16573-4214c2 & 17:00:33.38     &$-$42:25:18.0        \\ 
 17:07:58.78                  &$-$40:02:24.0                & 17040-3959c1 & 17:07:58.78     &$-$40:02:24.0        \\ 
 17:23:00.30                  &$-$38:13:54.0                & 17195-3811c1 & 17:23:00.98     &$-$38:13:54.0        \\ 
                {17:23:00.30} &$                -$38:14:58.0& 17195-3811c2 & 17:23:00.30     &$-$38:14:58.0        \\ 
                              &$                 $          & 17195-3811c3 & 17:23:00.98     &$-$38:15:38.0        \\ 
 17:38:49.87                  &$-$32:43:27.0                & 17355-3241c1 & 17:38:49.87     &$-$32:43:27.0        \\ 
\bottomrule                                                                          
\end{tabular}
\end{table}

A first attempt to exploit such large amount of data to define an evolutionary sequence for the clumps and their embedded sources was carried out by \citet{Molinari+08}, who distinguished three different types of objects, on the basis of their mm and IR properties: {\bf (Type 1)} objects with dominant mm emission, and not associated with a mid-IR source; {\bf (Type 2)} objects with both IR and mm emission; and {\bf (Type 3)} objects with clearly dominant IR emission. 
Using a simple model, the authors could explain these different types in terms of an evolutionary scenario, in order of increasing age: {(Type 1)} starless cores and/or high-mass proto-stars embedded in dusty clumps; {(Type 2)} deeply embedded Zero-Age Main Sequence (ZAMS) OB star(s) still accreting material from the parental clump, and {(Type 3)} OB stars surrounded only by the remnants of the molecular cloud. 
From our recent ATCA 1.3~cm continuum and line (H$_2$O maser at 22~GHz) observations \citep{Sanchez-Monge+13} of a large number ($\sim \,200$) of these massive clumps selected from the SEST mm-continuum observations, we found that Type 1 sources are rarely associated with cm-continuum emission (8\%), Types 2 (75\%) and 3 (28\%) more frequently. At the same time, H$_2$O maser emission was found associated with 13\%, 26\%, and 3\% of sources of Type 1, 2 and 3, respectively. These findings corroborate the evolutionary sequence derived by \citet{Molinari+08}. 

In this paper we will explore how the gas and dust properties in massive clumps depend on the evolutionary phase, as determined from the source type and the presence of signposts of high-mass star formation. 

\begin{figure}[tbp]
	\centering
	\includegraphics[angle=-90,width=0.75\columnwidth]{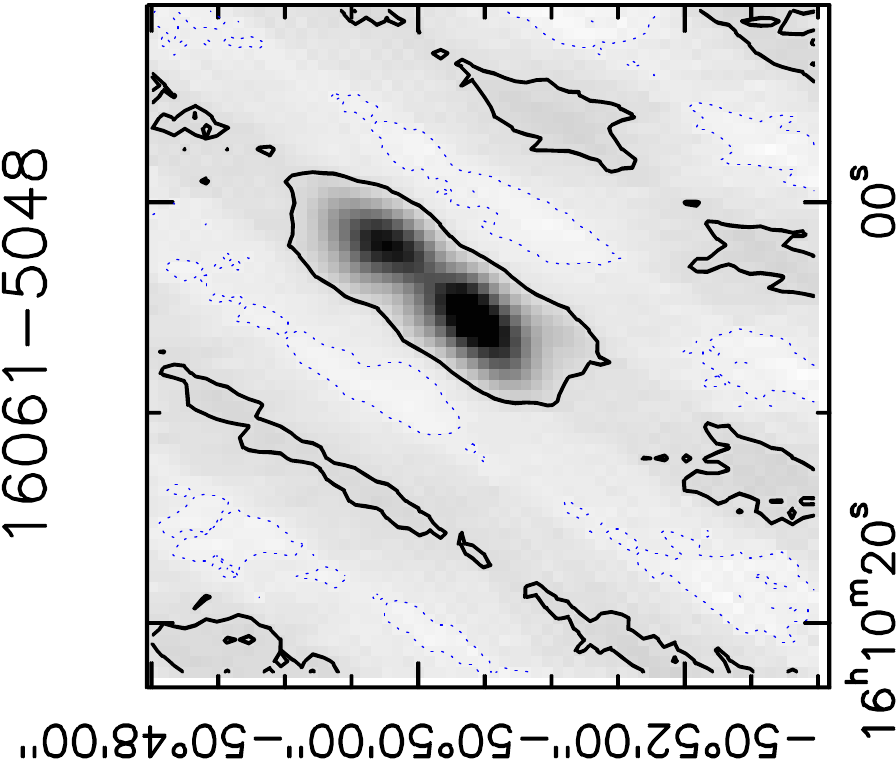}
	\caption{Typical map for a source in which we filter out the extended emission. The contours are $\pm3\%$ of the intensity peak of $628.3 \usk\milli\jytab$.}
	\label{fig:neg_lobes}
\end{figure}

The paper is organized as follows: in Sect.~\ref{sec:sample} we briefly describe the sample selection, in Sect.~\ref{sec:obs} we describe the observations performed with the Australia Telescope Compact Array, and describe the data reduction procedure; in Sect.~\ref{sec:results} we show the results for the quantities directly derived from the observations; in Sect.~\ref{sec:discussion} we discuss how the sample is divided into ``star-forming'' and ``quiescent'' clumps, and into clumps likely hosting a ZAMS star (Types 2 and 3) and clumps that are starless or with a deeply embedded protostar (Type 1). The mean clump properties are investigated to search for differences as a function of evolutionary phase. In Sect.~\ref{sec:sketch} we give a sketch of the different classes of objects identified and finally in Sect.~\ref{sec:summary} we summarize our findings.

\section{Sample and tracer}\label{sec:sample}

The 39 fields considered in this paper were selected from the \citet{Beltran+06} survey at $1.2\mm$, carried out with SEST/SIMBA towards IRAS sources, and contain 46 massive millimetre clumps. The coordinates of the field centres and of the clumps are listed in Table~\ref{tab:positions}. Each field contains at least one massive clump. 
The selection of fields was done according to simple criteria: (i) source declination $\delta<-30\degree$; (ii) comparable numbers of MSX-dark and -bright sources; (iii) clumps as isolated as possible, i.e. with a separation greater than 1 SEST beam ($24\arcsec$) between MSX and non-MSX emitters to limit confusion; (iv) masses in excess of $\sim40\msun$ in the \citet{Beltran+06} catalogue.
In this work we will use the gas temperature derived from ammonia observations and the dust temperature derived from a modified black-body fit to the SED to obtain a more accurate estimate of the mass and related quantities. For a spectroscopic tool we selected ammonia, which is an ideal tracer for cold, dense gas, not depleting up to high number densities ($\gtrsim\pot{6}\cm^{-3}$) and an excellent thermometer \citep{HoTownes83}. 
The ammonia inversion transitions are split into five electric quadrupole hyperfine components, a main one at the centre of the spectrum and four satellites, from the intensity-ratio of which one can derive the optical depth $\tau$. This allows a direct estimate of the column density. 
The five lines are further split into several closely spaced components by magnetic interactions between the nuclei; however, these lines are typically not resolved observationally.

\section{Observations and data reduction} \label{sec:obs}

The fields were observed in the NH$_3$(1,1) and (2,2) inversion transitions ($23694.50 \usk \mega\hertz$ and $23722.63 \usk \mega\hertz$, respectively) and in the H$_2$O($6_{16}-5_{23}$) maser line ($22235.08 \usk \mega\hertz$), with the Australia Telescope Compact Array (ATCA). 
The observations were performed between the 4th and 8th of March 2011, for a total telescope time of 48 hours. We used the array in configuration 750D, providing baselines from $31\usk \mathrm{m}$ to $4469\usk \mathrm{m}$. The primary beam of the telescope at these frequencies is $\sim 2\overset{\prime}{.} 5$. The flux density scale was determined by observing the standard primary calibrator PKS1934$-$638 ($0.78\jy$ at $23650 \mega\hertz$), with an uncertainty expected to be $\lesssim10\%$. Gain calibration was performed through frequent observations of nearby compact quasars; 0537$-$441 was used as the bandpass calibrator. Pointing corrections were derived from nearby quasars and applied online. Weather conditions were generally good, with a weather path noise $\sim 400\usk \micro\mathrm{m}$ or better.

The total time on source was divided into series of snapshots with a variable duration of between 3 and 5 minutes observed over a range as large as possible in hour angle, to improve the uv-coverage for each target. As a consequence of the observing strategy, the total on source integration time varies, and is typically between $\sim 30\usk$min and $\sim 1 \usk$hour. The CABB correlator provided two zoom bands of $64\usk \mega\hertz$ each, with a spectral resolution of $32\usk$kHz ($\sim 0.4 \kms$ at $\sim23.7\usk\mathrm{GHz}$). The two ammonia inversion transitions were observed in one band, and the other was centred on the H$_2$O maser line.

The data were edited and calibrated with the MIRIAD software package, following standard procedures. Deconvolution and imaging were performed in AIPS with the ``imagr'' task, applying natural weighting to the visibilities. Ammonia emission lines were visible only on the shortest baselines, thus we discarded all baselines $\gtrsim 30 \usk \mathrm{k \lambda}$. In order to obtain images with the same angular resolution, we reconstructed all of them with a clean circular beam of diameter $20\asec$, except for 17195$-$3811, 17040$-$3959c1 and 16428$-$4109c1. These sources have a poorer uv-coverage, resulting in a beam of roughly $20\asec$ $\times 40 \asec$. Moreover, 16254$-$4844c1 and 16573$-$4214c2 were observed with a very limited range for the hour angle, making the `clean' impossible.
Ammonia spectra were extracted from the data cubes in two different ways: from a circular area of diameter $20\asec$ centred at the peak emission, or averaged over the (larger) region enclosed in the $3\sigma$ contour of the NH$_3$(1,1) integrated emission. 
The spectra extracted from the data cube were imported in CLASS\footnote{Part of the GILDAS (Grenoble Image and Line Data Analysis Software \url{http://iram.fr/IRAMFR/GILDAS/}) package}, and the lines were fitted using METHOD NH3 for the NH$_3$(1,1) inversion transition, that takes into account the hyperfine splitting of the line, thus giving as output also the optical depth of the main line. This method was also used for the (2,2) transition, in order to obtain a better estimate of the full width at half maximum (FWHM) linewidth $\Delta V$ and $\tau$ for the $9$ sources for which we detected the (2,2) hyperfine structure. The spectral rms ranges from $3$ to $55\usk \milli\jytab$, with typical values around $10\usk \milli\jytab$. The value of the rms for each spectrum is given in Table~\ref{tab:line_prop}.

H$_2$O maser emission is detected on all baselines, allowing us to achieve the highest angular resolution allowed by the array configuration ($\sim1-2\asec$). For 16254$-$4844c1 and 16573$-$4214c2, we could only establish whether there is maser emission or not, and we do not derive positions for the maser spots.

\section{Results and analysis} \label{sec:results}

\subsection{Ammonia line profiles and properties}\label{ssec:line_prop}

The NH$_3$(1,1) integrated emission (zeroth moment) is shown in panels (a) and (b) of Fig.~\ref{fig:mom0_sest} together with the SEST $1.2\mm$ emission. 
Of the $46$ clumps listed in Table~\ref{tab:positions}, $36$ were detected in both NH$_3$(1,1) and (2,2); $43$ have been detected in NH$_3$(1,1).
Three clumps were not detected in NH$_3$ at all: 15454$-$5335c2, 14166$-$6118c1 and 16164$-$4929c2. We discuss them in more detail in Appendix~\ref{app:ind_sou}.
It is evident from the data that we filter out extended emission for some objects: Figure~\ref{fig:neg_lobes} shows that the lack of information on the largest spatial scales of emission causes the persistence of negative features in the corresponding maps.
The general morphology of the ammonia emission traces well the mm-continuum emission.
The peaks of NH$_3$(1,1) may show significant displacement with respect to the millimetre peak. For some sources this may be caused by a low signal-to-noise ratio of the NH$_3$ emission. Alternatively, the offset could be the result of optical depth or chemical effects. Twelve clumps have optical depths larger than $1.5$ in the main line and a reliable map, but only 3 of them show an offset. We also made maps of the emission of one of the satellite lines, (that are likely to be optically thin, as their optical depth is $\sim 4$ times smaller than that of the main line), for the 7 sources showing the largest displacement between ammonia and millimetre peak. In only one source (15470-5419c1) is the peak of the satellite line emission coincident with the millimetre peak; in the others the offset remains unchanged. Thus, optical depth effects cannot be the dominant cause for the offset.

\begin{figure}[tb]
 \centering
 \includegraphics[angle=-90,width=0.75\columnwidth]{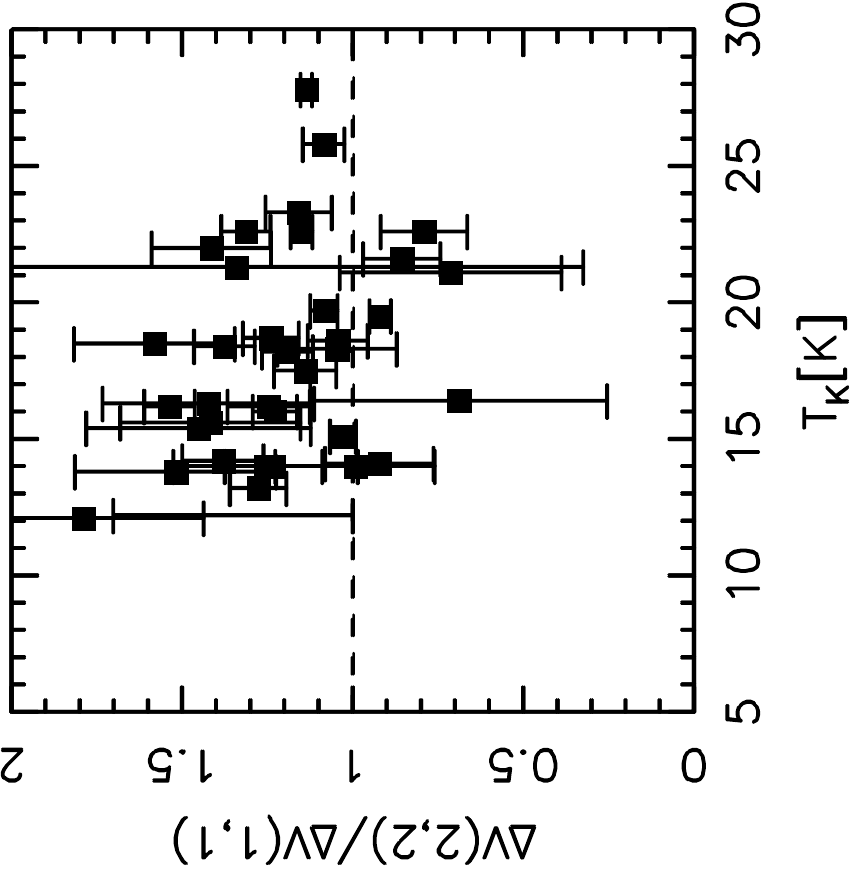}
 \caption{Ratio of the line FWHM $\Delta V$ for the NH$_3$(2,2) and (1,1) lines as a function of kinetic temperature, derived from our data assuming $\Delta V$(2,2)$=\Delta V$(1,1).}
 \label{fig:11/22}
\end{figure}

The peak flux of the two ammonia transitions, the rms of the spectra, $\Delta V$, the optical depth $\tau$, the systemic velocity $\vlsr$ of the line, the rotation temperature $\trot$, the kinetic temperature $\tk$ and the ammonia column density $N(\mathrm{NH_3})$ are listed in Table~\ref{tab:line_prop}. 
The optical depth of the (1,1) transition ranges from $\ll1$ to $\sim4$, showing that ammonia emission is moderately thick in these objects. 

Figure~\ref{fig:11/22} shows that the FWHM $\Delta V$ of the (2,2) transition is on average slightly larger than that of the (1,1), indicating that the two transitions do not trace exactly the same volume of gas, the (2,2) transition being more sensitive to regions with a higher degree of turbulence due to its higher energy. The same result is found by \citet{Rygl+10}.
To estimate a rotation temperature \citep[e.g., ][]{Mangum+92, Busquet+09} from the line ratio the two ammonia transitions must trace the same volume of gas. Thus, if this assumption is correct, the $\Delta V$ should be the equal. The difference in $\Delta V$ that we measure is sufficiently small not to invalidate our assumption that the emission from NH$_3$(1,1) and (2,2) comes from the same region. We thus consider our temperature estimates reliable.

The kinematic distances were recomputed for the clumps with the $\vlsr$ derived from NH$_3$ and the rotation curve of \citet{BrandBlitz93}, and were found to be in agreement with those given in \citet{Beltran+06}, except for 08477$-$4939c1, 16061$-$5048c1 and c2, 16093$-$5128c1, and 17040$-$3959c1. We choose to use the \citet{BrandBlitz93} instead of the more recent one of \citet{Reid+09} because it still the best sampled in terms of Heliocentric- and Galactocentric distances and Galactocentric azimuth. A comparison shows that the kinematic distances derived with the \citet{Reid+09} curve are systematically smaller by $< 10-15\%$ for virtually all of our sources.

In the inner Galaxy, objects along the line-of-sight on either side of the tangent point have the same radial velocity, which leads to an ambiguity in the kinematic distance (``near'' and ``far'') for several of our targets. Thus, we checked all our sources for associated $8 \mum$ absorption features in the Spitzer/GLIMPSE images, for H\textsc{i} self-absorption observations towards them in the literature and for the height with respect to the Galactic midplane. The near distance is chosen if the complex is observed in absorption against the Galactic mid-InfraRed (MIR) background or if the source at the far distance is further than $150\pc$ from the Galactic plane \citep[$\sim 2-3$ times the scaleheight of the molecular gas distribution; see][]{Dame+87, BrandBlitz93, DameThaddeus94}. 
Twenty-two of our sources meet the first criterion, and 8 targets would be located at more than $150\pc$ from the midplane of the Milky Way at the far distance.
Finally, \citet{GreenMcClureGriffiths11} report H\textsc{i} self-absorption measurements for 7 \hii\ regions near our observed fields (containing 12 clumps in total). They locate 3 \hii\ region/clump complexes at the far distance. However, we are confident that 2 (15557-5215 and 17040-3959) of those 3 are instead at the near distance as the $8\mum$ images show a clear absorption patch, and this is unlikely if the sources were on the far side of the Galactic centre. Hence the far distance was assigned to only 1 of our fields (containing 1 clump). The distances adopted are listed in Table~\ref{tab:line_prop}. Where the near-far ambiguity could not be resolved (8 sources), the near distance was assumed.

\subsection{Temperatures from ammonia}\label{ssec:temp}

We derive the rotation temperature ($\trot$), and the molecular column density, following the method described in \citet{Busquet+09}. This assumes that the transitions between the inversion doublets can be approximated as a two level system \citep[see ][]{HoTownes83}, and that the excitation temperature and line widths are the same for both the (1,1) and (2,2) transition (see Sect.~\ref{ssec:line_prop}). 
The kinetic temperature $\tk$ was then extrapolated from $\trot$ using the empirical method outlined in \citet{Tafalla+04}. This relation gives results accurate to a $5\%$ level for temperatures $\lesssim20\kel$. 
This procedure was used to derive the gas temperature both from the spectra extracted from an area equal to that of the beam around the peak of the ammonia emission, and from those averaged over the whole area of NH$_3$(1,1) emission.

In order to also have an estimate of the uncertainty, the method to derive $\trot$, $\tk$ and ammonia column density $N(\mathrm{NH_3})$ was implemented in JAGS\footnote{\url{http://mcmc-jags.sourceforge.net/}} (Just Another Gibbs Sampler). JAGS is a program for the analysis of Bayesian models, based on Markov Chain Monte Carlo simulations. It computes the \textit{posterior} probability distribution, summarizing our knowledge of the quantities considered, given a user-defined model (i.e. the equations and the assumptions of Gaussianity for the quantities directly derived from the fit in our case), the data and our prior knowledge of the quantities involved \citep{Andreon11}. This program was used to derive $\trot$, $\tk$ and $N(\mathrm{NH_3})$ and their uncertainty, propagating the Gaussian uncertainty of the parameters of the fit, as given by CLASS. Constant \textit{priors} were used on these parameters, i.e. $T\tau$ and $\tau$. To check the dependency of the results on the choice of the \textit{prior}, we used also a Gaussian \textit{prior} with a large $\sigma$. The results show that the derived parameters are virtually independent of the \textit{prior} choice.

The temperatures and $N(\mathrm{NH_3})$ derived from the peak spectrum in this way are listed in Table~\ref{tab:line_prop}, with their uncertainties. On the other hand, Table~\ref{tab:line_prop_mean} shows the observed spectral parameters for the spectra averaged over the whole NH$_3$(1,1) emission, the rotation and kinetic temperatures and the average ammonia column density, with the respective uncertainties. 
$\tk$ obtained from the spectra extracted from both the peak of the NH$_3$ and those obtained from the whole area of emission are in the range between $\sim10$ and $\sim28 \kel$. $\trot$ and $\tk$ calculated from the two sets of ammonia spectra agree very well in most cases. 
Few exceptions exist, where the $68\%$ credibility intervals for the kinetic temperature do not overlap (3 cases), but with differences of $\sim5\kel$ at most, possibly due to dilution of the NH$_3$(2,2) emission, averaged over the same area as the (1,1). Thus in the following we use the $\tk$ derived at the peak of ammonia emission

The gas temperatures derived from ammonia imply that the average $\Delta V$ of the ammonia lines (between $\sim0.7$ and $3.7\kms$) is well in excess of the thermal broadening in such cold gas ($\sim0.15\kms$ for $\tk=20\kel$), indicating that turbulence may play a major role in supporting the clumps.

\begin{figure}[tbp]
 \centering
 \includegraphics[angle=-90,width=0.75\columnwidth]{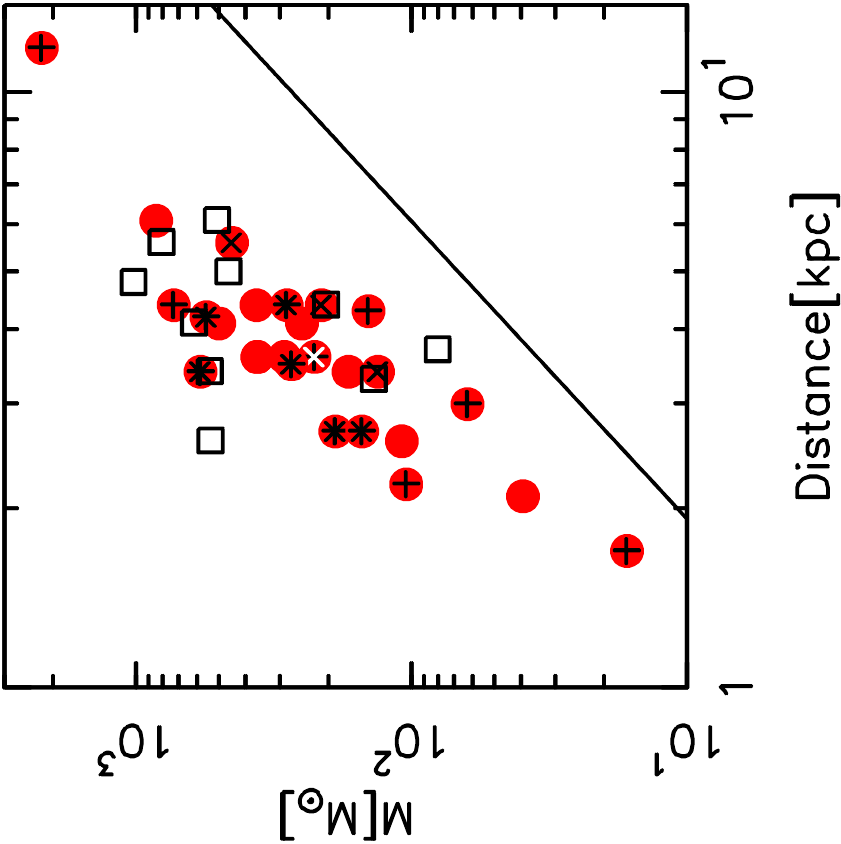}
 \caption{Mass of the clumps as a function of distance. The black solid line indicates the typical mass sensitivity for the SEST images (see text). The sources with signs of active star formation are shown as red filled circles, those without as black open squares (see Sect.~\ref{sec:discussion}). Associated MSX emission is indicated as a black plus, and radio-continuum emission as a black cross. The white cross indicates 17195-3811c1 (see text).} 
 \label{fig:m_dist}
\end{figure}

\subsection{Ammonia abundances}

To determine characteristic ammonia abundances we used the spectra averaged over all the NH$_3$ emission. 
We derived $\coldhtwo$ from the average $1.2\mm$ emission, collected over the same area as the ammonia, assuming that the clump is homogeneous (see Sect.~\ref{ssec:mass_dens_size}), and divided the NH$_3$ column density by $\coldhtwo$. The total range of abundances for all the sources in the sample lies between $\sim\pot{-9}$ and $\sim\pot{-7}$. For most of the objects the abundances are in the typical range of $\sim \pot{-8}-\pot{-7}$ \citep[cf.][and references therein]{Wienen+12}. We compared the NH$_3$ abundance derived in this way with the abundance derived at the peak of ammonia emission: we find this latter quantity is typically slightly greater than the former, with ratios in the range $\sim 0.6-10$ and mean and median values of $2.5$ and $1.2$, respectively. A sub-sample of the clumps observed in ammonia was also observed in \co\ and N$_2$H$^+$ with APEX \citep{Fontani+12}. The carbon monoxide was found to be heavily depleted in these sources, showing that they are in an early phase of evolution. On the contrary, the observed ammonia abundances indicate that NH$_3$ is not depleted on a large scale in these clumps, in agreement with studies of clumps in low-mass star-forming regions, whereas CO is also depleted \citep[e.g., ][]{Tafalla+02}.
\begin{table*}
 \centering
 \caption{Parameters of the ammonia spectra extracted from an area equal to that of the beam, around the NH$_3$ emission peak. The columns indicate the clump name, the peak flux of the (1,1) transition and the rms of the spectrum, the $\vlsr$ of the emission, the $\Delta V$ of NH$_3$(1,1), the opacity of the (1,1) line and its uncertainty, the peak flux of the (2,2) transition and the rms of the spectrum, and the $\Delta V$ of NH$_3$(2,2), $\trot$, $\tk$ and ammonia column density, with their uncertainties, the near and far kinematic distance. The clumps above the horizontal line are those classified as star-forming, while the clump below it are those classified as quiescent (see Sect.~\ref{sec:discussion}).}
 \label{tab:line_prop}
 \includegraphics[width=0.8\textwidth]{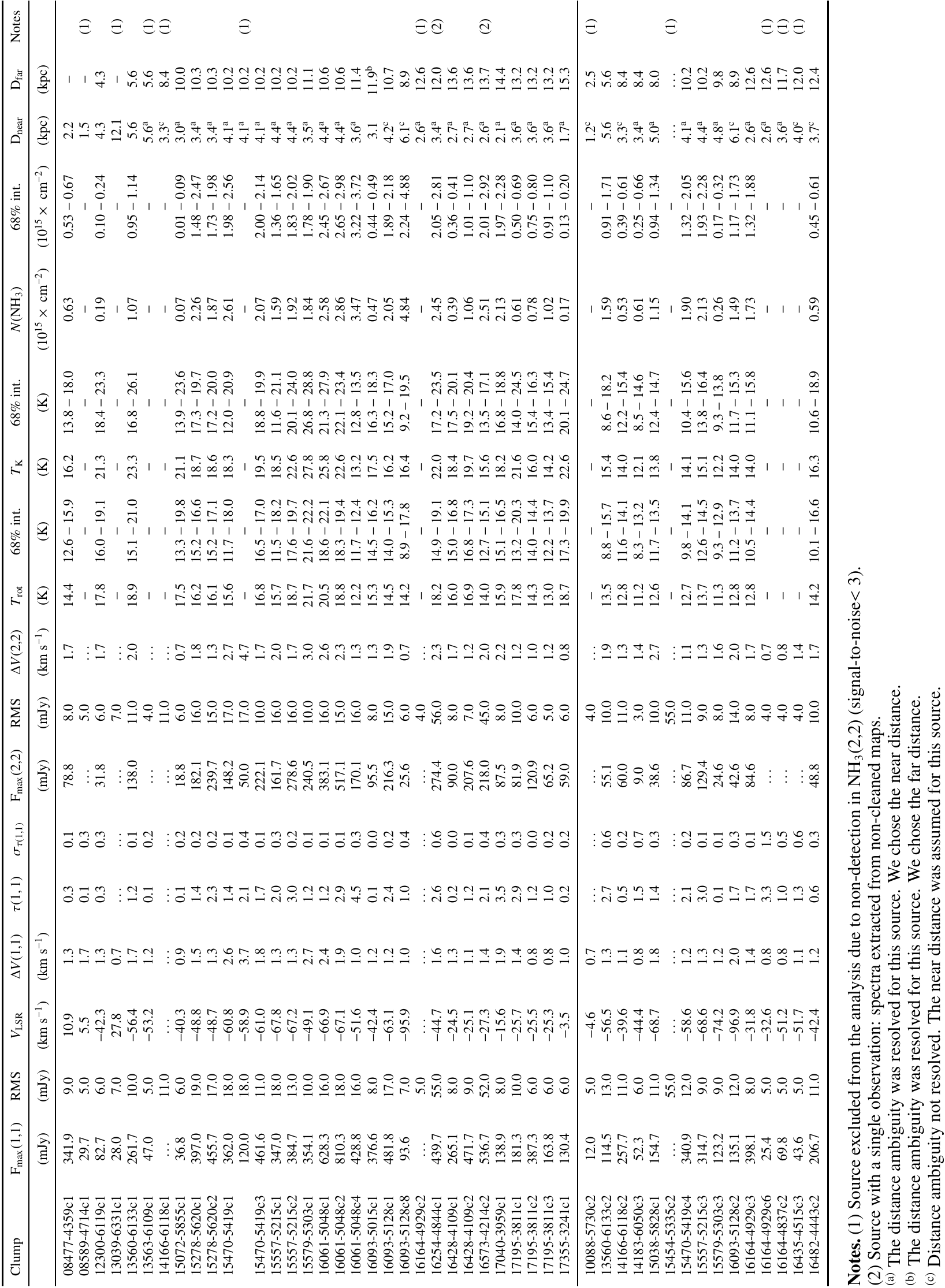}
\end{table*}

\subsection{Clump masses, diameters and gas densities}\label{ssec:mass_dens_size}

Determining accurate masses for the clumps is crucial to determine the evolutionary phase of the clump from a mass-luminosity plot \citep{Molinari+08}, and to see if the clump is massive enough to form high-mass stars. 
With our temperature determination (assuming that the gas, dust and kinetic temperatures $T_g$, $\td$ and $\tk$ are equal) we are able to compute more accurate masses than those listed in \citet{Beltran+06}, that were derived assuming $T_d=30\kel$. The clump masses are calculated from the integrated $1.2\mm$ ($250\,\mathrm{GHz}$) flux through \citep{Hildebrand83}:
\begin{equation}
  \mathrm{M_{gas}} = \gamma \frac{\mathrm{S_{250}} D^2}{\kappa_{250} \mathrm{B_{250}}(\td)},
 \label{eq:dustmass}
\end{equation}
where $\mathrm{S_{250}}$ is the total flux density at $250\usk\mathrm{GHz}$, $D$ is the distance, $\gamma$ is the gas-to-dust ratio, $\mathrm{B_{250}}(\td)$ is the emission of a black body with temperature equal to $\td$ at $250\usk\mathrm{GHz}$, and $\kappa_{250}\equiv\kappa_0(250\usk\giga\hertz/\nu_0[\giga\hertz])^\beta$ is the dust opacity per unit mass at the indicated frequency. We used $\kappa_0=0.8\usk\cm\usk\gram^{-1}$ at $\nu_0=230.6\usk\giga\hertz$, as recommended by \citet{OssenkopfHenning94}. 
The index $\beta$ was derived from the modified black-body fit to the spectral energy distribution of the clumps, using only the SEST and Hi-GAL fluxes (see Sect.~\ref{ssec:sed}), where the data were available, otherwise we chose $\beta=2$, as in \citet{Beltran+06}.

\begin{figure}[tbp]
 \centering
 \includegraphics[angle=-90,width=0.75\columnwidth]{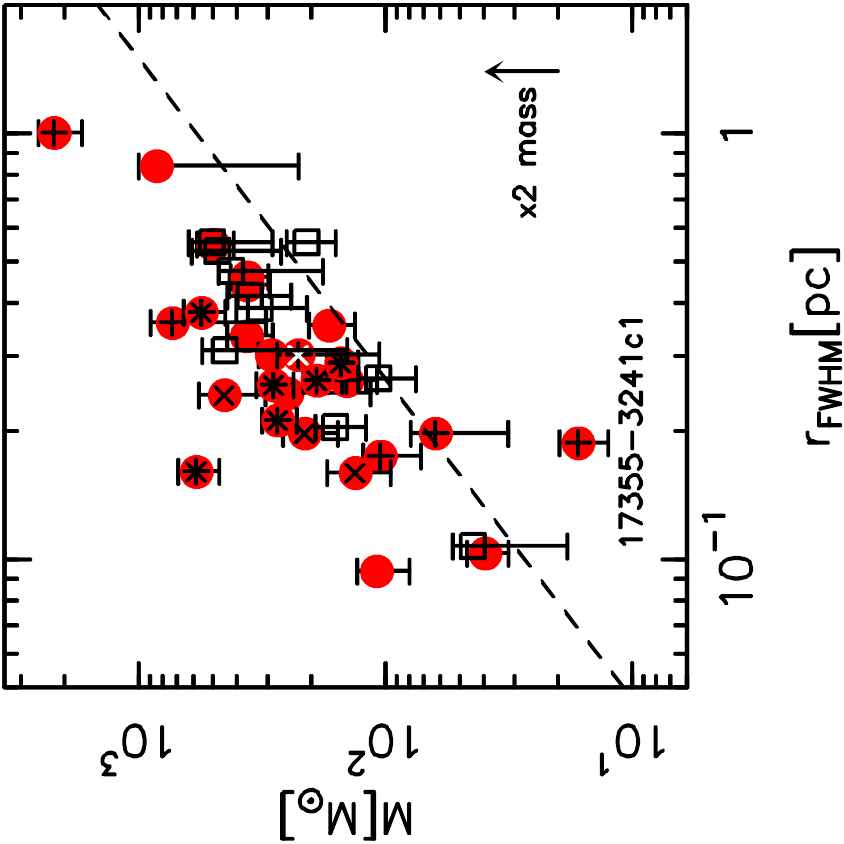}
 \caption{Mass-radius plot; $r_\mathrm{FWHM}$ is the beam-corrected radius at FWHM level. The symbols are the same as in Fig.~\ref{fig:m_dist}.
The boundary for massive-star formation derived by \citet{KauffmannPillai10} ($M[\msuntab]=870(r/\pc)^{1.33}$, rescaled to match our dust opacity) is indicated as a dashed line. Sources above this line are able to form massive stars. The uncertainty in mass is shown for each point. A variation of a factor of 2 in mass (see text) is indicated in the bottom right corner.} 
 \label{fig:m-r_diagram}
\end{figure}

The masses and their uncertainties are again estimated with JAGS, taking into account the probability distribution of $\tk$, as derived from the ammonia observations, the uncertainty of the integrated $1.2$-mm flux (determined with standard techniques from the flux density rms in the images) and a $15\%$ calibration uncertainty.
We find systematically higher masses than \citet{Beltran+06}, because the temperatures are always lower than $30\kel$. The masses of the clumps tend to increase with the distance (see Fig.~\ref{fig:m_dist}), as a result of the fact that nearby high-mass clumps are rare and that at large distances one cannot always separate individual clumps. In Fig.~\ref{fig:m_dist} we show the minimum detectable mass from the $1.2\mm$ maps, calculated from Eq.~\ref{eq:dustmass}, with a typical $3\sigma$ flux density of $100\usk$mJy/beam and a $\td=15\kel$.
In this work we adopted the mass computed within the FWHM contour, in order to consider only the inner regions of the clump, excluding the external envelope (see Sect.~\ref{sec:discussion}), and because the measured diameter depends on the signal-to-noise ratio. Therefore when we generically speak of the mass we refer to masses computed within the FWHM contour. In this way the mass could be underestimated by a factor of 2, if the source is Gaussian and the  envelope contribution is negligible. Our mass estimates are thus conservative, and the possible variation is indicated in figures~\ref{fig:m-r_diagram} and \ref{fig:m-l_plot} for comparison. The masses are listed in Table~\ref{tab:fwhm}. For completeness, masses and densities computed within the $3\sigma$ contour contour are shown in Table~\ref{tab:3s}.

Angular diameters were derived from $1.2\mm$ maps. The beam-corrected angular diameters $\theta$ of the clumps at FWHM level are estimated assuming that the sources are Gaussian, using the relation $\theta=\sqrt{\mathrm{FWHP^2}-\mathrm{HPBW^2}}$, with $\mathrm{FWHP}=2\sqrt{\mathrm{A}/\pi}$, where $A$ is the area within the contour at half peak intensity, and $\mathrm{HPBW}$ is the SEST half-power beam width. If the angular size derived in this way is less than half the beam size, the source is deemed unresolved and we set an upper limit to its size equal to half the HPBW \citep{Wilson09ToRA}.   
The linear diameters at FWHM level range from $\sim0.2$ to $\sim2.0\pc$ (Table~\ref{tab:fwhm}).

\citet{KauffmannPillai10} derived an empirical relation between mass and radius to separate the clumps that are able to form high-mass stars from those that are not. With the mass now much better constrained, we can use this relation to test if our clumps have the potential of forming massive stars. 
In Fig.~\ref{fig:m-r_diagram} we show the mass and size of our clumps, compared to the \citet{KauffmannPillai10} relation, scaled to the same dust opacity as used in the present work. From the figure we observe that the vast majority of our sources lie above the \citet{KauffmannPillai10} relation, indicated as a dashed line, corroborating the idea that the whole sample is constituted of similar objects and suggesting that virtually all of them could form massive stars. This makes our sample a good one to study the evolution of massive clumps potentially able to form massive stars.

The column- and volume-densities of molecular hydrogen were calculated using the mass and the diameters, assuming spherical symmetry and correcting for helium \citep[$\sim8\%$ in number; e.g.,][]{Allen73}.
These two quantities are found to lie between $\sim0.1-6\times\pot{23}\cm^{-2}$, and $0.2-20\times\pot{5}\cm^{-3}$, respectively: values like these are typical of IRDCs \citep[e.g., ][]{Egan+98, Carey+98, Carey+00, Pillai+06}.
The surface density $\Sigma$ was determined by averaging the mass over the deconvolved FWHM area of emission at $1.2\mm$.
$\Sigma$ for the clumps in this sample is found to lie between $0.03$ and $1.5\usk\gram\cm^{-2}$.

The mass, volume-, column-, and surface densities with their $68\%$ credibility intervals for all the clumps are listed in Table~\ref{tab:fwhm}.

\subsection{Spectral energy distribution} \label{ssec:sed}

Important insights in the evolutionary state of a source can be gained through its $L/M$ ratio, as proposed by \citet{Saraceno+96} for the low-mass regime and by \citet{Molinari+08} for the high-mass regime. We constructed the Spectral Energy Distribution (SED) for the sources in our sample complementing the SEST data with Herschel\footnote{\citealt{Pilbratt+10}. Here we use the PACS \citep{Poglitsch+10} and SPIRE \citep{Griffin+10} instruments.}/Hi-GAL \citep[$500\mum$, $350\mum$, $250\mum$, $160\mum$, $70\mum$; ][]{Molinari+10}, MIPSGAL \citep[$24\mum$; ][]{Carey+09}, MSX \citep[band A $8.28\mum$, C $12.13 \mum$, D $14.65 \mum$, E $21.30 \mum$; ][]{Price+01} and GLIMPSE \citep[$8.0\mum$, $5.8\mum$, $4.5\mum$, $3.6\mum$; ][]{Benjamin+03, Churchwell+09} data. We smoothed all the images to a common resolution of $25\asec$ (the approximate resolution of the $350 \mum$ image) except that at $500\mum$, which has a resolution of $\sim36\asec$. Two different polygons were defined for each wavelength: one to derive the flux density of the source, and the other for the background in the region.
The bolometric luminosity was calculated by integrating the fluxes over frequency, interpolating linearly between the measured fluxes at different frequencies in logarithmic space. The uncertainty was estimated by simply interpolating between the lower and upper limit of the $68\%$ credibility interval of the fluxes used to derive the luminosity, respectively.

The fluxes at the longest wavelengths in the SED can be used to infer the typical properties of the dusty envelope. With this in mind we fitted the SED fluxes, down to $70\mum$ with a modified black body. The point at $70\mum$ was included in the fit to better constrain the temperature in the case where the SED has its peak shortward of $160\mum$. The inclusion of the flux at $70\mum$ implies that we will measure a higher $\td$, because we are tracing the warm dust layers near the embedded (proto-)star. The fit procedure is described in Appendix~\ref{sec:app_SEDfit}. The results of the modified black-body fit, the luminosity derived integrating the SED from $1.2\mm$ down to $3.6\mum$, and the uncertainties in these quantities for each clump are listed in Table~\ref{tab:greybody}. In the table we list only the clumps with data in all the five Herschel/Hi-GAL bands. Figures~\ref{fig:sed_sfs} and \ref{fig:sed_qs} show the SED with the fit results. 17195-3811c1 is on the edge of the Herschel/HiGAL $160/70\mum$ maps, thus is not included here. However we use the lower limits on the fluxes at these wavelengths for the fit with the Robitaille models (see below) for this latter source.

\begin{table*}[tbp]
\centering
\caption{Parameters derived from the modified black-body fit of the SED down to $70\mum$, and luminosity of the clumps (integrated from $1.2\mm$ and $3.6\mum$). The clumps above the horizontal line are those classified as star-forming, while the clump below it are those classified as quiescent (see Sect.~\ref{sec:discussion}). The columns shows the dust temperature ($\td$), the mass ($M$) of the gas and the dust emissivity index $\beta$ from the modified black-body fit, and the luminosity of the clumps derived integrating the SED, with their uncertainties.}
\label{tab:greybody}
\begin{tabular}{lrrrrrrrr}
\toprule
Clump        & $\td$        & $68\%$ int.        & $M$        & $68\%$ int.        & $\beta$    & $68\%$ int.        & $L$       & $68\%$ int.          \\
\midrule
             & {\scriptsize (K)}        & {\scriptsize (K)}                   & {\scriptsize ($\pot{2}\times\msuntab$)}  & {\scriptsize ($\pot{2}\times\msuntab$)} &            &                       & {\scriptsize ($\pot{2}\times\lsuntab$)} & {\scriptsize ($\pot{2}\times\lsuntab$)} \\
\midrule
13560-6133c1 &$    24.3$&$    22.8-    25.5$&$   6.2  $&$   5.3  -   7.2  $&$     1.4$&$     1.2-     1.5$&$      30.1$&$      25.6-      34.3$ \\ 
13563-6109c1 &$    22.0$&$    20.8-    23.3$&$   2.4  $&$   2.0  -   2.9  $&$     1.8$&$     1.6-     2.0$&$      27.3$&$      24.2-      30.4$ \\ 
15072-5855c1 &$    26.5$&$    25.0-    28.0$&$   0.5  $&$   0.4  -   0.6  $&$     1.7$&$     1.5-     1.9$&$      11.4$&$      10.2-      12.6$ \\ 
15278-5620c1 &$    27.8$&$    26.0-    29.3$&$   6.0  $&$   5.1  -   7.0  $&$     1.8$&$     1.6-     1.9$&$     266.2$&$     241.1-     291.1$ \\ 
15278-5620c2 &$    11.2$&$     9.8-    12.3$&$   4.4  $&$   3.5  -   5.3  $&$     2.1$&$     1.7-     2.4$&$       1.4$&$       0.8-       1.8$ \\ 
15470-5419c1 &$    16.6$&$    15.8-    17.3$&$   4.0  $&$   3.5  -   4.8  $&$     1.5$&$     1.3-     1.7$&$       4.0$&$       3.2-       4.5$ \\ 
15470-5419c3 &$    19.2$&$    18.3-    20.0$&$   3.6  $&$   3.1  -   4.3  $&$     1.6$&$     1.4-     1.8$&$       7.9$&$       6.7-       8.9$ \\ 
15557-5215c1 &$    23.8$&$    22.3-    25.0$&$   5.3  $&$   4.5  -   6.3  $&$     1.8$&$     1.6-     2.0$&$      76.2$&$      68.0-      84.4$ \\ 
15557-5215c2 &$    15.6$&$    14.8-    16.8$&$   3.6  $&$   2.9  -   4.5  $&$     2.1$&$     1.8-     2.3$&$       7.4$&$       5.9-       8.5$ \\ 
15579-5303c1 &$    24.5$&$    23.0-    25.5$&$   5.3  $&$   4.6  -   6.1  $&$     1.9$&$     1.7-     2.0$&$      84.8$&$      75.9-      93.6$ \\ 
16061-5048c1 &$    19.9$&$    19.0-    20.8$&$   6.0  $&$   5.1  -   7.0  $&$     2.0$&$     1.8-     2.1$&$      33.8$&$      29.7-      37.6$ \\ 
16061-5048c2 &$    21.3$&$    20.3-    22.3$&$   5.9  $&$   5.1  -   7.0  $&$     2.2$&$     2.0-     2.3$&$      86.3$&$      77.0-      95.5$ \\ 
16061-5048c4 &$     9.5$&$     8.8-    10.3$&$   8.5  $&$   7.0  -  10.2  $&$     2.5$&$     2.2-     2.8$&$       2.1$&$       1.4-       2.6$ \\ 
16093-5015c1 &$    20.8$&$    19.8-    21.8$&$  20.9  $&$  17.8  -  24.3  $&$     1.6$&$     1.4-     1.8$&$      90.0$&$      77.5-     101.5$ \\ 
16093-5128c1 &$    23.8$&$    22.5-    25.0$&$   4.6  $&$   4.0  -   5.3  $&$     2.2$&$     2.0-     2.4$&$     256.9$&$     234.4-     279.3$ \\ 
16093-5128c8 &$    13.8$&$    12.8-    15.3$&$   1.8  $&$   1.5  -   2.2  $&$     2.3$&$     2.0-     2.6$&$       3.5$&$       2.4-       4.2$ \\ 
16254-4844c1 &$    17.6$&$    16.8-    18.3$&$   2.6  $&$   2.2  -   3.1  $&$     1.6$&$     1.5-     1.8$&$       4.1$&$       3.4-       4.6$ \\ 
16573-4214c2 &$    16.0$&$    15.3-    16.5$&$   2.1  $&$   1.8  -   2.4  $&$     1.8$&$     1.6-     1.9$&$       2.6$&$       2.2-       3.0$ \\ 
17040-3959c1 &$    17.2$&$    16.5-    17.8$&$   0.5  $&$   0.4  -   0.6  $&$     2.3$&$     2.2-     2.5$&$       2.5$&$       2.2-       2.8$ \\ 
17355-3241c1 &$    23.8$&$    22.5-    24.8$&$   0.35 $&$   0.3  -   0.4  $&$     2.1$&$     1.9-     2.2$&$      19.0$&$      17.1-      21.0$ \\ 
\midrule
14166-6118c2 &$    18.1$&$    15.3-    20.8$&$   0.5  $&$   0.4  -   0.6  $&$     1.7$&$     1.2-     2.0$&$       1.3$&$       0.8-       1.7$ \\ 
14183-6050c3 &$    16.1$&$    15.0-    17.0$&$   1.0  $&$   0.7  -   1.2  $&$     1.9$&$     1.6-     2.3$&$       2.3$&$       1.5-       2.7$ \\ 
15038-5828c1 &$    12.2$&$    11.3-    13.3$&$   5.4  $&$   4.5  -   6.5  $&$     2.1$&$     1.8-     2.4$&$       3.3$&$       1.9-       3.8$ \\ 
15470-5419c4 &$    11.1$&$    10.3-    12.0$&$   6.2  $&$   5.1  -   7.5  $&$     2.4$&$     2.0-     2.6$&$       2.9$&$       1.9-       3.7$ \\ 
15557-5215c3 &$     9.6$&$     8.5-    10.5$&$   3.6  $&$   2.9  -   4.5  $&$     2.9$&$     2.5-     3.3$&$       1.8$&$       1.1-       2.6$ \\ 
15579-5303c3 &$    15.6$&$    14.5-    17.0$&$   3.4  $&$   2.7  -   4.3  $&$     1.6$&$     1.3-     1.9$&$       3.9$&$       2.1-       4.8$ \\ 
16093-5128c2 &$    10.4$&$     9.0-    11.5$&$   5.1  $&$   4.0  -   6.1  $&$     2.7$&$     2.3-     3.0$&$       3.2$&$       1.7-       4.0$ \\ 
16164-4929c3 &$     9.2$&$     8.3-    10.3$&$   4.7  $&$   3.8  -   5.7  $&$     2.6$&$     2.2-     2.9$&$       1.0$&$       0.6-       1.7$ \\ 
16164-4929c6 &$    11.0$&$    10.3-    12.0$&$   1.1  $&$   0.9  -   1.4  $&$     2.4$&$     2.1-     2.7$&$       0.7$&$       0.4-       1.0$ \\ 
16164-4837c2 &$     8.1$&$     7.5-     8.8$&$   3.2  $&$   2.7  -   3.9  $&$     3.0$&$     2.7-     3.4$&$       0.8$&$       0.5-       1.0$ \\ 
16435-4515c3 &$    10.5$&$     9.8-    11.3$&$   4.3  $&$   3.5  -   5.1  $&$     2.7$&$     2.4-     2.9$&$       2.9$&$       1.8-       3.6$ \\ 
16482-4443c2 &$     8.7$&$     8.0-     9.5$&$   1.9  $&$   1.6  -   2.3  $&$     3.1$&$     2.8-     3.5$&$       0.8$&$       0.5-       1.0$ \\ 
\bottomrule                                                                          
\end{tabular}
\end{table*}

\begin{figure}[tbp]
 \centering
 \includegraphics[angle=-90,width=0.75\columnwidth]{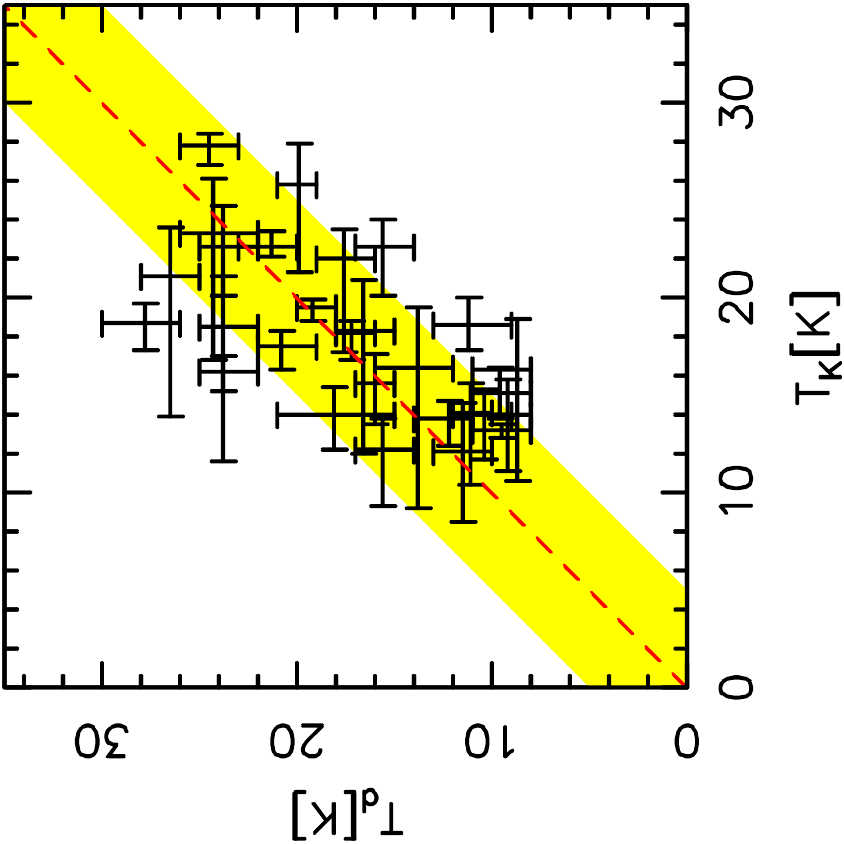}
 \caption{Comparison between the kinetic temperature ($\tk$) derived from ammonia and the dust temperature ($\td$) from the SED-fit. The dashed line indicates equal temperatures and the yellow-shaded region shows a difference of $\pm 5 \kel$ between the two temperatures.}
 \label{fig:tk_td}
\end{figure}

From Fig.~\ref{fig:tk_td} we can see that the characteristic $\td$ obtained from the fit of a modified black body to the SED down to $70\mum$ is usually in good agreement with $\tk$ derived from the ammonia observations. Seven sources have $|\tk-\td|\ge5\kel$ and uncertainties not large enough to explain this difference, implying a statistically significant discrepancy. Four of these objects have $\td>\tk$: these are also the cases where $\td$ is high, always above $20\kel$. The discrepancy may arise from a combination of different causes: the fact that the ratio of the two lowest transitions of ammonia is optimal to derive temperatures only up to $20-25\kel$, that ammonia and dust emission are probing different regions of the clump, and that the strong emission in these clumps from warm dust, heated by the central star and visible at $70\mum$, is biasing the modified black-body fit towards higher $\td$. 

\begin{figure}[tbp]
 \centering
 \includegraphics[angle=-90,width=0.75\columnwidth]{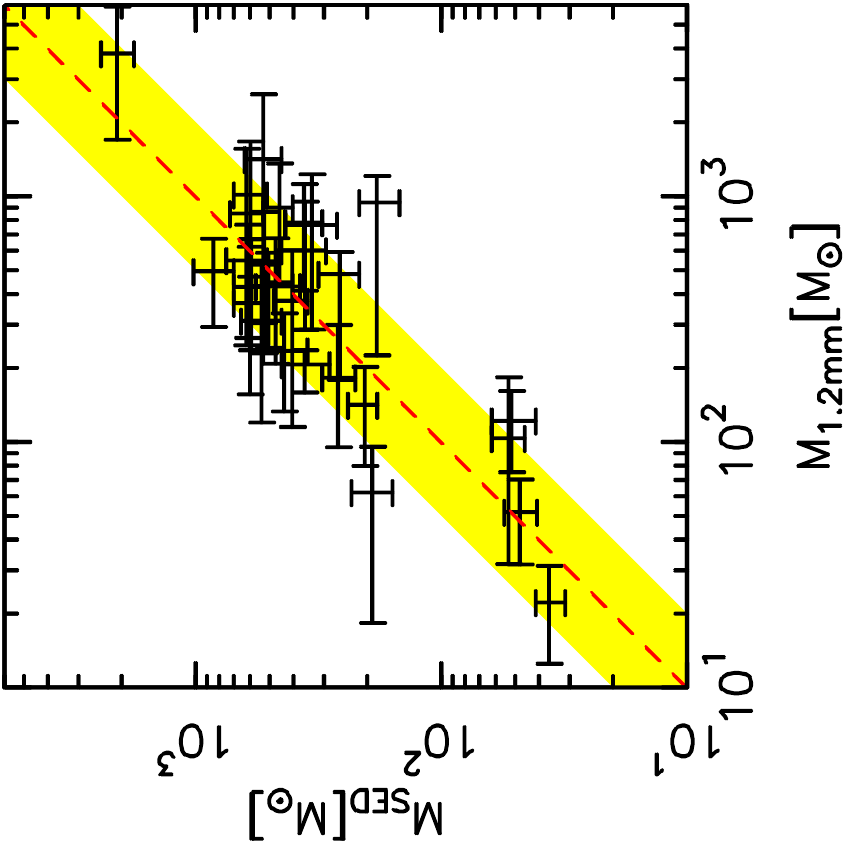}
 \caption{Comparison between the mass derived from the $1.2\mm$ continuum and from the SED fit. The uncertainties in $M_\mathrm{SED}$ are indicated. The bars for the $1.2\mm$ continuum range from the lower limit of the mass within the FWHM contour to the upper limit of the $3\sigma$ contour. The dashed line indicates equal masses and the yellow-shaded region shows a difference of a factor of two.}
 \label{fig:msed_msest}
\end{figure}

The gas masses obtained from the modified black-body fit usually agree, within the uncertainties, with those derived simply from the $1.2\mm$ continuum, and lie between the mass within the $3\sigma$ and that within the FWHM contour (Fig.~\ref{fig:msed_msest}). As not  all sources have a complete SED, and considering the reasonable agreement between masses determined from the $1.2\mm$ integrated flux and from the modified black-body fit to the SED, we decided to use the former in the following analysis, so that we could also assign a size to the source consistently.

\smallskip

\citet{Robitaille+06} developed a code to compute the SED of axisymmetric YSOs. This code considers a central YSO with a rotationally flattened infalling envelope, the presence of bipolar cavities and a flared accretion disk,
making use of a Monte Carlo radiation transfer algorithm to compute the flux of the object at wavelengths from the mm- to the near-IR  regimes. 

A vast range of stellar masses and of evolutionary phases are covered, from the earliest stages of strong infall to the late phase where only the circumstellar disk remains around the central object, and the envelope is completely dispersed.
Following the discussion in \citet{Molinari+08}, we use these models only for clumps that clearly contain an embedded source, and with detectable fluxes shortward of $70\mum$ in the smoothed images. Otherwise, only the modified black-body fit is done. To fit the SED with the Robitaille models, we make use of the online SED fitting tool\footnote{\url{http://caravan.astro.wisc.edu/protostars/}} \citep{Robitaille+07}.
In the tool, we allowed the foreground interstellar extinction to range between $1$ and $2\usk\mathrm{mag}\usk\kilo\pctab^{-1}$ \citep[e.g., ][]{Allen73, Lynga82, Scheffler82}. 
The stellar masses obtained from the fit range between $\sim5$ and $\sim30\msun$, corresponding to spectral types approximately B7-O8. The luminosities derived vary between $\sim600$ and $65000\lsun$, agreeing with those derived by simply integrating the SED, interpolating linearly in the log-log space. Our estimate of $L$ tends to be lower, as the linear interpolation in the log-log space gives a lower limit for the luminosity and because of the model assumptions. However, for consistency, in the following we will use our estimate of the luminosity for all sources.

The envelope mass from the fit of the Robitaille models is greater than that derived either from the modified black-body fit or from the $1.2\mm$-continuum. 
In this regard, \citet{Offner+12} compared synthetic SEDs of deeply embedded proto-stars, derived from simulated observation obtained with a 3D radiative transfer code for dust emission, for a vast range of parameters, with the best fit obtained from the standard grid of Robitaille models. These authors showed that usually the fit recovers the true luminosity and stellar mass, although with large uncertainties, but systematically overestimates the mass of the envelope, mainly due to the assumption of a different dust model. The difference is more pronounced in the mm-regime, strongly influencing the mass determination. Our assumption of dust opacity is similar to that used by \citet{Offner+12} in the mm-regime, explaining the discrepancy between our mass estimates and the envelope mass from the fit with the online SED fitting tool.

A summary of the results of the fits with the Robitaille models is shown in Table~\ref{tab:robitaille}.

\begin{table}
\centering
\caption{Summary of the properties derived from the fit of the Robitaille models to the SED of the objects with significant mid-IR emission. Our estimated range in $L$ is shown for comparison. The ranges in $M_*$, $L_{Rob}$ and $M_\mathrm{env}$ are those spanned by the best ten models. The masses are derived with a different dust model, and thus deviate from our estimate.}
\tiny
\label{tab:robitaille}
\begin{tabular}{lcccc}
\toprule
Clump        & $M_*$       & $L_{Rob}$            & $L$     & $M_\mathrm{env}$ \\
             & ${\scriptsize (\msuntab)}$ & ${\scriptsize (\pot{2}\times\lsuntab)}$ & ${\scriptsize (\pot{2}\times\lsuntab)}$ & ${\scriptsize (\pot{2}\times\msuntab)}$          \\
\midrule
13560-6133c1 & $9.6-15.4$  & $25-43$              & $26-34$   & $15.0-36.0$      \\
13563-6109c1 & $9.7-13.8$  & $27-49$              & $24-30$   & $5.5-18.0$       \\
15072-5855c1 & $5.9-8.8$   & $6-28$               & $10-13$   & $1.6-15.0$       \\
15278-5620c1 & $14.7-22.9$ & $307-665$            & $241-291$ & $16.0-28.0$      \\
15557-5215c1 & $10.1-27.4$ & $37-234$             & $68-84$   & $5.2-23.0$       \\
15557-5215c2 & $6.0-9.6$   & $6-15$               & $6-9$     & $7.3-15.0$       \\
15579-5303c1 & $13.4-17.8$ & $149-334$            & $76-94$   & $14.0-34.0$      \\
16061-5048c1 & $12.5-18.5$ & $52-116$             & $30-38$   & $23.0-40.0$      \\
16061-5048c2 & $11.0-24.4$ & $83-781$             & $77-96$   & $16.0-21.0$      \\
16093-5015c1 & $12.8-21.9$ & $45-209$             & $78-102$  & $19.0-52.0$      \\
16093-5128c1 & $14.8-24.2$ & $104-638$            & $234-279$ & $9.8-40.0$       \\
17195-3811c1 & $9.3-12.9$  & $23-138$             & $-$       & $7.2-31.0$       \\
17355-3241c1 & $7.6-8.2$   & $21-34$              & $17-21$   & $2.2-2.5$        \\
\bottomrule
\end{tabular}
\end{table}

\subsection{Stability and dynamics of the clumps}\label{ssec:dynamics}

To investigate the stability of the clumps, we performed the simplest virial analysis, without taking into account magnetic or rotational support. 

\begin{figure}[tbp]
 \centering
 \includegraphics[angle=-90,width=0.75\columnwidth]{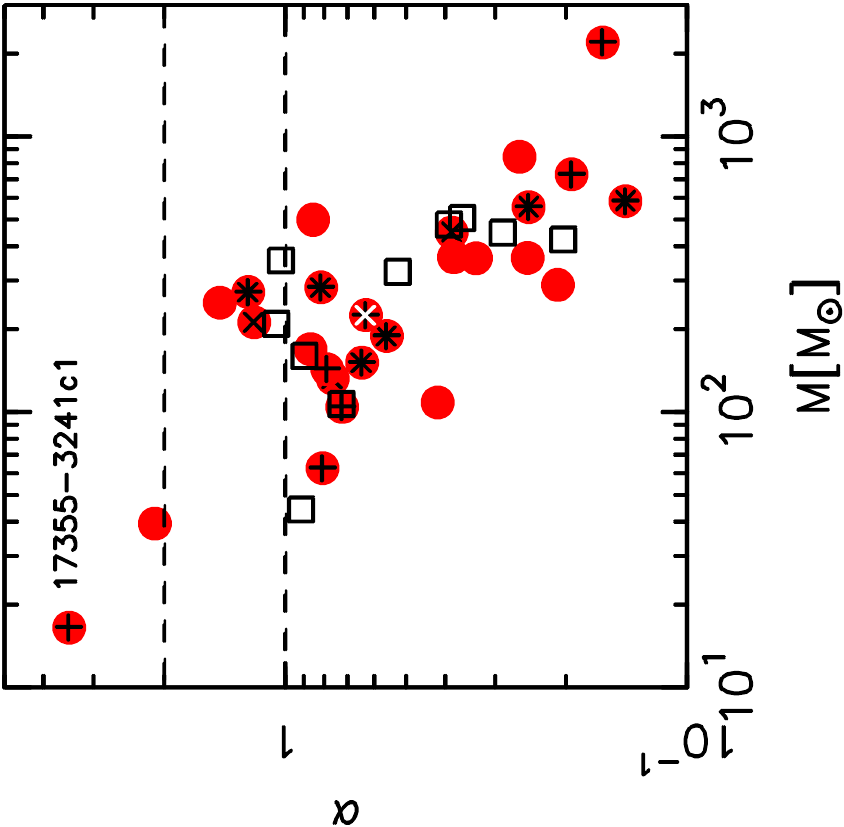}
 \caption{Virial parameter $\alpha=\mvir/M$ as a function of $M$. The symbols are the same as in Fig.~\ref{fig:m_dist}. The dashed lines indicates $\alpha=2$ (clump gravitationally bound), and $\alpha=1$ (clump in virial equilibrium).}
 \label{fig:virial}
\end{figure}

To derive the virial mass we assumed a constant density profile, using Eq.~$(3)$ in \citet{MacLaren88}, which implies that $\mvir[\msuntab]=210 \usk R\mathrm{[\pctab]} \usk \Delta V\mathrm{[\kmstab]}^2$, where $R$ is the radius and $\Delta V$ is the FWHM of the line. In this way we obtain an upper limit for the virial mass. Power law radial profiles for the gas volume density in the clumps can reduce the virial mass: the steeper the profile, the lower the virial mass. A radial dependence like $\nhtwo\propto\left(  r/r_0  \right)^{-2}$ reduces the virial mass by about a factor of $2$ \citep[cf. ][]{MacLaren88}. The virial parameter $\alpha=\mvir/\msest$ is used as an indicator for gravitational stability; $\alpha<2$ implies that the clumps are gravitationally bound, and $\alpha=1$ indicates virial equilibrium. Due to our choice of homogeneous clumps to derive the virial mass, the values that we derive for the virial parameter $\alpha$ are upper limits. Figure~\ref{fig:virial} shows that for virtually all the clumps we find $\alpha\lesssim 1$, implying that they are dominated by gravity. In Sect.~\ref{ssec:alpha} we discuss in detail the observed values of $\alpha$.

First order moment maps reveal the presence of velocity gradients, typically around $1-2\kms\pc^{-1}$. A more detailed analysis for the most interesting clumps is the subject of a forthcoming paper.

\subsection{Water maser emission}

\begin{figure}[tbp]
 \centering
 \includegraphics[angle=-90,width=0.75\columnwidth]{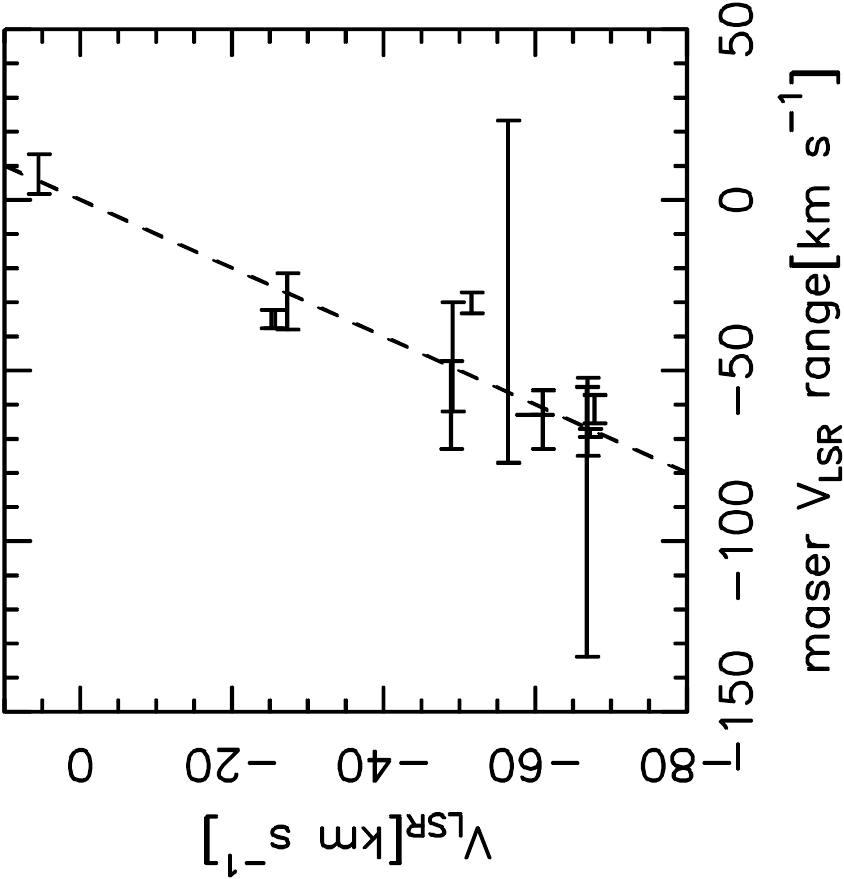}
 \caption{Range of velocity of the water maser \textit{vs.} $\vlsr$ of the clump. The dashed line indicates $V_\mathrm{H_2O}=\vlsr$, i.e. a maser velocity equal to the systemic velocity of the clump.}
 \label{fig:maser_v}
\end{figure}

Thirteen clumps in our sample show maser emission. The spectra were extracted from the data cubes at each position where emission was detected (see Fig.~\ref{fig:maser_spec}) within a polygon comparable to the beam dimensions (between $\sim 1\asec$ and $\sim 2\asec$), and imported in CLASS. The lines were then fitted with Gaussians.
Two sources have very strong lines, reaching nearly $\sim100\usk \jy$. We typically find multiple velocity components (up to 22) towards a single clump. A summary of the maser emission properties is shown in Table~\ref{tab:maser}. The water maser range of velocities usually straddles the systemic velocity of the clump, as shown in Fig.~\ref{fig:maser_v} (cf. \citealt{Brand+03}). The positions of the maser spots are indicated in Fig.~\ref{fig:sest+24mum} as white open squares. A comparison with \citet{Sanchez-Monge+13} shows that 2 sources detected in their study are not detected in our observations, while 2 targets that we detect, were not detected in \citet{Sanchez-Monge+13} (cf. Table~\ref{tab:sf_signs}), as expected because of the well-known variability of water masers \citep[e.g., ][]{Felli+07}. All of these sources show other signs of active star formation.

\begin{table*}[tbp]
\centering
\caption{Summary of the emission characteristics of H$_2$O masers. The columns show the clump name, the number of Gaussian components in the spectrum, the range of $\vlsr$ over which we detect emission, the flux density peak, the integrated emission, the water maser luminosity, the $\vlsr$ and $\Delta V$ of the strongest component (S.C.), and the offset of the maser spot with respect to the phase centre (Ph.C.). No correction was performed for the primary beam.}
\tiny
\label{tab:maser}
\begin{tabular}{lrrrrrrrr}
\toprule
Clump        &  Maser & $\vlsr(\mathrm{min,max})$ & $F_\mathrm{peak}$        & $\int F d\nu$                  & $L_\mathrm{H_2O}$ & $\vlsr$ S.C. & $\Delta V$ S.C. & Offset (Ph.C.) \\
\midrule
             &        & {\scriptsize ($\kmstab $)}& {\scriptsize ($\jytab$)} &  {\scriptsize ($\jytab \kms$)} & {\scriptsize $\pot{-7}\times\lsuntab$} & {\scriptsize ($\kmstab$)}       &  {\scriptsize ($\kmstab$)} & {\scriptsize ($\asec        $)}   \\
\midrule  
08589-4714c1 &$   3$&$   0.5;  14.8$& $1.67$ & $2.4$  & $1.3$   & $4.8$   & $1.2$ & $2,-10$    \\
13560-6133c1 &$  22$&$ -78.2;  26.0$& $5.78$ & $37.4$ & $272.0$ & $-53.3$ & $1.1$ & $-15,35$   \\
15278-5620c1 &$   6$&$ -68.2; -43.5$& $84.5$ & $85.4$ & $228.9$ & $-47.5$ & $0.8$ & $10,79$    \\
15470-5419c1 &$   1$&$ -64.6; -61.2$& $1.95$ & $2.1$  & $8.2$   & $-63.0$ & $1.0$ & $26,-8$    \\
15470-5419c3 &$   4$&$ -63.7; -51.9$& $3.93$ & $9.3$  & $36.3$  & $-60.0$ & $1.1$ & $-2,10$    \\
             &$   2$&$ -75.5; -64.1$& $0.57$ & $1.2$  & $4.7$   & $-66.4$ & $1.3$ & $-31,-50$  \\
15557-5215c1 &$   2$&$ -68.4; -52.4$& $3.54$ & $7.5$  & $33.7$  & $-57.2$ & $1.6$ & $42,-7$    \\
             &$   2$&$ -67.9; -60.4$& $9.03$ & $11.3$ & $50.7$  & $-65.4$ & $1.0$ & $6,-57$    \\
15557-5215c2 &$   1$&$ -71.0; -67.8$& $2.56$ & $2.3$  & $9.4$   & $-69.4$ & $0.8$ & $32,33$    \\
             &$   1$&$ -69.4; -67.1$& $0.18$ & $0.15$ & $0.7$   & $-67.1$ & $0.8$ & $7,4$      \\
15579-5303c1 &$   9$&$ -63.1; -27.7$& $75.4$ & $120.4$& $643.3$ & $-47.7$ & $0.9$ & $-53,13$   \\
             &$   1$&$ -38.5; -33.7$& $0.59$ & $0.67$ & $3.6$   & $-35.8$ & $1.0$ & $-54,15$   \\
16061-5048c1 &$   6$&$ -78.7; -50.9$& $19.6$ & $23.3$ & $104.6$ & $-67.0$ & $0.8$ & $-3,2$     \\
             &$   2$&$ -71.2; -68.1$& $0.47$ & $0.51$ & $2.3$   & $-69.0$ & $0.8$ & $-9,4$     \\
16061-5048c2 &$   2$&$ -79.5; -69.6$& $1.50$ & $2.7$  & $12.1$  & $-77.7$ & $1.6$ & $45,-27$   \\
             &$   3$&$-135.8; -63.5$& $0.54$ & $1.8$  & $8.1$   & $-65.1$ & $0.9$ & $35,-20$   \\
             &$   3$&$ -79.6; -52.0$& $0.46$ & $0.75$ & $3.4$   & $-54.8$ & $1.8$ & $42,-24$   \\
16061-5048c4 &$   3$&$ -35.6; -25.1$& $0.70$ & $2.1$  & $6.3$   & $-33.2$ & $1.5$ & $-5,-2$    \\
16573-4214c2 &$   5$&$ -39.6; -20.2$& $1.30$ & $2.5$  & $3.9$   & $-29.9$ & $0.7$ & $-$\tablefootmark{a}        \\
17195-3811   &$   5$&$ -39.7; -29.5$& $8.39$ & $10.4$ & $31.3$  & $-32.2$ & $1.0$ & $-$\tablefootmark{a}        \\
\bottomrule                                                                          
\end{tabular}
\tablefoot{
\tablefoottext{a}{These sources have poor uv-coverage, thus no position for the maser spots is given.}
}
\end{table*}

\section{Discussion} \label{sec:discussion}

\begin{figure}[tbp]
 \centering
	\includegraphics[width=\columnwidth]{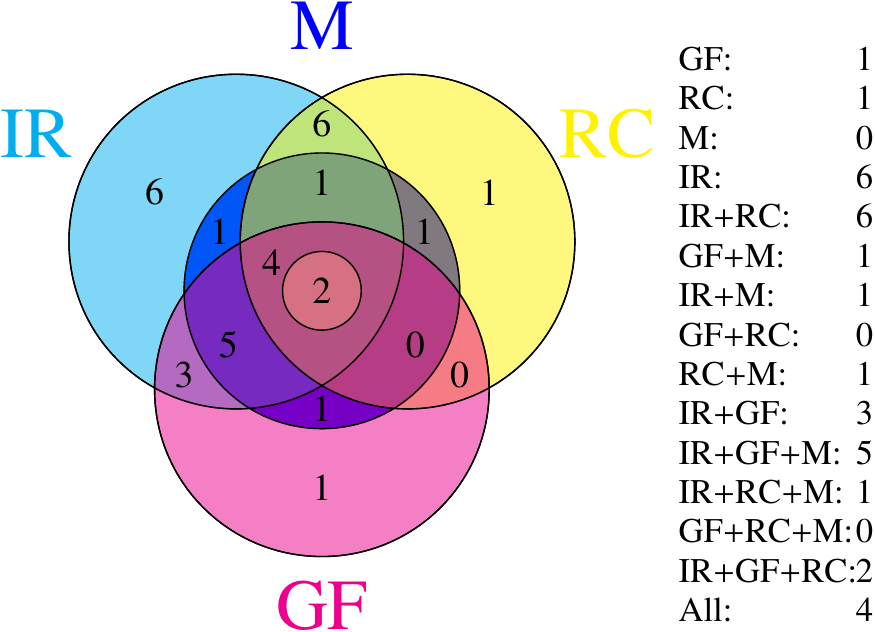}
 \caption{Summary of specific star formation indicators in the clumps. The labels correspond to green fuzzies (GF), mid-IR emission (IR), H$_2$O maser (M), and radio continuum emission (RC) (see Sect.~\ref{sec:discussion} for details). For example, the small central circle shows the combination IR+GF+RC, while the ``triangular'' area with a darker shade surrounding the small circle represents the presence of all four signposts of star formation.}
 \label{fig:summ_sf}
\end{figure}

In order to investigate how the clump properties depend on their evolutionary state, the sources were  first separated into two sub-samples, according to the presence or absence of signposts of active star formation. In particular, we considered the presence of water maser(s), ``green fuzzies'',  $24\usk\mu\mathrm{m}$ and radio-continuum emission.

\begin{figure}[tbp]
 \centering
	\includegraphics[width=0.75\columnwidth]{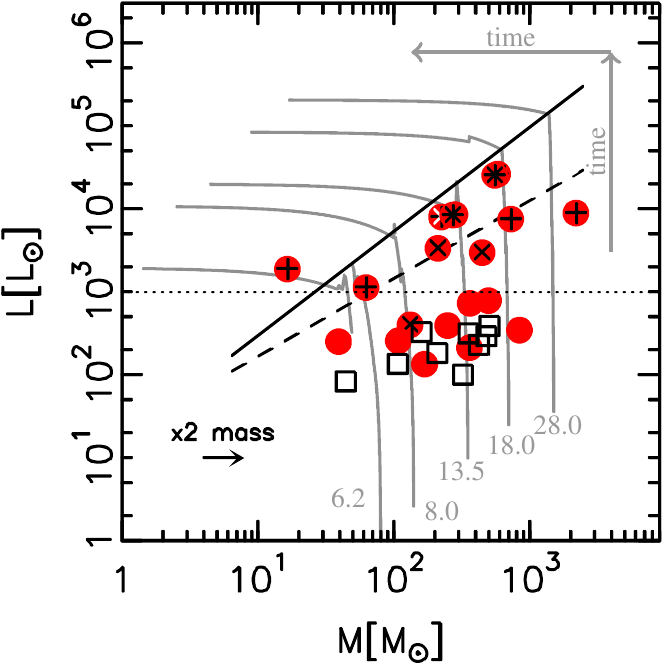}
 \caption{Mass-Luminosity plot for the sources in our sample with Hi-GAL observations. The mass is computed within the FWHM contour of $1.2\mm$ emission. The symbols are the same as in Fig.~\ref{fig:m_dist}. The black solid line indicates the ZAMS locus, according to \citet{Molinari+08}, while the dashed line indicates the ZAMS locus as determined by \citet[][]{Urquhart+13}. The grey lines show the evolution of cores of different masses; the lines are labeled with the final mass of the most massive star (in $\msun$). Time increases from bottom to top and from right to left, as indicated. Radio-continuum emission and MSX emission are found nearly exclusively in clumps near the ZAMS. A variation of a factor of 2 in mass (see text) is indicated in the bottom left corner.}
 \label{fig:m-l_plot}
\end{figure}

\begin{itemize}
\item $24\mum$ emission: We overlaid the Spitzer MIPSGAL images at $24\usk\mu\mathrm{m}$ and the SEST $1.2\mm$ maps of \citet{Beltran+06}, in order to identify those clumps with and without IR emission. If a $24\mum$ source is found at the location of the $1.2\mm$ emission peak, then it is considered as being associated with it. 
MIPSGAL images cover 41 of the clumps in this sample, the remaining 5 are covered by MSX images. Twenty-five clumps are IR-bright at $24\mum$ as shown in Fig.~\ref{fig:sest+24mum}, and 3 among those without Spitzer $24\mum$ data show emission in the $21\mum$ MSX image.

\item ``Green Fuzzies'': extended $4.5\mum$ emission produced by shock-excited molecular lines, commonly associated with Class II CH$_3$OH masers \citep{Cyganowski+09}. To identify the ``green fuzzies'', we followed the procedure described in \citet{Chambers+09}. Again, only 41 clumps out of the 46 are covered by GLIMPSE data. 
Sixteen clumps show the presence of extended, excess $4.5\mum$ emission. 

\item Water masers: 13 clumps in our sample show maser emission. The water maser is a known indicator of star formation, thought to appear in the early stages of the process \citep{BreenEllingsen11}, and is observed both in low- and high-mass star formation regions.

\item Radio-continuum: 40 of the clumps in our sample were observed with ATCA at $22\usk\giga\hertz$ and $18\usk\giga\hertz$ \citep{Sanchez-Monge+13}. Twelve sources were detected, and 3 more have a tentative detection at about $3\sigma$ at one of the frequencies. The radio-continuum alone is not always considered sufficient to classify a source as star forming. This is because we find one source (16061$-$5048c4) in the sample where the radio emission is likely to come from an ionization front in the outer layers of the clump, based on the morphology of the emission or the lack of an IR source in the Spitzer/Hi-GAL images. This special case is discussed in the Appendix. 
\end{itemize}

If any of these signposts is observed (except radio continuum in the special case discussed in the Appendix), a clump is indicated as star-forming.
Based on these criteria, of the 46 objects observed, 31 were classified as star-forming and 15 as quiescent. A summary of the clumps with specific indicators of ongoing star formation is given in Fig.~\ref{fig:summ_sf}. All the star formation signposts in each clump are presented in Table~\ref{tab:sf_signs}. Figure~\ref{fig:sest+24mum} shows all the observed fields of view and clumps (dashed and solid red circles, respectively), with the SEST emission (contours) superimposed on the MIPS $24\mum$ image, and the position of maser spots (white open squares) and ``green fuzzies'' (green open circles).
Removing those clumps with no NH$_3$(2,2) detection, $26$ and $10$ clumps remain in the star-forming and in the quiescent sub-samples, respectively.

\smallskip

In the following, we compare the average properties of the clumps using the parameters we have derived, for the two sub-samples, to look for systematic differences in the physical properties characterizing the two classes. Note that hereafter the quiescent and star-forming sub-samples will be denoted with the acronyms \textbf{QS} and \textbf{SFS}, respectively.

\begin{figure}[tbp]
 \centering
 \includegraphics[angle=-90,width=0.75\columnwidth]{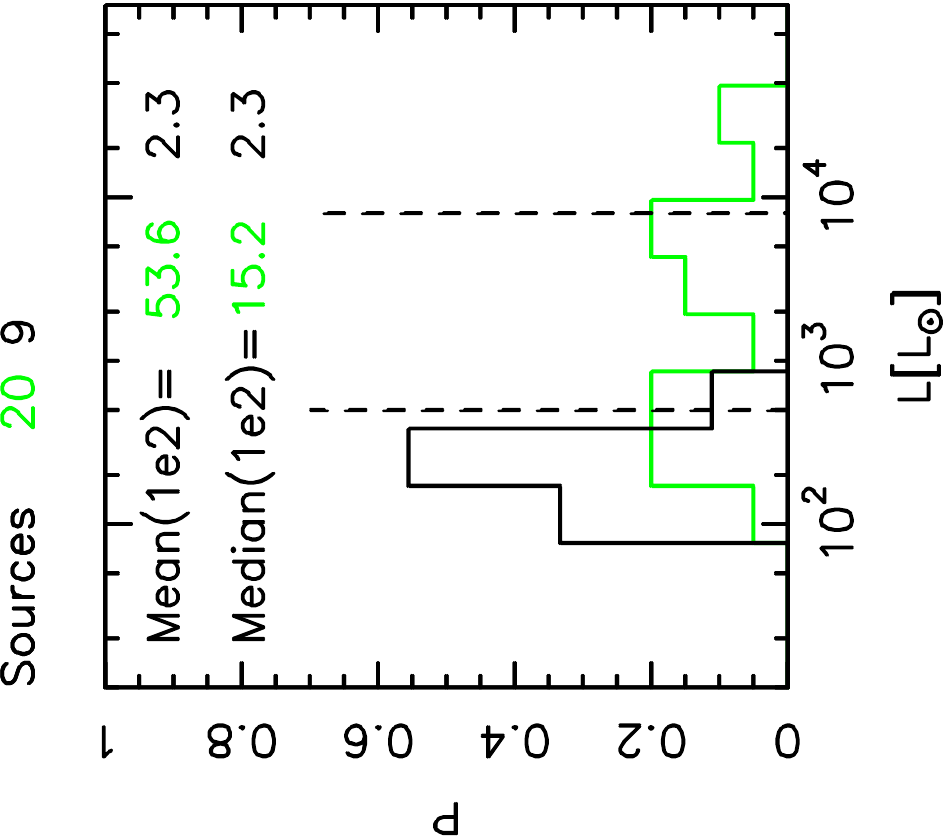}
 \caption{Normalized histogram of the luminosities for the SFS (green) and for the QS (black). $500\lsun$ and $8000\lsun$ are indicated by the dashed lines.}
 \label{fig:luminosities}
\end{figure}

\subsection{The mass-luminosity plot}\label{ssec:ml_plot}

To refine the separation in different evolutionary phases we make use of the mass-luminosity ($M-L$) plot, which is an efficient and well-established diagnostic tool to disentangle the different evolutionary phases of star formation in the low-mass regime \citep{Saraceno+96}. \citet{Molinari+08} proposed that it could also be used for high-mass stars, under the hypothesis that star formation at high- and low-mass proceeds in a similar fashion, with accretion from the surrounding environment playing a major role \citep[e.g., ][]{Krumholz+09}. \citet{Molinari+08} built a simple model for the evolution of a clump, based on the turbulent core prescriptions of \citet{McKeeTan03}, ranging from the early collapse phase to the complete disruption of the dusty envelope by the central object.

Figure~\ref{fig:m-l_plot} shows the $M-L$ plot for the sources in our sample for which we were able to derive the luminosity (see Sect.~\ref{ssec:sed} and Table~\ref{tab:greybody}) from the integration of the SED (with the linear interpolation in the log-log space) and the mass derived from the $1.2\mm$ emission, for which we used the temperature determination obtained from the ammonia observations (Sect.~\ref{ssec:temp}). The black solid line indicates the ZAMS locus, according to \citet{Molinari+08}, while the dashed line indicates the ZAMS locus as determined by \citet[][]{Urquhart+13}. The latter authors consider clumps showing methanol maser emission and with a luminosity from the Red MSX Sources survey \citep{Urquhart+08}. $90\%$ of the sources studied by \citet{Urquhart+13} have $L>\pot{3}\lsun$, the remaining $10\%$, with $L<\pot{3}\lsun$, have low gas masses ($\pot{1}-\pot{2}\msun$), and could be forming intermediate-mass stars. Objects classified as either YSO and UC\hii\ regions were used to determine the ZAMS locus, as no statistically significant difference is found between the two classes. The difference in slope between the two ZAMS-lines could be due to the fact that in \citet{Molinari+08} the luminosities were determined from the Robitaille models, using IRAS fluxes in the FIR, and could thus be overestimated.
The grey curves show the evolution of the source predicted by the simple model, for different final masses, from $\sim 6$ to $\sim30\msun$. Time increases in the direction of the arrows shown in the upper right part of the plot. In the collapse phase, before the central object reaches the ZAMS, the mass of the core envelope does not change by much, but the luminosity increases rapidly. After the central star reaches the ZAMS, the luminosity does not vary much, and the energetic radiation and wind begin to destroy the parental clump, visible as a steady decrease in envelope mass.

Comparing the distribution of sources in the plot with the evolutionary tracks of the \citet{Molinari+08} model, our objects span the total range in envelope masses, for final masses of the star between about $6$ and $30\msun$.
The stellar masses are in agreement with those derived from the fit of the SED with the Robitaille models for the objects near the ZAMS, for which the final mass is similar to the current mass, as the main accretion phase is over.  Massive stars appear to form virtually always in clusters \citep[e.g., ][]{LadaLada03}. In both the \citet{Molinari+08} and the Robitaille models the luminosity is considered as being dominated by the most massive object formed in the cluster. The derived bolometric luminosity and stellar mass could thus be overestimated. A detailed study of the stellar population in these clumps will be carried out in a subsequent paper, by means of mid-IR and near-IR data.

\begin{figure*}[tbp]
 \centering
	\includegraphics[width=\textwidth]{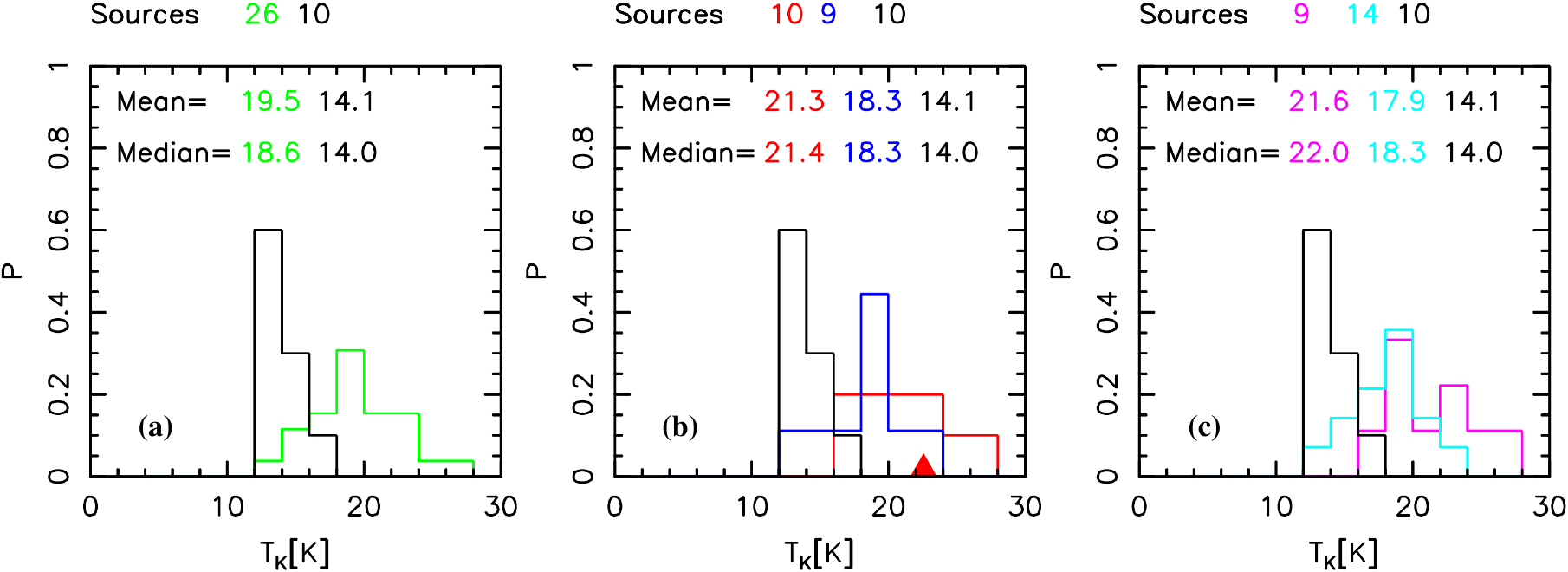}
 \caption{Normalized histogram (to the total number of sources in each class) of kinetic temperature for \textbf{(a)} the SFS (green), and the QS (black); \textbf{(b)} the same as \textbf{(a)}, but the SFS-2, with $L>\pot{3}\lsun$ (in red) are separated from the SFS-1 (in blue), the red triangle shows the temperature for the Type 3 source; \textbf{(c)} the same as \textbf{(a)}, but the SFS was divided into clumps with (magenta) and without (cyan) radio-continuum emission. Mean and median values of the temperature are indicated in each panel. The total number of sources in each class is shown above each panel.}
 \label{fig:histo_tk}
\end{figure*}

A first macroscopic difference between the SFS and the QS sources is the luminosity. As can be seen in Fig.~\ref{fig:luminosities}, the clumps in the QS have low luminosities, distributed between $\sim100$ and $500\lsun$. On the other hand, the sources in the SFS show a peak at $\sim500\lsun$, but also a second peak at $L\sim8\times\pot{3}\lsun$ (both indicated as dashed lines in the figure). 
The distribution of the sources in the $M-L$ plot indicates that part of the sources in the SFS (10 out of 20, including 17195-3811c1, with the mean luminosity derived with the online SED fitting tool) are likely hosting a ZAMS star and have stopped the accelerating accretion phase, according to the model of \citet{Molinari+08}. The sources with signs of active star formation, but well below the ZAMS loci, are essentially indistinguishable from the quiescent ones in terms of luminosities. 
Figure~\ref{fig:m-l_plot} shows that in our sample all sources with $L>\pot{3}\lsun$ have strong IR (MSX) and/or radio continuum emission, while those below this threshold do not.
The radio emission from the sources near the ZAMS locus, with final masses greater than $8\msun$ shows that the interpretation of the $M-L$ plot is essentially correct, and that the prediction of the end of the accelerating accretion phase is reasonably good. 

The small range of luminosity and its low average value for the QS shows that this is a homogeneous sample, with all the clumps in an early phase of evolution. 
On the other hand, the SFS appear to include clumps in widely different evolutionary stages: the points in the diagram go from clumps similar to those of the QS, to clumps containing a ZAMS star and beyond, where the star is dispersing the envelope. 
For our sample, a simple criterion in luminosity is sufficient to separate the sources that likely have an embedded ZAMS star from the rest. Thus, in the following we will refer to objects containing a ZAMS star as those with $L>\pot{3}\lsun$. In Fig.~\ref{fig:m-l_plot} these objects fall in the region encompassed by the ZAMS loci, except 13560$-$6133c1 and 16093$-$5015c1, slightly below the \citet[][]{Urquhart+13} ZAMS locus. 

We can easily divide the clumps in this sample in Type 1, 2 and 3 according to the $M-L$ plot \citep[see ][]{Molinari+08}; the classification of sources in one of the three types is shown in Table~\ref{tab:sf_signs}. The clumps between the two ZAMS loci, with $L>\pot{3}\lsun$ would be Type 2 clumps (including the two sources slightly below the Urquhart et al. ZAMS locus, 13560$-$6133c1 and 16093$-$5015c1, the former of which shows also radio continuum emission), while the rest of the SFS and the QS would be Type 1.
Type 1 sources include both pre-stellar and proto-stellar sources in early stages of evolution, for which the development of an \hii\ region may be quenched by the high accretion rates.
The leftmost point in the mass luminosity plot (Fig.~\ref{fig:m-l_plot}) identifies the most evolved source in our sample, 17355$-$3241c1, with a relatively low mass and a high luminosity. This clump is the only Type 3 source in our sample, and has very strong emission even in the IRAC bands. 17355$-$3241c1 exemplifies the last phase of the evolution in the $M-L$ plot, when the parent cloud is dispersed by the destructive action of the central star. This source is discussed in more detail in the Appendix. 
\begin{figure}[tbp]
 \centering
 \includegraphics[angle=-90,width=0.35\textwidth]{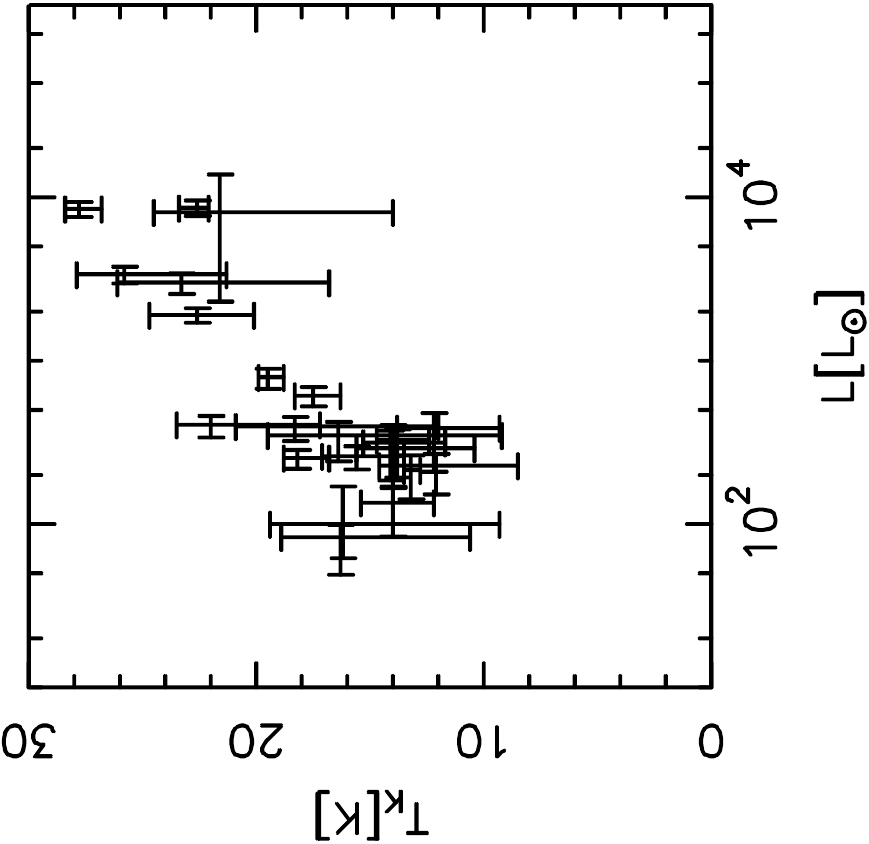}
 \caption{Correlation between luminosity and kinetic temperature. The uncertainties for $L$ and $\tk$ are indicated in the figure.}
 \label{fig:tk_l}
\end{figure}

\begin{figure*}[tbp]
 \centering
	\includegraphics[width=0.7\textwidth]{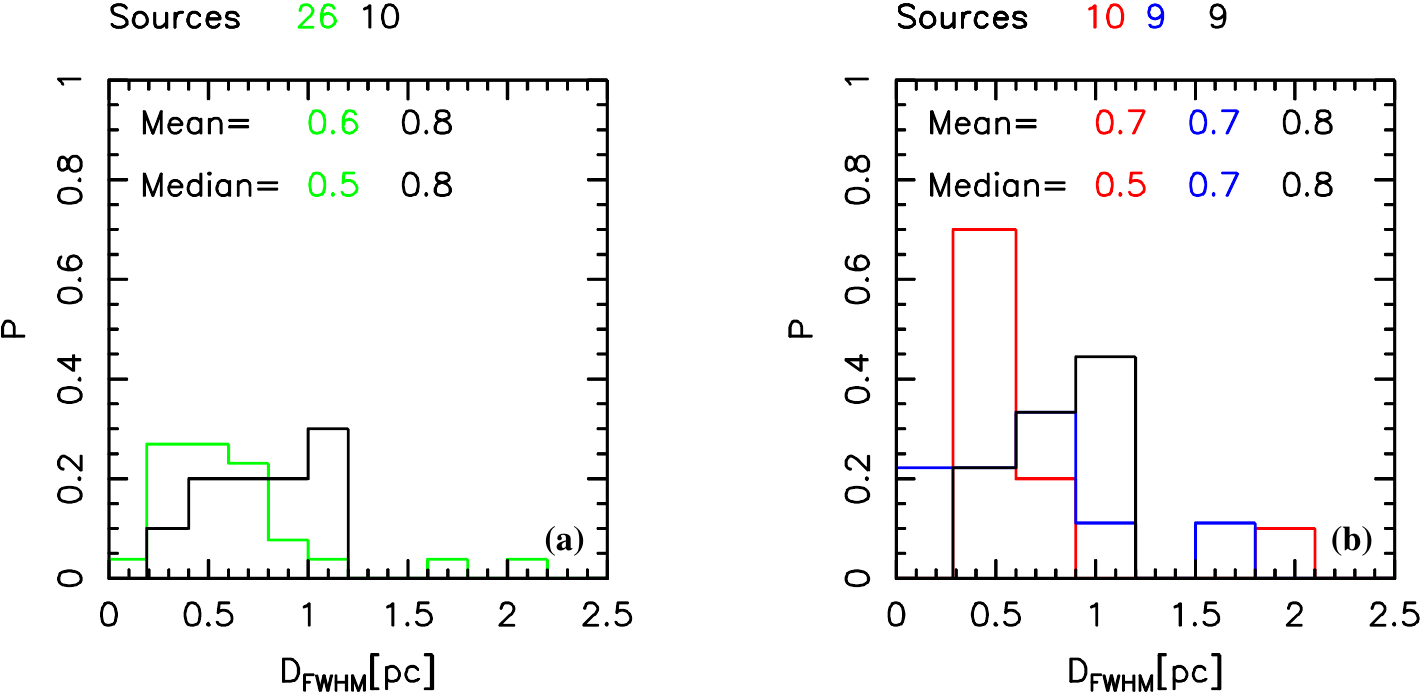}
 \caption{Histogram of the diameters of the clumps. Panel \textbf{(a)} shows the diameters of the SFS (green) \textit{vs.} the QS (black). Panel \textbf{(b)} shows the diameters for the SFS-2 is indicated in red, the SFS-1 is indicated in blue, and the QS in black. The total number of sources in each class is shown above each panel.}
 \label{fig:lin_diam}
\end{figure*}

Thus, the $M-L$ plot and the signposts of active star formation give complementary information about the evolutionary state of the clump, allowing us to refine the \citet{Molinari+08} classification, separating objects likely hosting a ZAMS star from the other sources in the SFS. Our original sample is thus finally divided into three different classes (without considering Type~3 objects): Type 1 quiescent clumps, apparently starless, Type 1 with signs of active star formation, but still a low luminosity and Type 2 sources, hereafter QS, SFS-1 and SFS-2, respectively. Table~\ref{tab:mean_zams_nozams} shows explicitly that SFS-1 and SFS-2 have very different luminosity-to-mass ratios. The clumps in the SFS-1 can be in a very early phase of the process of formation of a high-mass object; alternatively, the signposts of active star formation could be generated by more evolved lower-mass stars.

\subsection{Properties of sources in different stages of evolution}

\subsubsection{Temperatures}

The {temperature} of gas and dust may be influenced by the presence of a (proto-)star deeply embedded in a clump. 
Figure~\ref{fig:histo_tk}a shows the histogram of temperatures for the two samples. The normalized counts of the SFS are shown in green, and those of the QS in black. We find that it is possible to observe temperature differences on a large scale, comparing the average values of $\tk$ and $\td$ of the QS and the SFS. The typical $\tk$ of the star-forming clumps is greater than that of the QS, with mean values of $\tk = 19.5\unc{-2.9}{+1.5}\kel$ and $14.1\unc{-3.2}{+1.8}\kel$, for the SFS and QS, respectively. Thus, the average temperature increases as evolution proceeds. 

In Fig.~\ref{fig:histo_tk}b we can see that the SFS-2 (in red) show a slightly higher $\tk$ ($\tk = 21.4\unc{-3.8}{+1.7}\kel$) than the SFS-1 (in blue) ($\tk=18.3\unc{-3.0}{+1.4}\kel$). Figure~\ref{fig:histo_tk}b shows also the QS, to underline that both the SFS-1 and SFS-2 sources on average have a higher $\tk$. Figure~\ref{fig:histo_tk}c shows that the SFS with $1.3$~cm radio-continuum emission are hotter than those without it.
A similar conclusion is found by \citet{Sanchez-Monge+13b}, studying NH$_3$(1,1) and (2,2) at high angular resolution in cores in clustered high- and intermediate-mass star forming regions. They find that starless cores have an average $T\sim15\kel$, lower than $T\sim21\kel$ found for proto-stellar cores. These values are very close to those found in this work. \citet{Sanchez-Monge+13b} show that the higher temperatures in starless cores in clustered environments with respect to more isolated cases can be explained considering external heating from the nearby massive stars. Also \citet{Rygl+10} and \citet{Urquhart+11} find that actively star-forming clumps are slightly hotter than the quiescent ones.
A behaviour similar to that of the kinetic temperature is observed for $\td$ as a function of evolutionary phase, not unexpected given the good agreement of the two temperatures (see Sect.~\ref{ssec:sed}; Fig.~\ref{fig:tk_td}).
The mean values are $\td=24.0\unc{-1.6}{+1.5}; 15.8\unc{-1.6}{+1.4}; 11.8\unc{-1.5}{+1.7} \kel$ for SFS-2, SFS-1 and QS, respectively.
 
Disregarding the 7 points with $|\tk-\td|\ge5\kel$ and non-overlapping $68\%$ credibility intervals, we find a correlation between the luminosity and the kinetic temperature, shown in Fig.~\ref{fig:tk_l}. A correlation between $\tk$ and $\lbol$ was also found by e.g., \citet{Churchwell+90}, \citet{Wu+06}, \citet{Urquhart+11}, and by \citet{Sanchez-Monge+13b}, for cores in clustered environments, using the luminosity of the whole region. 

\smallskip

\begin{figure}[tbp]
 \centering
	\includegraphics[angle=-90, width=0.35\textwidth]{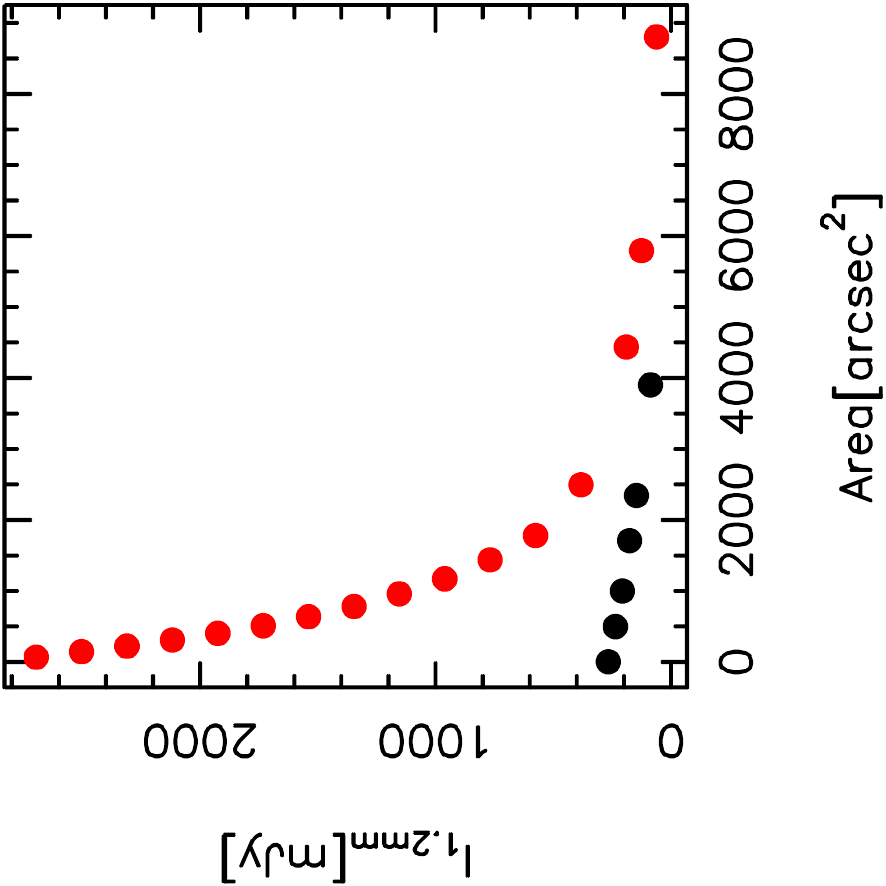}
 \caption{$1.2\mm$ intensity level as a function of the area within the contour for a typical QS (black) and SFS-2 (red) source, respectively.}
 \label{fig:12mm_extrap}
\end{figure}

\subsubsection{Sizes}\label{sssec:sizes}

From the panel (a) of Fig.~\ref{fig:lin_diam} we note that the SFS tend to have smaller FWHM diameters than the QS. From the distribution of sizes shown in panel (b) we note that the SFS-2 have a peak at the smallest linear dimensions. 
Plotting the area within a specific intensity contour of $1.2\mm$ emission as a function of the intensity level and extrapolating linearly to zero intensity we get an idea of the source extent if we could observe with infinite sensitivity \citep[cf. ][]{BrandWouterloot94}. Figure~\ref{fig:12mm_extrap} shows the $1.2\mm$ intensity level as a function of the area within the contour for representative sources in QS and SFS-2. For the SFS-2 we ignore the central emission peak for the fit. This method allow us to  estimate the effect of the lower temperature on the clump size at the typical noise levels of the SEST maps.

This procedure shows that with our noise levels we miss $\sim 30\%$ of the emission area for a typical QS, while only $\sim10-15\%$ is lost for a typical SFS-2. The SFS-1 usually shows an intermediate behaviour. The larger fraction of emission area below the noise level for the QS confirms that we are not able to detect the external envelope of the coldest clumps, and that the actual linear size of QS sources is very similar to that of SFS-2 objects. On the other hand, the area within the FWHM contour is smaller for SFS-2 sources, possibly indicating that sources hosting a ZAMS object are more compact and centrally concentrated. We investigate an alternative possibility, namely that the observed FWHM size may be $T$-dependent, performing a simple test using a 1D simulation with RATRAN \citep{HogerheijdeVanDerTak00}, constructing a clump with a typical radial dependence of the density \citep[$\propto r^{-1.7}$, cf. ][]{Beuther+02b, Mueller+02}, and comparing the continuum at $250 \usk \giga\hertz$ with and without a central luminous heating source. The radial temperature dependence is assumed to be $\propto (r/r_0)^{-0.4}$ \citep[e.g., ][]{WolfireCassinelli86}. We observe that the clump with the embedded source, has a smaller (by $20-30\%$) FWHM. This could explain the smaller sizes derived for SFS-2.

\citet[][]{Urquhart+13}, also conclude that high-mass star forming clumps showing methanol maser emission are more compact and centrally concentrated than the rest of sources in the ATLASGAL survey \citep{Schuller+09}, comparing the ``compactness'' of the sources by means of the ratio of the peak and integrated sub-mm flux, for a much larger sample.

\subsubsection{Densities}

\begin{figure}[btp]
 \centering
 \includegraphics[angle=-90,width=0.33\textwidth]{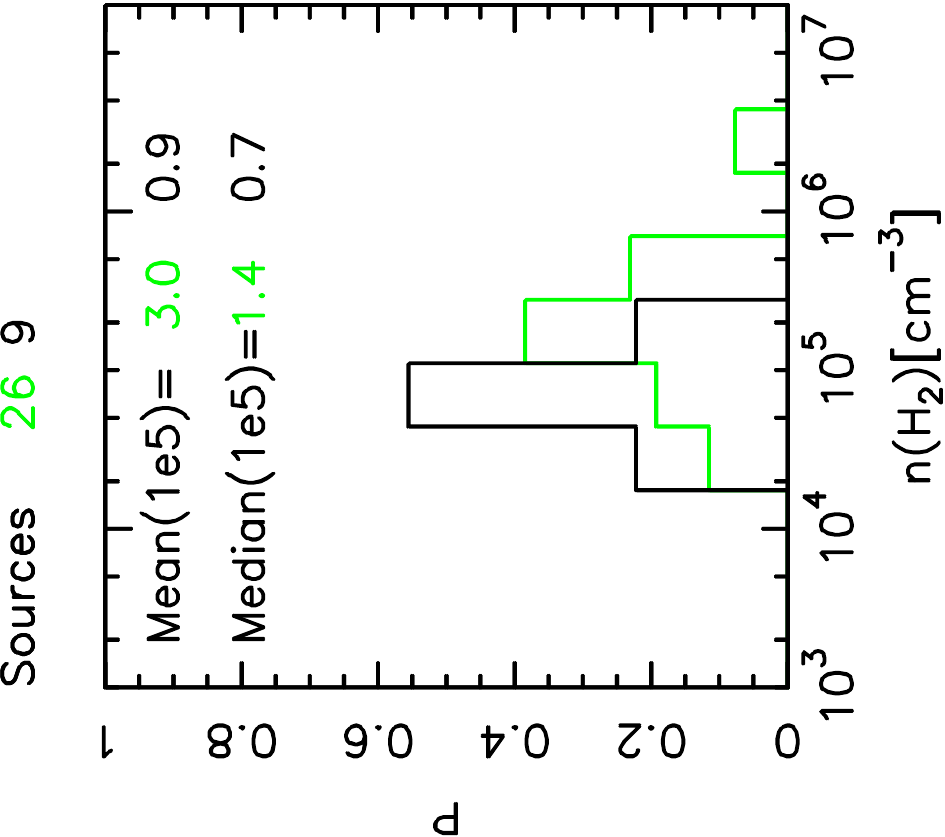}
 \caption{Normalized histogram of volume density of molecular hydrogen averaged within the FWHM contour. We show the SFS in green and the QS in black. The total number of sources in each class is shown above the panel.}
 \label{fig:histo_dens_fwhm}
\end{figure}

\begin{figure*}[btp]
 \centering
	\includegraphics[width=\textwidth]{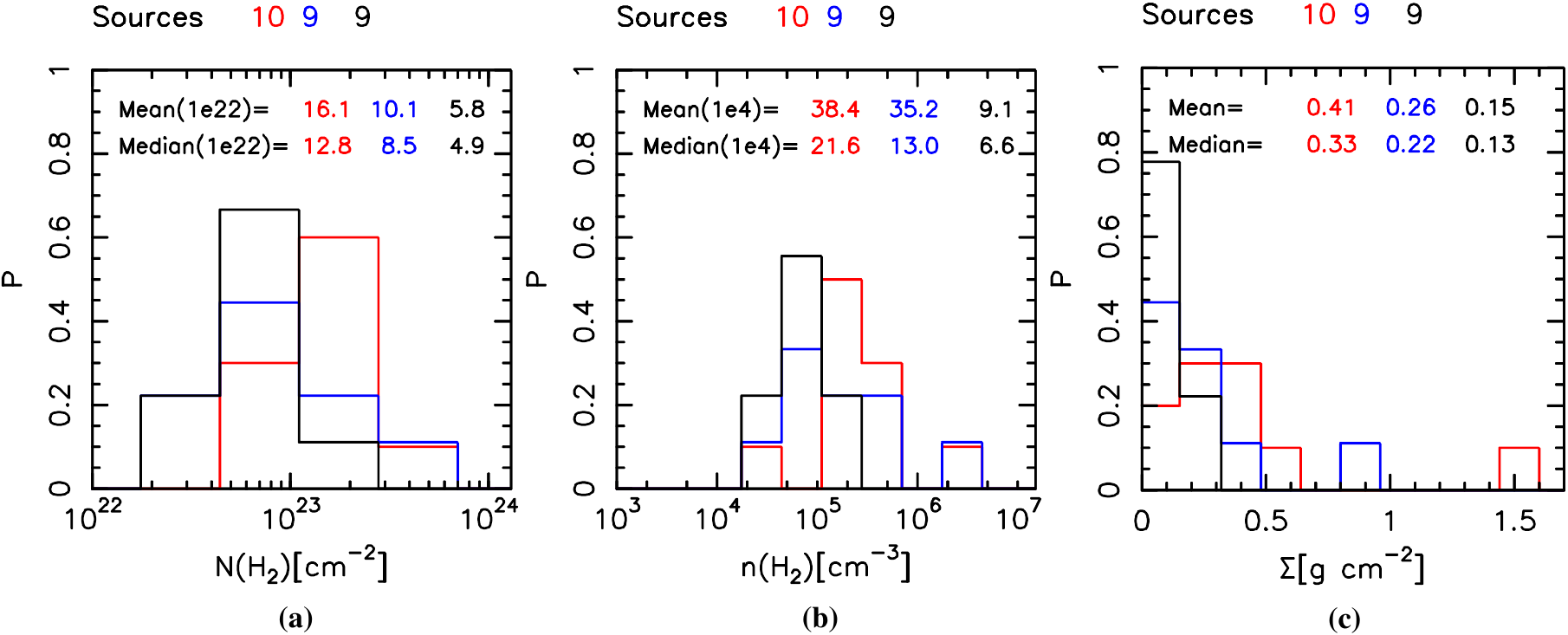}
 \caption{Normalized histograms of \textbf{(a)} column-, \textbf{(b)} volume- and \textbf{(c)} surface-densities of molecular hydrogen averaged within the FWHM contour. We show the SFS-2 in red, SFS-1 in blue and the QS in black. The total number of sources in each class is shown above the panels.}
 \label{fig:histo_dens_fwhm_zams}
\end{figure*}

The mean {column-, volume- and surface-densities} are compared for QS and SFS, and for SFS-1 and SFS-2 in Fig.~\ref{fig:histo_dens_fwhm} and \ref{fig:histo_dens_fwhm_zams}, respectively.

The histogram of the SFS is shifted towards higher densities; \citet{Chambers+09} obtain the same result, with star-forming sources on average denser than the quiescent ones. Taking into account the separation into SFS-1 and SFS-2, the density histograms show that the SFS-2 usually have higher values of column-, volume- and surface-densities than either QS or SFS-1; the latter two classes have density distributions peaking at similar values, with that of the SFS-1 showing a tail with values similar to those of the SFS-2. Thus, the clumps hosting a ZAMS star have higher densities, indicating that as star formation proceeds, the clumps appear to become denser in the central parts. The same is found by \citet{ButlerTan12}, comparing their sample of starless cores to a sample of more evolved objects \citep[from ][]{Mueller+02}. We caution that, as the source size could be underestimated due to the presence of a central heating source, causing the clump to appear more centrally peaked (see Sect.~\ref{sssec:sizes}), the densities could be overestimated for the SFS-2, thus explaining the observed differences with both QS and SFS-1.

\citet{Rygl+13} argue that star formation signposts are not present in clumps with a column density below a value of $4\times\pot{22}\cm^{-2}$. No such threshold effect for the onset of star formation is observed in our sample, but only two of our clumps of any type have column densities well below $4\times\pot{22}\cm^{-2}$ (cf. Table~\ref{tab:fwhm}).

We note that the values of the mass surface density are typically lower than the theoretical threshold of $\Sigma=1\usk\gram\cm^{-2}$ for massive star formation, on average by a factor $2-6$. The theoretical threshold for $\Sigma$ given by \citet{KrumholzMcKee08} holds for a single core, stabilized against fragmentation only by radiative heating. 
\citet{LopezSepulcre+10} show that massive star formation, indicated  by the presence of massive molecular outflows, most probably driven by massive YSOs, is occurring also in clumps with a much lower mass surface density, of the order of $\Sigma\sim0.3\usk\gram\cm^{-2}$. \citet{ButlerTan12} show that similar values of the mass surface density $\Sigma$ are typical also of massive starless cores, assuming a gas-to-dust ratio of $150$, thus consistent with our average value of $\sim0.2\usk\gram\cm^{-2}$ for the QS, for a gas-to-dust ratio of $100$. 

\begin{figure*}[tbp]
 \centering
	\includegraphics[angle=-90, width=0.35\textwidth]{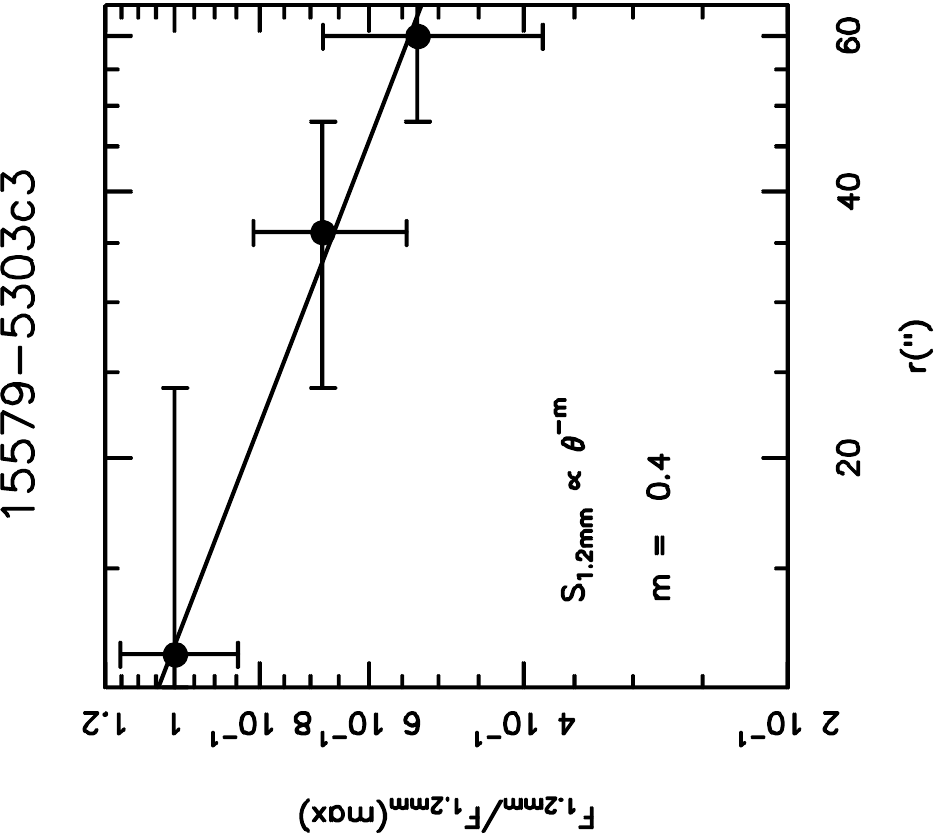} \hspace*{1cm}
	\includegraphics[angle=-90, width=0.35\textwidth]{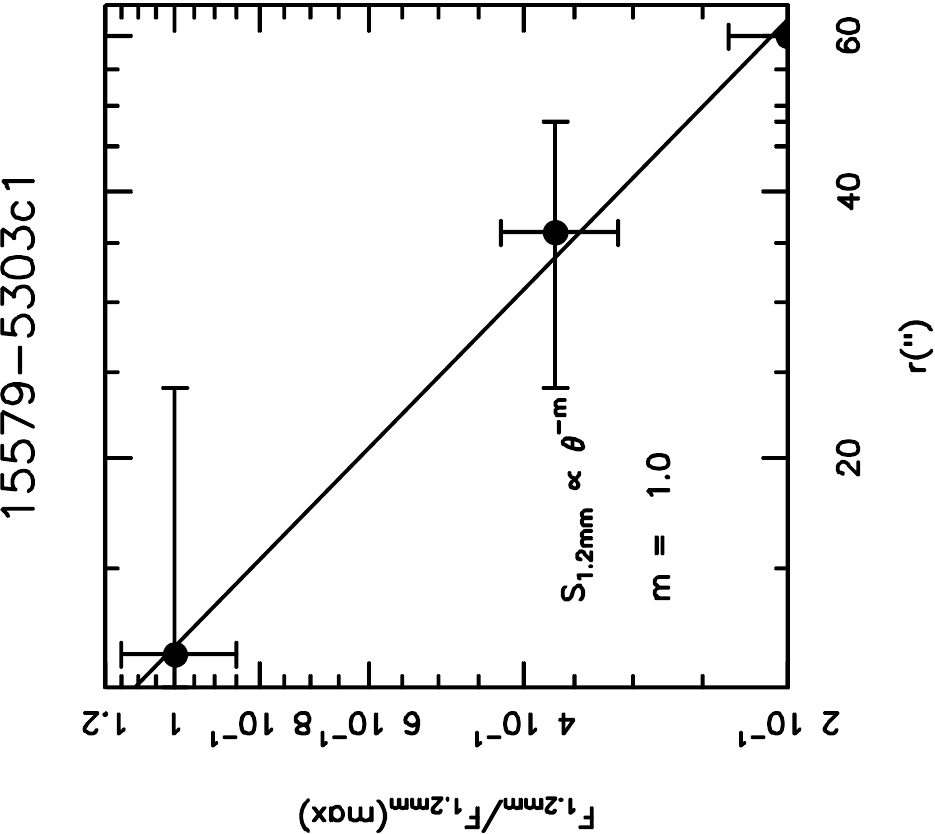}
 \caption{Average $1.2\mm$ flux, normalized to the maximum, as a function of the radius of the largest circle of the annulus for a typical QS (left) and SFS-2 (right) source. The uncertainty on the average flux in indicated. The beam FWHM size is indicated as an errorbar in $x$. In the bottom left corner the slope $m$ is indicated.}
 \label{fig:density_pl}
\end{figure*}

We investigate in more detail the possibility that the radial dependence of the density changes in different classes of objects, by fitting a power law to the average radial $1.2\mm$ intensity profile calculated in concentric annuli centered on the clump. Figure~\ref{fig:density_pl} shows the normalized $1.2\mm$ emission as a function of the angular radius of the largest circle of the annulus. The beam FWHM size is indicated as an errorbar in $x$. Following the notation in \citet{Ward-Thompson+94} (and references therein), the power law indices of the $\mm$ emission $m$, of the density $p$ and of the temperature $q$ are related by $m=p+Q(\nu,T)q-1$, where $Q(\nu,T)$ is a coefficient depending on the wavelength and the temperature, near to unity for $h\nu/(k_B T)\ll1$ \citep[see ][]{Adams91}. We find typical power law indices for mm emission $m \simeq 0.2-0.4$ for QS and $\sim 1.0-1.3$ for SFS-2 clumps, respectively. We assumed that quiescent clumps are isothermal and that Type 2 sources are likely to have a radial gradient in temperature, described by $T\propto (r/r_0)^{q}$, with $q=-0.4$ \citep[e.g., ][]{WolfireCassinelli86}. This implies that the power law index $p$ for volume density is $\sim 1.5-1.8$ in SFS-2 sources, steeper than in QS sources, with $p\sim 1.2-1.4$. The value of $p$ is highly uncertain, because of the assumptions made and because of the very limited number of points for each source, however we are mainly interested in the difference between the QS and SFS-2, and not the specific value of $p$. As the clumps in QS and SFS have similar masses and total sizes (see Sect.~\ref{ssec:mass_dens_size} and \ref{sssec:sizes}), this difference in density profile suggests that SFS-2 clumps may indeed be denser in the central regions. Other authors investigated the differences in the density power law index for sources in different stages of evolution. \citet{Beuther+02b} also derive an average radial power law dependence for density with $p \sim 1.5-2.0$. These authors find that the density distribution is flatter for sources in the very early stages ($p\sim1.5$), then it becomes steeper during the collapse and accretion phase ($p\sim1.9$), and finally it flattens again ($p\sim1.5$) in the dispersal phase. A similar result is reported by \citet{ButlerTan12}, who compared power law indices for density in starless clumps ($p \sim 1.1$), obtained with a very different technique, with the same quantity derived by \citet{Mueller+02} for more evolved objects ($p \sim 1.8$). 

\smallskip

\subsubsection{Mass}

The masses measured for our clumps are in the range $10-2000\msun$, which indicates that we are probing the low-mass end of the \citet{KauffmannPillai10} relation (cf. their Fig.~2b and Fig.~\ref{fig:m-r_diagram} in this paper). 
With respect to more massive  clumps, those in this sample will probably form a very limited number of stars with $M>8\msun$, making the assumption of a single massive star in the clump plausible. On the other hand, we know that massive stars are being formed in some objects, as we observe compact radio continuum emission, likely arising in \hii\ regions \citep{Sanchez-Monge+13}. 

Comparing the source size and densities (Fig.~\ref{fig:lin_diam} and \ref{fig:histo_dens_fwhm_zams}) SFS-2 sources are typically the most compact, suggesting that an evolution of the clumps towards more compact entities with evolution might indeed occur. However, as time proceeds and the massive stars dissociate the molecular gas they move down and to the right in the $M-r$ plot (Fig.~\ref{fig:m-r_diagram}), as shown by 17355$-$3241c1.

\smallskip

\subsubsection{Velocity and linewidth}

The velocity gradients found for the clumps ($1-2\kms\pc^{-1}$; Sect.~\ref{ssec:dynamics}) are comparable for the SFS and QS, and also the fraction of clumps showing gradients is similar: about $50\%$ of the objects.

\begin{figure}[tbp]
 \centering
	\includegraphics[angle=-90, width=0.45\textwidth]{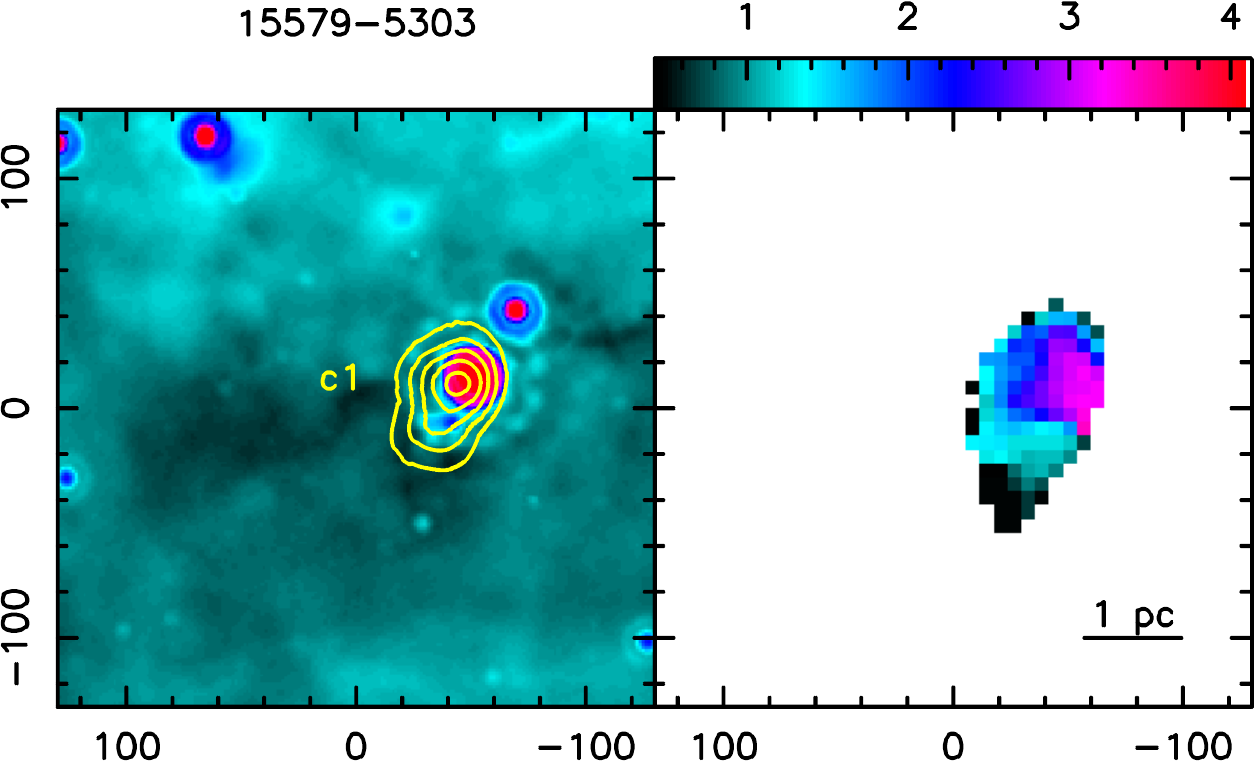}
 \caption{Example of NH$_3$(1,1) second moment map. Left panel shows Spitzer/MIPSGAL $24\mum$ image in colourscale and NH$_3$(1,1) integrated emission in contours, while the right panel shows the second moment map.}
 \label{fig:ex_2nd_mom}
\end{figure}

From the second moment map we find that the clumps in the SFS have larger linewidths, with an average value of $2.2\kms$ compared to $1.6\kms$ for the sources in the QS. For 08477$-$4359c1, 13560$-$6133c1, 15557$-$5215c1 and c2, 15579$-$5303c1 and 16093$-$5015c1 we find an increase of the linewidth between the position of a $24\mum$ source and the rest of the clump of $\sim10-30\%$. Figure~\ref{fig:ex_2nd_mom} shows an example of this increase in linewidth at the location of a $24\mum$ source. These sources always have a $\Delta V$ of the (2,2) line larger than that of the (1,1) at this position. This indicates that we are probing the regions where the embedded source is injecting turbulence. All these sources have a luminosity in excess of $\pot{3}\lsun$, except 15557$-$5215c2 ($L=740\lsun$), and 08477$-$4359c1, for which we do not have Herschel data. Similar linewidths are found in dense cores located within clustered massive star forming regions \citep{Sanchez-Monge+13b}: $\sim1.2\kms$ and $\sim2.0\kms$ for starless and proto-stellar cores, respectively.

\subsubsection{Virial parameter} \label{ssec:alpha}

Figure~\ref{fig:virial} presents the {virial parameter} $\alpha\equiv \mvir/\msest$ as a function of $\msest$. As already noted, all clumps in our sample appear dominated by gravity. Moreover, these are upper limits for $\alpha$ as the virial mass should be reduced by up to a factor of 2 \citep[see ][]{MacLaren88} as a consequence of the density gradients found in the clumps; another factor of 2 may arise because the mass was was computed within the FWHM contour. Two sources in the SFS has $\alpha\gtrsim2$: this could  be due to the action of the embedded YSO(s) disrupting the parental cloud. 
The value of the virial parameter tend to decrease as the clump mass increases. 

Let's explore a few possibilities to explain $\alpha < 1$ for such a large number of sources.
The first is that we might be underestimating the gas/dust temperature, and are thus overestimating the mass from the $1.2\mm$ continuum. As NH$_3$(1,1) and (2,2) essentially trace cold gas and they could be optically thick, this may be a possibility. An independent determination of the temperature in the clumps is given by the dust temperature derived from the modified black-body fit of the mm-FIR part of the SED. Dust emission is optically thin in this regime, thus probing even the inner regions of the clump. Comparing $\tk$ and $\td$ we find that the two temperatures are usually in good agreement, as discussed in Sect.~\ref{sec:results} and shown in Fig.~\ref{fig:tk_td}. Therefore we do not expect the bulk of the gas in the clump to have temperatures systematically higher than those adopted here, and consequently lower masses. Moreover, also quiescent clumps have $\alpha \ll 1$, and the temperature in this case should not be underestimated, as no heating source is present in these sources. Even if it were the case, it is difficult to account for a factor of  $3-10$. The independent estimate of the mass through the SED essentially confirms that the mass of the clumps is correct.

Another possibility is that we are underestimating the radius of the clump, leading us to underestimate the virial mass. \citet{PanagiaWalmsley78} show that if the source is not Gaussian, we may underestimate the radius by even a factor of 2. 
However, it is not clear why the most massive clumps should be different from the others. The same holds for the uncertainty connected to the gas-to-dust ratio, assumed to be 100. 

\citet{Fontani+02} and \citet{LopezSepulcre+10} reach opposite conclusions regarding the stability of the clumps in their samples. In the former work the authors show that the ratio between $\mvir/M$ reaches values as low as ours, using as a tracer CH$_3$CCH, while \citet{LopezSepulcre+10} obtain $\alpha\sim1$ using the mass derived from dust emission and the virial mass from the C$^{18}$O line emission. Also \citet{Hofner+00} find $\alpha<1$ using the less abundant C$^{17}$O.
Some of the clumps in our sample have been observed in C$^{18}$O$(3-2)$ by \citet{Fontani+12}, and these authors suggest that the clumps are in virial equilibrium or even have virial masses greater than the clump mass (i.e., $\alpha\gtrsim1$). Comparing the C$^{18}$O and the NH$_3$ linewidths we find that the former are larger, typically by a factor of $1.6$ and even up to $\sim2.6$. As a consequence, the virial masses calculated using the C$^{18}$O linewidth would be larger by a factor $2.7$ (up to $6.8$), and $\alpha$ would increase accordingly. 
\citet{Sanchez-Monge+13b} find linewidths for cores similar to those found in this study.
The difference in linewidth between NH$_3$ and C$^{18}$O may be explained by the fact that C$^{18}$O is tracing a more extended and diffuse region, where $\Delta V$ may be larger due to the effects of the environment in which the clump is embedded, or due to the presence of several gas concentrations along the line of sight (i.e., additional indistinguishable velocity components in the spectra), not dense/massive enough to be visible in NH$_3$, or by the presence of velocity gradients across the clumps. Moreover, C$^{18}$O is more prone to be entrained in outflows driven by objects already formed in the clump. In conclusion, $\alpha$ may be underestimated, nevertheless, even taking into account the possibilities discussed above, $\alpha$ would still be $<1$ for the sources with the highest masses. Moreover, the contribution of external pressure helps gravity.

If these clumps were supported only by turbulence and thermal pressure, they would collapse on the timescale of the free-fall time. Given the short timescales set by the free-fall time $t_{ff}=\sqrt{3\pi/(32  G \rho)} \sim 5\times\pot{4}\yr$, this may suggest that magnetic field plays a significant role in stabilizing the clumps. 
Following the relation for virial equilibrium given in \citet{McKee+93}, neglecting the surface pressure term, we get the expression for the equilibrium magnetic field
\begin{equation}
 \begin{split}
	B = & 1.4\times\pot{-5} \left[ \left( \frac{M}{100\msun} \right) \left( \frac{R}{1\pc} \right)^{-4} \right]^{0.5} \times \\
	    & \left[ \left( \frac{M}{100\msun} \right) - 0.71 \left( \frac{R}{1\pc} \right) \left( \frac{\Delta V}{2\kms} \right)^2 \right]^{0.5} [G],
 \end{split}
\end{equation}
where $M$ is the clump mass, $R$ is the clump radius and $\Delta V$ is the FWHM of the line.
Using appropriate numbers for these parameters we find that $|\overrightarrow{B}|\approx 0.1-1\usk\milli\mathrm{G}$ is sufficient to stabilize the clumps. Such values of the magnetic field have been observed towards regions undergoing massive star formation \citep[e.g., ][]{Crutcher05, Girart+09}.

\begin{figure}[tbp]
 \centering
 \includegraphics[angle=-90,width=0.75\columnwidth]{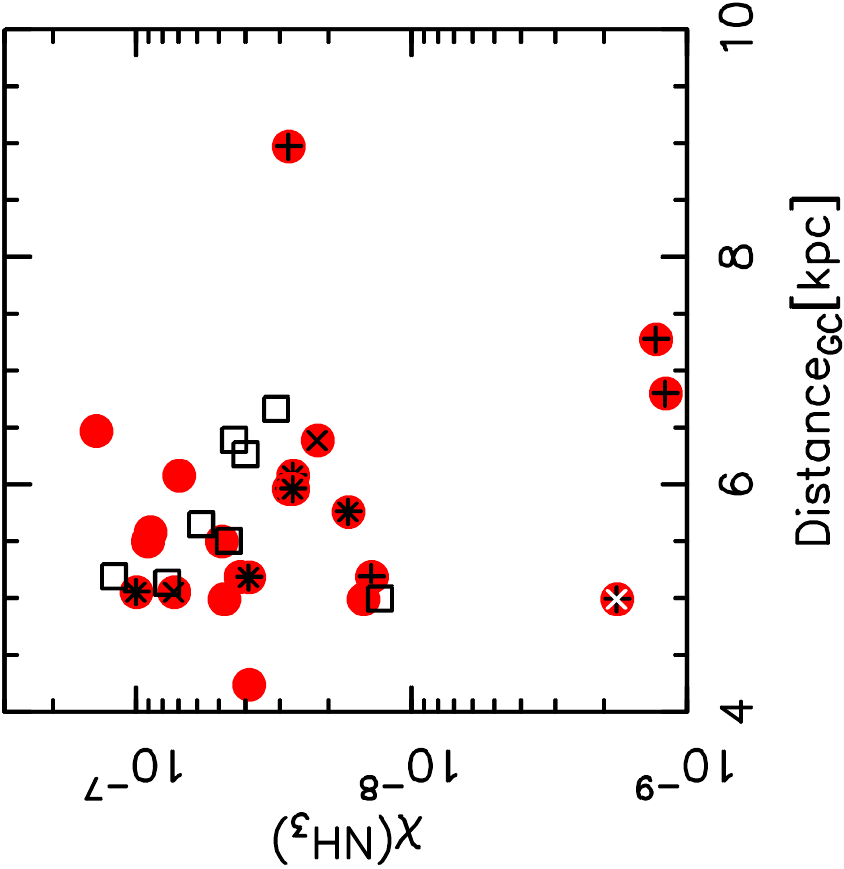}
 \caption{Ammonia abundance as a function of Galactocentric distance. The symbols are the same as in Fig.~\ref{fig:m_dist}.}
 \label{fig:dgc_abund}
\end{figure}

\section{Observational classification of high-mass clumps} \label{sec:sketch}

\begin{figure*}[tbp]
 \centering
	\includegraphics[width=\textwidth]{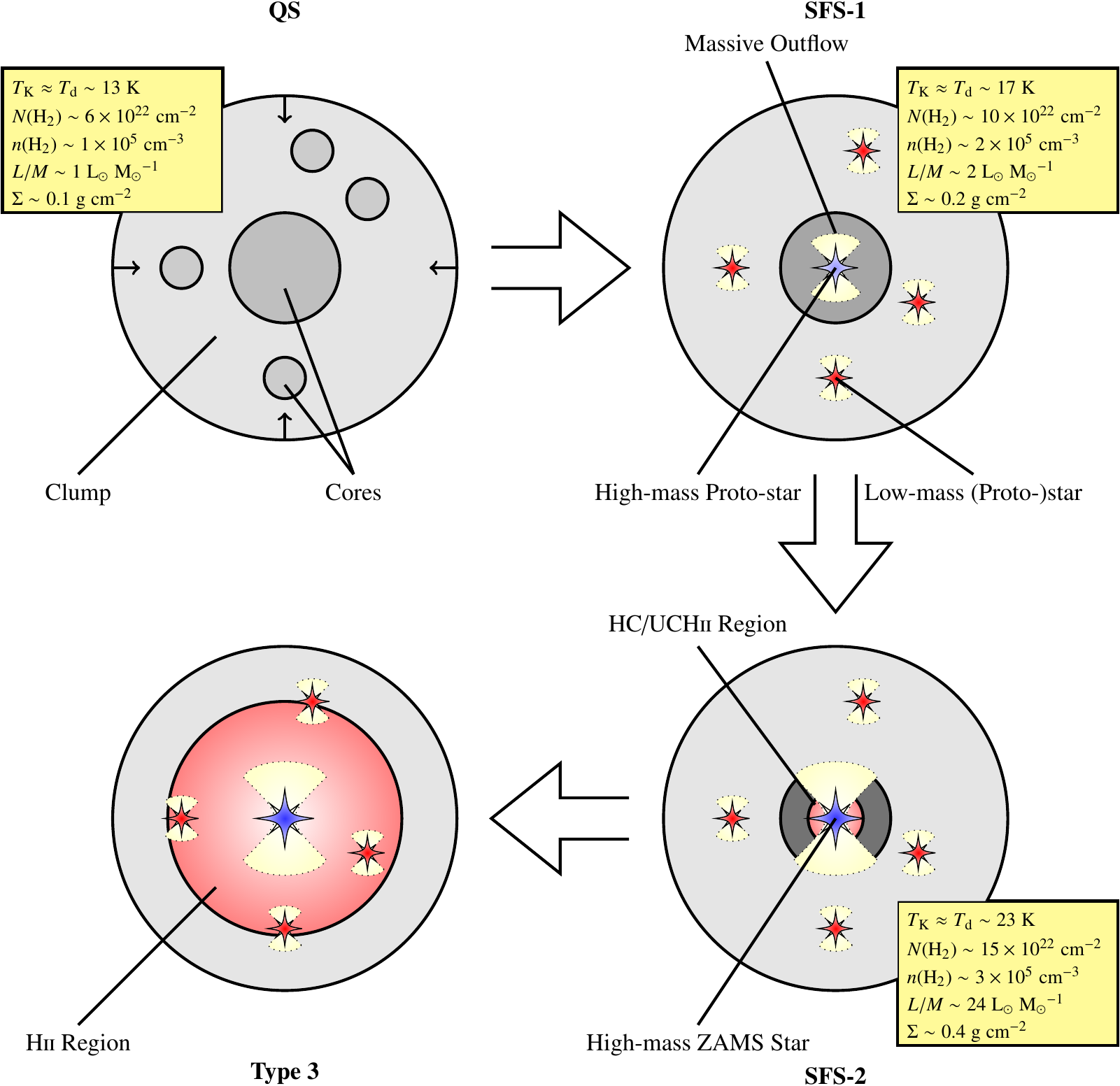}
\caption{A simple sketch of the evolutionary phases considered in this paper for massive star formation. The representative properties of clumps in the different evolutionary stages as derived in this work are listed in the yellow rectangles.}
\label{fig:sketch_ev_ph}
\end{figure*}

Figure~\ref{fig:sketch_ev_ph} shows a sketch of the different evolutionary phases identified in this work, with representative values for the observed properties indicated in the yellow rectangles. The arrows connecting the different source types show how the evolution proceeds. In addition to the properties listed in the figure, we find that the SFS-2 have smaller FWHM diameters, that can be due to an intrinsically smaller size and/or just an observational effect caused by the presence of a temperature gradient generated by the heating of the embedded massive (proto-)stars (see Sect.~\ref{sssec:sizes}). The density profile seems to be steeper in the SFS-2 clumps than in the QS, even when allowing for a temperature gradient in the former sub-sample. Thus the central density may indeed be higher for clumps with a similar size and mass. More detailed studies are needed to confirm this result.

Mean and median values for various parameters of the QS and SFS, and of the SFS-1 and SFS-2 are listed in Tables~\ref{tab:mean_sfs_qs_fwhm} and \ref{tab:mean_zams_nozams}. 

\section{Summary and conclusions} \label{sec:summary}

From ATCA NH$_3$ observations of 46 clumps previously observed with the SEST in the $1.2$-mm continuum we derived the average properties of the gas for a sample of 36 of these, detected in both NH$_3$(1,1) and (2,2). 
With a reliable and independent temperature estimate through the NH$_3$ (1,1) and (2,2) line ratio, we determined the mass of the gas.

We performed the simplest virial analysis to investigate the stability of the clumps against gravity. All sources, but one, show a virial parameter $\alpha\lesssim2$ (Fig.~\ref{fig:virial}), showing that gravity is the dominating force. The most massive clumps typically show $\alpha < 1$, and we showed that this is likely to be real (see Sect.~\ref{ssec:alpha}). The role of the magnetic field in stabilizing the sources against gravitational collapse is thus a major one. The required strength of the field was estimated to be $|\overrightarrow{B}|\approx 0.1-1\usk\milli\mathrm{G}$, in agreement with the sparse measurements in regions of high-mass star formation (Sect.~\ref{ssec:alpha}).

We find ammonia abundances to be in the range $\pot{-7}-\pot{-9}$, but within the canonical values of $\sim\pot{-7}-\pot{-8}$ for the vast majority of the sample, showing that this molecule is not depleted, as opposed to CO in these clumps \citep{Fontani+12}.

These data were complemented with Herschel/Hi-GAL, Spitzer/MIPSGAL, MSX and Spitzer/IRAC data, to construct the SEDs of the sources (Sect.~\ref{ssec:sed}). 
From the SEDs we derived the luminosity of 32 sources (i.e. those with Herschel/Hi-GAL data), out of which we have 29 sources with a reliable ammonia detection in both lines, so that we could locate the clumps in a $M-L$ plot (Fig.~\ref{fig:m-l_plot}).
The sample was divided into sub-samples of clumps in different phases of evolution on the basis of the presence or absence of signs of ongoing star formation and on the location of the sources in the $M-L$ plot. To classify a clump as star-forming, we considered the presence of at least one of the following tracers: $24\mum$ emission, $1.3\cm$ continuum emission, water masers, and ``green fuzzies'' (excess emission at $4.5\mum$). The star-forming sub-sample (SFS) includes these sources, while the quiescent sub-sample (QS) contains those without detectable signs of ongoing star formation.

We used the $M-r$ relation found by \citet{KauffmannPillai10} to assess if also the clumps without clear signs of ongoing massive star formation in our sample are potentially able to form high-mass stars. Virtually all sources lie above of the empirical relation, confirming that our sample is a good one to study evolution in the first stages of the formation of high-mass stars.

We explored if and how the average properties of a clump depend on the presence of active star formation and on its evolutionary phase (Sect.~\ref{sec:discussion}). The information from the $M-L$ plot and that on the presence of ongoing star formation are complementary: we find that Type 1 sources include both the star forming clumps with a low luminosity, well below the ZAMS loci in the $M-L$ plot, and the clumps in the QS, while Type 2 sources are the clumps in the SFS hosting a ZAMS star. For our sample, a convenient criterion based on $L$ was enough to separate Type 1 and Type 2 objects (cf. Fig.~\ref{fig:m-l_plot}): sources with $L>\pot{3}\lsun$ are  likely to host a ZAMS star. This idea is corroborated by the large fraction of these sources that are encompassed by the two determinations of the ZAMS locus \citep[][]{Molinari+08, Urquhart+13}, and show radio-continuum and strong mid-IR emission, i.e. an UC\hii\ region. Therefore we define the following classes of objects using both the information obtained from the $M-L$ plot and from classical signposts of star formation: QS for quiescent Type 1 sources, SFS-1 for Type 1 sources with signs of active star formation, and SFS-2 for Type 2 sources.
A sketch of these phases with the typical values for the physical parameters derived in this work is shown in Fig.~\ref{fig:sketch_ev_ph}.
Analyzing the typical properties of the clumps in our sample we find that they depend on the evolutionary phase of the source. The differences found can be summarized as follows:

\begin{itemize}
 \item SFS-2 sources always show radio continuum or strong mid-IR emission, suggesting the presence of an \hii\ region, while QS and SFS-1 objects do not.
 \item The average temperatures (both kinetic and dust) of the three evolutionary classes slowly increases from the QS, to SFS-1, to SFS-2 sources, with typical values of $\sim13\kel$, $17\kel$ and $\sim23\kel$, respectively. 
 \item The temperature of the clumps appears to be correlated with the luminosity of the source (Fig.~\ref{fig:tk_l}).
 \item SFS-2 objects have smaller FWHM diameters (median values of $0.5$ \textit{vs.} $0.8\pc$ for SFS-2 and QS, respectively), due to the presence of strong and compact peaks of emission at mm wavelengths. This could be caused by the presence of a temperature gradient in the SFS-2. 
 \item As a consequence, clumps classified as SFS-2 on average have higher volume-, column- and surface-densities inside the FWHM intensity contour of the $1.2\mm$ continuum emission.
 \item Assuming that density (for all clumps) and temperature (for SFS-2 clumps) both vary as power laws as a function of clump radius, we derived the power law indices for density, and found them to be steeper in more evolved sources. Typical power law indices for the molecular hydrogen volume density are $p\sim1.2-1.4$ for QS and $\sim1.5-1.8$ for SFS-2 sources. These results indicate that more evolved sources are indeed denser and more centrally concentrated. 
 \item The fact that SFS-2 sources are the most extreme in terms of compactness suggests that QS sources are still contracting.
\end{itemize}

\begin{acknowledgements}
The Australia Telescope is funded by the Commonwealth of Australia for operation as a National Facility managed by CSIRO. This work is based in part on observations made with the Spitzer Space Telescope, which is operated by the Jet Propulsion Laboratory, California Institute of Technology under a contract with NASA. This research made use of data products from the Midcourse Space Experiment. Processing of the data was funded by the Ballistic Missile Defense Organization with additional support from NASA Office of Space Science. This research has also made use of the NASA/ IPAC Infrared Science Archive, which is operated by the Jet Propulsion Laboratory, California Institute of Technology, under contract with the National Aeronautics and Space Administration. This research made use of the NASA ADS database.
\end{acknowledgements}

\bibliographystyle{bibtex/aa}
\bibliography{bibtex/biblio.bib}

\appendix
\section{Comments on individual sources} \label{app:ind_sou}

\subsection{16061$-$5048c4 and 16435$-$4515c3}

16061$-$5048c4 is included in the SFS due to the presence of a water maser. No clear mid-IR emission can be seen in the Spitzer images, suggesting a clump in very early stages of evolution or a very high extinction. 

For example, \citet{BreenEllingsen11} suggest that a maser is one of the first sign of active star formation to appear. This is because the maser is likely excited by the outflow, the onset of which happens in a phase where the mid-IR emission may still be too weak to be detected. This seems to be consistent with the position of this source in the M-L plot. This is, in fact, one of the objects with the lowest kinetic temperature in the SFS and with a very low luminosity. 

However, 16061$-$5048c4, in addition to the maser, shows a clear detection in the radio-continuum emission. 
Visual inspection of the radio image shows an elongated morphology of the emission at this wavelength, following closely the shape of the mm-clump, and displaced towards South-East.
We conclude that this is not a compact \hii\ region produced by the central YSO, but an ionization front generated by a nearby massive star (most likely the corresponding IRAS source), rather than by the YSO.

A similar situation is observed in 16435$-$4515c3. This clump shows no signs of active star formation, apart from radio emission. It could be that also in this case this emission comes from an ionization front outside the clump. In any case, this source was excluded from the analysis because the NH$_3$(2,2) was not detected.

\subsection{17355$-$3241c1}

As evolution proceeds, the massive ZAMS star delivers huge quantities of ionizing photons and energetic particles into the parent molecular cloud. This process disperses the clump, leaving only remnants of the original cloud, around an \hii\ region.

17355$-$3241c1 is the most evolved source in our sample, as  can be seen in the mass-luminosity plot. This is the only Type 3 source in our sample, with still a detectable $1.2\mm$ flux, but with dominant IR-emission. 

The properties of this clump appear  to confirm the evolutionary scenario: it has the lowest mass, column and surface density of the entire sample, and a high temperature and luminosity. Moreover, the L/M ratio is the highest among all the observed clumps, above $120\lsun\msun^{-1}$. The virial parameter $\alpha$ ($\sim 3.5$) is the highest. This could be due to the destructive action of the central ZAMS star, dispersing the clump from which it was formed. On the other hand, this object does not show compact radio-continuum emission. The $M-L$ plot shows that the mass of the central ZAMS star should be $M\sim6\msun$, corresponding to a B5 star, comparing its position in the $M-L$ plot with the evolutionary tracks. The output of Lyman continuum photons of such a star is low, and below the detection limit of the radio observations \citep[cf. ][]{Thompson84, Sanchez-Monge+13}. However the uncertainty on the stellar mass is quite large. The SED fit with Robitaille models suggests a stellar mass of $\sim8\msun$. Another possibility to explain the lack of radio continuum emission is that it is extended and filtered out by the interferometer. Finally, we remind that in both cases the stellar mass is an upper limit, under the assumption that the luminosity is dominated by the most massive object.

The SED of this object is different from that typical of the other objects with appreciable emission in the mid-IR, being almost flat in terms of energy up to $3.6\mum$. Also the Robitaille model indicates that this is an older object.

\subsection{Undetected sources}

A number of sources do not have detections in ammonia, especially in the (2,2) transition. Among the SFS, they are weak also in (1,1), suggesting low beam-averaged NH$_3$ column densities.
16164$-$4929c2 and 14166$-$6118c1 have a clear detection at $1.2\mm$, but no ammonia counterpart. NIR images show  that these two objects contain a star cluster. The SED, the morphology of the IR emission and radio-continuum observations of these two clumps show that they host a massive ZAMS star, that has already developed an \hii\ region, different from the nearby clumps detected in ammonia. 13563$-$6109c1 also shows signs of the presence of a recognizable embedded cluster in NIR images, even if not as clear as 16164$-$4929c2 and 14166$-$6118c1. 

On the other hand, four more sources with similar properties are found in the QS, but without clear signs of an embedded cluster. Two of these have quite strong NH$_3$(1,1) detections, indicating very low gas temperatures. These clumps might be just too cold to be detected in (2,2) or simply not massive enough to form stars with $M>8\msun$.

Finally, also 15454$-$5335c2 was not detected in ammonia. This source was identified at the edge of the SEST map. No IR source was found to be associated with it. The non-detection in NH$_3$ and visual inspection of the SEST map suggest that this might be just a noise spike in the SEST $1.2\mm$ continuum map.

\section{SED fit}\label{sec:app_SEDfit}

The modified black-body fit was done with a simple Bayesian approach, considering Gaussian uncertainties on the fluxes, taking into account the rms of the image and the calibration uncertainty at each wavelength \citep[$15\%$ for SEST and SPIRE, $20\%$ for PACS red and $10\%$ for PACS blue fluxes: ][]{Beltran+06, Swinyard+10, Poglitsch+10}. We adopted a modified black body as a model, comparing the observed and predicted fluxes at each wavelength, and multiplying the probability for each point, to obtain the total probability of the model. We took a constant \textit{prior} on the mass, while we used a Gaussian \textit{prior} on $\beta$, with mean value $\mu=2$ and standard deviation $\sigma=2$, and on the dust temperature $\td$, with $\mu=20\kel$ and $\sigma=15\kel$. We considered temperatures between $5$ and $50\kel$ in 200 equal linear steps, masses between $10$ and $10000\msun$ in 200 equal logarithmic steps, and $\beta$ between $0.1$ and $4$, in 40 equal linear steps. To test the dependence of the fit on the choice of \textit{prior}, we also tried to use a constant \textit{prior} on all the parameters of the fit. The results show that $\td$, $M$ and $\beta$ are not sensitive to this choice, within the uncertainties.
To derive the probability distribution, and thus the uncertainty on the single parameters we integrated over the other two parameters. 

\section{Tables and figures}\label{sec:figtab}

\setlength{\tabcolsep}{4pt}
\begin{table*}[tbp]
\centering
\caption{Properties of each clump, averaged within the FWHM contour. The columns show respectively: clump name, mass, volume- and column-density of H$_2$, mass surface density, with their $68\%$ credibility interval, and the beam-corrected diameters of the clumps. The clumps above the horizontal line are those classified as star-forming, while the clumps below it are those classified as quiescent.}
\label{tab:fwhm}
\tiny
\begin{tabular}{l*{10}c}
\toprule
Clump        & $M$        & $68\%$ int.        & $\nhtwo$       & $68\%$ int.         & $\coldhtwo$       & $68\%$ int.          & $\Sigma$ & $68\%$ int.        & Diameter  & Notes \\ 
\midrule
             & \multicolumn{2}{c}{\scriptsize ($\pot{2}\times\msuntab$)}          & \multicolumn{2}{c}{\scriptsize ($\pot{4}\times\cm^{-3}$)} & \multicolumn{2}{c}{\scriptsize ($\pot{22}\times\cm^{-2}$)} & \multicolumn{2}{c}{\scriptsize ($\pot{-1}\times$g$\cm^{-2}$)} & {\scriptsize ($\pc$)} & \\
\midrule
08477$-$4359c1 &$       1.1$&$    0.7-       1.2$&$         26.0$&$     17.8-   30.5$&$             9.0$&$       6.1-   10.5$&$    2.29$&$    1.57-    2.69$ &$       0.3$&       \\ 
12300$-$6119c1 &$       1.4$&$    1.1-       1.7$&$         10.5$&$      7.9-   12.4$&$             5.4$&$       4.1-    6.4$&$    1.39$&$    1.04-    1.64$ &$       0.5$&       \\ 
13560$-$6133c1 &$       4.5$&$    2.5-       5.7$&$         41.5$&$     22.7-   52.5$&$            19.9$&$      10.9-   25.1$&$    5.09$&$    2.79-    6.43$ &$       0.5$&       \\ 
15072$-$5855c1 &$       0.6$&$    0.3-       0.8$&$         10.8$&$      5.5-   13.5$&$             4.2$&$       2.1-    5.3$&$    1.07$&$    0.54-    1.35$ &$       0.4$&       \\ 
15278$-$5620c1 &$       5.9$&$    4.7-       6.9$&$        187.0$&$    150.0-  220.0$&$            59.0$&$      47.4-   69.7$&$   15.10$&$   12.10-   17.80$ &$       0.3$&       \\ 
15278$-$5620c2 &$       1.7$&$    1.3-       2.0$&$          5.1$&$      4.0-    6.1$&$             3.5$&$       2.8-    4.2$&$    0.90$&$    0.71-    1.08$ &$       0.7$&       \\ 
15470$-$5419c1 &$       2.5$&$    1.2-       3.1$&$         22.4$&$     10.3-   27.3$&$            10.8$&$       5.0-   13.2$&$    2.77$&$    1.27-    3.38$ &$       0.5$&       \\ 
15470$-$5419c3 &$       5.0$&$    4.1-       5.8$&$          4.0$&$      3.3-    4.6$&$             4.3$&$       3.6-    5.0$&$    1.11$&$    0.91-    1.28$ &$       1.1$&       \\ 
15557$-$5215c1 &$       7.3$&$    3.1-       9.0$&$         20.9$&$      8.7-   25.5$&$            14.8$&$       6.1-   18.0$&$    3.78$&$    1.57-    4.62$ &$       0.7$&       \\ 
15557$-$5215c2 &$       3.7$&$    2.9-       4.4$&$         13.0$&$     10.1-   15.5$&$             8.5$&$       6.7-   10.2$&$    2.19$&$    1.70-    2.61$ &$       0.7$&       \\ 
15579$-$5303c1 &$       2.7$&$    2.3-       3.2$&$         38.0$&$     31.7-   43.8$&$            15.8$&$      13.2-   18.3$&$    4.06$&$    3.39-    4.68$ &$       0.4$&       \\ 
16061$-$5048c1 &$       2.1$&$    1.6-       2.6$&$         36.5$&$     26.7-   44.5$&$            14.2$&$      10.4-   17.3$&$    3.63$&$    2.66-    4.43$ &$       0.4$&       \\ 
16061$-$5048c2 &$       2.9$&$    2.4-       3.3$&$         22.3$&$     18.5-   26.1$&$            11.3$&$       9.3-   13.2$&$    2.89$&$    2.39-    3.37$ &$       0.5$&       \\ 
16061$-$5048c4 &$       3.6$&$    2.9-       4.3$&$          5.6$&$      4.6-    6.6$&$             4.9$&$       4.0-    5.8$&$    1.25$&$    1.01-    1.47$ &$       0.9$&       \\ 
16093$-$5015c1 &$      22.1$&$   17.0-      25.5$&$          2.9$&$      2.2-    3.3$&$             5.7$&$       4.4-    6.6$&$    1.46$&$    1.13-    1.69$ &$       2.0$&       \\ 
16093$-$5128c1 &$       5.6$&$    4.5-       6.6$&$         13.5$&$     10.8-   15.9$&$            10.1$&$       8.0-   11.8$&$    2.58$&$    2.06-    3.03$ &$       0.8$&       \\ 
16093$-$5128c8 &$       8.5$&$    2.2-      10.0$&$          1.9$&$      0.5-    2.2$&$             3.1$&$       0.8-    3.7$&$    0.81$&$    0.21-    0.95$ &$       1.7$&       \\ 
16254$-$4844c1 &$       1.3$&$    0.9-       1.7$&$         43.2$&$     30.9-   55.8$&$            13.6$&$       9.7-   17.5$&$    3.47$&$    2.49-    4.49$ &$       0.3$& (1)   \\ 
16428$-$4109c1 &$       1.9$&$    1.5-       2.2$&$         13.8$&$     10.9-   15.6$&$             7.2$&$       5.6-    8.1$&$    1.83$&$    1.44-    2.07$ &$       0.5$&       \\ 
16428$-$4109c2 &$       1.5$&$    1.3-       1.7$&$          8.3$&$      7.1-    9.5$&$             4.7$&$       4.0-    5.4$&$    1.21$&$    1.03-    1.38$ &$       0.6$&       \\ 
16573$-$4214c2 &$       1.1$&$    0.8-       1.3$&$        174.0$&$    128.0-  208.0$&$            32.2$&$      23.6-   38.4$&$    8.25$&$    6.04-    9.84$ &$       0.2$& (1)   \\ 
17040$-$3959c1 &$       0.4$&$    0.3-       0.5$&$         47.5$&$     38.0-   56.0$&$             9.7$&$       7.7-   11.4$&$    2.47$&$    1.98-    2.91$ &$       0.2$& (2)   \\ 
17195$-$3811c1 &$       2.3$&$    1.1-       2.7$&$         10.9$&$      5.1-   13.2$&$             6.5$&$       3.0-    7.8$&$    1.65$&$    0.78-    2.00$ &$       0.6$&       \\ 
17195$-$3811c2 &$       2.9$&$    2.4-       3.3$&$         14.0$&$     11.7-   15.7$&$             8.3$&$       6.9-    9.3$&$    2.13$&$    1.78-    2.38$ &$       0.6$&       \\ 
17195$-$3811c3 &$       3.7$&$    3.0-       4.2$&$          5.0$&$      4.0-    5.7$&$             4.5$&$       3.7-    5.2$&$    1.15$&$    0.94-    1.33$ &$       0.9$&       \\ 
17355$-$3241c1 &$      0.15$&$    0.1-       0.2$&$          3.3$&$      2.5-    3.9$&$             1.2$&$       0.9-    1.5$&$    0.31$&$    0.24-    0.37$ &$       0.4$&       \\ 
\midrule                                                                                                                                                                              
13560$-$6133c2 &$       4.5$&$    1.4-       5.5$&$         20.1$&$      6.4-   24.8$&$            12.2$&$       3.9-   15.1$&$    3.13$&$    1.00-    3.86$ &$       0.6$&       \\ 
14166$-$6118c2 &$       1.1$&$    0.7-       1.3$&$          7.6$&$      5.3-    9.1$&$             4.0$&$       2.8-    4.8$&$    1.01$&$    0.71-    1.22$ &$       0.5$&       \\ 
14183$-$6050c3 &$       4.2$&$    1.8-       4.7$&$          5.2$&$      2.2-    5.9$&$             4.9$&$       2.1-    5.5$&$    1.25$&$    0.53-    1.40$ &$       1.0$&       \\ 
15038$-$5828c1 &$       1.6$&$    1.2-       1.9$&$         24.9$&$     18.7-   29.8$&$            10.0$&$       7.5-   12.0$&$    2.56$&$    1.92-    3.07$ &$       0.4$&       \\ 
15470$-$5419c4 &$       4.8$&$    2.6-       6.1$&$          4.3$&$      2.4-    5.4$&$             4.5$&$       2.5-    5.7$&$    1.14$&$    0.63-    1.45$ &$       1.1$&       \\ 
15557$-$5215c3 &$       2.1$&$    1.6-       2.5$&$          1.6$&$      1.2-    2.0$&$             1.8$&$       1.4-    2.1$&$    0.45$&$    0.34-    0.55$ &$       1.1$&       \\ 
15579$-$5303c3 &$       5.1$&$    2.9-       6.3$&$         10.2$&$      5.8-   12.7$&$             8.1$&$       4.6-   10.1$&$    2.08$&$    1.18-    2.57$ &$       1.1$&       \\ 
16093$-$5128c2 &$       3.6$&$    2.4-       4.4$&$          6.6$&$      4.5-    8.1$&$             5.4$&$       3.7-    6.6$&$    1.37$&$    0.93-    1.69$ &$       0.8$&       \\ 
16164$-$4929c3 &$       3.2$&$    2.1-       4.0$&$          7.3$&$      4.7-    9.1$&$             5.6$&$       3.6-    6.9$&$    1.43$&$    0.92-    1.78$ &$       0.8$&       \\ 
16482$-$4443c2 &$       0.4$&$    0.2-       0.5$&$         47.4$&$     19.6-   57.0$&$            10.0$&$       4.1-   12.0$&$    2.56$&$    1.06-    3.08$ &$      <0.2$& (3)   \\ 
\bottomrule                                                                          
\end{tabular}
     \tablefoot{
     (1) Source with very poor uv-coverage in NH$_3$: spectrum extracted from non-cleaned maps.\\
     (2) Lower limit for the mass and the derived quantities.\\
     (3) Volume-, column- and surface-densities are lower limits, as the source is unresolved.
     }
\end{table*}

\begin{table*}[tbp]
\centering
\caption{Properties of each clump, averaged within the $3\sigma$ contour. The columns show respectively: clump name, mass, volume- and column-density of H$_2$, mass surface density, with their $68\%$ credibility interval, and the diameters of the clumps. The clumps above the horizontal line are those classified as star-forming, while the clumps below it are those classified as quiescent.}
\label{tab:3s}
\tiny
\begin{tabular}{l*{10}c}
\toprule
Clump        & $M$             & $68\%$ int.     & $\nhtwo$               & $68\%$ int.          & $\coldhtwo$  & $68\%$ int.     & $\Sigma$ & $68\%$ int.     & Diameter & Notes \\
\midrule
             & \multicolumn{2}{c}{\scriptsize ($\pot{2}\times\msuntab$)}          & \multicolumn{2}{c}{\scriptsize ($\pot{4}\times\cm^{-3}$)} & \multicolumn{2}{c}{\scriptsize ($\pot{22}\times\cm^{-2}$)} & \multicolumn{2}{c}{\scriptsize ($\pot{-1}\times$g$\cm^{-2}$)} & {\scriptsize ($\pc$)} & \\
\midrule
08477$-$4359c1    &$	       2.3  $&$       1.7    -       2.6  $&$	       4.1  $&$       3.1    -       4.6  $&$	       3.4  $&$       2.5    -       3.8  $&$	      0.87  $&$      0.65    -      0.98  $&$	       0.8 $&       \\ 
12300$-$6119c1    &$	       2.7  $&$       2.0    -       3.2  $&$	       1.8  $&$       1.4    -       2.1  $&$	       2.1  $&$       1.6    -       2.5  $&$	      0.53  $&$      0.40    -      0.63  $&$	       1.2 $&       \\ 
13560$-$6133c1    &$	      12.0  $&$       8.2    -      16.0  $&$	       1.7  $&$       1.1    -       2.1  $&$	       3.3  $&$       2.2    -       4.2  $&$	      0.85  $&$      0.56    -      1.10  $&$	       2.0 $&       \\ 
15072$-$5855c1    &$	       1.4  $&$       0.8    -       1.8  $&$	       2.0  $&$       1.0    -       2.5  $&$	       1.8  $&$       0.9    -       2.3  $&$	      0.46  $&$      0.24    -      0.58  $&$	       0.9 $&       \\ 
15278$-$5620c1    &$	      14.0  $&$      12.0    -      17.0  $&$	       2.8  $&$       2.3    -       3.3  $&$	       4.9  $&$       4.0    -       5.6  $&$	      1.20  $&$      1.00    -      1.40  $&$	       1.8 $&       \\ 
15278$-$5620c2    &$	       3.0  $&$       2.6    -       3.3  $&$	       1.6  $&$       1.4    -       1.8  $&$	       2.0  $&$       1.8    -       2.2  $&$	      0.51  $&$      0.45    -      0.57  $&$	       1.3 $&       \\ 
15470$-$5419c1    &$	       5.0  $&$       2.5    -       6.0  $&$	       3.0  $&$       1.5    -       3.6  $&$	       3.6  $&$       1.8    -       4.3  $&$	      0.91  $&$      0.45    -      1.10  $&$	       1.2 $&       \\ 
15470$-$5419c3    &$	      11.0  $&$      10.0    -      11.0  $&$	       1.0  $&$       1.0    -       1.1  $&$	       2.2  $&$       2.2    -       2.4  $&$	      0.57  $&$      0.55    -      0.61  $&$	       2.2 $&       \\ 
15557$-$5215c1    &$	      21.0  $&$      10.0    -      26.0  $&$	       2.4  $&$       1.2    -       2.9  $&$	       4.9  $&$       2.5    -       6.1  $&$	      1.30  $&$      0.63    -      1.60  $&$	       2.1 $&       \\ 
15557$-$5215c2    &$	       8.4  $&$       7.5    -       9.5  $&$	       1.4  $&$       1.2    -       1.6  $&$	       2.5  $&$       2.3    -       2.9  $&$	      0.65  $&$      0.58    -      0.73  $&$	       1.9 $&       \\ 
15579$-$5303c1    &$	       8.1  $&$       7.8    -       8.7  $&$	       1.2  $&$       1.2    -       1.3  $&$	       2.3  $&$       2.2    -       2.5  $&$	      0.60  $&$      0.57    -      0.63  $&$	       1.9 $&       \\ 
16061$-$5048c1    &$	       5.2  $&$       4.3    -       6.2  $&$	       1.7  $&$       1.4    -       2.0  $&$	       2.5  $&$       2.1    -       3.0  $&$	      0.64  $&$      0.53    -      0.76  $&$	       1.5 $&       \\ 
16061$-$5048c2    &$	       5.7  $&$       4.9    -       6.7  $&$	       2.8  $&$       2.4    -       3.3  $&$	       3.6  $&$       3.1    -       4.2  $&$	      0.92  $&$      0.79    -      1.10  $&$	       1.3 $&       \\ 
16061$-$5048c4    &$	       6.3  $&$       5.7    -       6.7  $&$	       1.9  $&$       1.7    -       2.0  $&$	       2.8  $&$       2.5    -       3.0  $&$	      0.71  $&$      0.65    -      0.77  $&$	       1.5 $&       \\ 
16093$-$5015c1    &$	      54.2  $&$      48.2    -      59.1  $&$	       0.4  $&$       0.3    -       0.5  $&$	       2.0  $&$       1.8    -       2.2  $&$	      0.51  $&$      0.45    -      0.56  $&$	       5.3 $&       \\ 
16093$-$5128c1    &$	      12.0  $&$      11.0    -      14.0  $&$	       2.1  $&$       1.9    -       2.3  $&$	       3.8  $&$       3.4    -       4.2  $&$	      0.97  $&$      0.86    -      1.10  $&$	       1.9 $&       \\ 
16093$-$5128c8    &$	      10.0  $&$       3.4    -      12.0  $&$	       1.2  $&$       0.4    -       1.4  $&$	       2.5  $&$       0.8    -       3.0  $&$	      0.65  $&$      0.21    -      0.76  $&$	       2.1 $&       \\ 
16254$-$4844c1    &$	       2.3  $&$       1.9    -       3.0  $&$	       2.3  $&$       1.9    -       3.0  $&$	       2.3  $&$       1.9    -       3.0  $&$	      0.60  $&$      0.48    -      0.77  $&$	       1.0 $& (1)   \\ 
16428$-$4109c1    &$	       8.1  $&$       6.6    -       9.3  $&$	       1.1  $&$       0.9    -       1.2  $&$	       2.1  $&$       1.7    -       2.4  $&$	      0.53  $&$      0.43    -      0.62  $&$	       2.0 $&       \\ 
16428$-$4109c2    &$	       4.8  $&$       4.2    -       5.6  $&$	       0.8  $&$       0.7    -       1.0  $&$	       1.5  $&$       1.3    -       1.8  $&$	      0.39  $&$      0.34    -      0.45  $&$	       1.8 $&       \\ 
16573$-$4214c2    &$	       1.7  $&$       1.4    -       2.0  $&$	       8.8  $&$       7.2    -      10.0  $&$	       5.1  $&$       4.2    -       6.0  $&$	      1.30  $&$      1.10    -      1.50  $&$	       0.6 $& (1)   \\ 
17040$-$3959c1    &$	       0.6  $&$       0.6    -       0.7  $&$	       6.3  $&$       5.7    -       6.9  $&$	       3.0  $&$       2.6    -       3.2  $&$	      0.75  $&$      0.68    -      0.83  $&$	       0.5 $& (2)   \\ 
17195$-$3811c1    &$	       6.5  $&$       3.6    -       7.8  $&$	       1.6  $&$       0.9    -       1.9  $&$	       2.6  $&$       1.4    -       3.1  $&$	      0.66  $&$      0.36    -      0.80  $&$	       1.6 $&       \\ 
17195$-$3811c2    &$	       8.3  $&$       8.0    -       8.5  $&$	       2.1  $&$       2.0    -       2.1  $&$	       3.3  $&$       3.2    -       3.4  $&$	      0.85  $&$      0.81    -      0.86  $&$	       1.6 $&       \\ 
17195$-$3811c3    &$	       6.3  $&$       5.5    -       6.8  $&$	       1.8  $&$       1.6    -       2.0  $&$	       2.8  $&$       2.4    -       3.0  $&$	      0.71  $&$      0.63    -      0.77  $&$	       1.5 $&       \\ 
17355$-$3241c1    &$	       0.3  $&$       0.2    -       0.3  $&$	       1.7  $&$       1.4    -       1.9  $&$	       0.9  $&$       0.8    -       1.0  $&$	      0.24  $&$      0.20    -      0.27  $&$	       0.6 $&       \\ 
\midrule                                                                                                                                                                                                                             
13560$-$6133c2    &$	       8.1  $&$       2.9    -       9.1  $&$	       3.5  $&$       1.2    -       3.9  $&$	       4.6  $&$       1.7    -       5.2  $&$	      1.20  $&$      0.43    -      1.30  $&$	       1.4 $&       \\ 
14166$-$6118c2    &$	       1.4  $&$       1.1    -       1.6  $&$	       5.1  $&$       3.9    -       6.0  $&$	       3.3  $&$       2.5    -       3.9  $&$	      0.85  $&$      0.65    -      0.99  $&$	       0.7 $&       \\ 
14183$-$6050c3    &$	       5.4  $&$       2.3    -       5.9  $&$	       3.2  $&$       1.4    -       3.6  $&$	       3.9  $&$       1.6    -       4.2  $&$	      0.99  $&$      0.41    -      1.10  $&$	       1.2 $&       \\ 
15038$-$5828c1    &$	       4.6  $&$       3.8    -       5.3  $&$	       2.2  $&$       1.9    -       2.5  $&$	       2.9  $&$       2.4    -       3.2  $&$	      0.73  $&$      0.61    -      0.83  $&$	       1.3 $&       \\ 
15470$-$5419c4    &$	       6.2  $&$       3.7    -       7.7  $&$	       2.6  $&$       1.6    -       3.2  $&$	       3.5  $&$       2.1    -       4.3  $&$	      0.89  $&$      0.53    -      1.10  $&$	       1.4 $&       \\ 
15557$-$5215c3    &$	       2.0  $&$       1.7    -       2.4  $&$	       1.6  $&$       1.3    -       1.8  $&$	       1.7  $&$       1.4    -       2.0  $&$	      0.44  $&$      0.37    -      0.51  $&$	       1.1 $&       \\ 
15579$-$5303c3    &$	      10.0  $&$       6.1    -      12.0  $&$	       1.4  $&$       0.9    -       1.7  $&$	       2.8  $&$       1.7    -       3.3  $&$	      0.71  $&$      0.42    -      0.86  $&$	       2.0 $&       \\ 
16093$-$5128c2    &$	       5.1  $&$       3.7    -       5.9  $&$	       2.7  $&$       2.0    -       3.1  $&$	       3.3  $&$       2.5    -       3.9  $&$	      0.86  $&$      0.63    -      0.99  $&$	       1.3 $&       \\ 
16164$-$4929c3    &$	       5.4  $&$       3.4    -       6.8  $&$	       2.5  $&$       1.6    -       3.2  $&$	       3.3  $&$       2.0    -       4.1  $&$	      0.84  $&$      0.52    -      1.05  $&$	       1.3 $&       \\ 
16482$-$4443c2    &$	       0.8  $&$       0.3    -       1.0  $&$	       7.8  $&$       3.1    -       9.3  $&$	       3.7  $&$       1.5    -       4.4  $&$	      0.94  $&$      0.38    -      1.10  $&$	       0.5 $&       \\ 
\bottomrule                                                                                                                                                                        
\end{tabular}                                                                                                                                                                
     \tablefoot{
     (1) Source with very poor uv-coverage in NH$_3$: spectrum extracted from non-cleaned maps.\\
     (2) Lower limit for the mass and the derived quantities.
     }
\end{table*}

\begin{table*}[tbp]
\centering
\caption{Summary of star formation signposts for all the observed clumps. The clumps above the horizontal line are those classified as star-forming, while the clump below it are those classified as quiescent. For the ``green fuzzies'', the notation Y+ indicates multiple ``green fuzzies'' associated with the clump. In the last column we show the Type of the source (see text).}
\label{tab:sf_signs}
\tiny
\begin{tabular}{lcccccc}
\toprule
Clump        & $24\mum$ Emission & MSX Emission   & GF     & $1.3\cm$-Continuum & H$_2$O Maser & Type \\
\midrule
             &                   &                &        &       &                    &       \\
\midrule
08477$-$4359c1 &                 Y &             Y? &  \dots             & N     & N                  & \dots \\ 
08589$-$4714c1 &                 Y &              Y &  \dots             & N     & Y                  & \dots \\ 
12300$-$6119c1 &             \dots &              Y &  \dots             & N     & N                  & \dots \\ 
13039$-$6331c1 &                 Y &              Y &     N              & N     & N                  & \dots \\       
13560$-$6133c1 &                 Y &              N &     N              & Y     & Y                  & 2     \\ 
13563$-$6109c1 &                 Y &              Y &     N              & Y?    & N                  & \dots \\ 
14166$-$6118c1 &                 Y &              Y &     N              & Y     & N                  & \dots \\ 
15072$-$5855c1 &                 Y &              Y &     Y              & N     & N                  & 2     \\ 
15278$-$5620c1 &                 Y &              Y &     Y              & Y     & Y                  & 2     \\ 
15278$-$5620c2 &                 Y &              N &     Y+             & N     & N                  & 1     \\ 
15470$-$5419c1 &                 N &              N &     Y              & N     & Y                  & 1     \\ 
15470$-$5419c3 &                 Y &              N &     Y+             & N     & Y                  & 1     \\ 
15557$-$5215c1 &                 Y &              Y &     Y              & N     & Y\tablefootmark{c} & 2     \\ 
15557$-$5215c2 &                 Y &              N &     Y              & N     & Y                  & 1     \\ 
15579$-$5303c1 &                 Y &              Y &     Y+             & Y     & Y                  & 2     \\ 
16061$-$5048c1 &                 Y &              N &     Y+             & Y     & Y                  & 2     \\ 
16061$-$5048c2 &                 Y &              Y &     Y+             & Y     & Y\tablefootmark{c} & 2     \\ 
16061$-$5048c4 &                 N &              N &     N              & Y     & Y                  & 1     \\ 
16093$-$5015c1 &                 Y &              Y &     Y+             & N     & N\tablefootmark{d} & 1     \\ 
16093$-$5128c1 &                 Y &              Y &     N              & Y     & N                  & 2     \\ 
16093$-$5128c8 &                 N &              N & Y\tablefootmark{b} & N     & N                  & 1     \\ 
16164$-$4929c2 &                 Y &              Y &     N              & Y     & N                  & \dots \\     
16254$-$4844c1 &                 Y &              N &     Y              & Y     & N\tablefootmark{d} & 1     \\ 
16428$-$4109c1 &             \dots &              Y &  \dots             & Y     & N                  & \dots \\ 
16428$-$4109c2 &             \dots &              Y &  \dots             & Y     & N                  & \dots \\ 
16573$-$4214c2 &                 Y &              N &     Y              & N     & Y\tablefootmark{a} & 1     \\ 
17040$-$3959c1 &                 Y &              N &     N              & N     & N                  & 1     \\ 
17195$-$3811c1 &                 Y &              Y &     Y              & N     & Y\tablefootmark{a} & 2     \\ 
17195$-$3811c2 &                 Y &              N &     Y              & Y?    & Y\tablefootmark{a} & \dots \\ 
17195$-$3811c3 &                 Y &              N &     N              & N     & Y\tablefootmark{a} & \dots \\ 
17355$-$3241c1 &                 Y &              Y &     N              & N     & N                  & 3     \\ 
\midrule                                                               
10088$-$5730c2 &             \dots &              N &  \dots             & N     & N                  & \dots \\ 
13560$-$6133c2 &                 N &              N &     N              & N     & N                  & 1     \\ 
14166$-$6118c2 &                 N &              N &     N              & N     & N                  & 1     \\ 
14183$-$6050c3 &                 N &              N &     N              & N     & N                  & 1     \\ 
15038$-$5828c1 &                 N &              N &     N              & N     & N                  & 1     \\ 
15454$-$5335c2 &                 N &              N &     N              & \dots & N                  & \dots \\      
15470$-$5419c4 &                 N &              N &     N              & N     & N                  & 1     \\ 
15557$-$5215c3 &                 N &              N &     N              & \dots & N                  & 1     \\ 
15579$-$5303c3 &                 N &              N &     N              & \dots & N                  & 1     \\ 
16093$-$5128c2 &                 N &              N &     N              & N     & N                  & 1     \\ 
16164$-$4929c3 &                 N &              N &     N              & N     & N                  & 1     \\ 
16164$-$4929c6 &                 N &              N &     N              & \dots & N                  & \dots \\ 
16164$-$4837c2 &             \dots &              N &     N              & \dots & N                  & \dots \\ 
16435$-$4515c3 &                 N &              N &     N              & Y?    & N                  & \dots \\ 
16482$-$4443c2 &                 N &              N &     N              & \dots & N                  & 1     \\ 
\bottomrule                                                                          
\end{tabular}
\tablefoot{
\tablefoottext{a}{These sources have poor uv-coverage, thus a firm assignment to a clump in the region was not possible.}
\tablefoottext{b}{Visual inspection of the Mid-IR images showed that this GF could be generated by the Spitzer PSF. However, we still consider the clump as star forming to be conservative, and because of the presence of a strong $24\mum$ source within the $3\sigma$ contour of the $1.2\mm$ continuum emission, at the location of the GF.}
\tablefoottext{c}{Source detected in our work, but not in that of \citet{Sanchez-Monge+13}.}
\tablefoottext{d}{Source detected in the work of \citet{Sanchez-Monge+13}, but not in our work.}
}
\end{table*}

\begin{landscape}

\setlength{\tabcolsep}{4pt}

\begin{table}[tbp]
\centering
\tiny
\caption{Mean and median values for temperatures, luminosity, dust emissivity index $\beta$, $L/M$ ratio, gas properties and diameters of the clumps in the QS and the SFS averaged within the FWHM contour of the $1.2\mm$ continuum emission. The pairs of columns show respectively: kinetic temperature, dust temperature, mass, diameter, column density, volume density, surface density, luminosity, dust emissivity index and $L/M$ ratio.}
\label{tab:mean_sfs_qs_fwhm}
\begin{tabular}{lcccccccccccccccccccccccccccccc}
\toprule
Parameter    & \multicolumn{2}{c}{$\tk$}                       & \multicolumn{2}{c}{$\td$}                       & \multicolumn{2}{c}{$M$}                     & \multicolumn{2}{c}{$D$}               & \multicolumn{2}{c}{$N_\mathrm{H_2}$}           & \multicolumn{2}{c}{$n_{\mathrm{H_2}}$}        & \multicolumn{2}{c}{$\Sigma$}                        & \multicolumn{2}{c}{$L$}                        & \multicolumn{2}{c}{$\beta$}                   & \multicolumn{2}{c}{$L/M$}                       \\
             & \multicolumn{2}{c}{\scriptsize (K)}                         & \multicolumn{2}{c}{\scriptsize (K)}                         & \multicolumn{2}{c}{\scriptsize ($\msuntab$)}                & \multicolumn{2}{c}{\scriptsize ($\pctab$)}           & \multicolumn{2}{c}{\scriptsize ($\pot{22}\times\cm^{-2}$)} & \multicolumn{2}{c}{\scriptsize ($\pot{5}\times\cm^{-3}$)} & \multicolumn{2}{c}{\scriptsize ($\gram\usk\cm^{-2}$)}           & \multicolumn{2}{c}{\scriptsize ($\pot{2}\times\lsuntab$)} &                       &                       & \multicolumn{2}{c}{\scriptsize ($\lsuntab/\msuntab$)}             \\
\midrule
             & SFS                    & QS                     & SFS                    & QS                     & SFS                  & QS                   & SFS                  & QS             & SFS                    & QS                    & SFS                   & QS                    & SFS                      &  QS                      & SFS                    & QS                    & SFS                   & QS                    & SFS                     & QS                    \\
\midrule
Mean         & $19.5\unc{-2.9}{+1.5}$ & $14.1\unc{-3.2}{+1.8}$ & $19.6\unc{-1.6}{+1.4}$ & $11.8\unc{-1.5}{+1.7}$ & $370\unc{-120}{+70}$ & $310\unc{-140}{+70}$ & $0.6$                & $0.8$          & $11.2\unc{-3.2}{+2.2}$ & $5.8\unc{-2.5}{+1.3}$ & $3.0\unc{-0.8}{+0.6}$ & $0.9\unc{-0.4}{+0.2}$ & $0.29\unc{-0.08}{+0.06}$ & $0.15\unc{-0.06}{+0.03}$ & $53.6\unc{-8.2}{+7.9}$ & $2.3\unc{-0.9}{+0.6}$ & $1.9\unc{-0.2}{+0.2}$ & $2.4\unc{-0.4}{+0.3}$ & $18.9\unc{-5.6}{+13.4}$ & $0.9\unc{-0.5}{+1.0}$ \\
Median       & $18.7$                 & $14.0$                 & $19.9$                 & $11.1$                 & $260$                & $340$                & $0.5$                & $0.8$          & $8.4$                  & $4.9$                 & $1.4$                 & $0.7$                 & $0.22$                   & $0.13$                   & $15.2$                 & $2.3$                 & $1.9$                 & $2.6$                 & $6.5$                   & $0.8$                 \\
\bottomrule  
\end{tabular}
\end{table}

\begin{table}[tbp]
\centering
\caption{Mean and median values for temperatures, luminosity, $L/M$ ratio, gas properties and diameters averaged within the FWHM contour of the $1.2\mm$ continuum emission of the clumps in the SFS, divided in SFS-1 and SFS-2. The pairs of columns show respectively: kinetic temperature, dust temperature, diameter, column density, volume density, surface density, luminosity and $L/M$ ratio.}
\label{tab:mean_zams_nozams}
\tiny
\begin{tabular}{lcccccccccccccccc}
\toprule
Parameter    & \multicolumn{2}{c}{$\tk$}                       & \multicolumn{2}{c}{$\td$}                       & \multicolumn{2}{c}{$D$}               & \multicolumn{2}{c}{$N_\mathrm{H_2}$}           & \multicolumn{2}{c}{$n_{\mathrm{H_2}}$}        & \multicolumn{2}{c}{$\Sigma$}                        & \multicolumn{2}{c}{$L$}                           & \multicolumn{2}{c}{$L/M$}                        \\
             & \multicolumn{2}{c}{\scriptsize (K)}                         & \multicolumn{2}{c}{\scriptsize (K)}                         & \multicolumn{2}{c}{\scriptsize ($\pctab$)}           & \multicolumn{2}{c}{\scriptsize ($\pot{22}\times\cm^{-2}$)} & \multicolumn{2}{c}{\scriptsize ($\pot{5}\times\cm^{-3}$)} & \multicolumn{2}{c}{\scriptsize ($\gram\usk\cm^{-2}$)}           & \multicolumn{2}{c}{\scriptsize ($\pot{2}\times\lsuntab$)} & \multicolumn{2}{c}{\scriptsize ($\lsuntab/\msuntab$)}              \\
\midrule
             & SFS-2                  & SFS-1                  & SFS-2                  & SFS-1                  & SFS-2                & SFS-1          & SFS-2                  & SFS-1                 & SFS-2                 & SFS-1                 & SFS-2                    &  SFS-1                   & SFS-2                     & SFS-1                 & SFS-2                    & SFS-1                 \\
\midrule
Mean         & $21.3\unc{-3.8}{+1.7}$ & $18.3\unc{-3.0}{+1.4}$ & $23.6\unc{-1.6}{+1.5}$ & $15.2\unc{-1.5}{+1.4}$ & $0.7$                & $0.7$          & $16.1\unc{-4.6}{+3.2}$ & $10.1\unc{-3.0}{+2.1}$& $3.8\unc{-1.0}{+0.7}$ & $3.5\unc{-1.0}{+0.7}$ & $0.41\unc{-0.12}{+0.08}$ & $0.26\unc{-0.08}{+0.05}$ & $101.7\unc{-15.0}{+15.0}$ & $3.9\unc{-0.8}{+0.6}$ & $24.4\unc{-7.8}{+20.3}$  & $2.1\unc{-0.7}{+1.3}$ \\
Median       & $21.4$                 & $18.3$                 & $23.8$                 & $16.0$                 & $0.5$                & $0.7$          & $12.8$                 & $8.5$                 & $2.2$                 & $1.3$                 & $0.33$                   & $0.22$                   & $83.0$                    & $3.5$                 & $24.2$                   & $1.6$                 \\
\bottomrule  
\end{tabular}
\end{table}
\end{landscape}

\begin{table*}[p]
	 \centering 
	 \scriptsize 
	 \caption{Parameters of the ammonia spectra averaged over the whole NH$_3$(1,1) area of emission. The columns indicate the clump name, the peak flux of the (1,1) transition and the rms of the spectrum, the $\vlsr$ of the emission, the $\Delta V$ of NH$_3$(1,1), the peak flux of the (2,2) transition and the rms of the spectrum, and the $\Delta V$ of NH$_3$(2,2), $\trot$, $\tk$ and ammonia column density, with their uncertainties. The clumps above the horizontal line are those classified as star-forming, while the clump below it are those classified as quiescent (see Sect.~\ref{sec:discussion}). Note that, contrary to Table~\ref{tab:line_prop}, we do not include clumps not detected in NH$_3$(1,1).} \label{tab:line_prop_mean}
	 \includegraphics{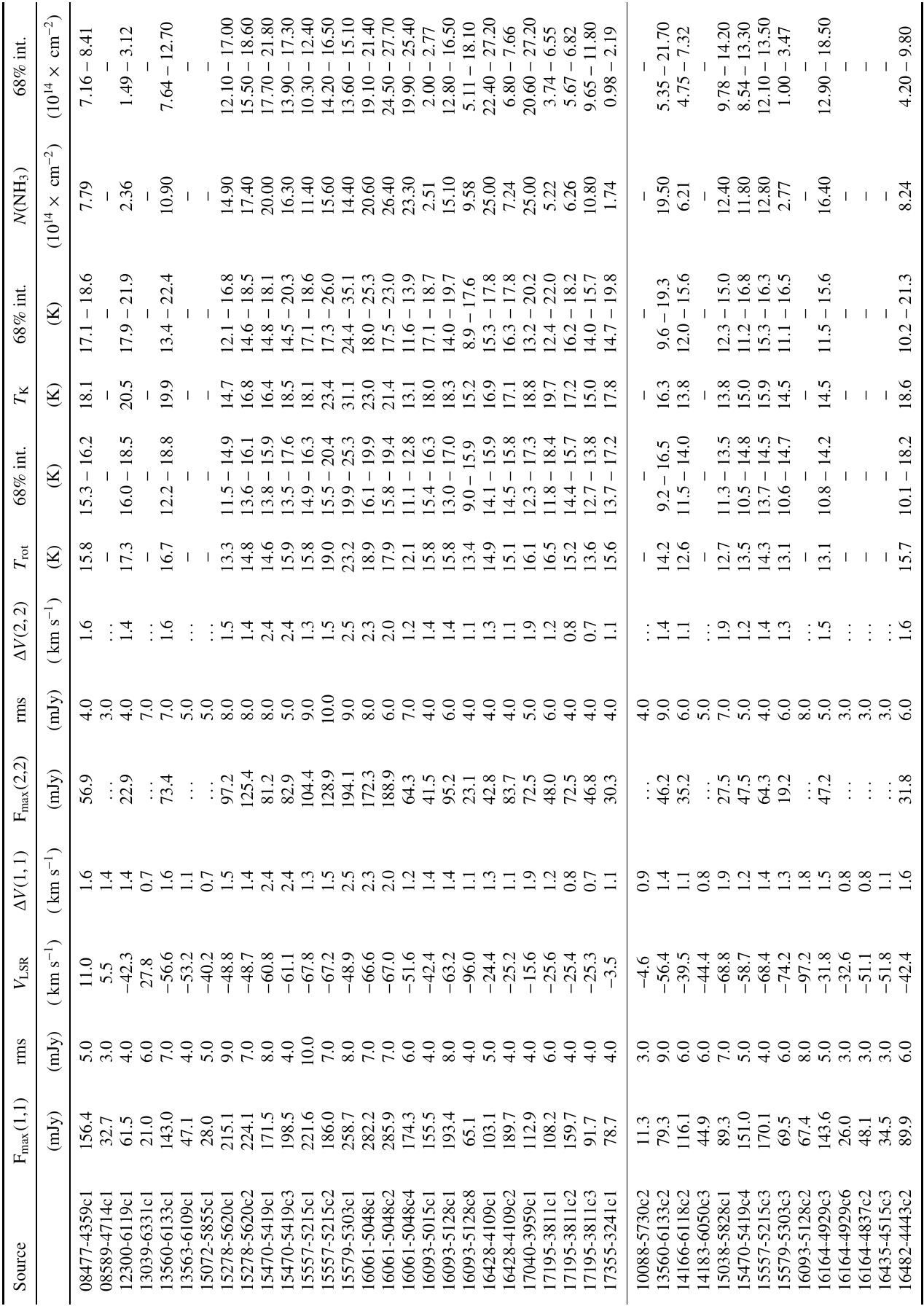}
\end{table*}

\begin{figure*}[tbp]
 \centering
 \includegraphics[angle=-90,width=0.225\textwidth]{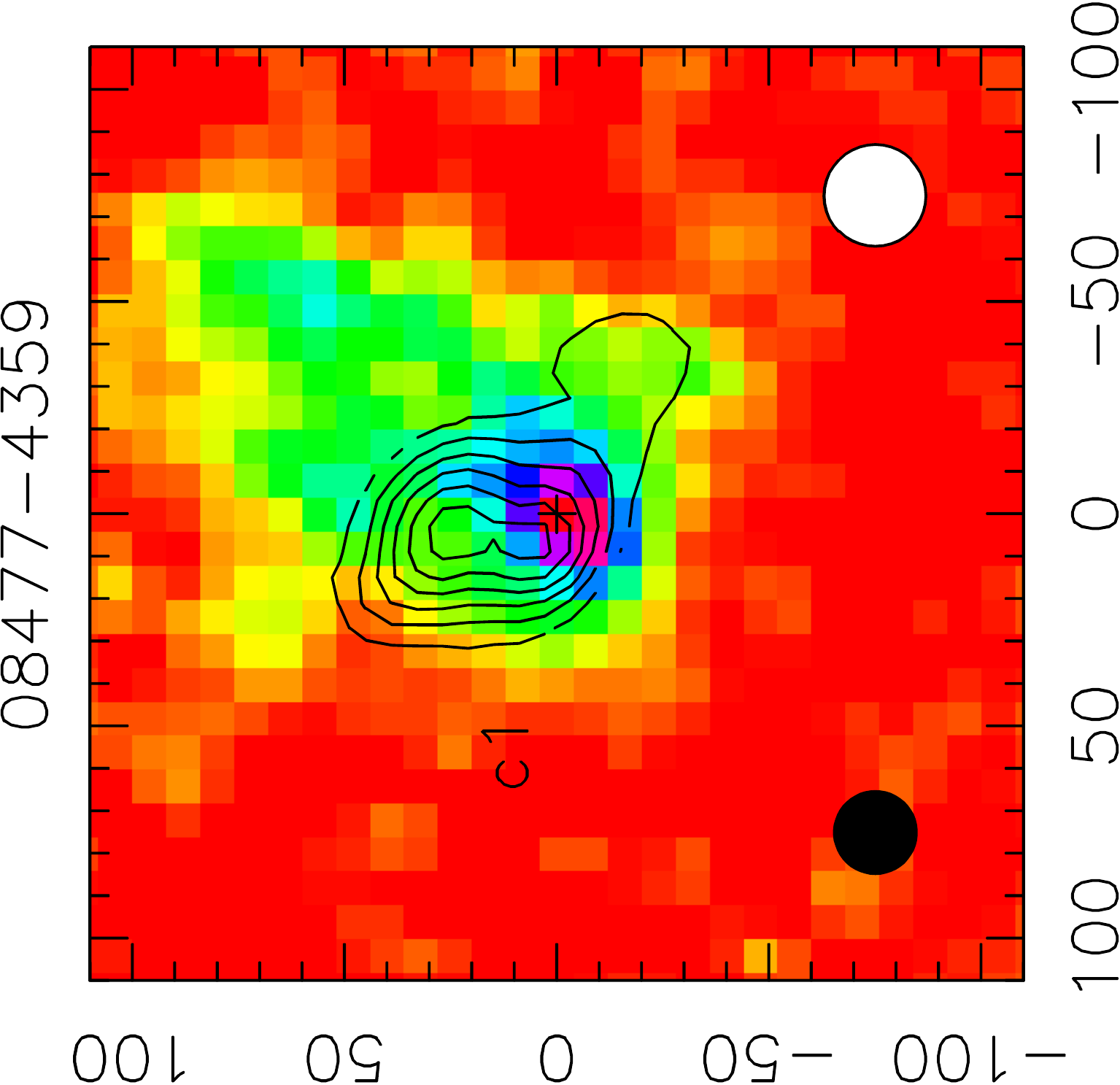}
 \includegraphics[angle=-90,width=0.225\textwidth]{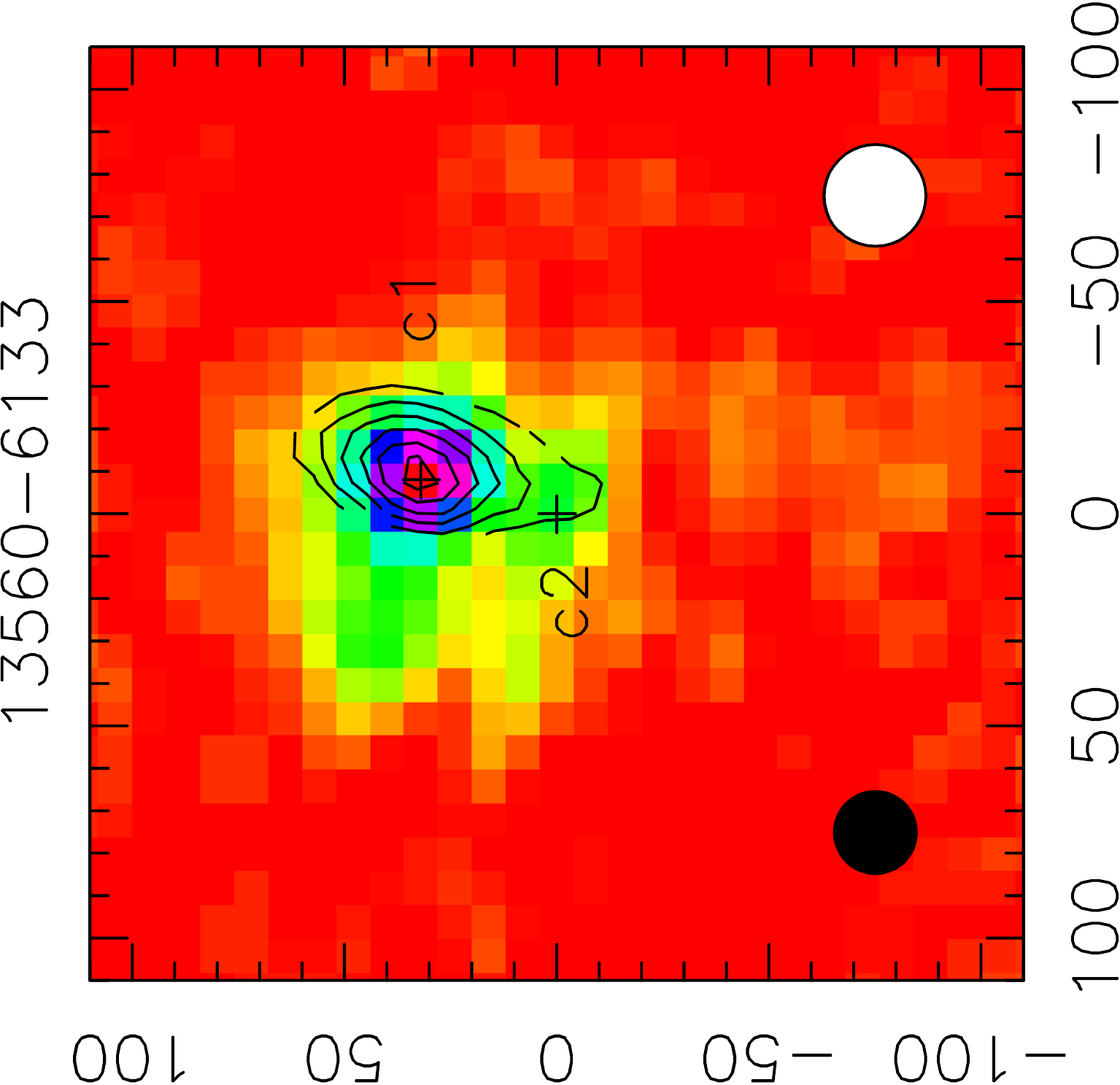}
 \includegraphics[angle=-90,width=0.225\textwidth]{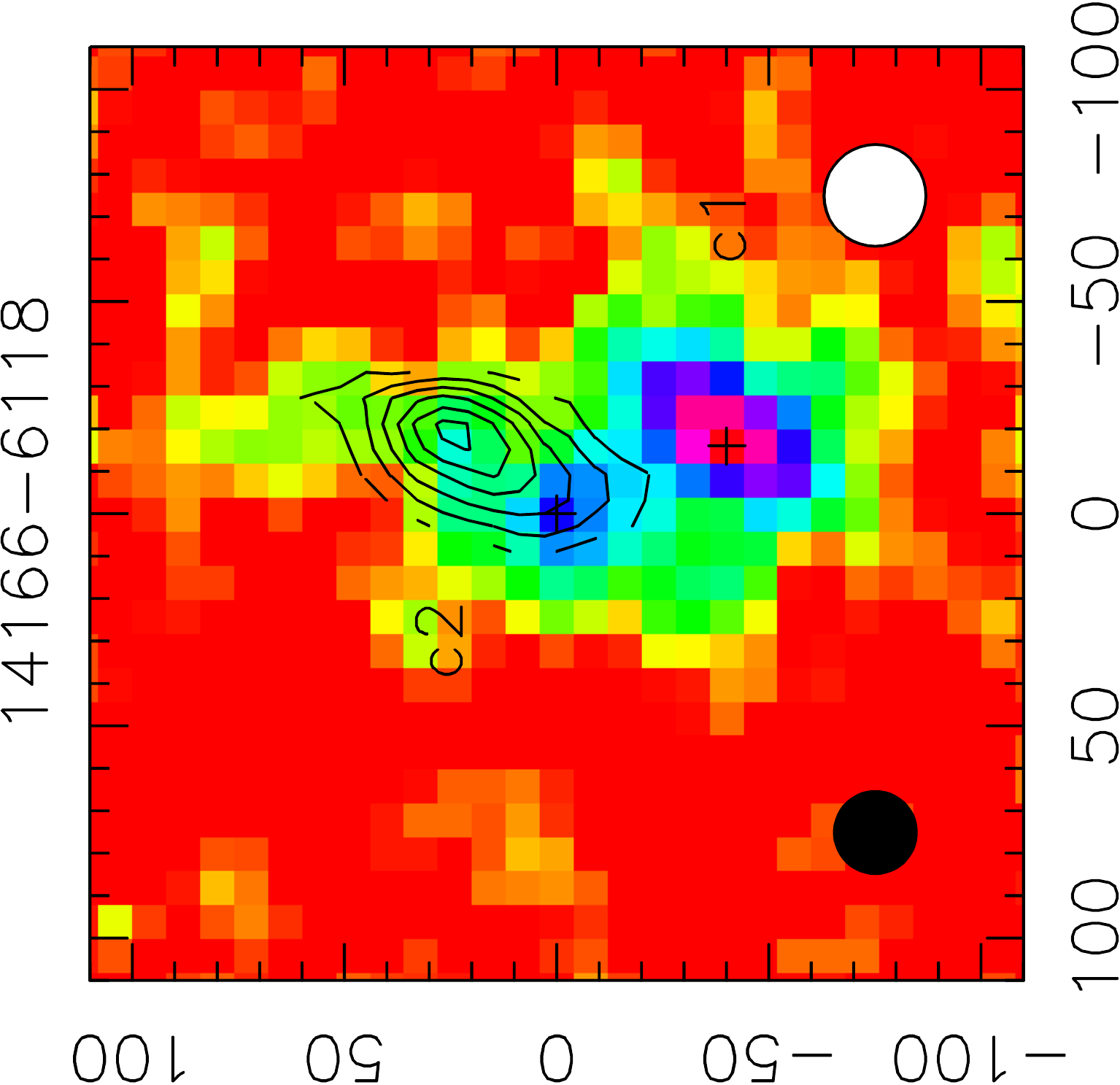}
 \includegraphics[angle=-90,width=0.225\textwidth]{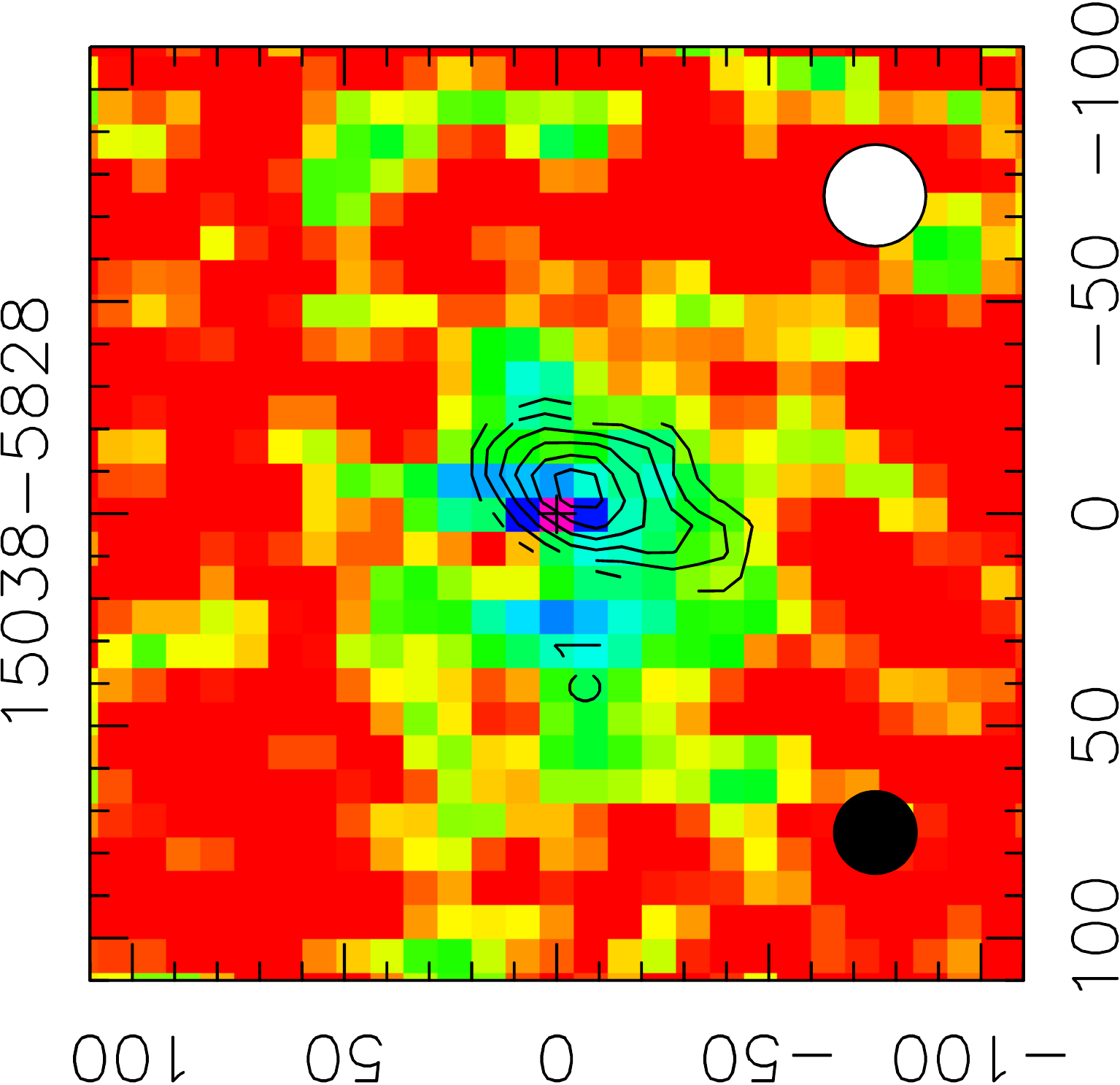}
 \includegraphics[angle=-90,width=0.225\textwidth]{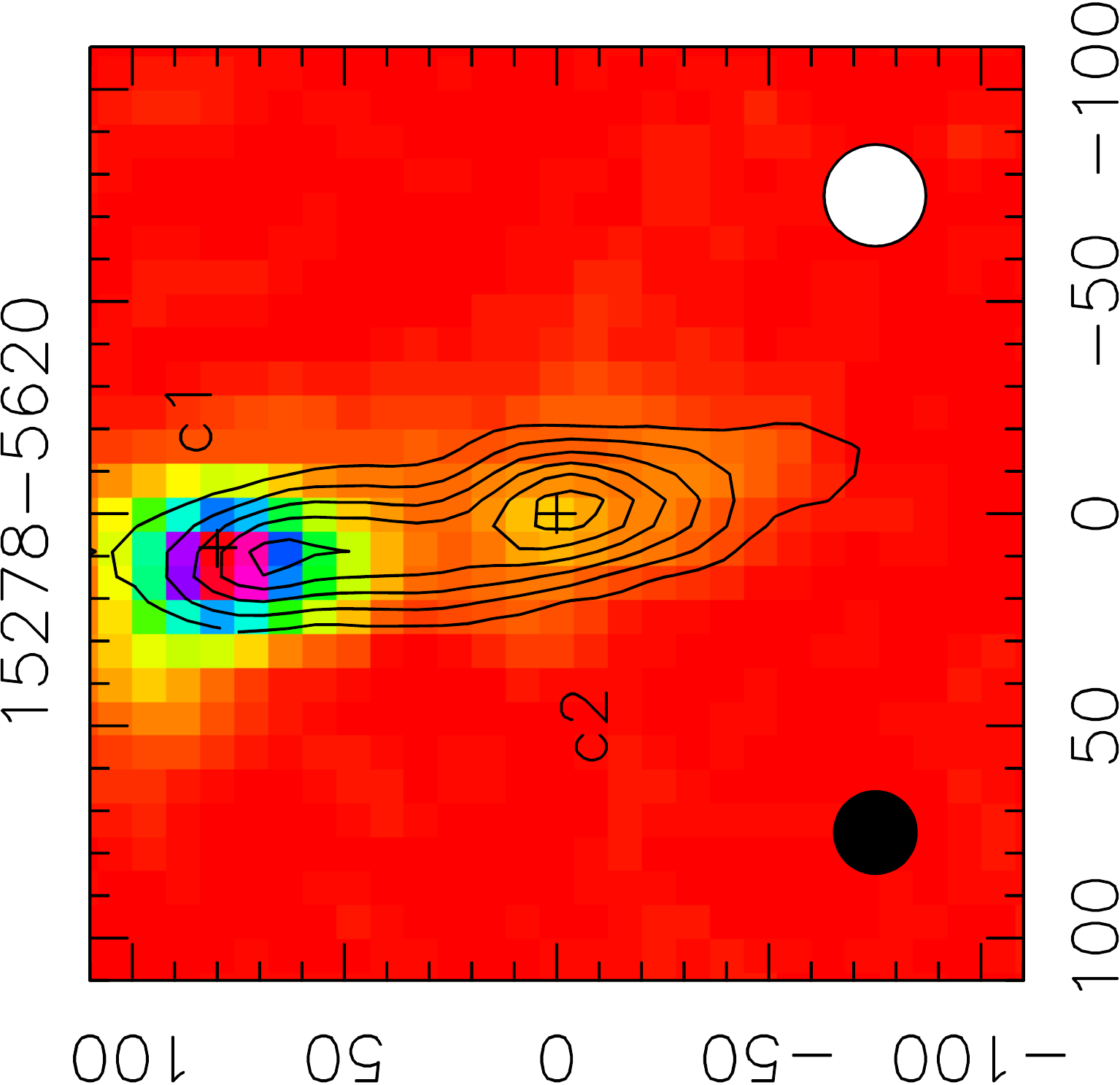}
 \includegraphics[angle=-90,width=0.225\textwidth]{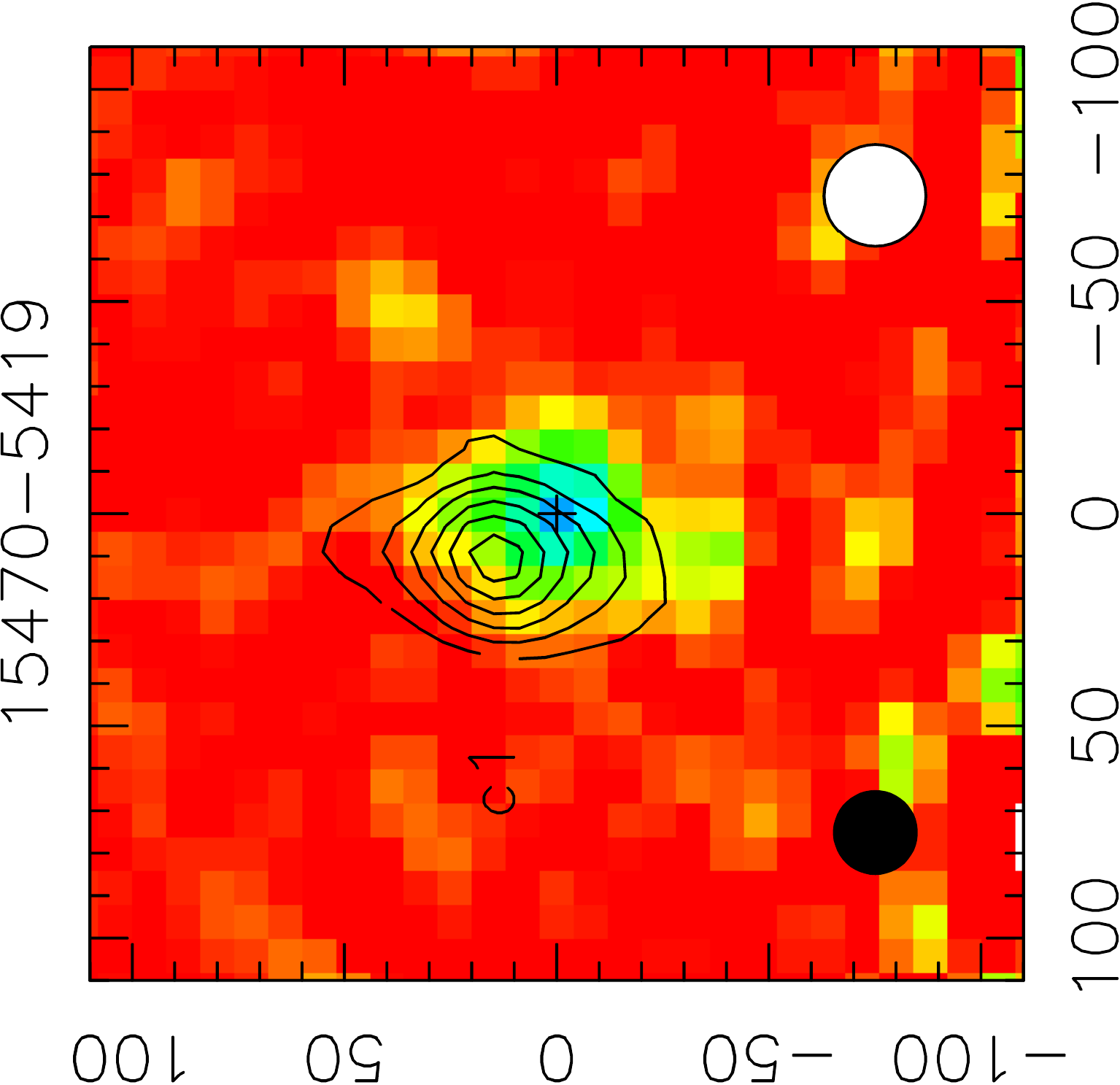}
 \includegraphics[angle=-90,width=0.225\textwidth]{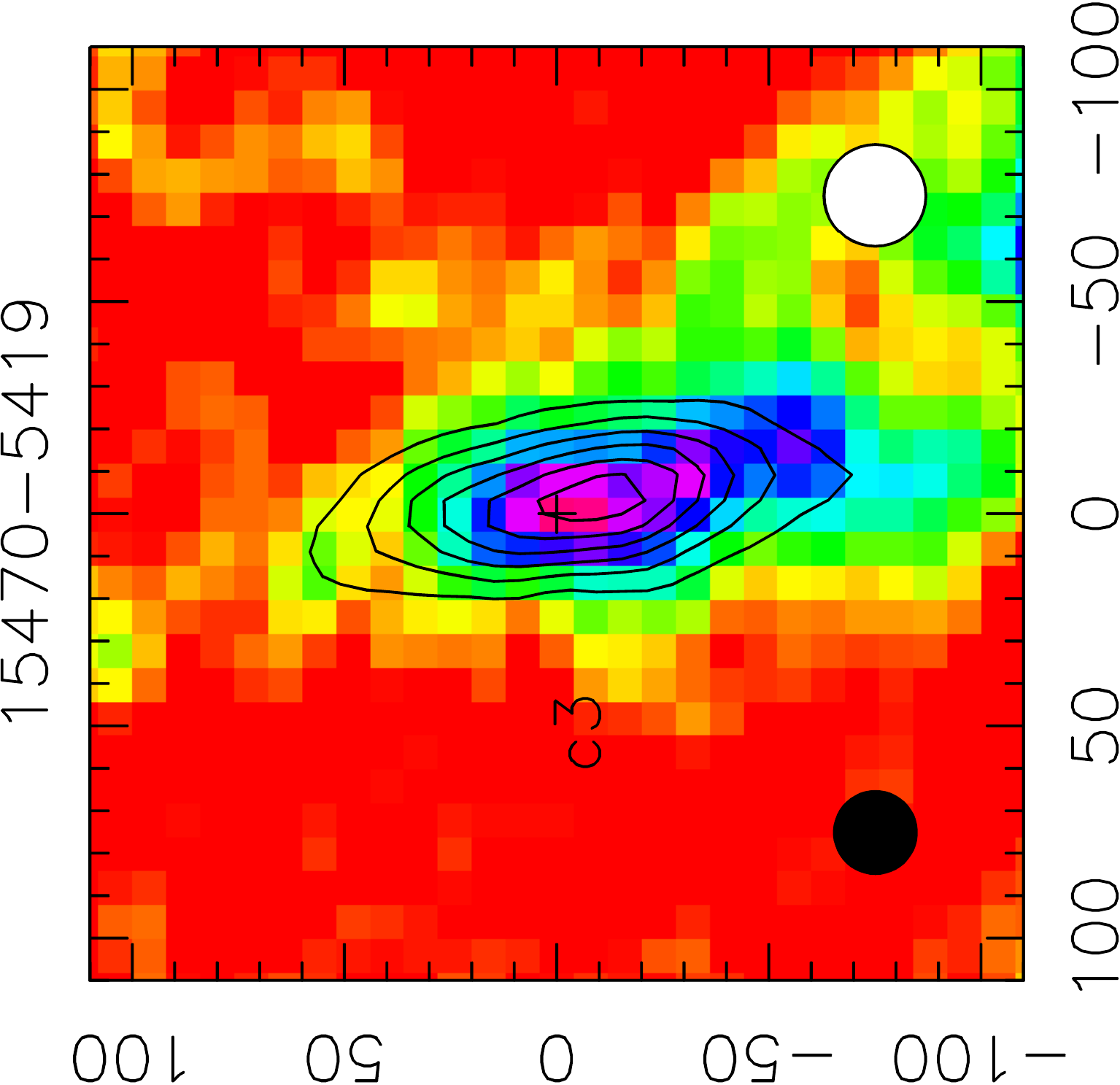}
 \includegraphics[angle=-90,width=0.225\textwidth]{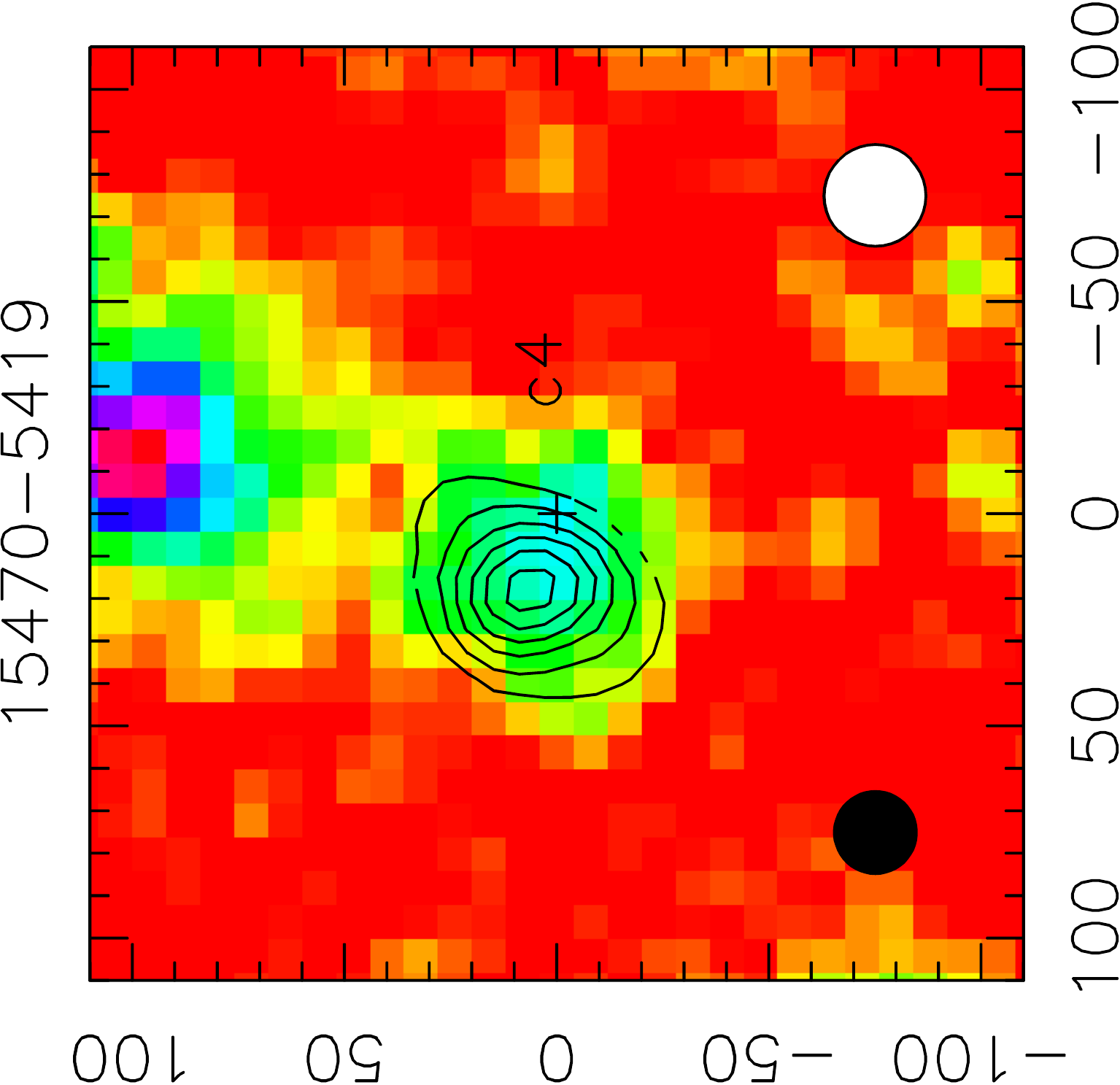}
 \includegraphics[angle=-90,width=0.225\textwidth]{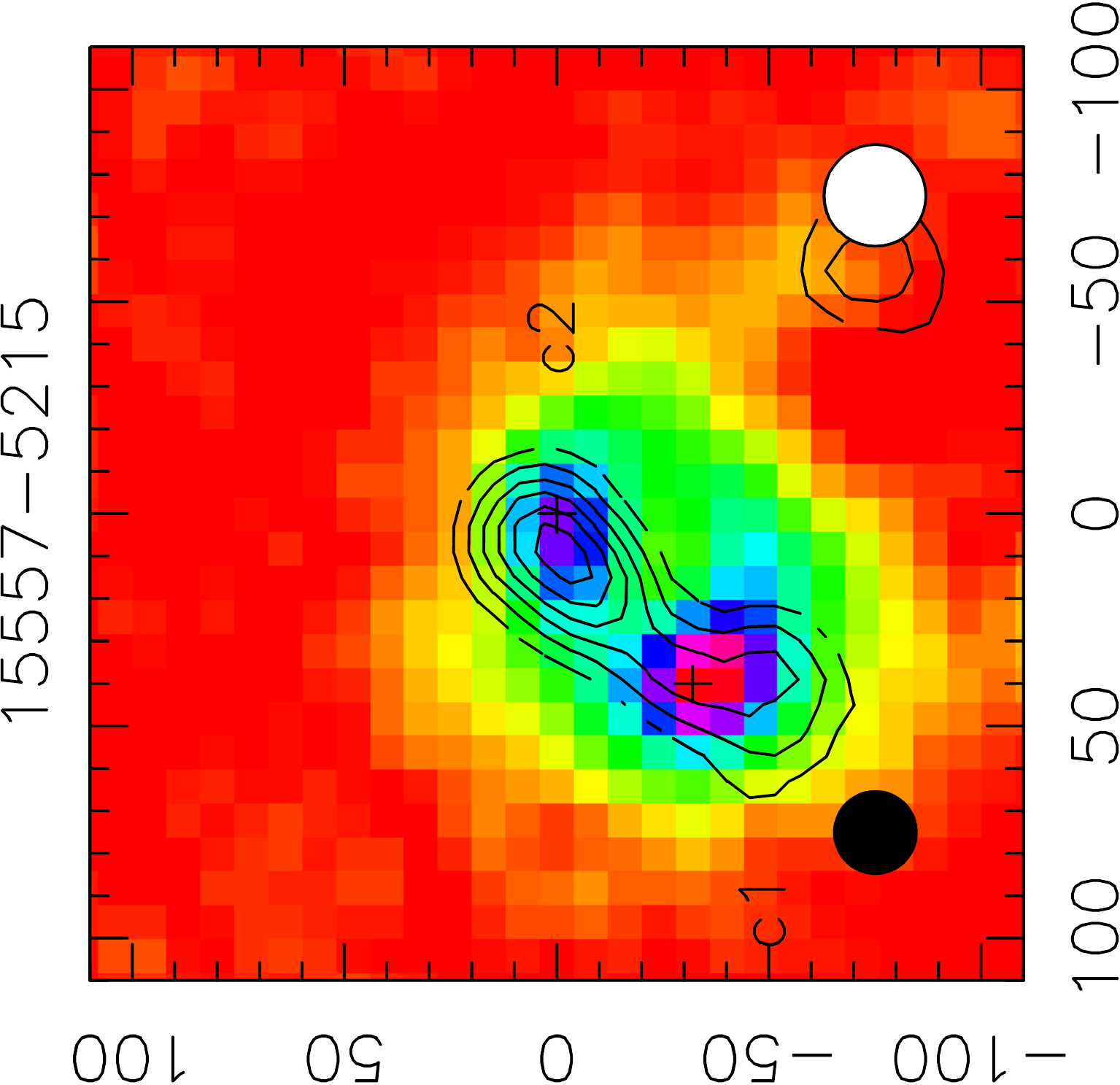}
 \includegraphics[angle=-90,width=0.225\textwidth]{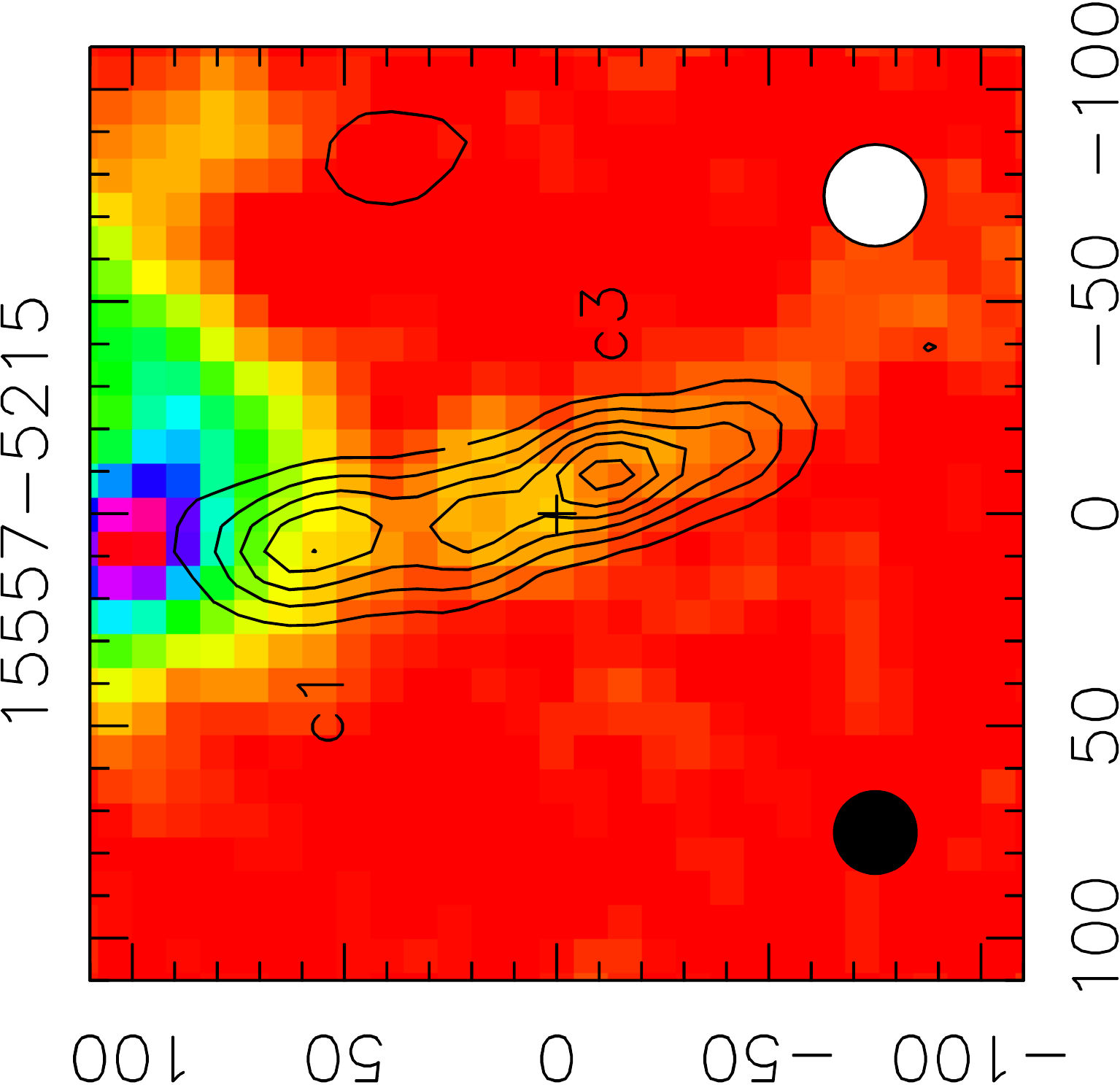}
 \includegraphics[angle=-90,width=0.225\textwidth]{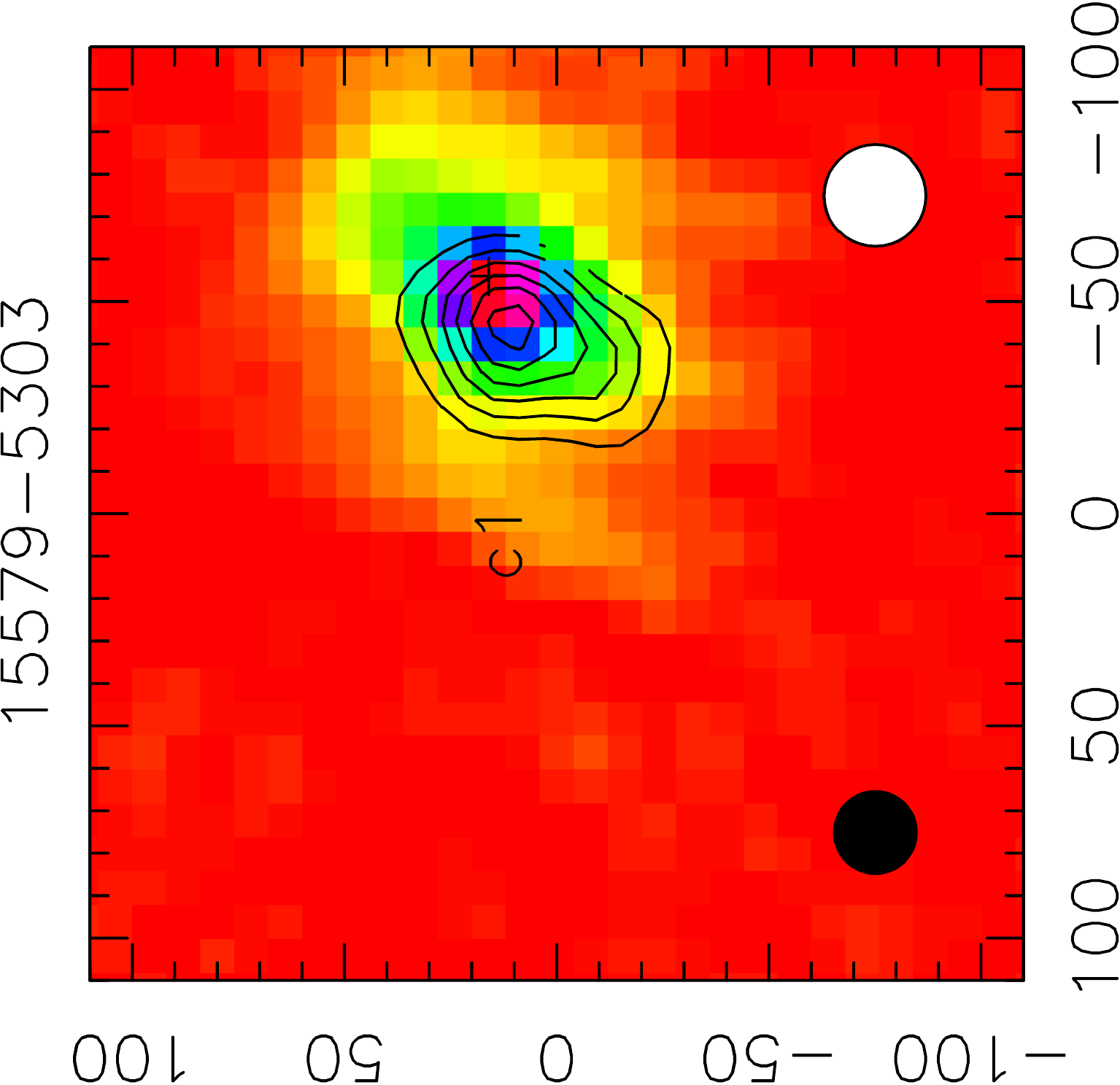}
 \includegraphics[angle=-90,width=0.225\textwidth]{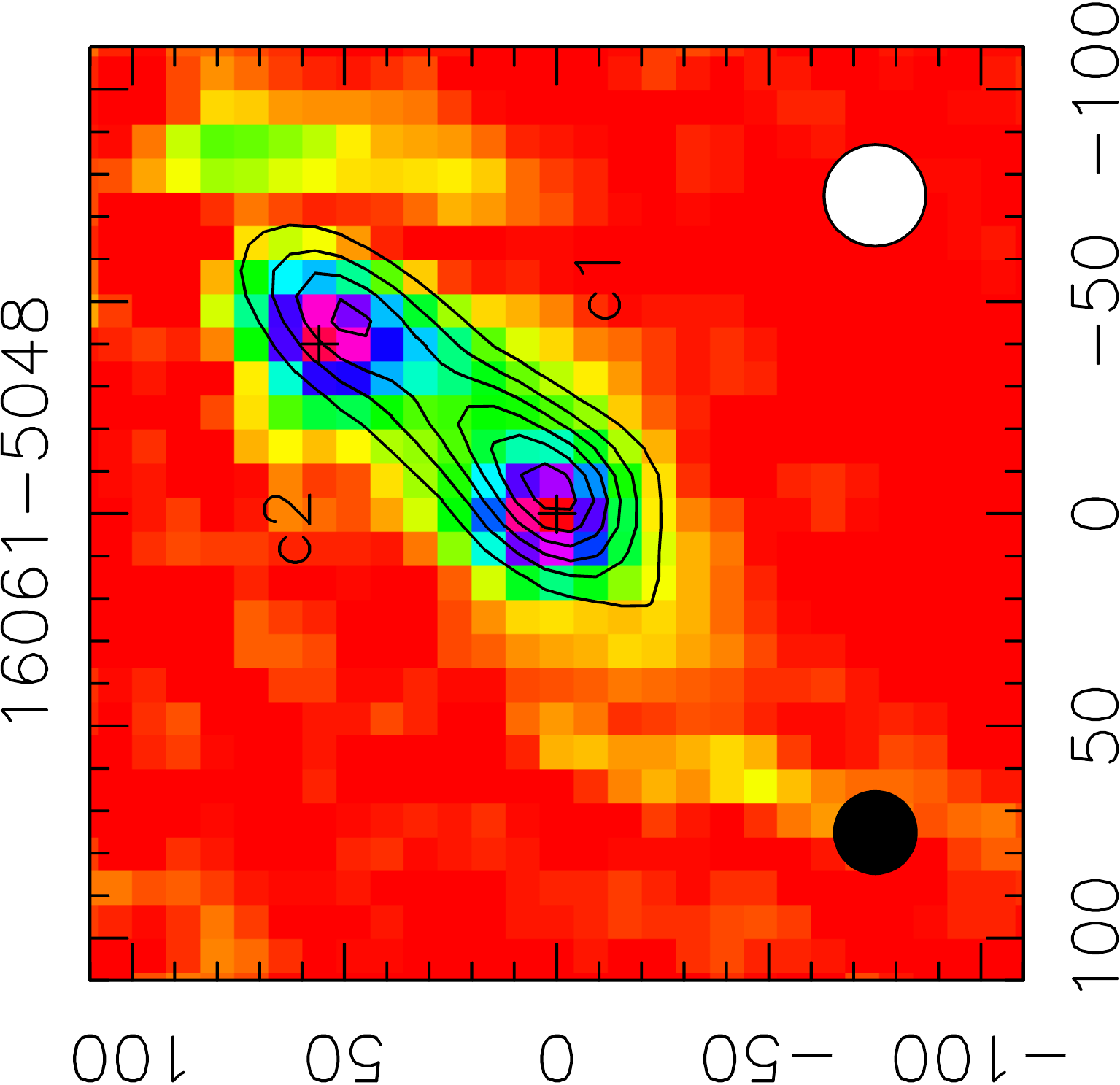}
 \includegraphics[angle=-90,width=0.225\textwidth]{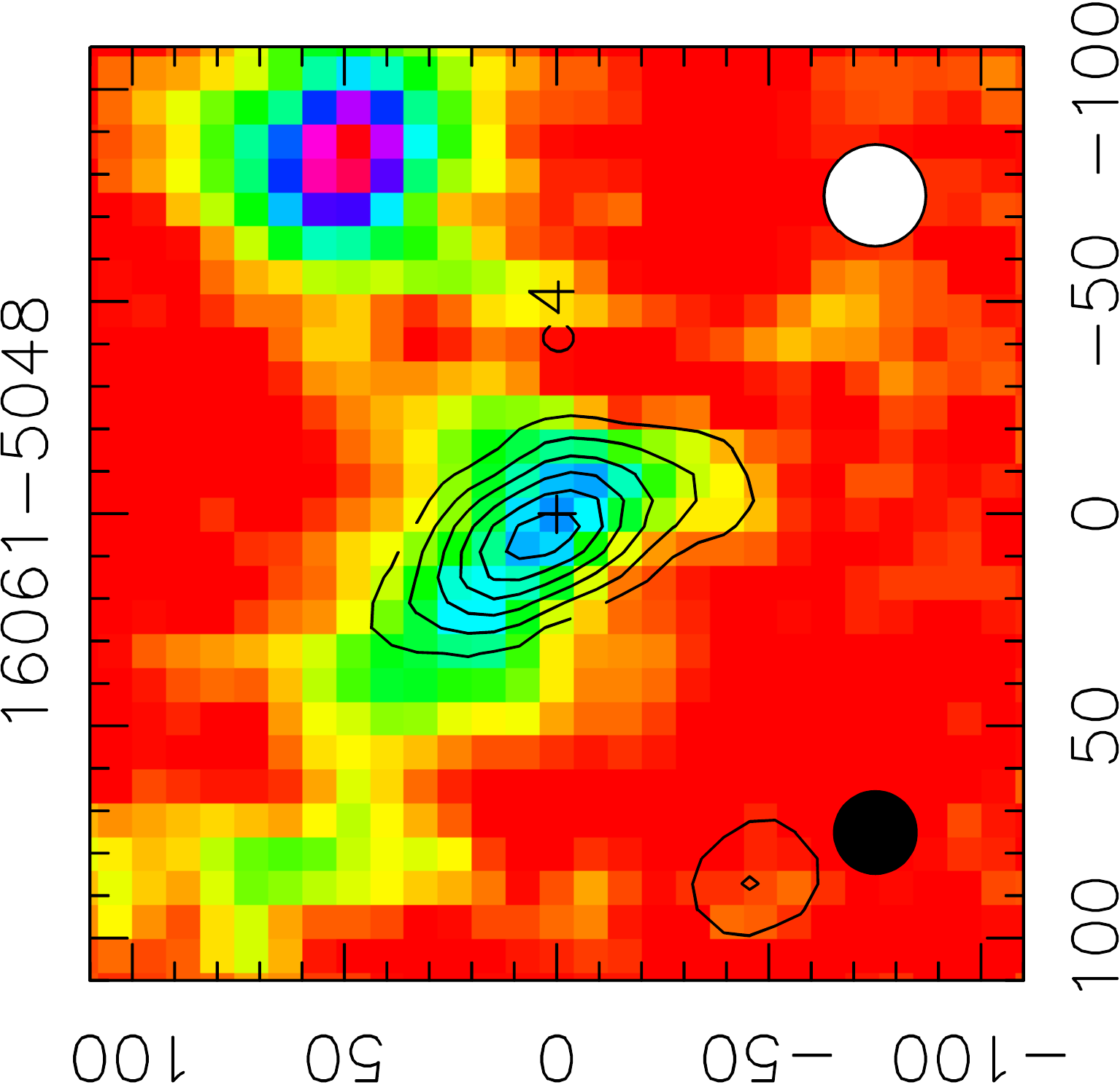}
 \includegraphics[angle=-90,width=0.225\textwidth]{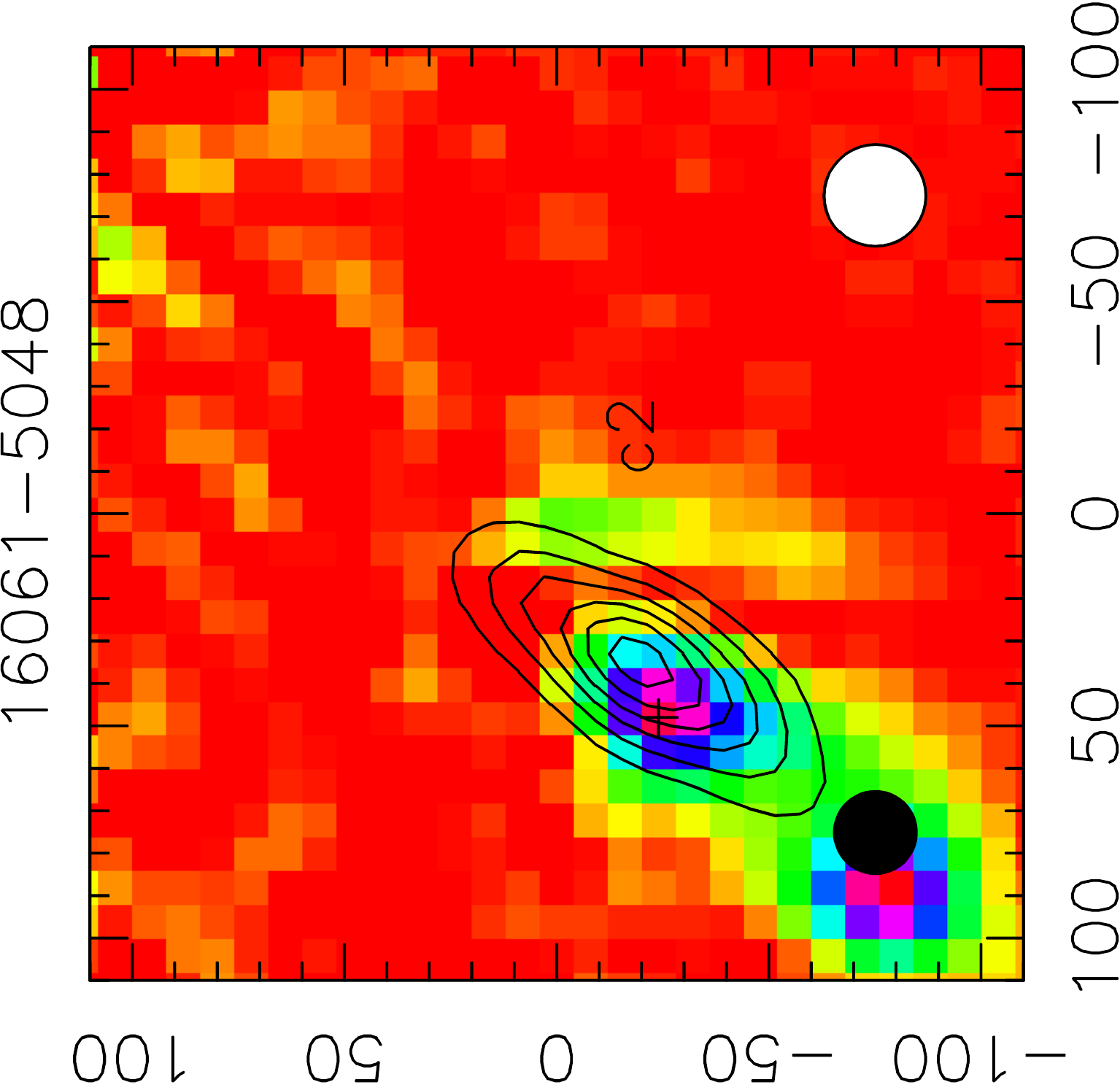}
 \includegraphics[angle=-90,width=0.225\textwidth]{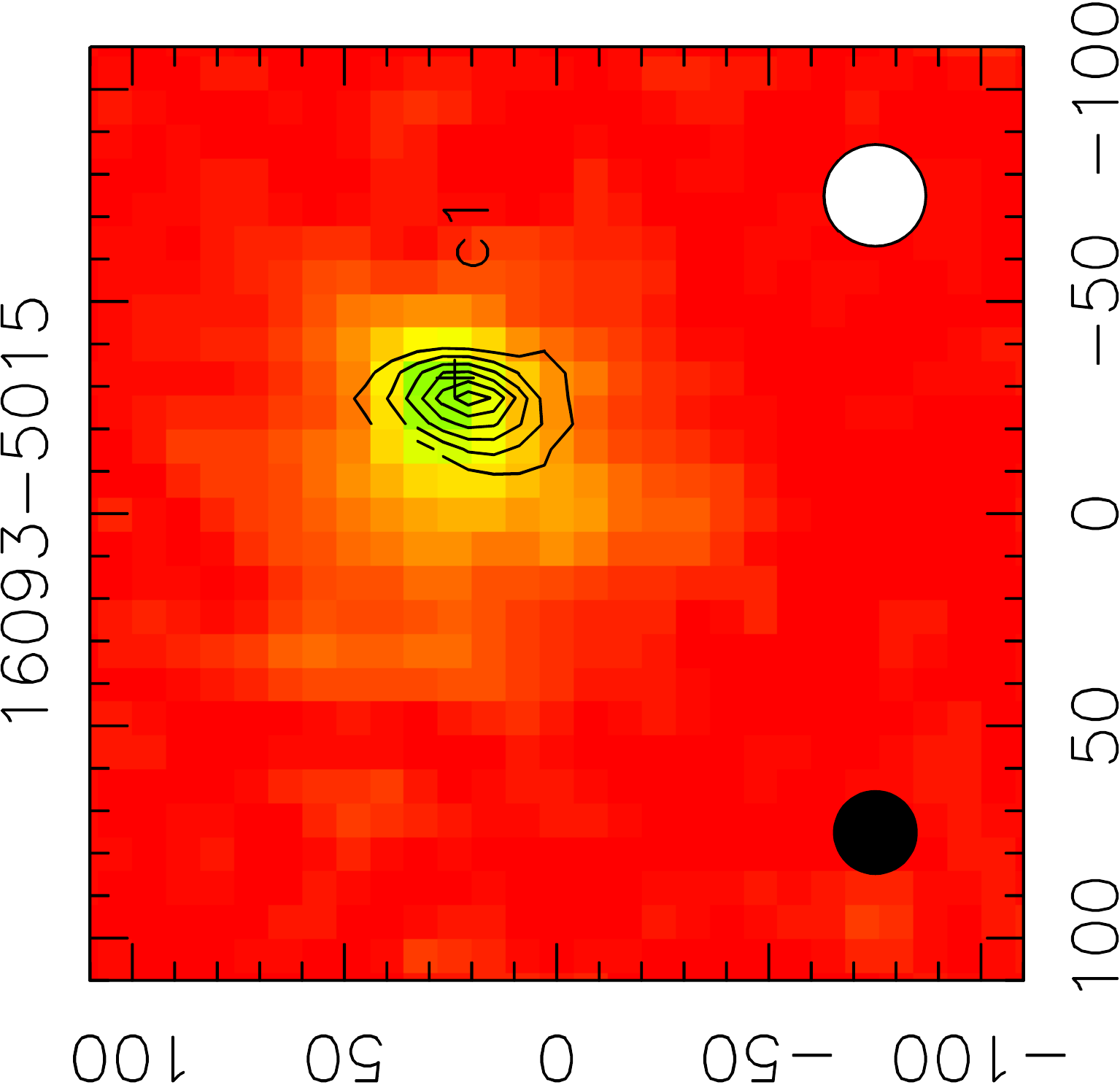}
 \includegraphics[angle=-90,width=0.225\textwidth]{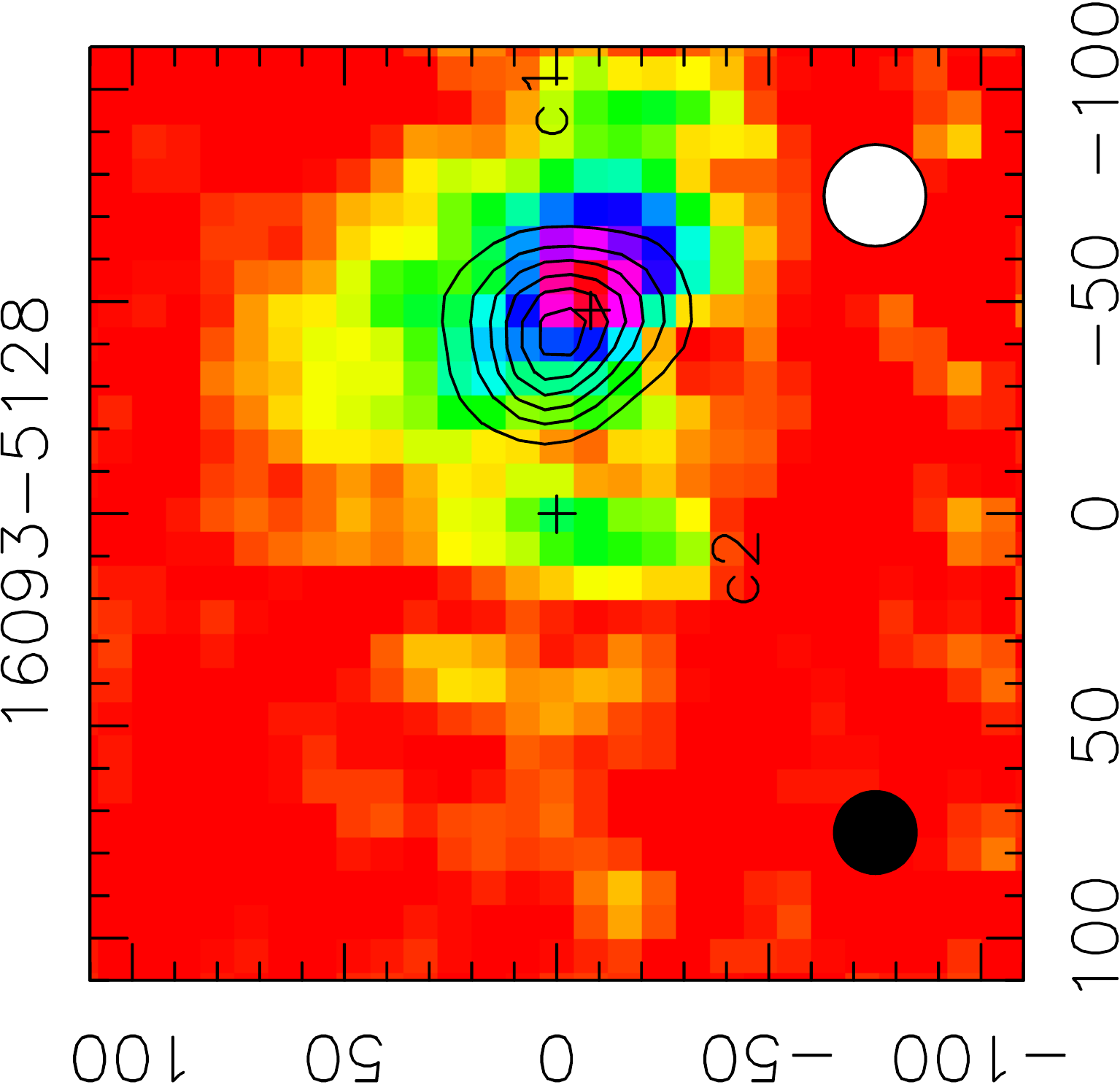}
 \includegraphics[angle=-90,width=0.225\textwidth]{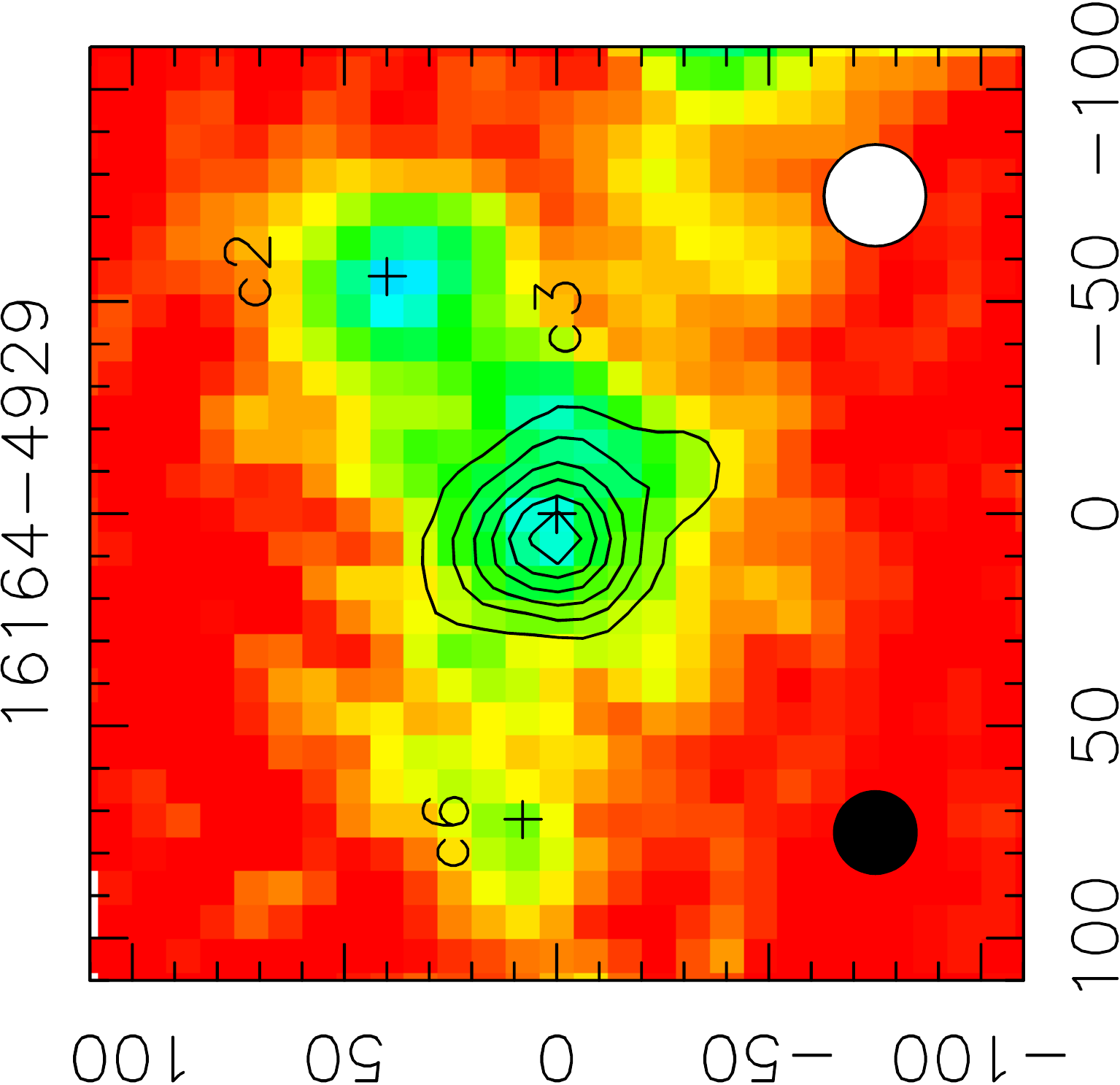}
 \includegraphics[angle=-90,width=0.225\textwidth]{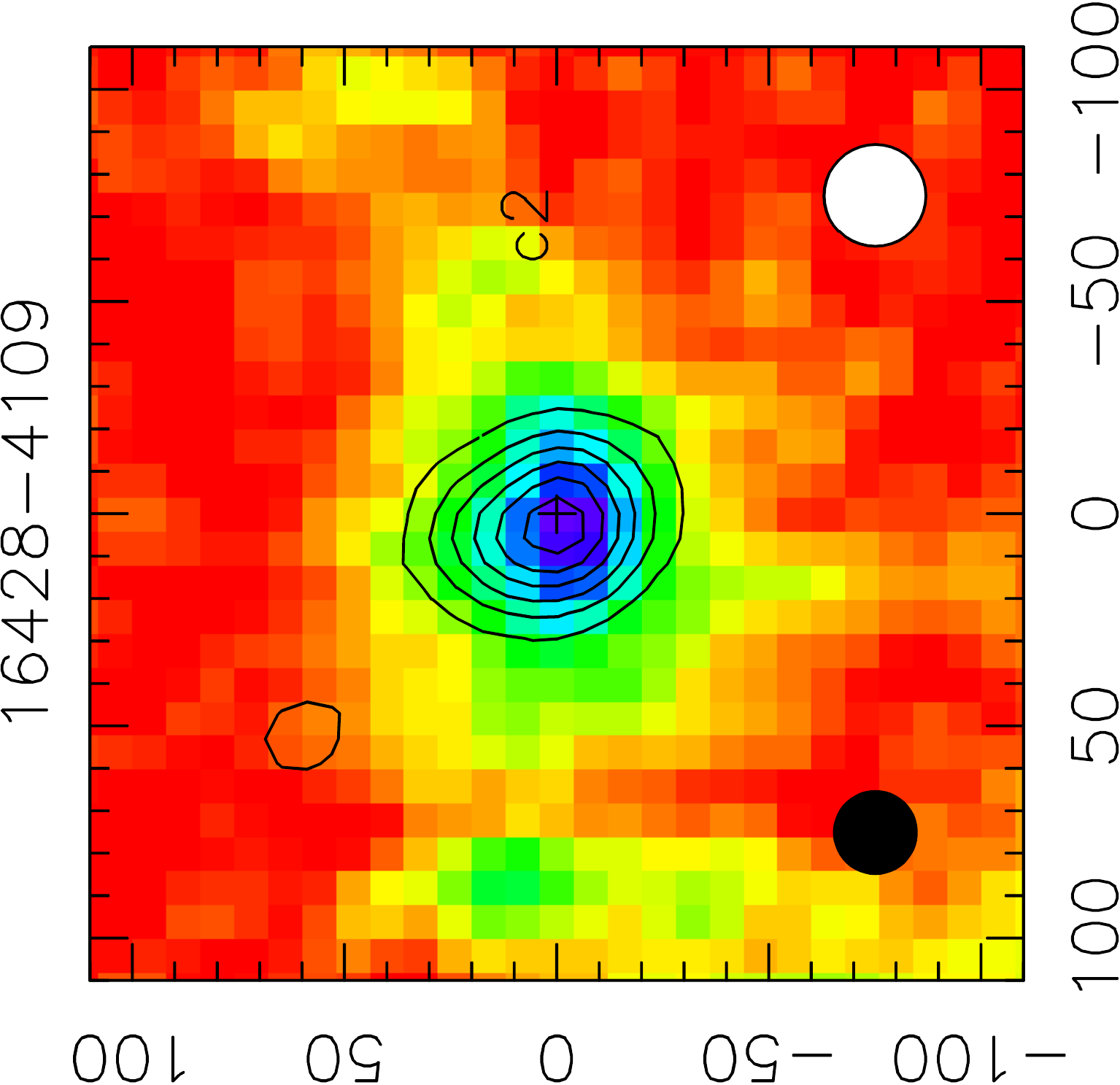}
 \includegraphics[angle=-90,width=0.225\textwidth]{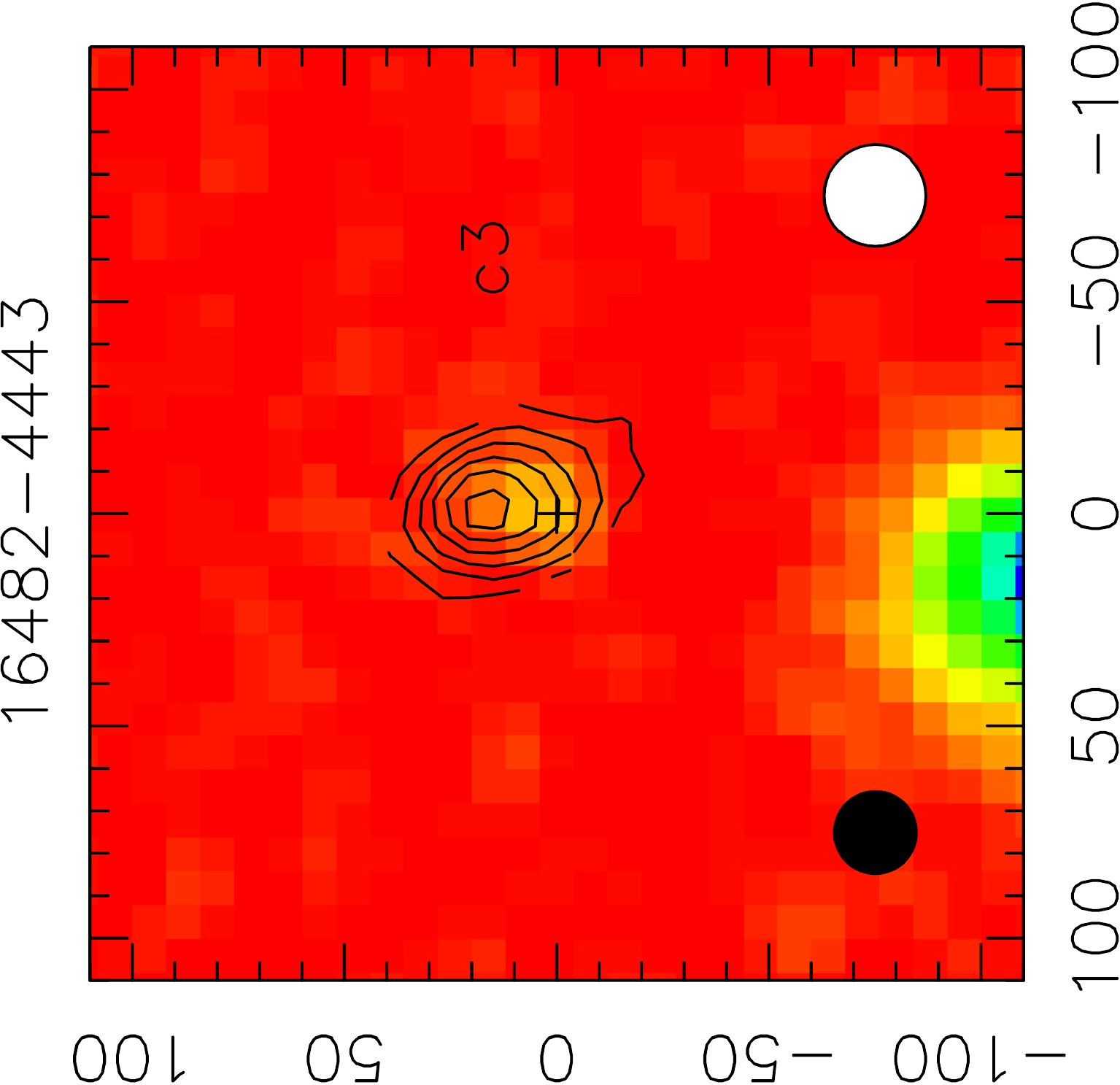}
 \includegraphics[angle=-90,width=0.225\textwidth]{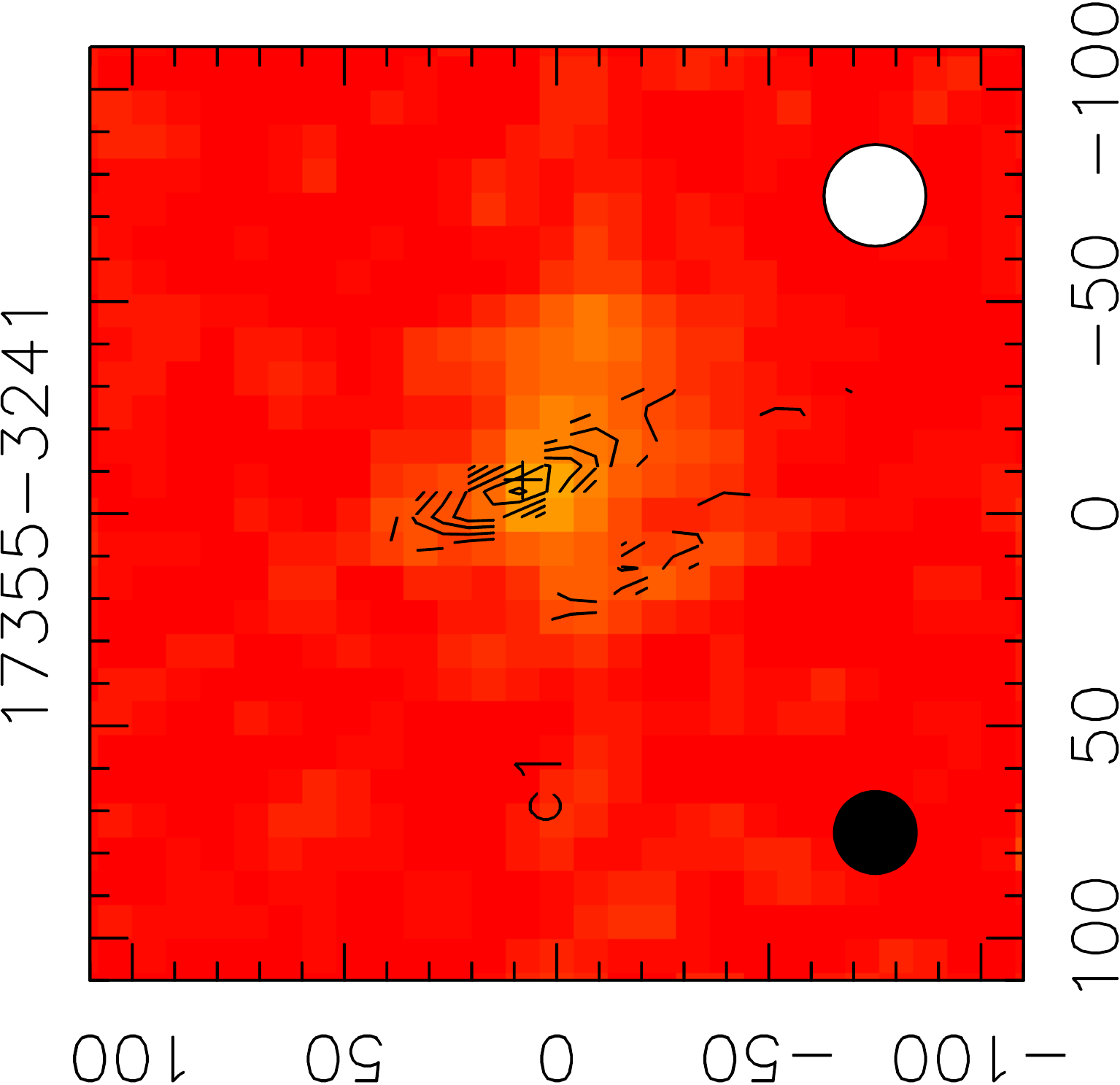}
 \caption{\textbf{(a)}:Colorscale: SEST $1.2\,$mm; contours: NH$_3$(1,1) zeroth moment. The contours start from $15\%$ of the peak, in step of $15\%$ of the peak integrated emission. The peak flux and the $\Delta V$ of the line are given in Table~\ref{tab:line_prop}. The position of each clump in our sample is indicated by a black cross. The SEST beam is indicated as a white circle, and the ATCA beam is shown as a black circle. The IRAS name is indicated above each panel, and the clump names are shown in the figures. The fields with poor uv-coverage (see Sect.~\ref{sec:obs}) are not used to produce overlays, as the maps produced are not as reliable as the others.}
 \label{fig:mom0_sest}
\end{figure*}

\begin{figure*}[tbp]
 \ContinuedFloat
 \centering
 \includegraphics[angle=-90,width=0.225\textwidth]{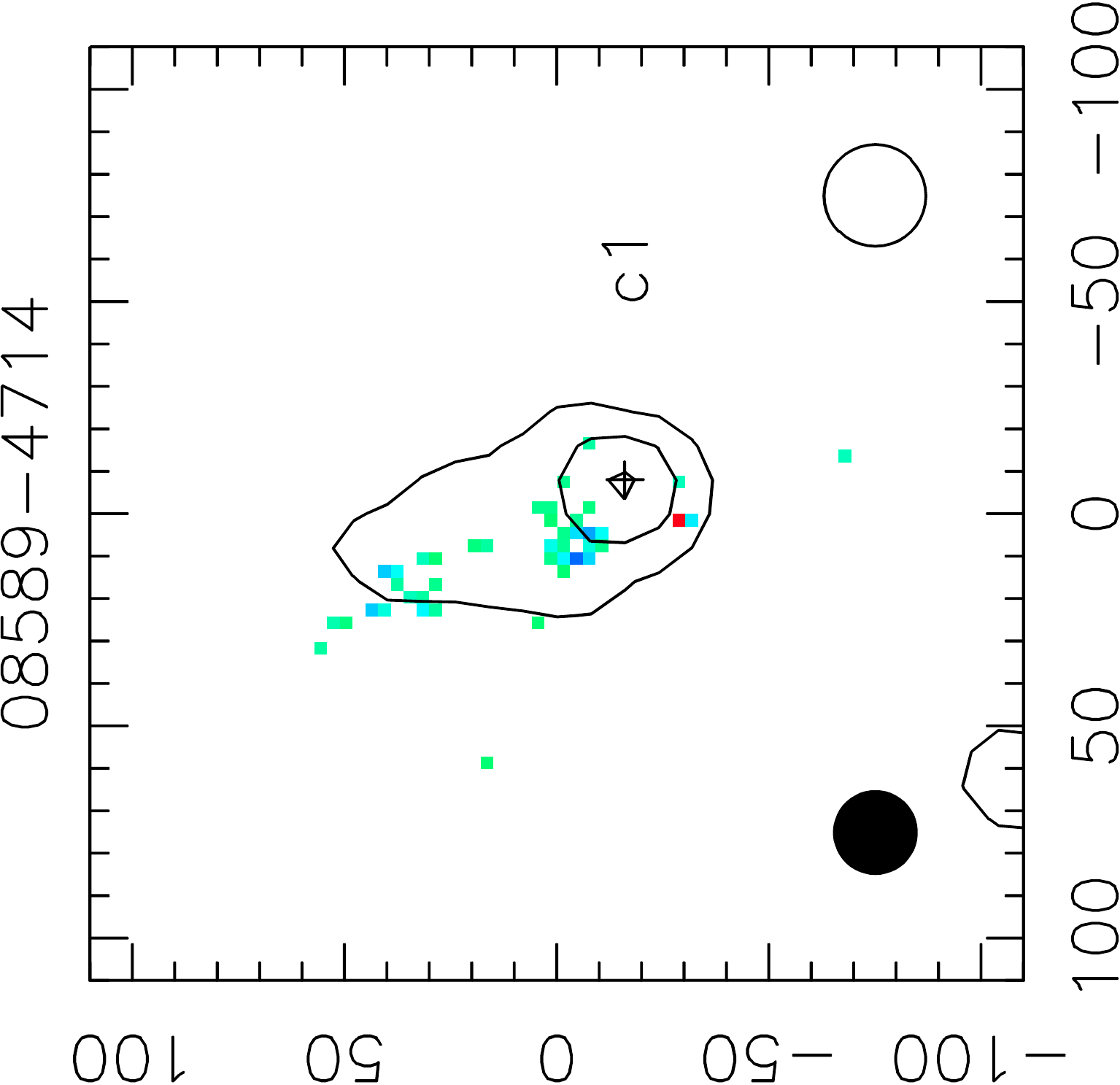}
 \includegraphics[angle=-90,width=0.225\textwidth]{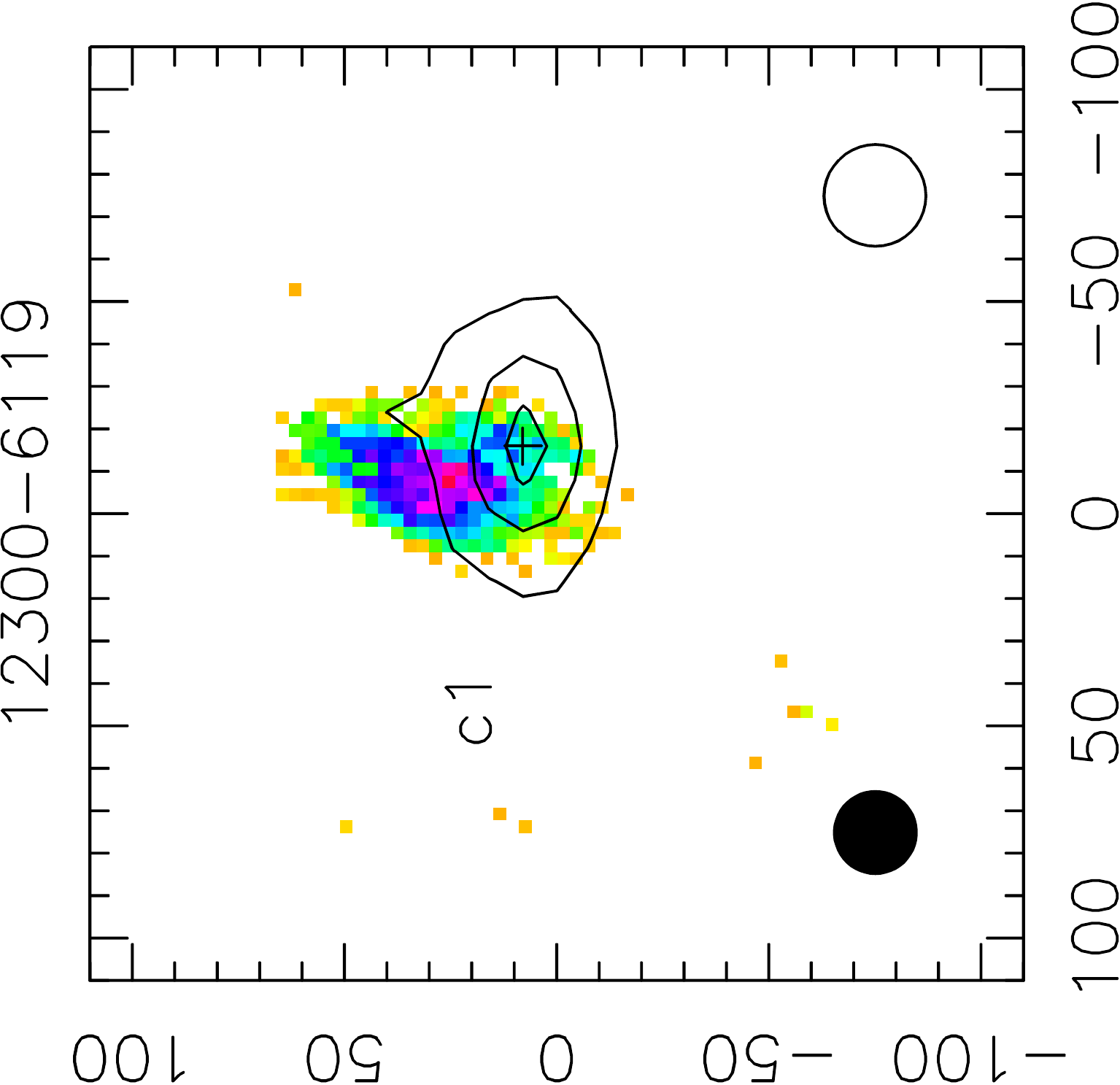}
 \includegraphics[angle=-90,width=0.225\textwidth]{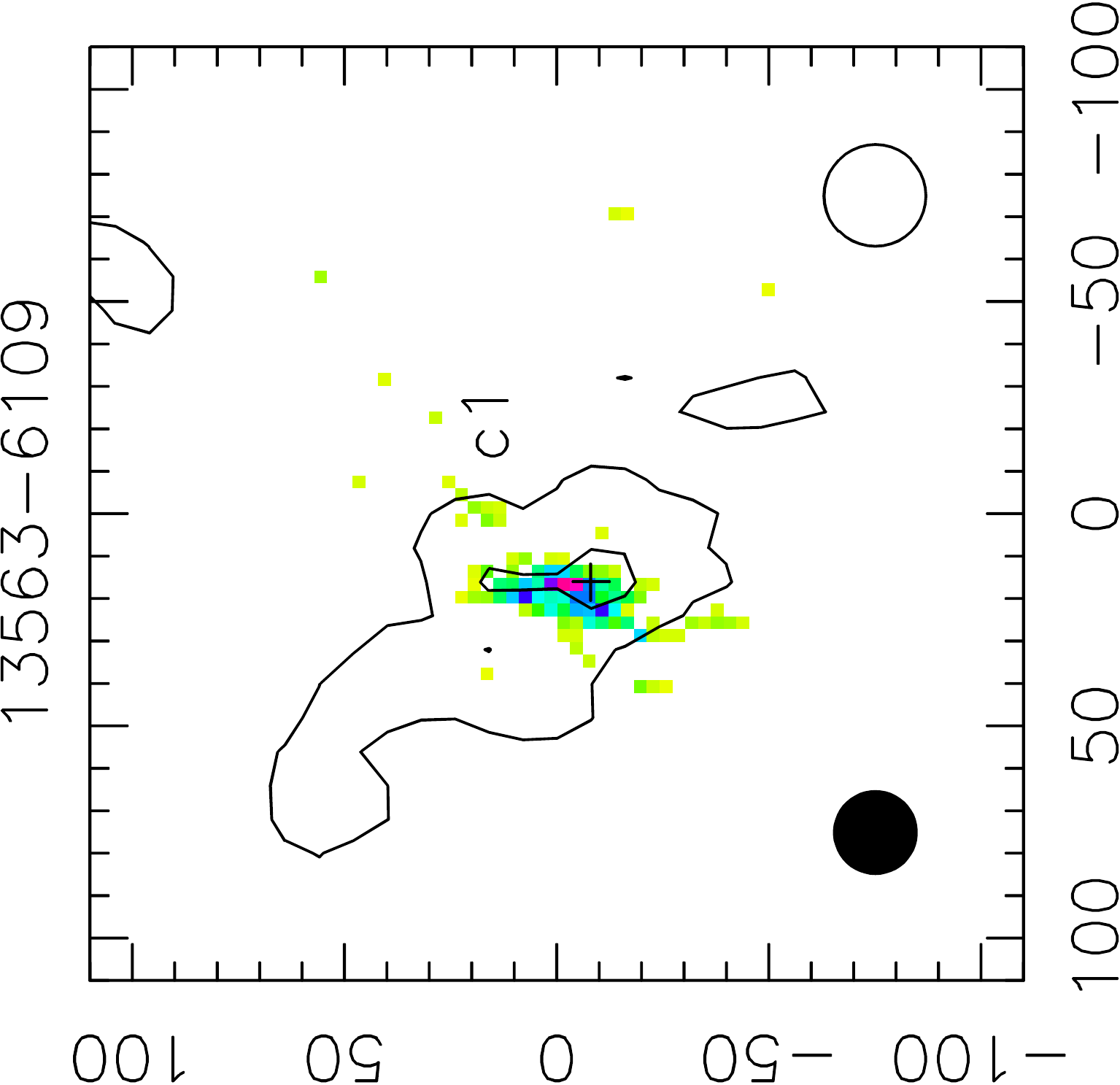}
 \includegraphics[angle=-90,width=0.225\textwidth]{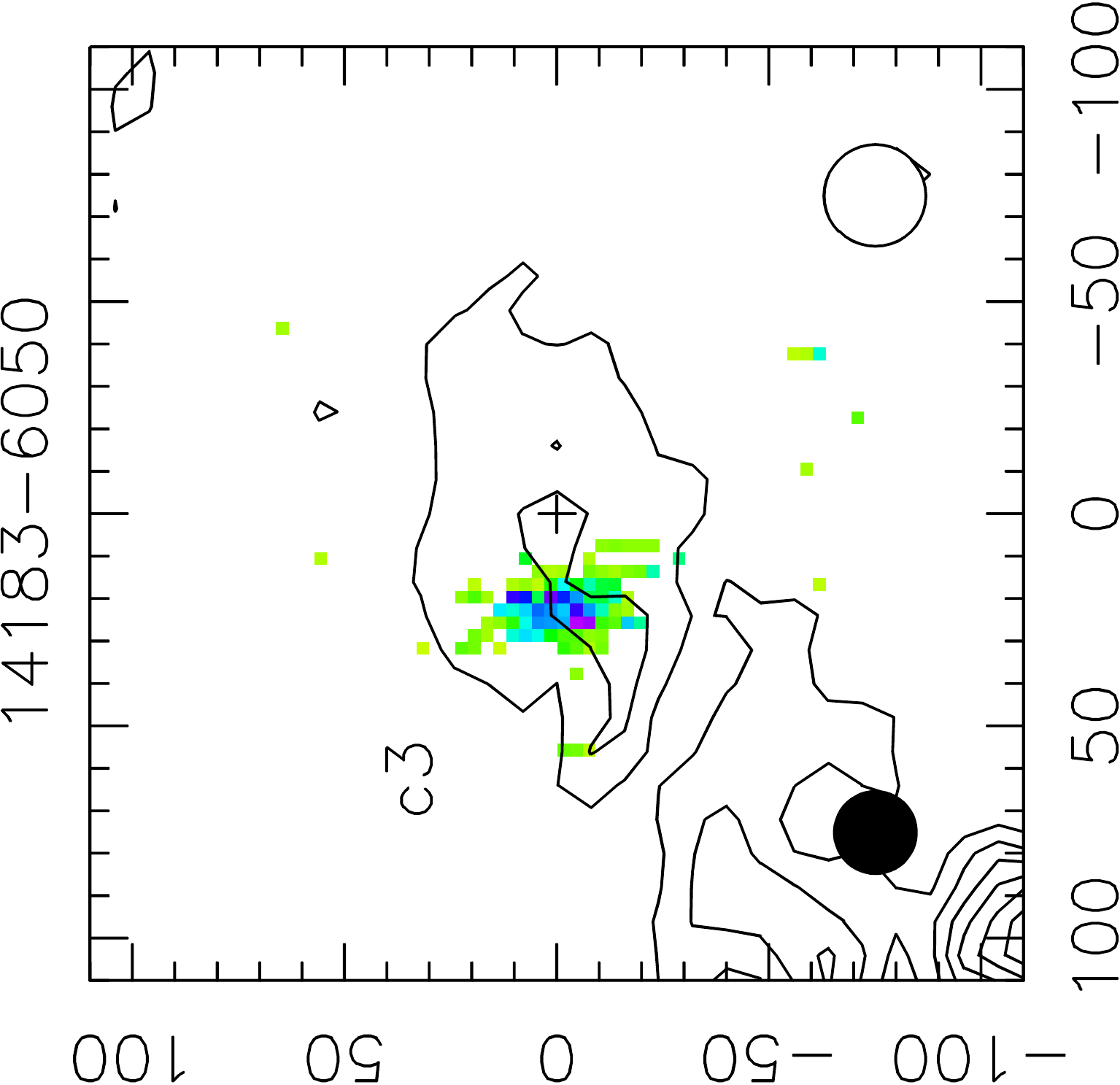}
 \includegraphics[angle=-90,width=0.225\textwidth]{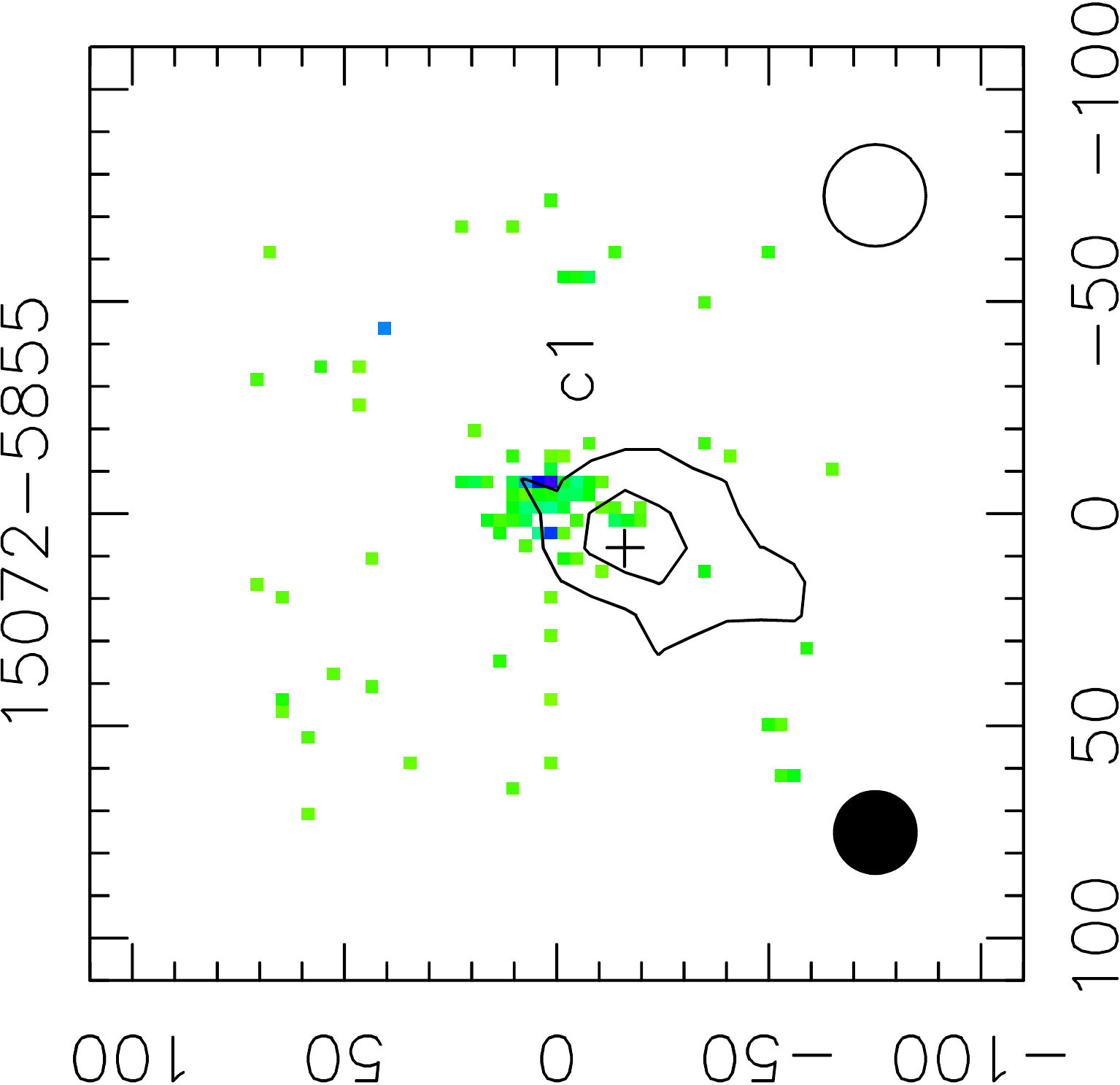}
 \includegraphics[angle=-90,width=0.225\textwidth]{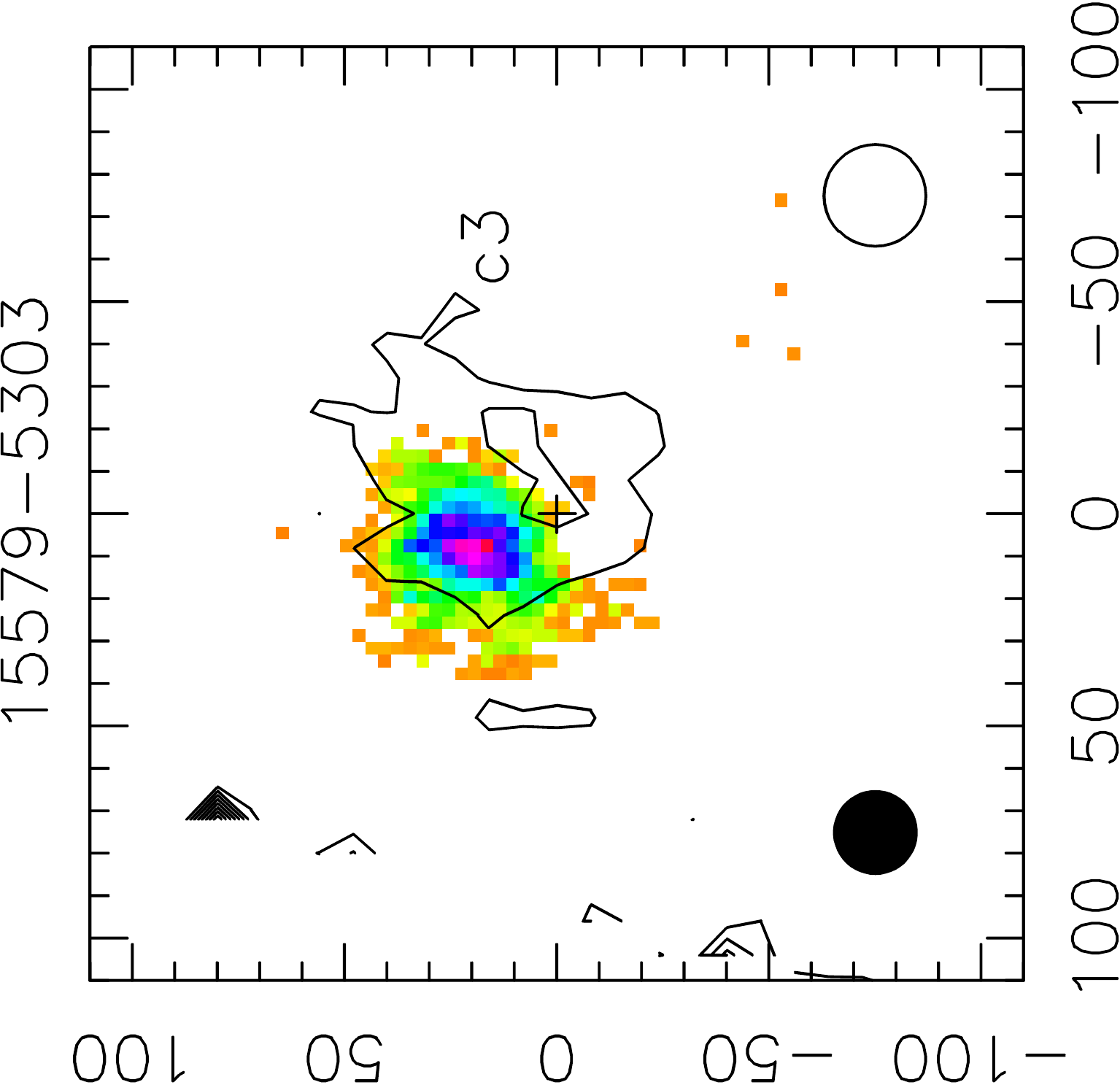}
 \includegraphics[angle=-90,width=0.225\textwidth]{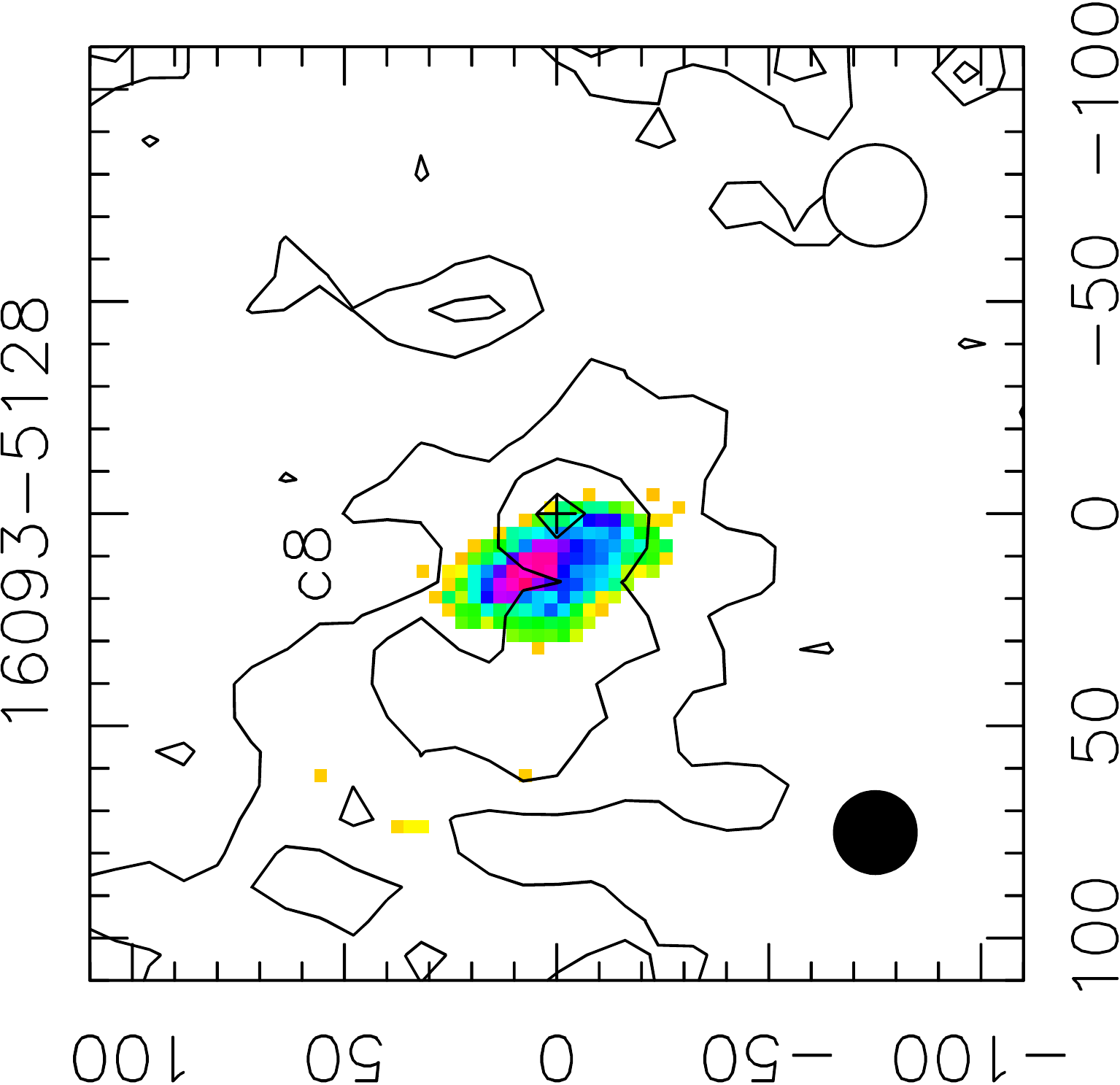}
 \includegraphics[angle=-90,width=0.225\textwidth]{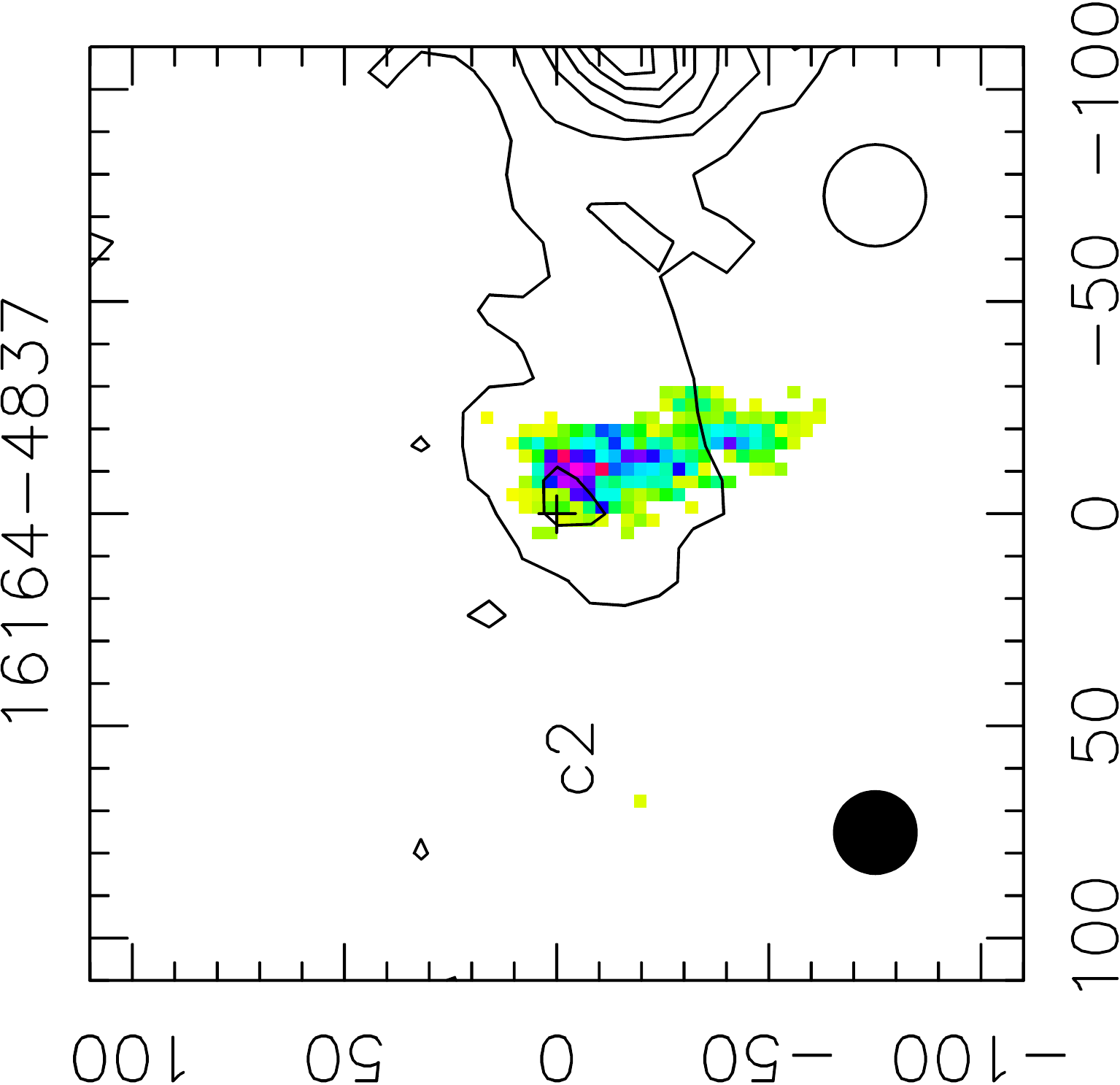}
 \caption{\textbf{(b)}: Colorscale: NH$_3$(1,1) zeroth moment; contours: SEST $1.2\,$mm. In this panel the colorscale and contours have been swapped in order to make the figure intelligible. The SEST contours are optimized to identify the clump. The position of each clump in our sample is indicated by a black cross. The SEST beam is indicated as a white circle, and the ATCA beam is shown as a black circle. The IRAS name is indicated above each panel, and the clump names are shown in the figures. The fields with poor uv-coverage (see Sect.~\ref{sec:obs}) are not used to produce overlays, as the maps produced are not as reliable as the others.}
\end{figure*}

\clearpage

\begin{figure*}[tbp]
 \centering
 \includegraphics[angle=-90,width=0.29\textwidth]{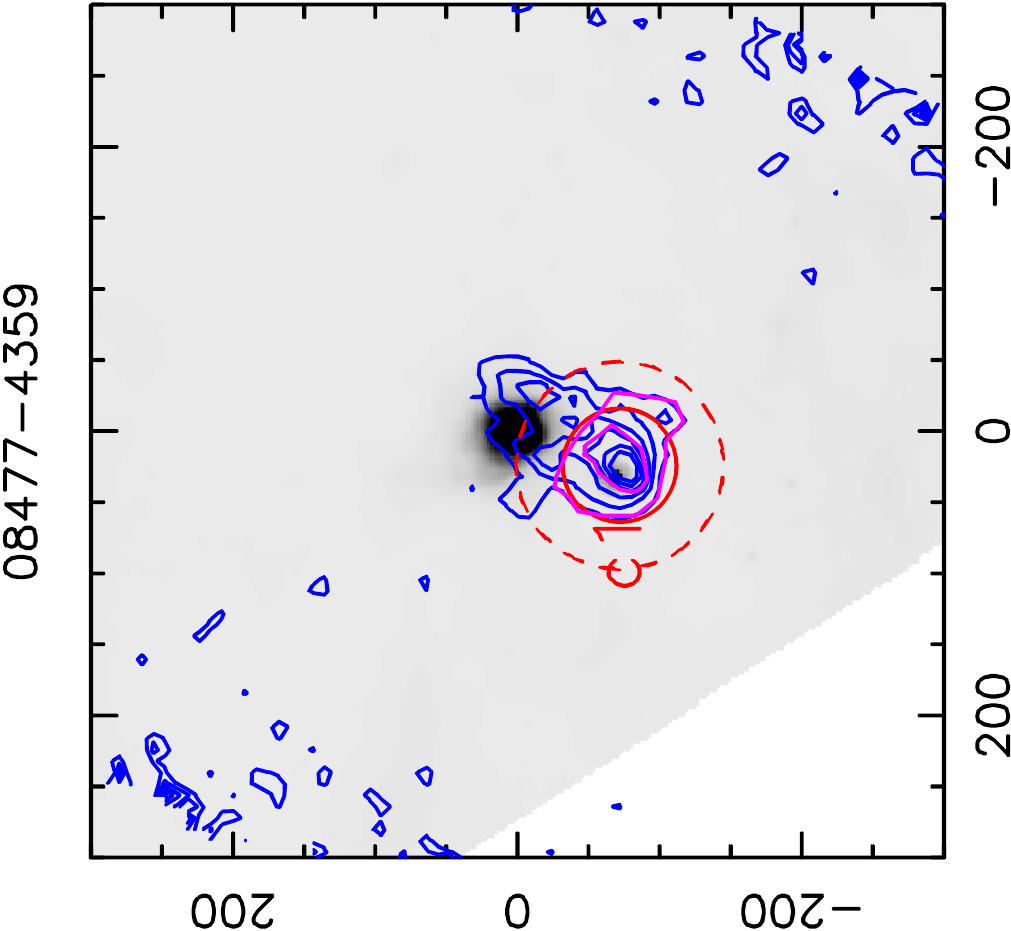}
 \includegraphics[angle=-90,width=0.29\textwidth]{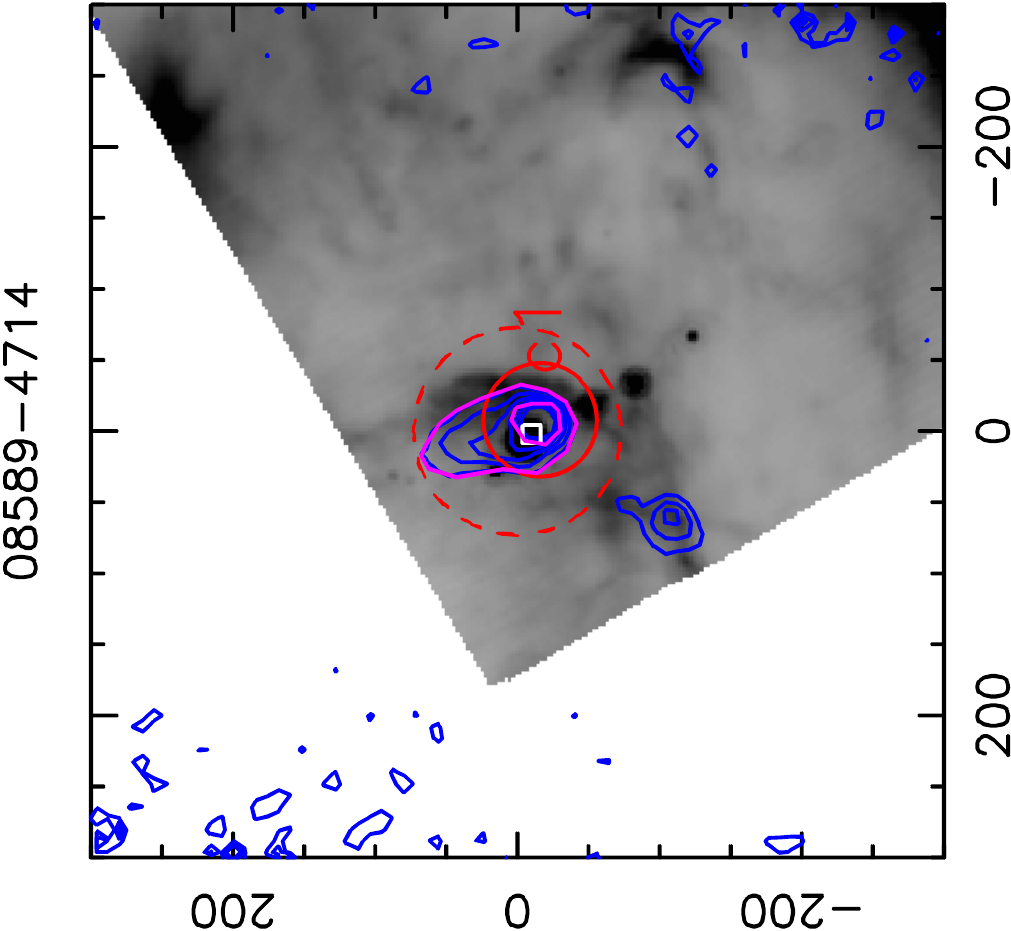}
 \includegraphics[angle=-90,width=0.29\textwidth]{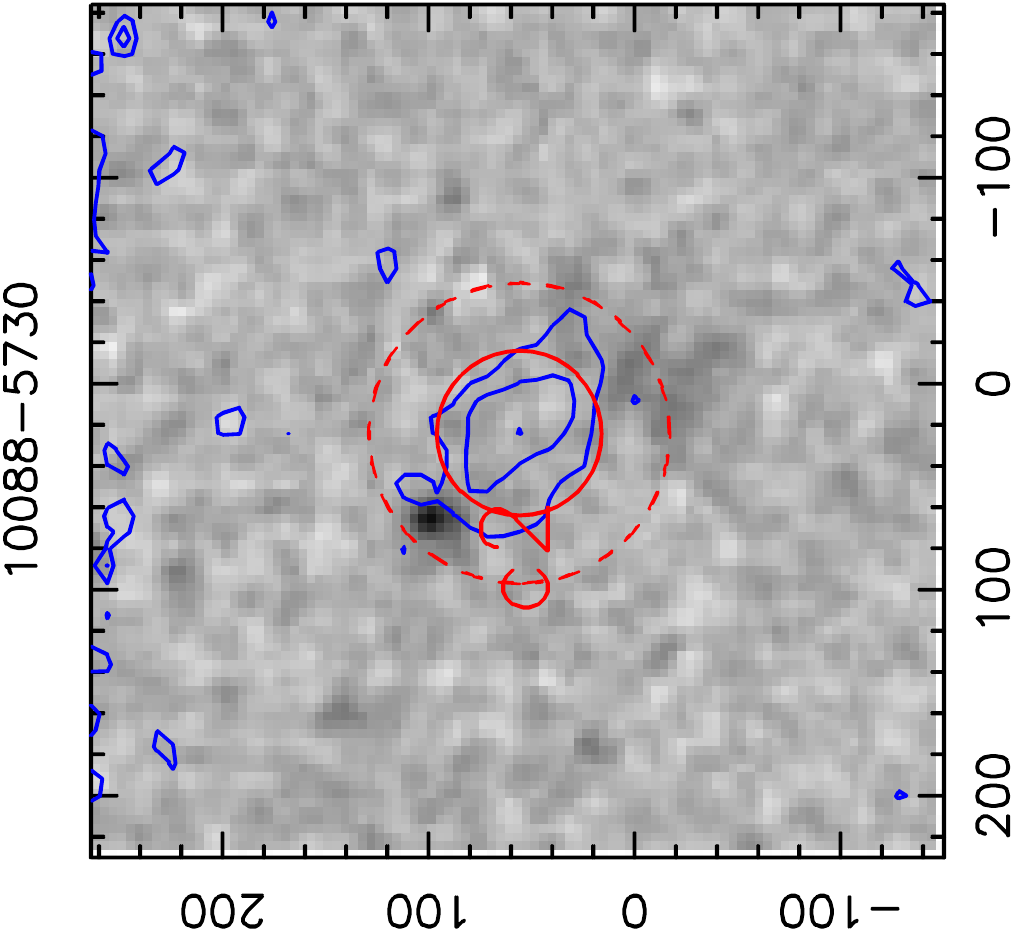}

 \includegraphics[angle=-90,width=0.29\textwidth]{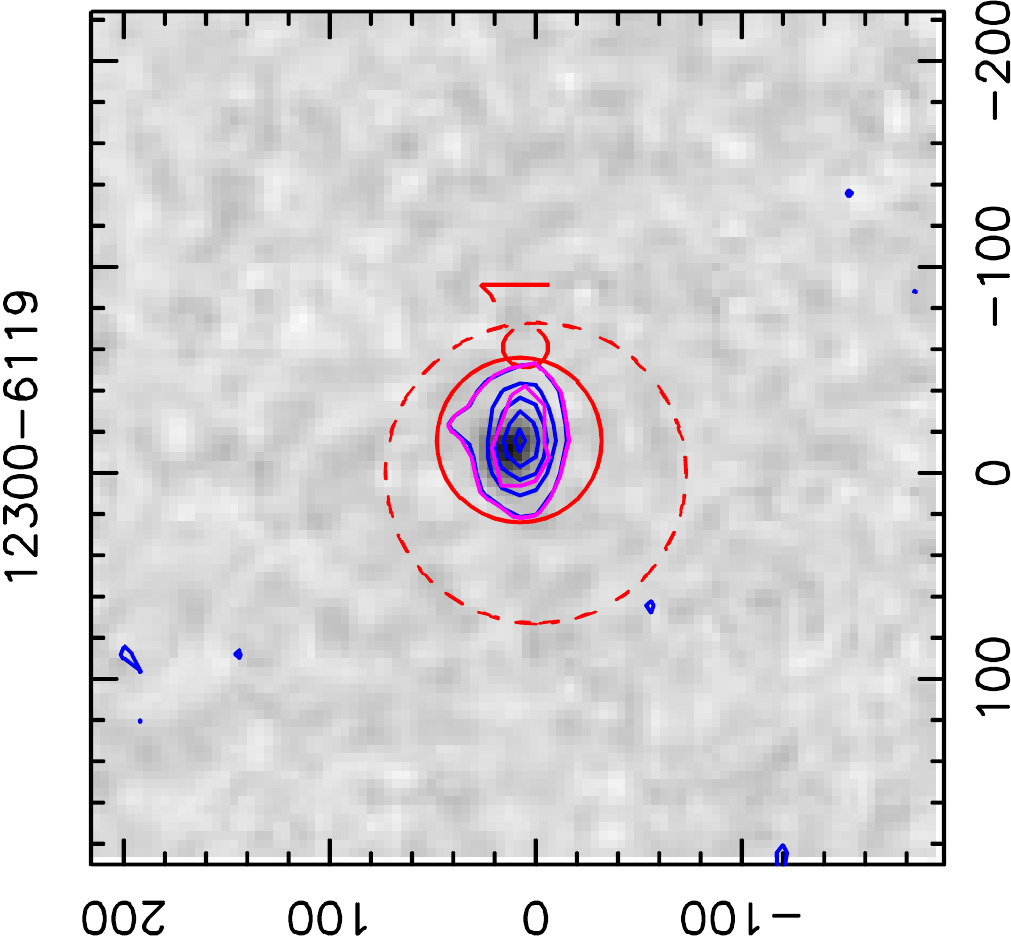}
 \includegraphics[angle=-90,width=0.29\textwidth]{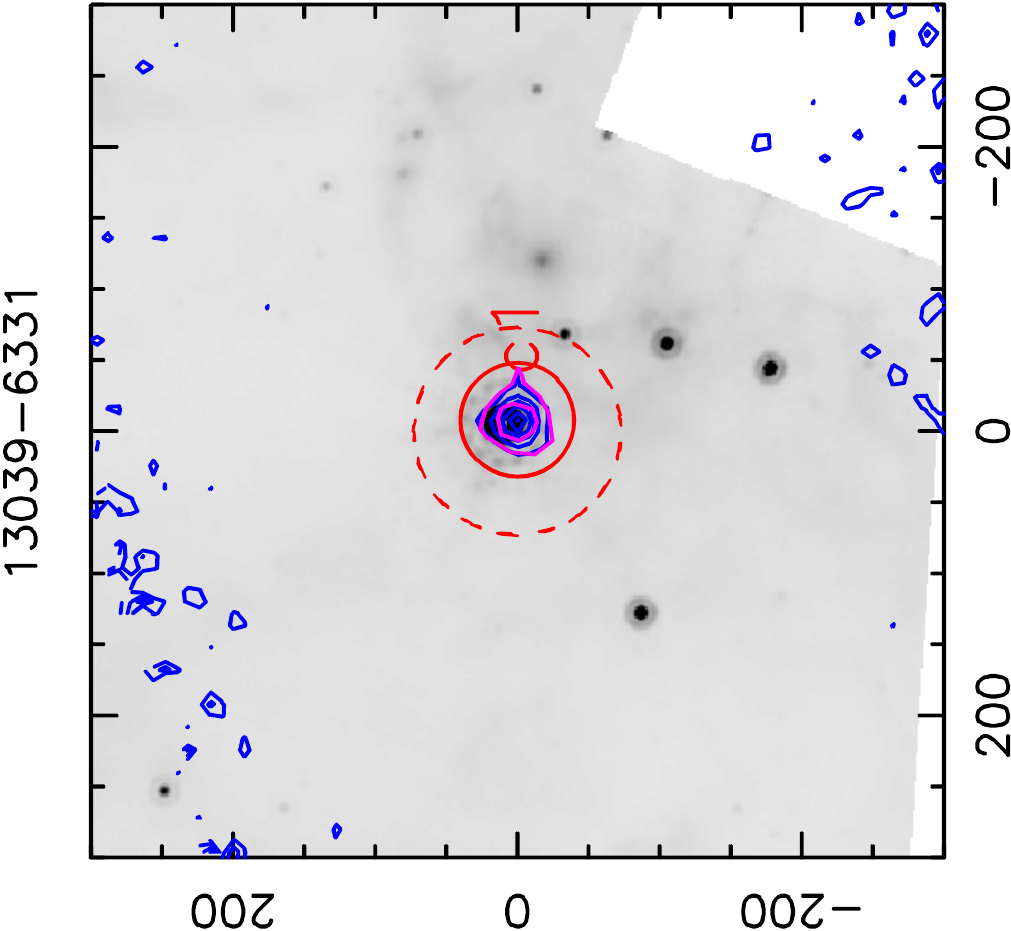}
 \includegraphics[angle=-90,width=0.29\textwidth]{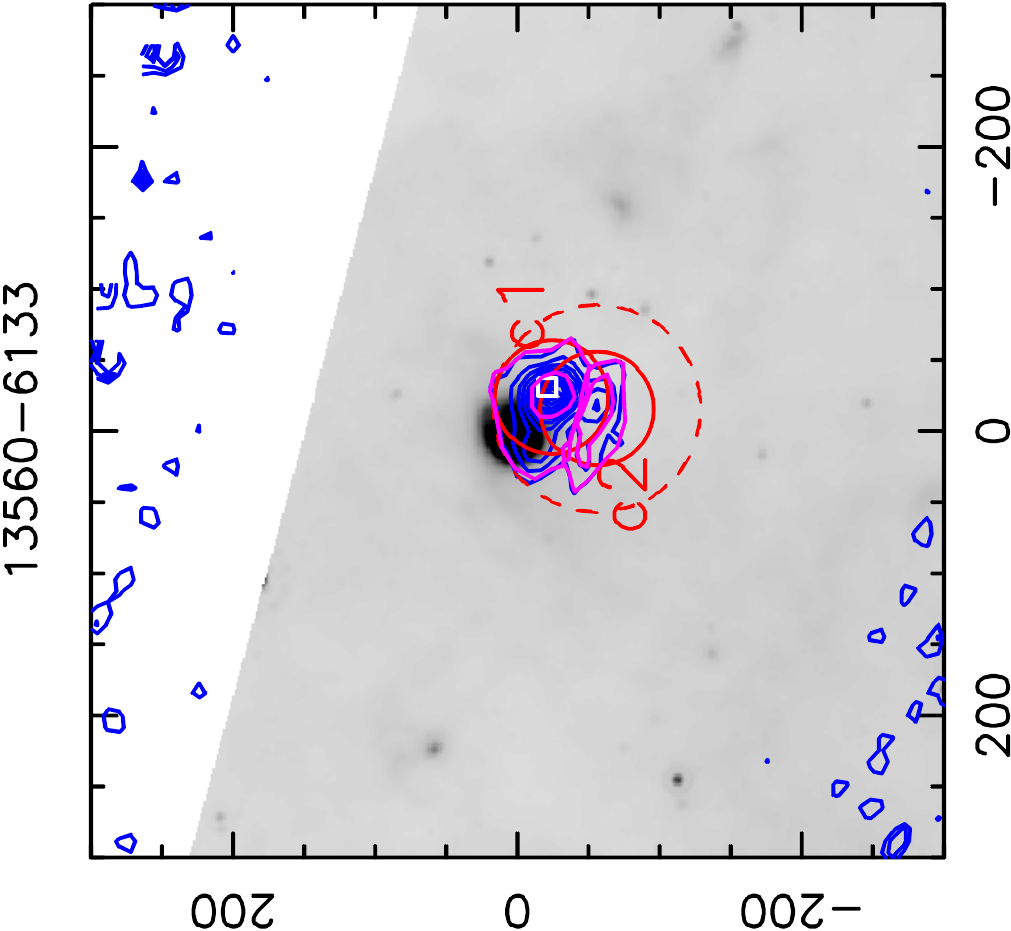}

 \includegraphics[angle=-90,width=0.29\textwidth]{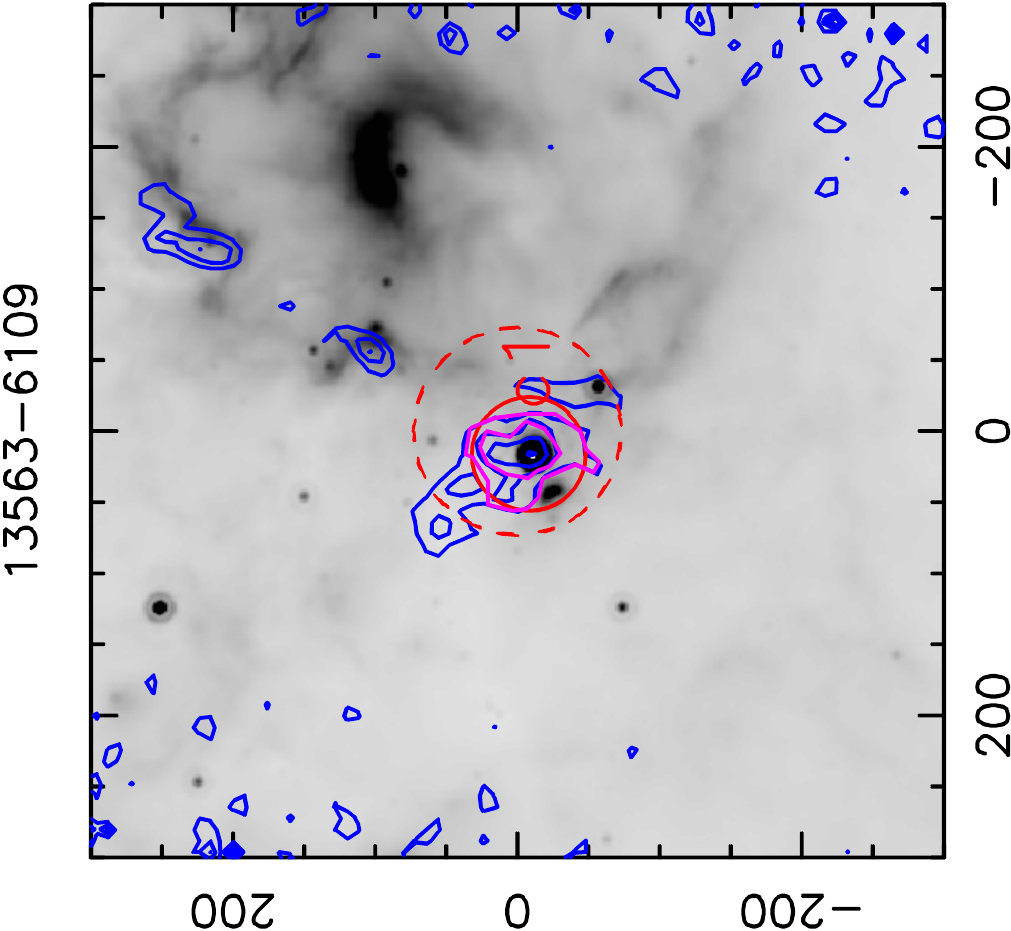}
 \includegraphics[angle=-90,width=0.29\textwidth]{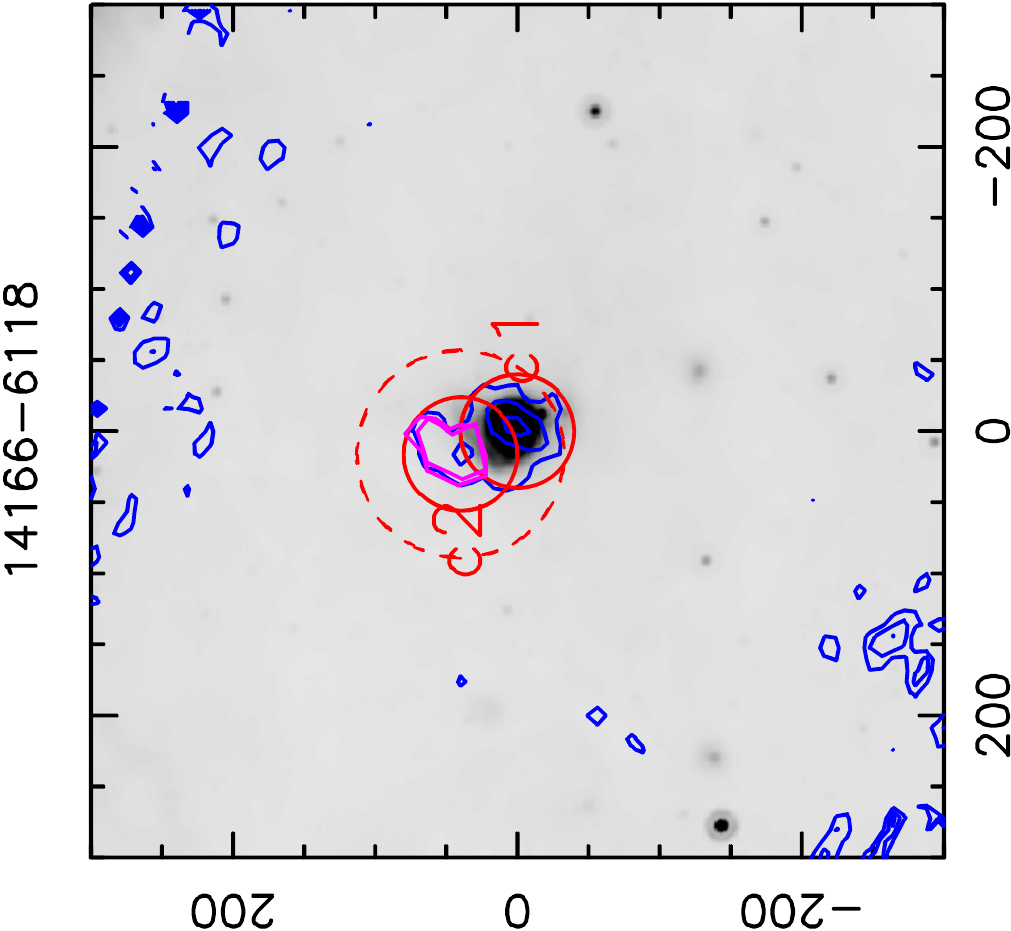}
 \includegraphics[angle=-90,width=0.29\textwidth]{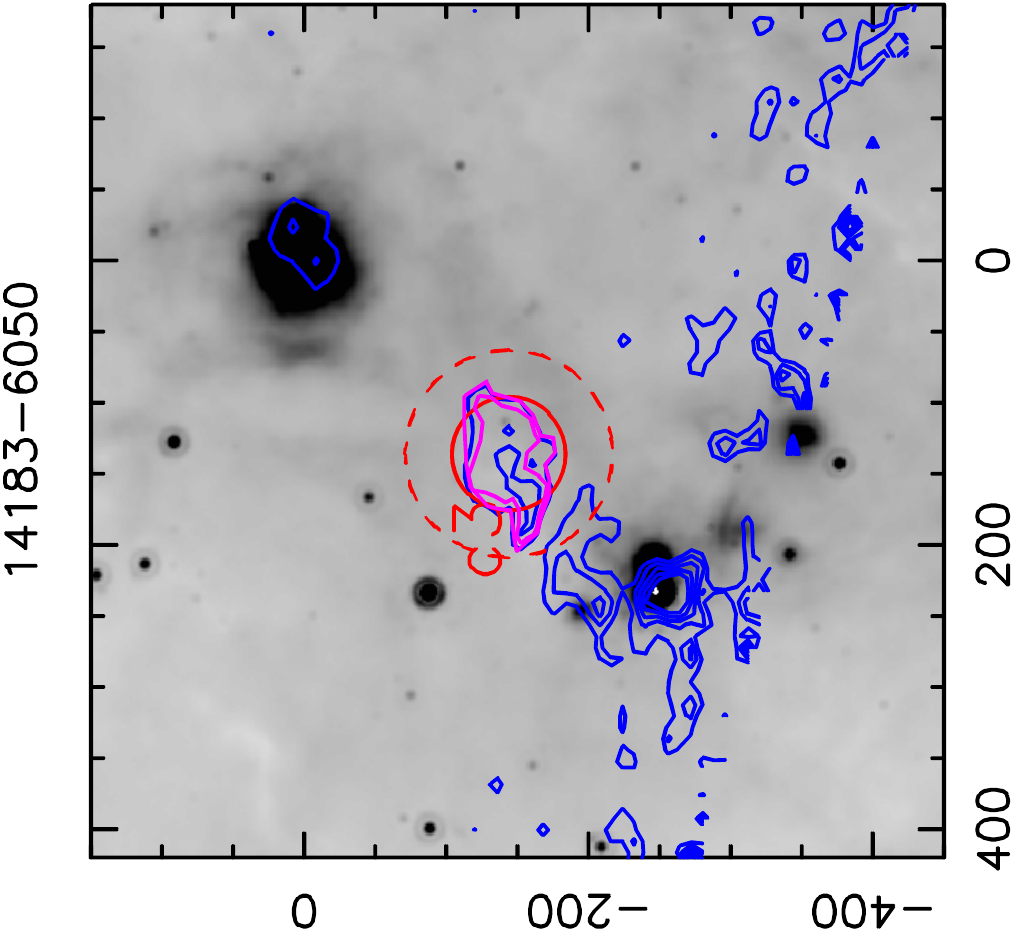}

 \includegraphics[angle=-90,width=0.29\textwidth]{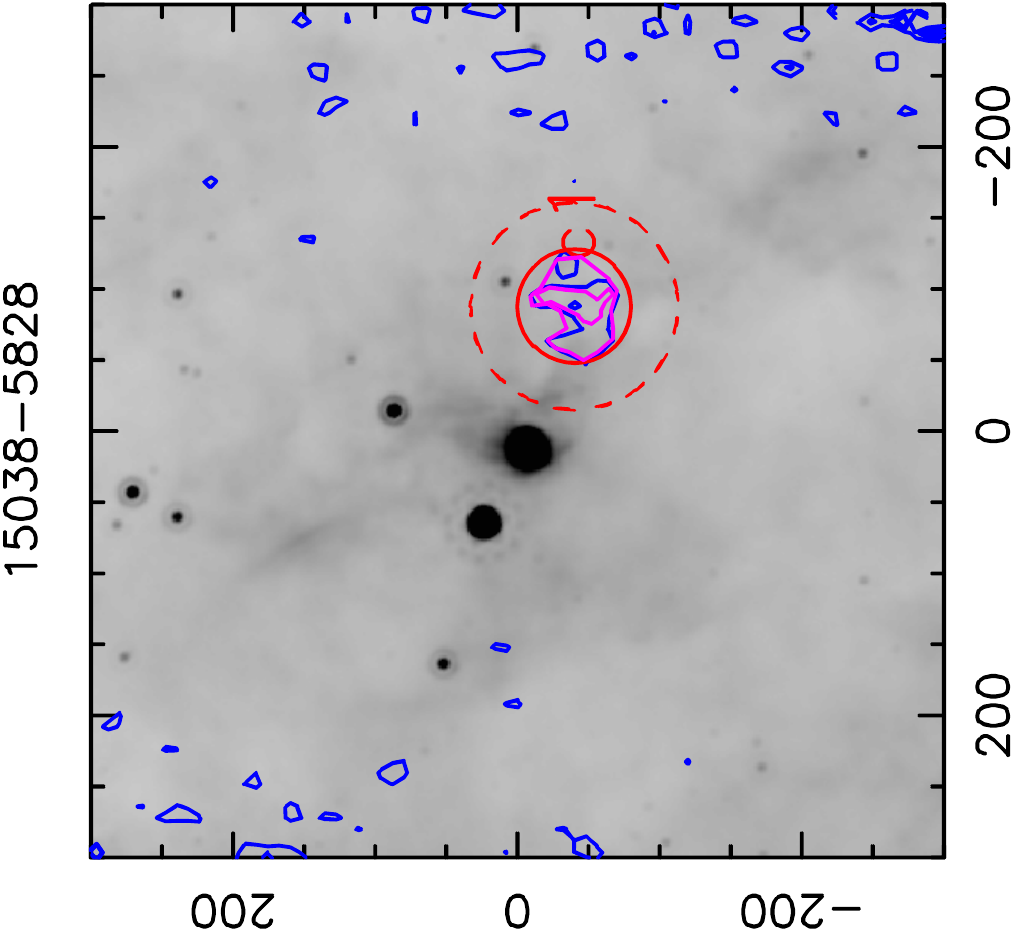}
 \includegraphics[angle=-90,width=0.29\textwidth]{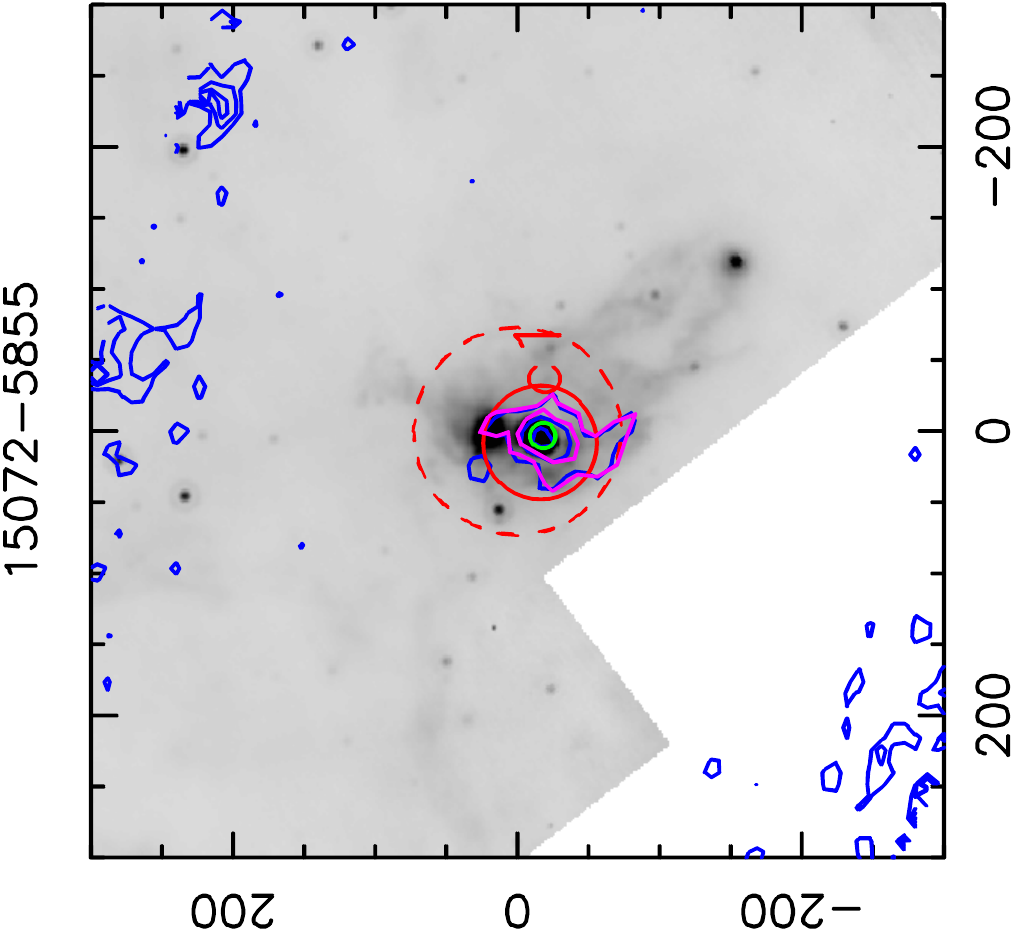}
 \includegraphics[angle=-90,width=0.29\textwidth]{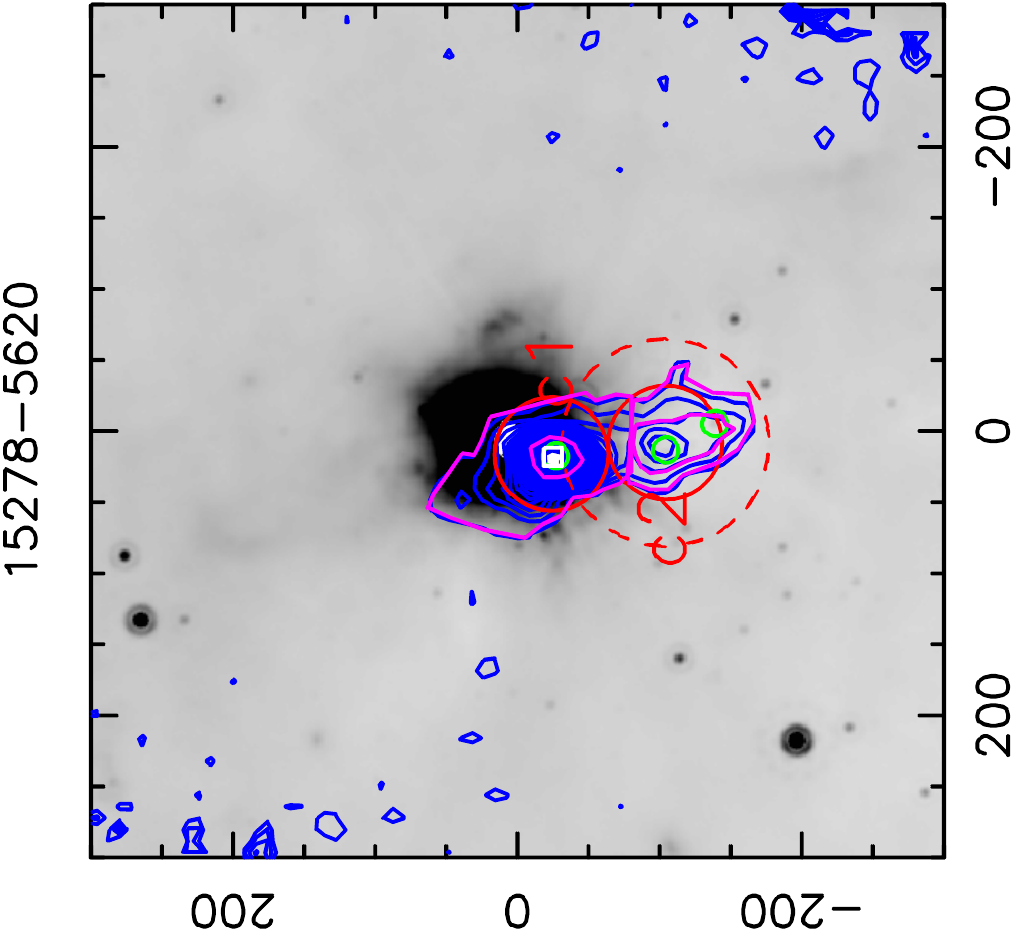}

 \caption{SEST $1.2\mm$ emission (contours) superimposed on the Spitzer/MIPS $24\mum$ images. The IRAS name is indicated above each panel, and the clump names are shown in the figures. Dashed and solid red lines indicate the ATCA field and the observed clumps, respectively. H$_2$O maser spots are shown as white open squares, while ``green fuzzies'' are indicated as green open circles. In the last four panels the SEST emission is superimposed on MSX images. Purple lines indicate the $3\sigma$ and FWHM polygons used to extract fluxes from the SEST images.}
 \label{fig:sest+24mum}
\end{figure*}

\begin{figure*}[tbp]
 \ContinuedFloat
 \centering
 \includegraphics[angle=-90,width=0.29\textwidth]{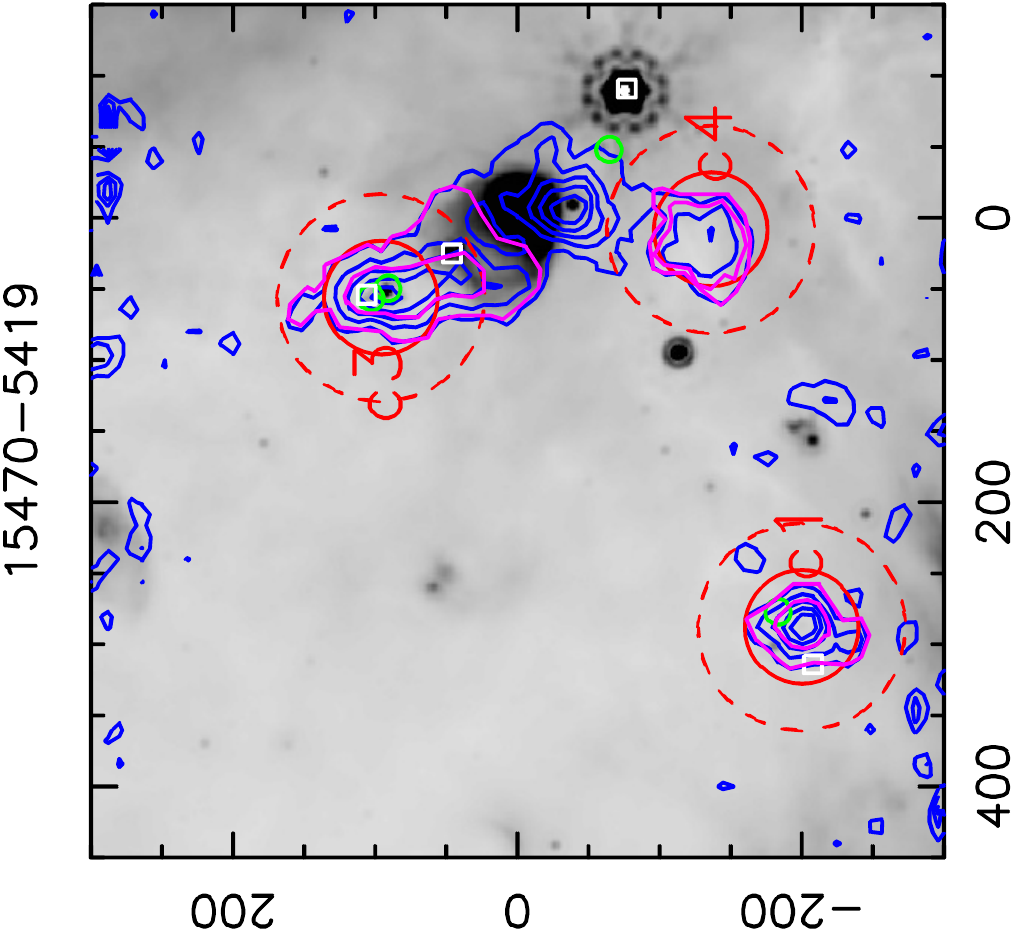}
 \includegraphics[angle=-90,width=0.29\textwidth]{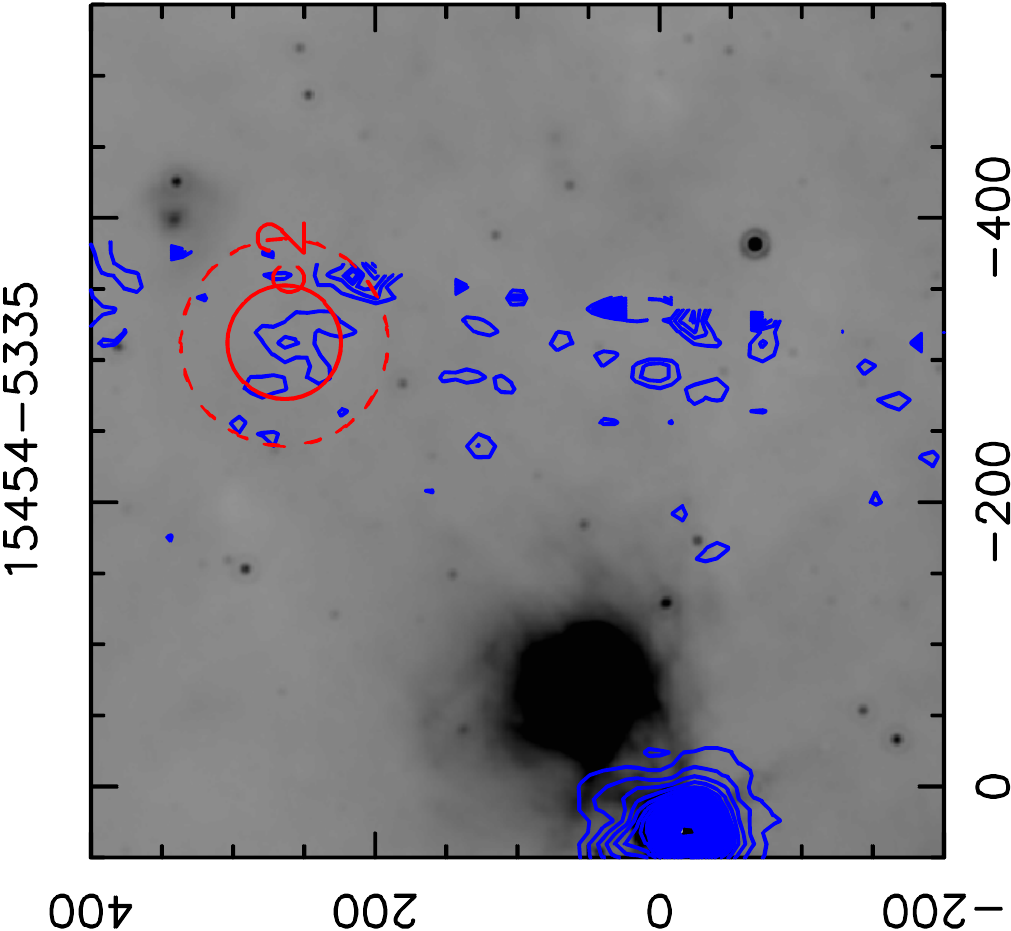}
 \includegraphics[angle=-90,width=0.29\textwidth]{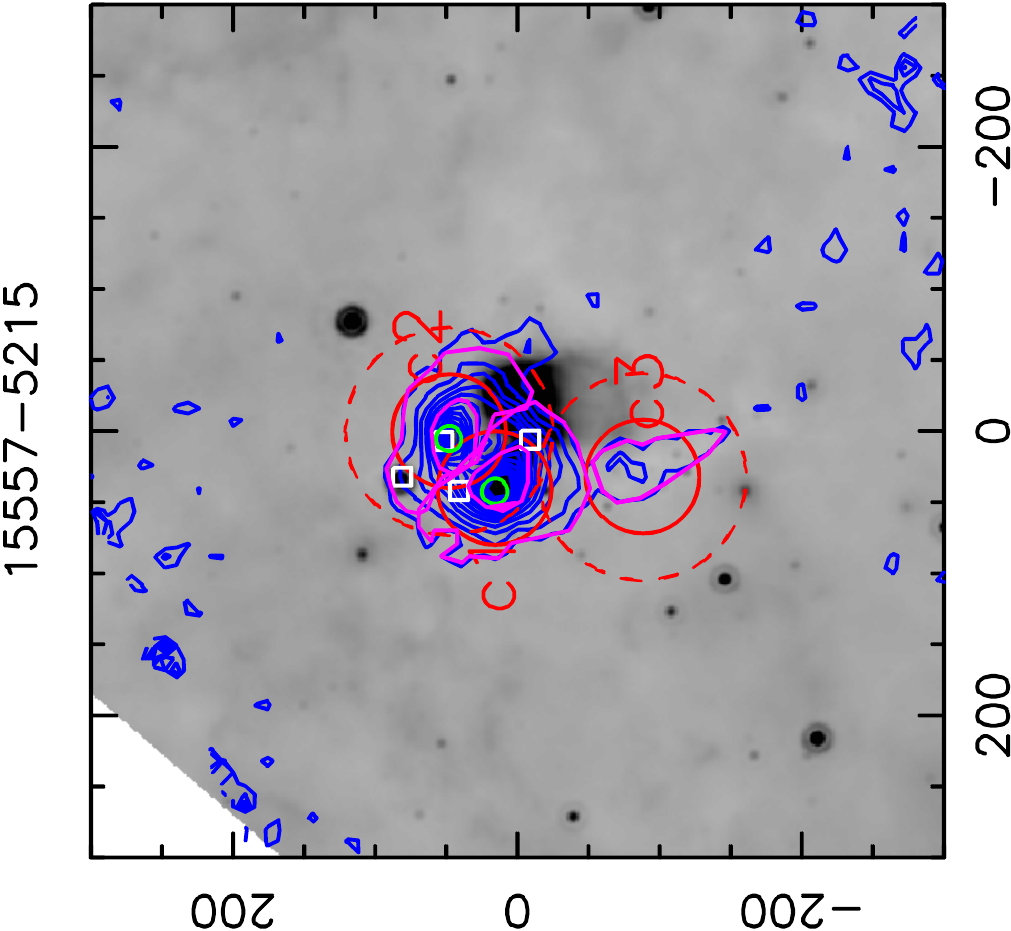}

 \includegraphics[angle=-90,width=0.29\textwidth]{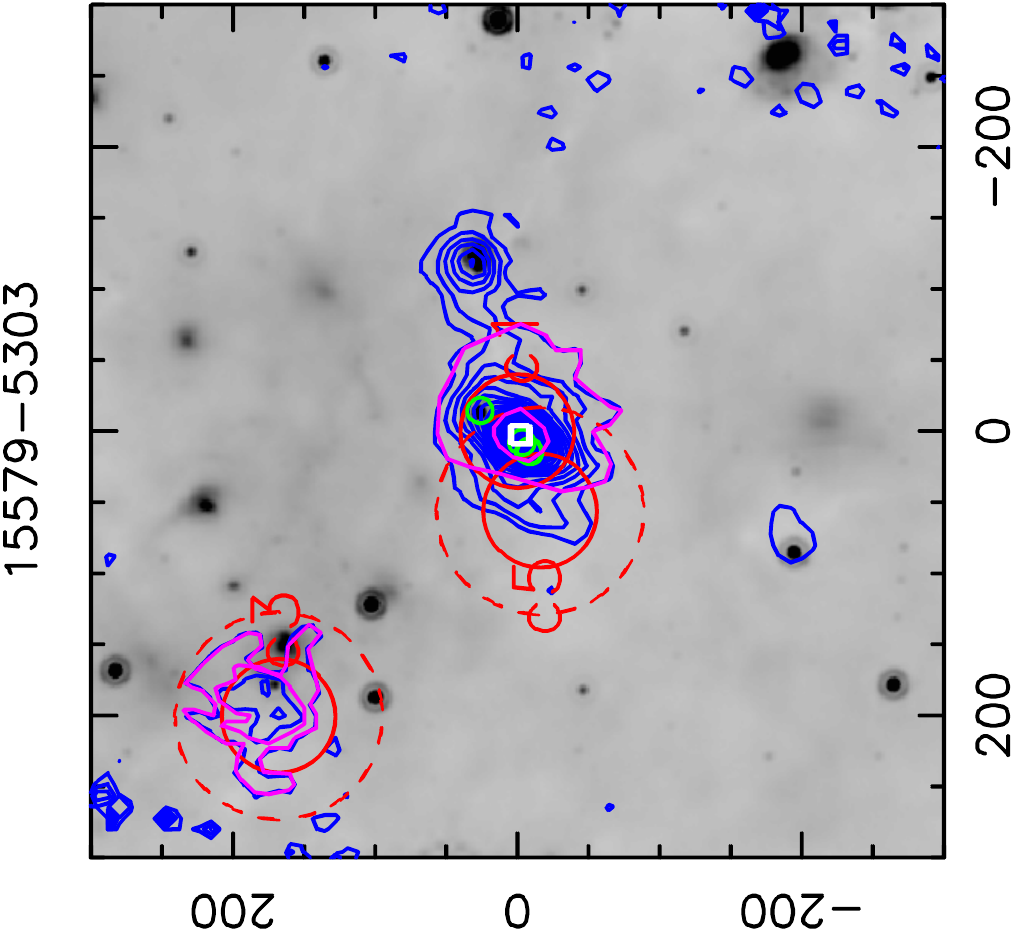}
 \includegraphics[angle=-90,width=0.29\textwidth]{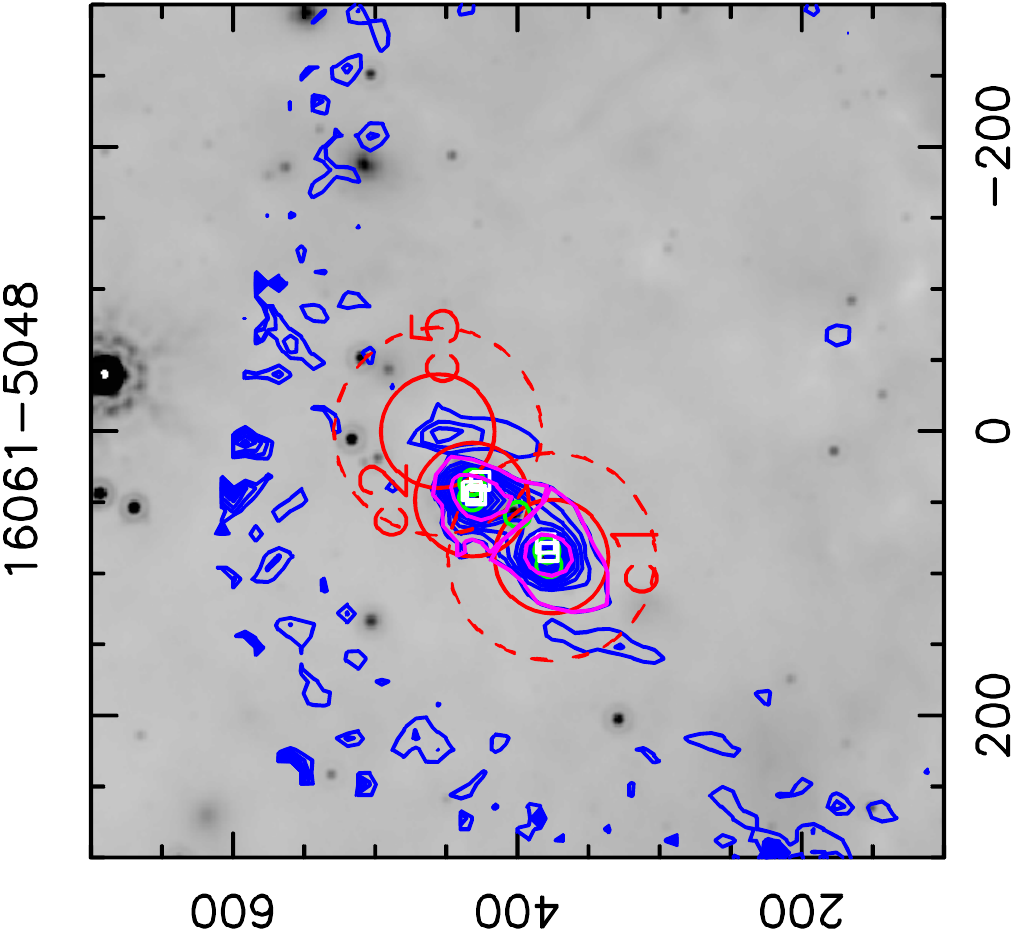}
 \includegraphics[angle=-90,width=0.29\textwidth]{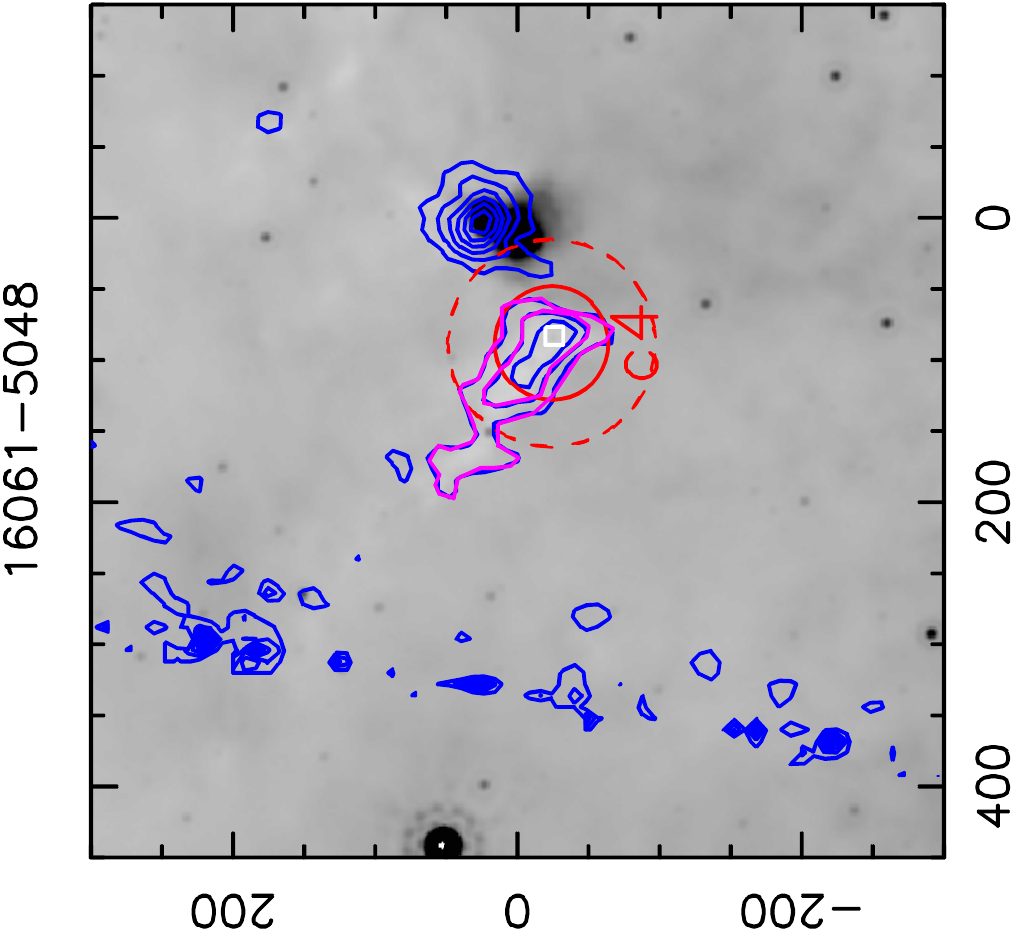}

 \includegraphics[angle=-90,width=0.29\textwidth]{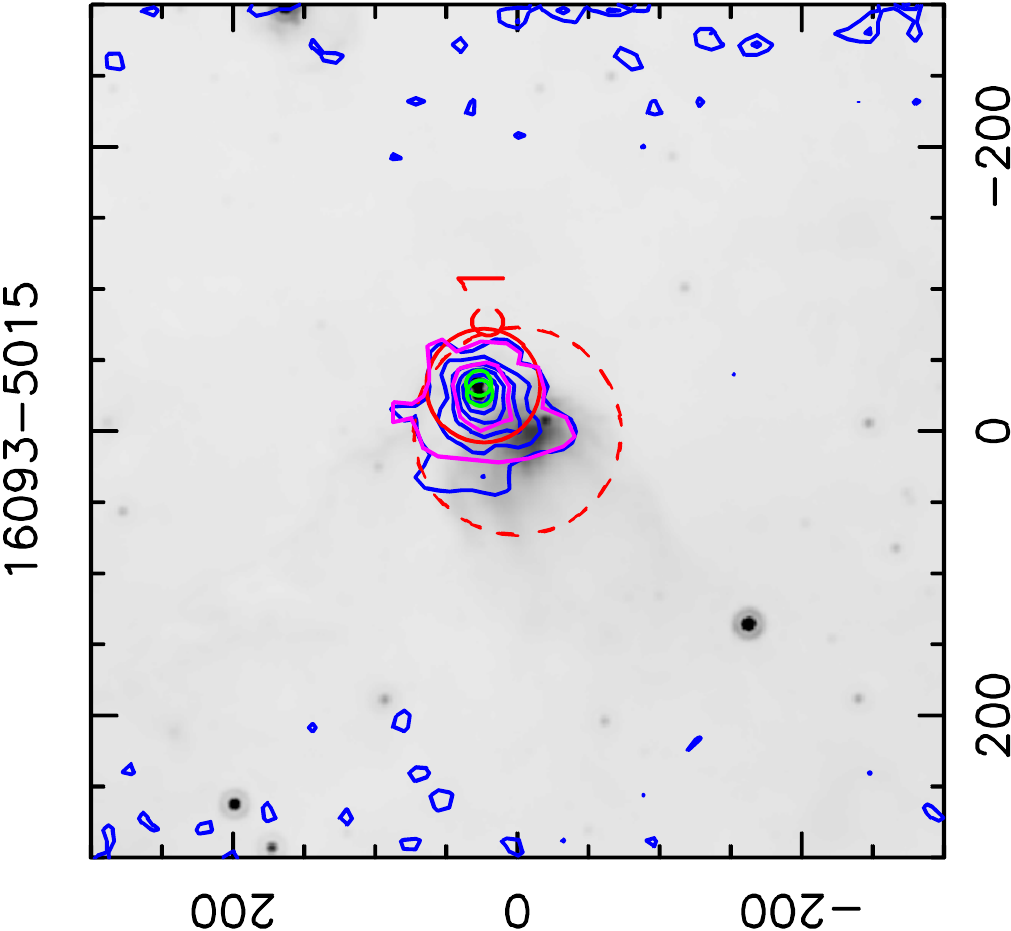}
 \includegraphics[angle=-90,width=0.29\textwidth]{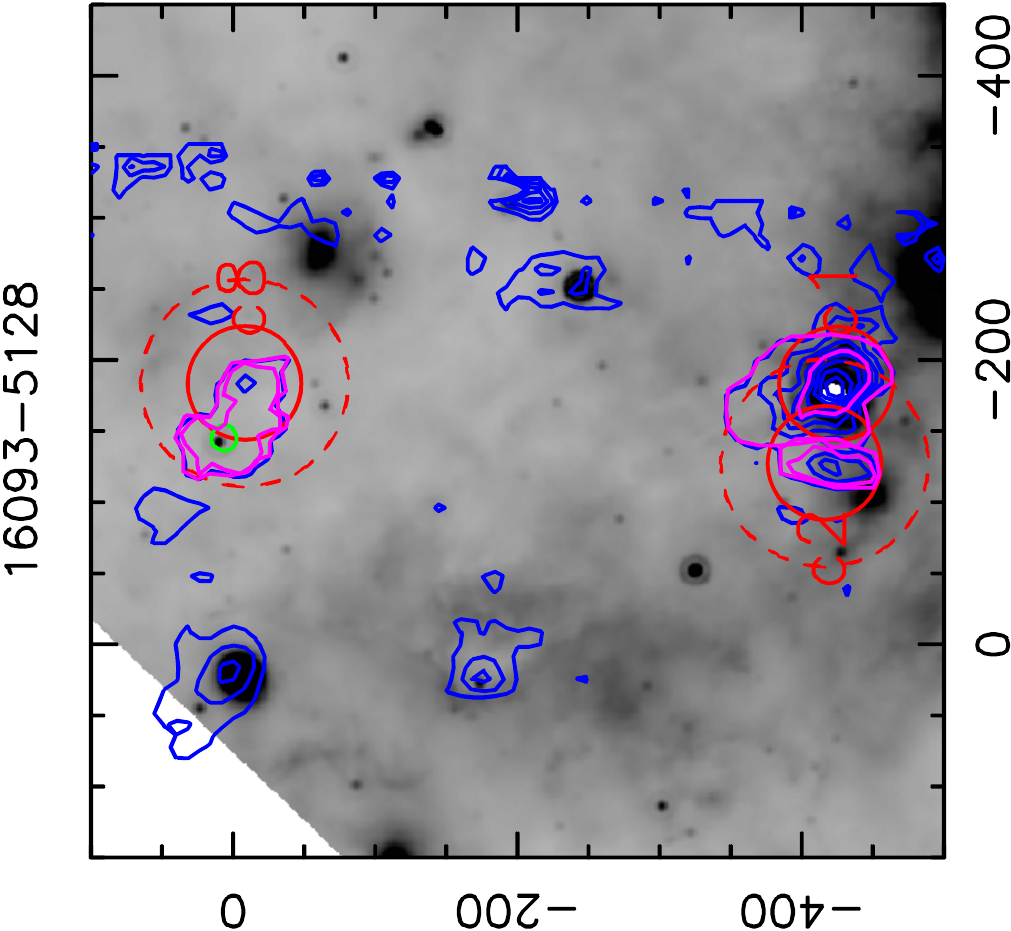}
 \includegraphics[angle=-90,width=0.29\textwidth]{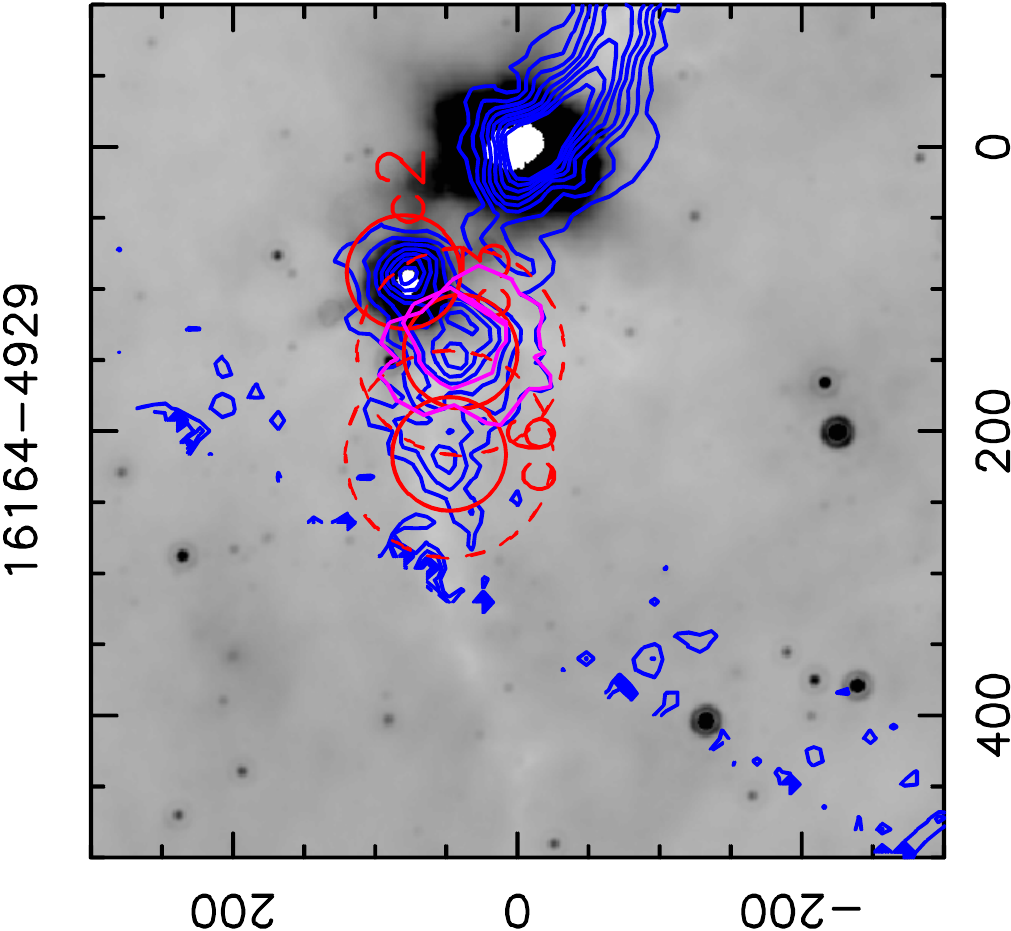}

 \includegraphics[angle=-90,width=0.29\textwidth]{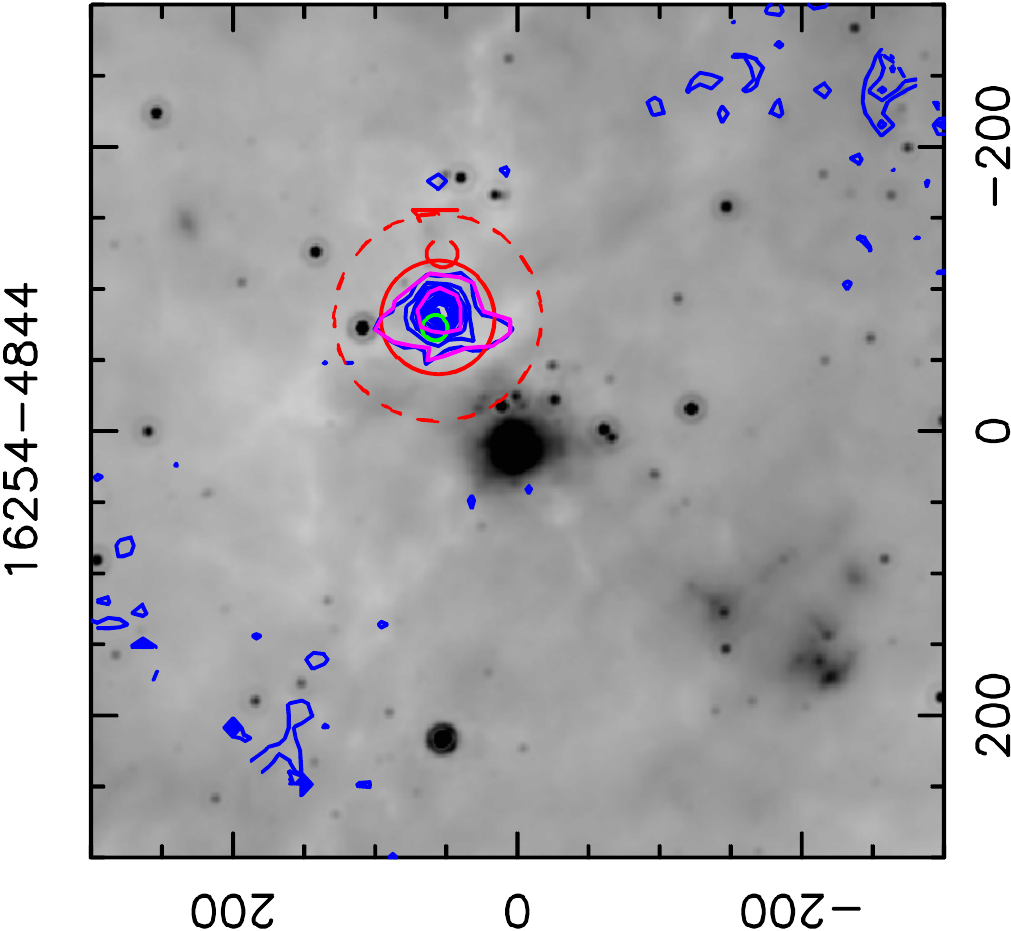}
 \includegraphics[angle=-90,width=0.29\textwidth]{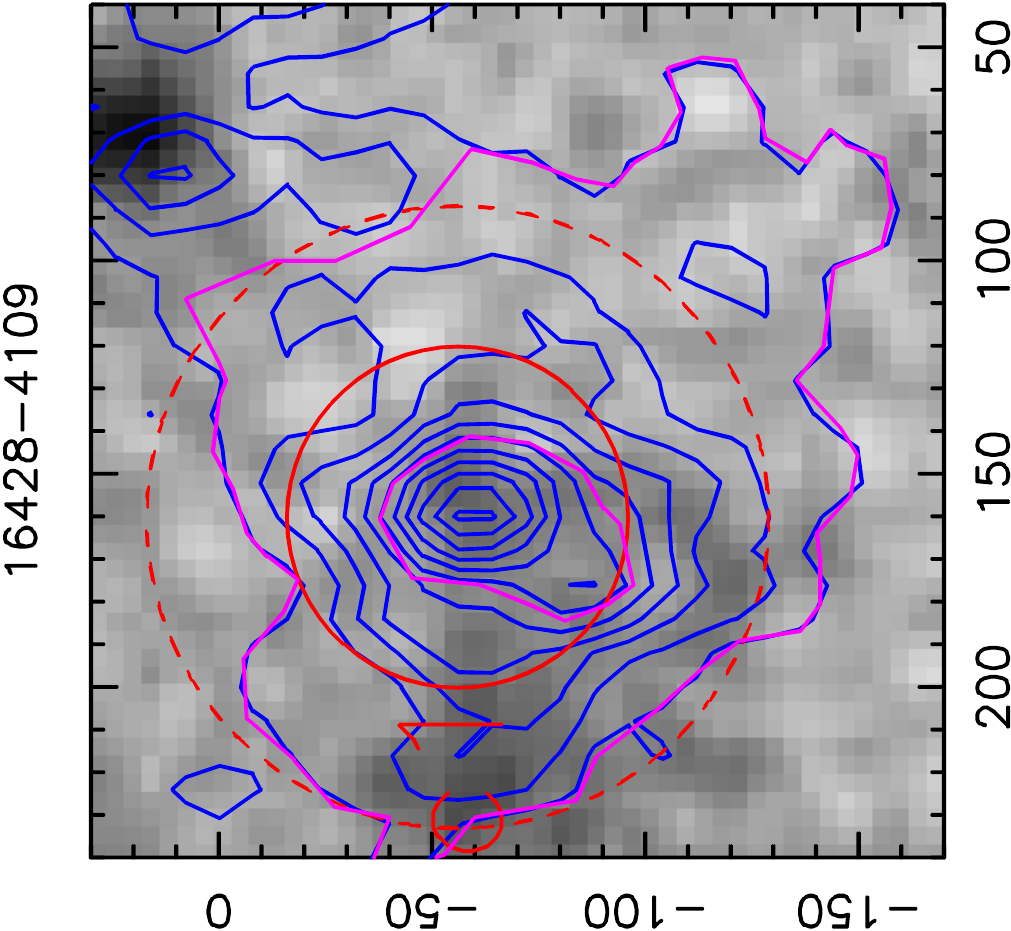}
 \includegraphics[angle=-90,width=0.29\textwidth]{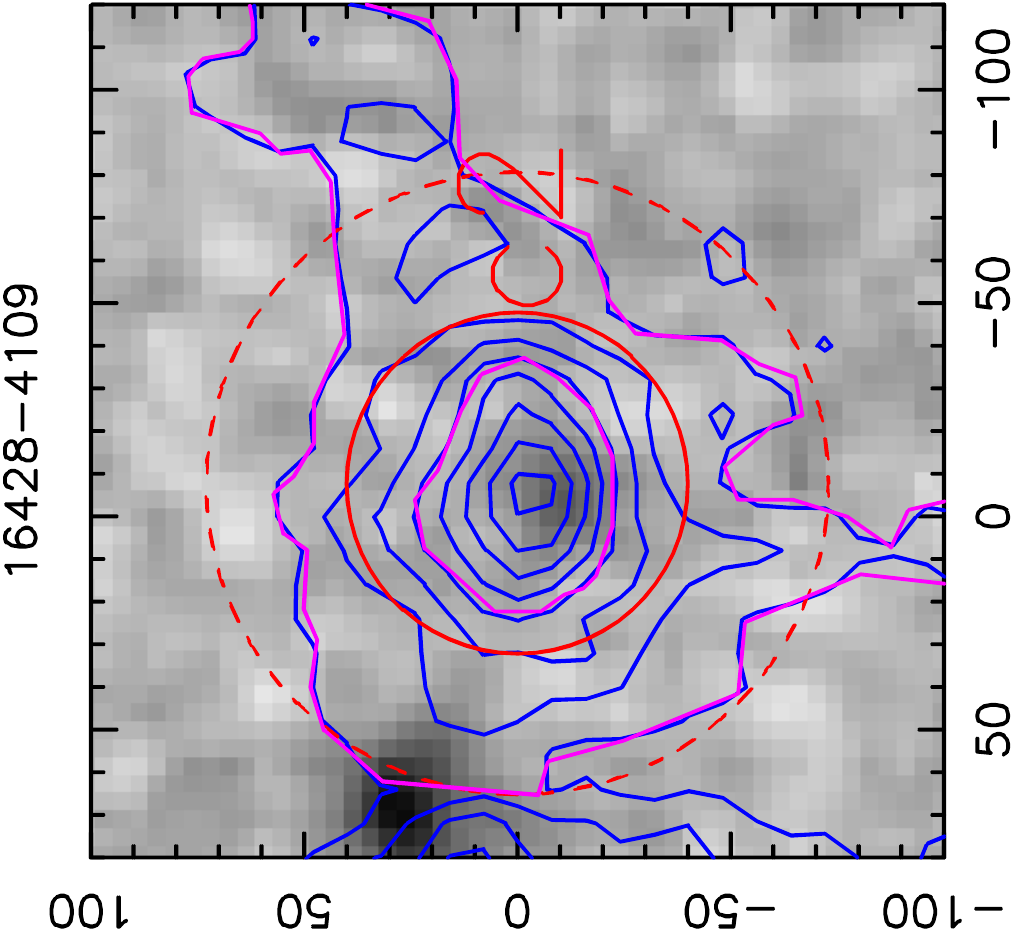}

 \caption{Continued.}
\end{figure*}

\begin{figure*}[tbp]
 \ContinuedFloat
 \centering
 \includegraphics[angle=-90,width=0.29\textwidth]{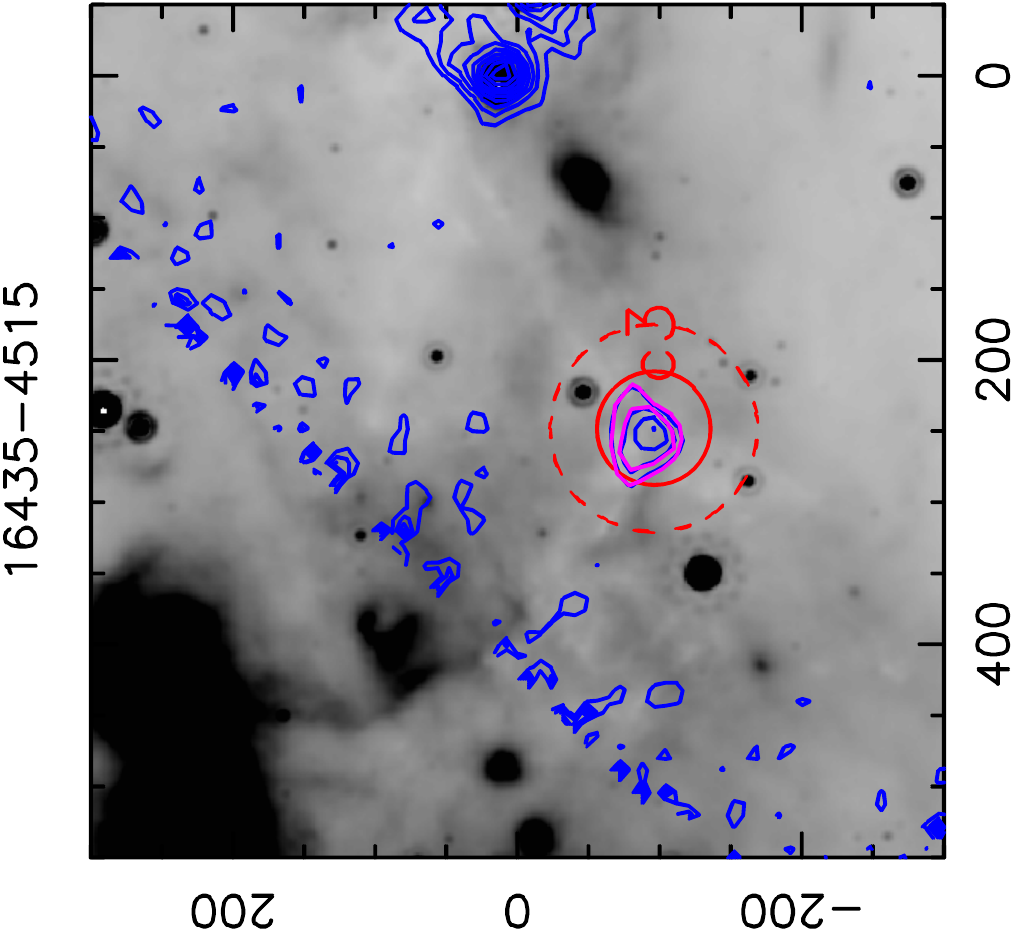}
 \includegraphics[angle=-90,width=0.29\textwidth]{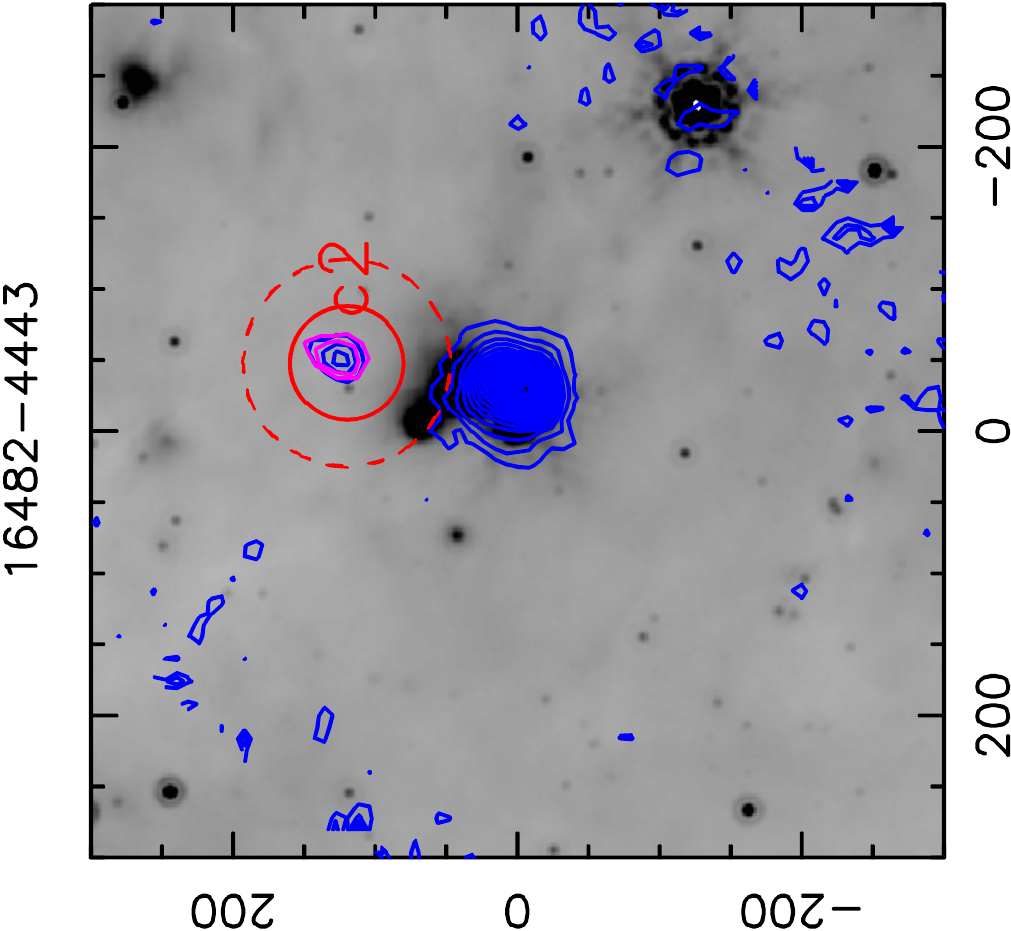}
 \includegraphics[angle=-90,width=0.29\textwidth]{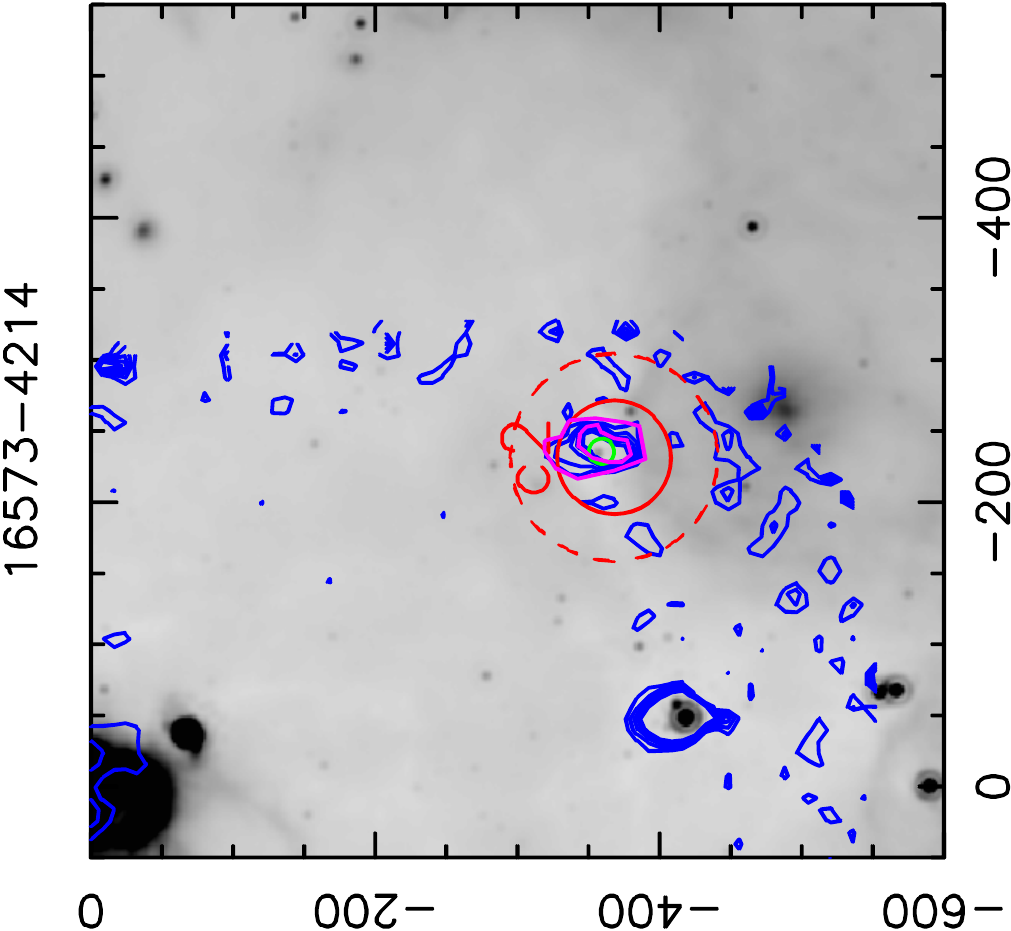}

 \includegraphics[angle=-90,width=0.29\textwidth]{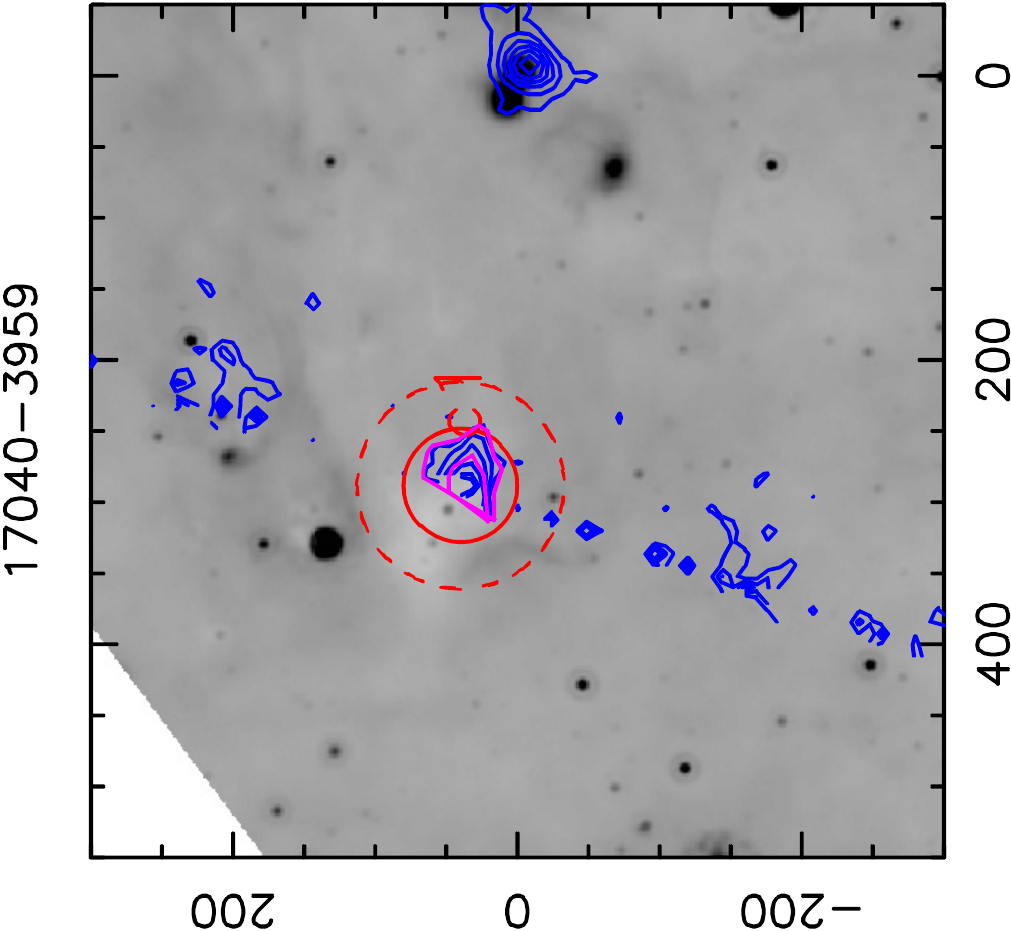}
 \includegraphics[angle=-90,width=0.29\textwidth]{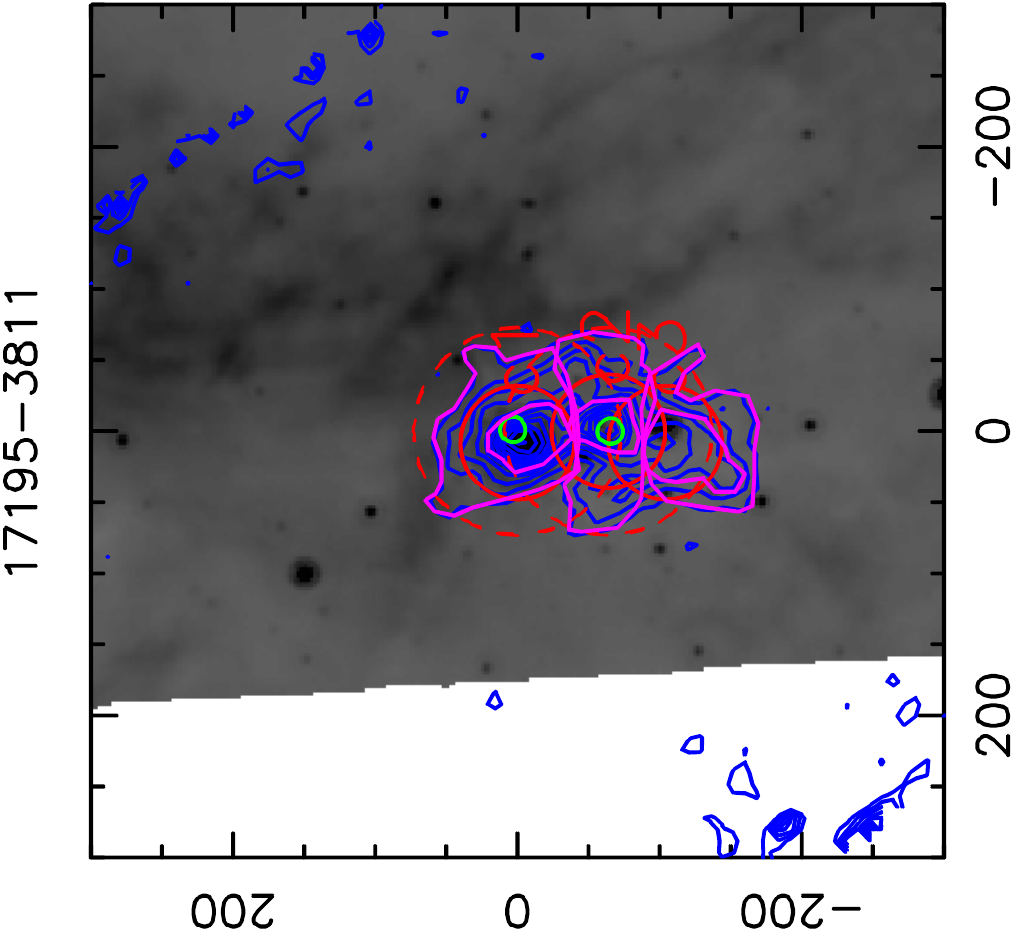}
 \includegraphics[angle=-90,width=0.29\textwidth]{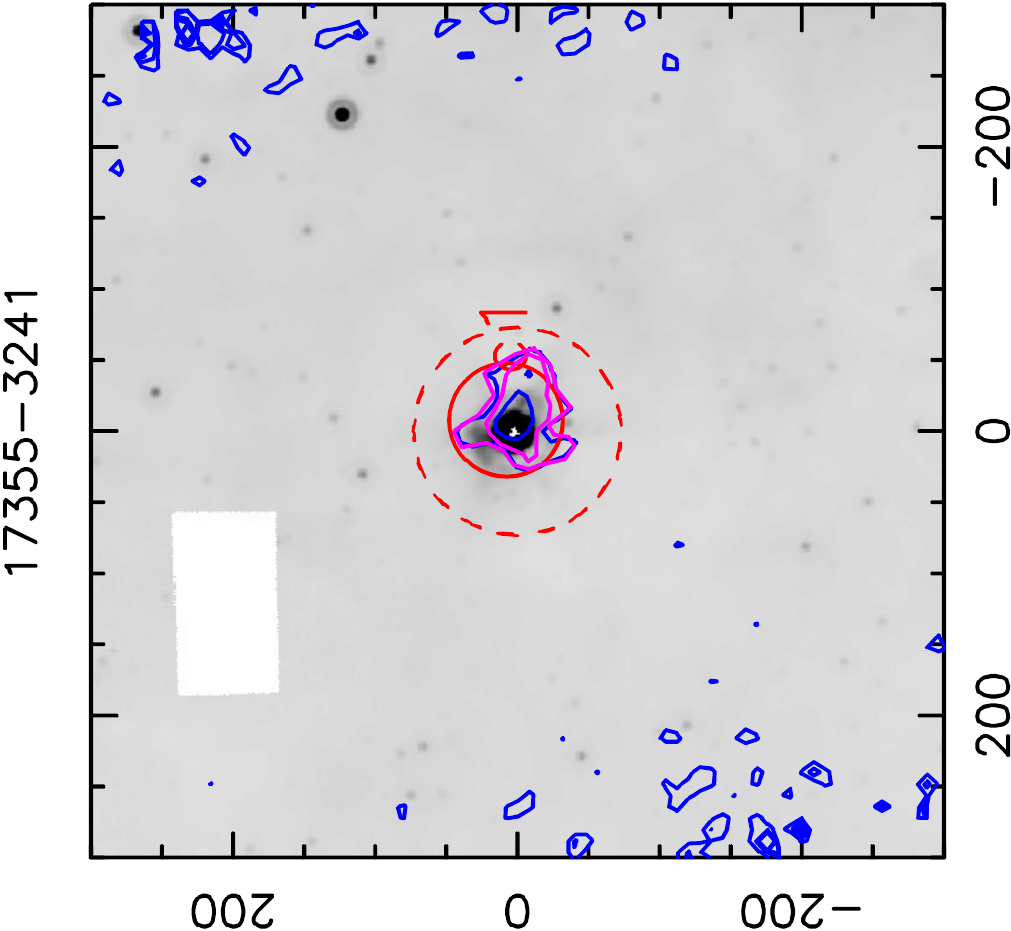}
 \caption{Continued.}
\end{figure*}

\clearpage

\begin{figure*}[tbp]
 \centering
 \includegraphics[angle=-90,width=0.24\textwidth]{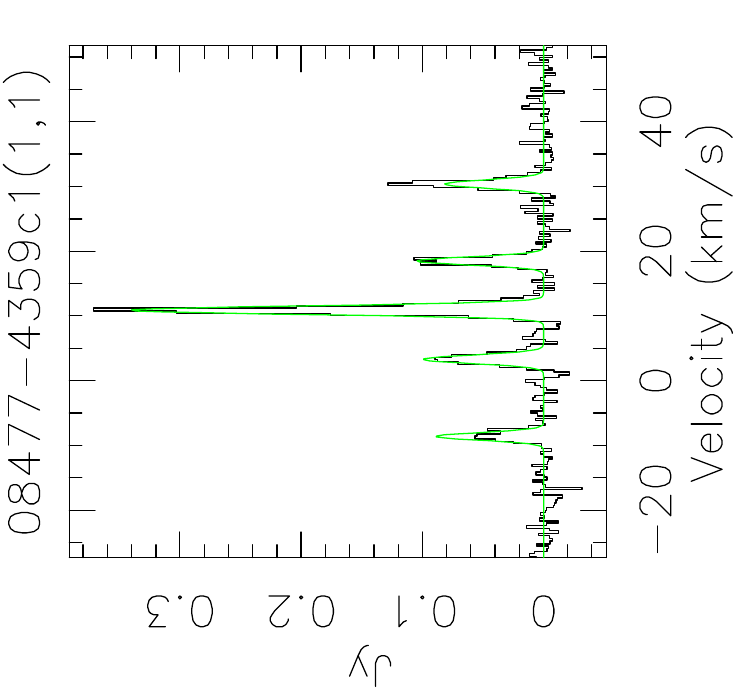}
 \includegraphics[angle=-90,width=0.24\textwidth]{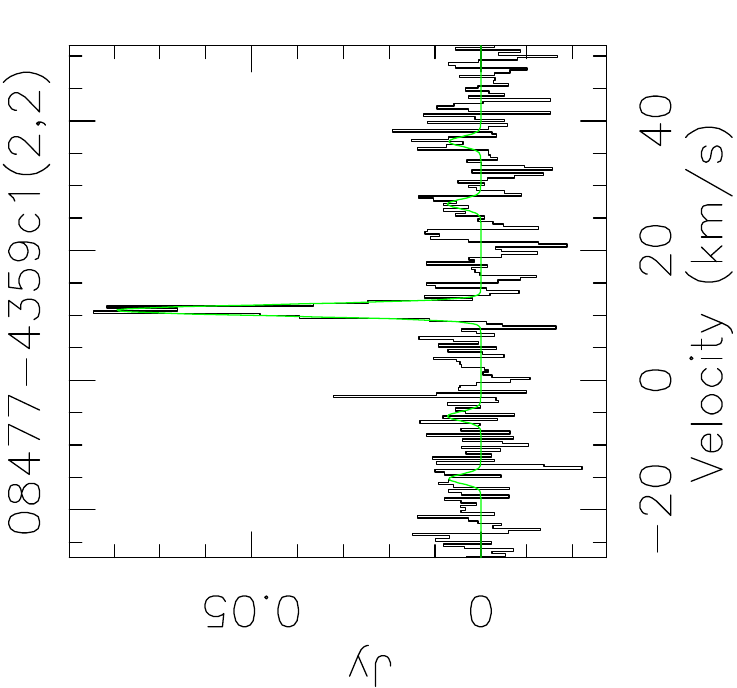}
 \includegraphics[angle=-90,width=0.24\textwidth]{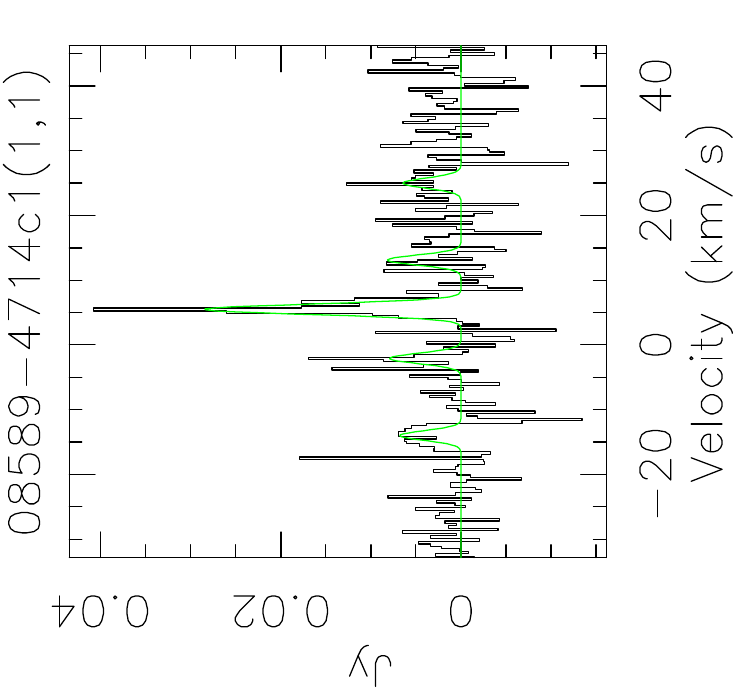}
 \includegraphics[angle=-90,width=0.24\textwidth]{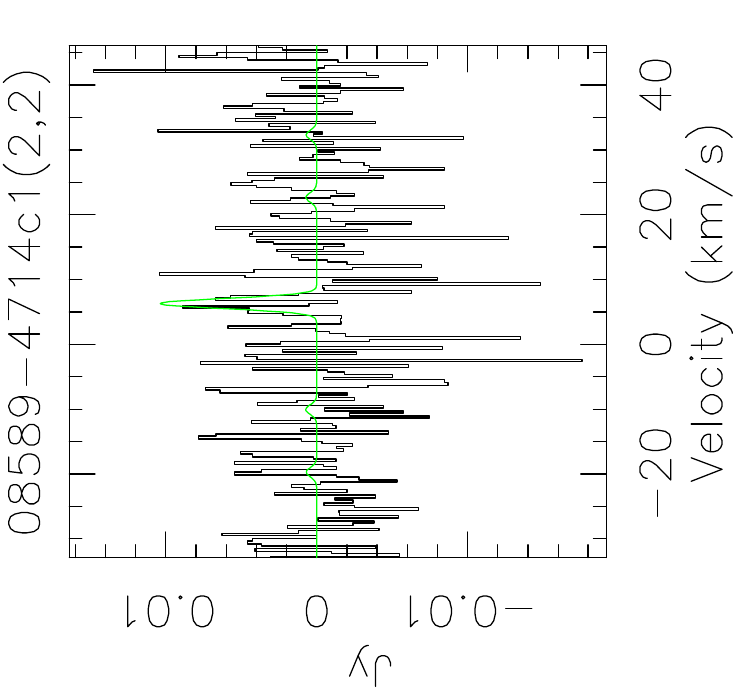}

 \includegraphics[angle=-90,width=0.24\textwidth]{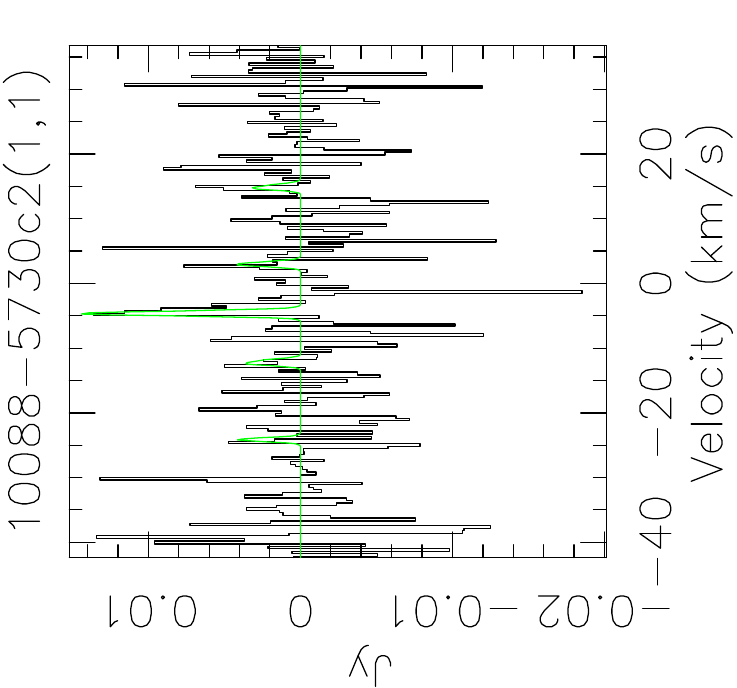}
 \includegraphics[angle=-90,width=0.24\textwidth]{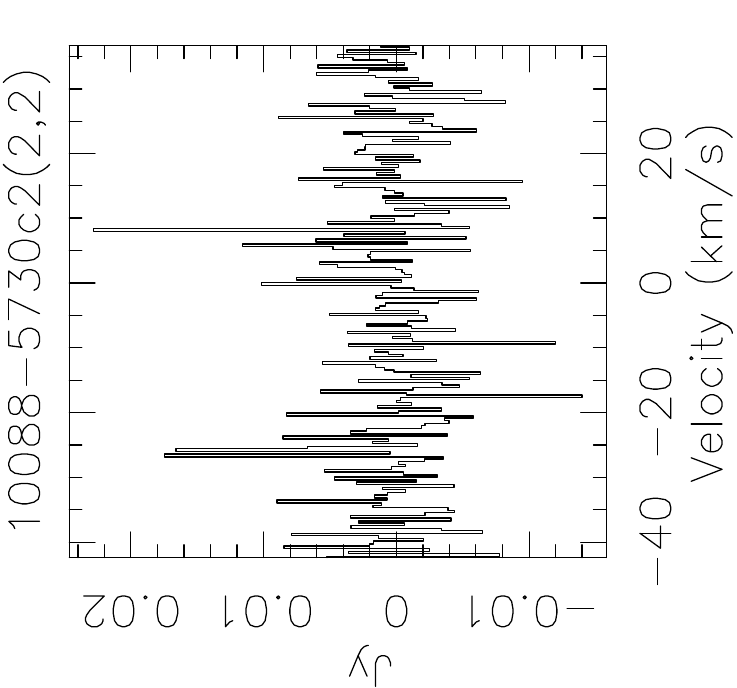}
 \includegraphics[angle=-90,width=0.24\textwidth]{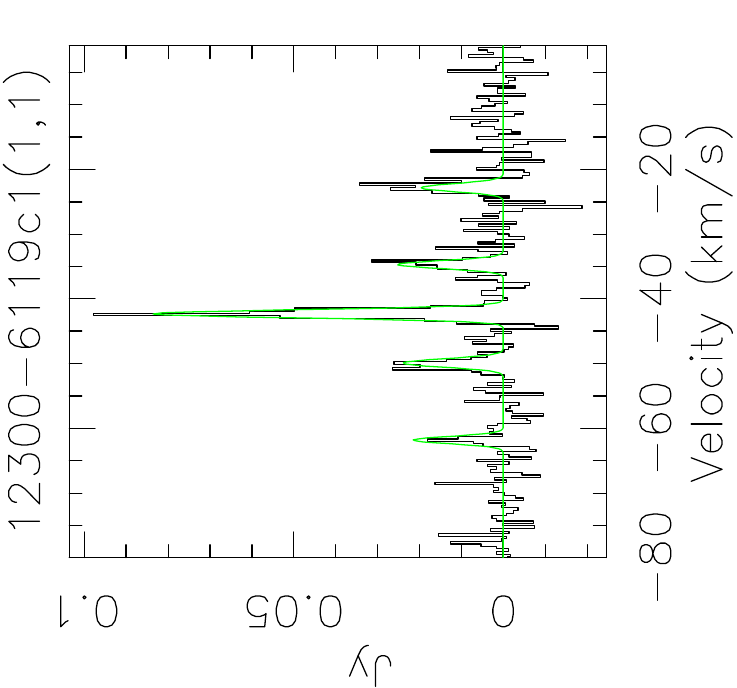}
 \includegraphics[angle=-90,width=0.24\textwidth]{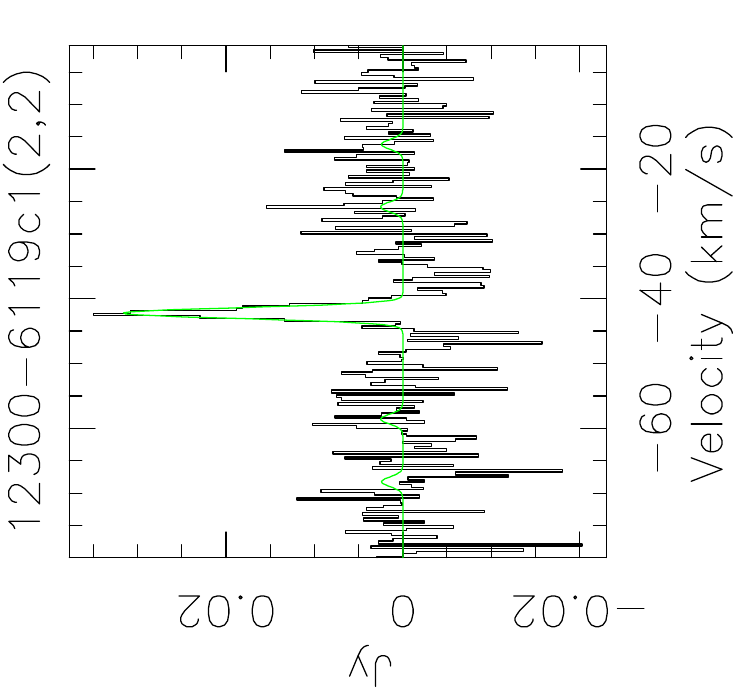}

 \includegraphics[angle=-90,width=0.24\textwidth]{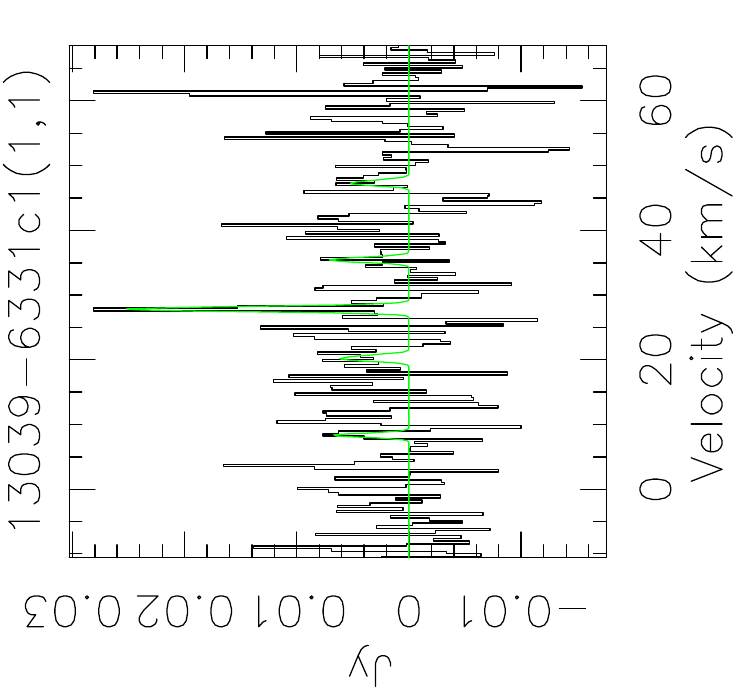}
 \includegraphics[angle=-90,width=0.24\textwidth]{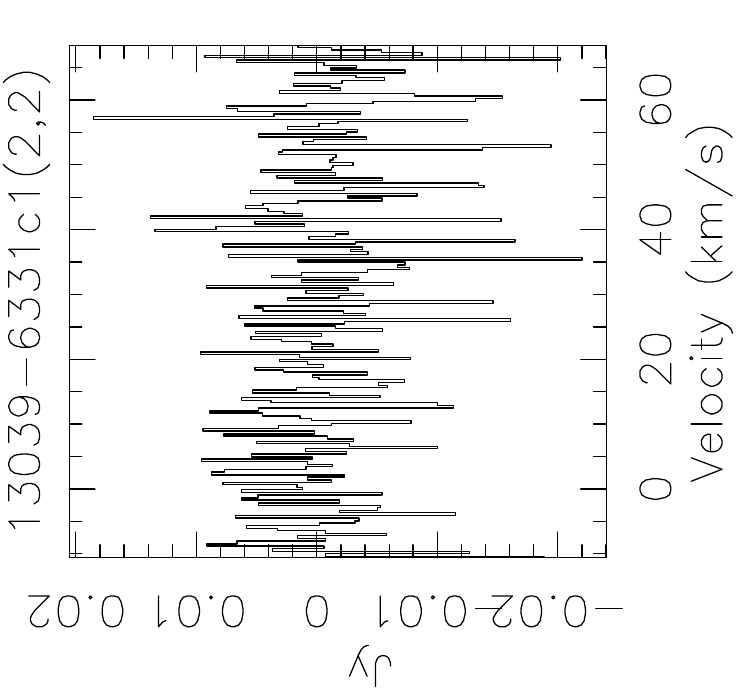}
 \includegraphics[angle=-90,width=0.24\textwidth]{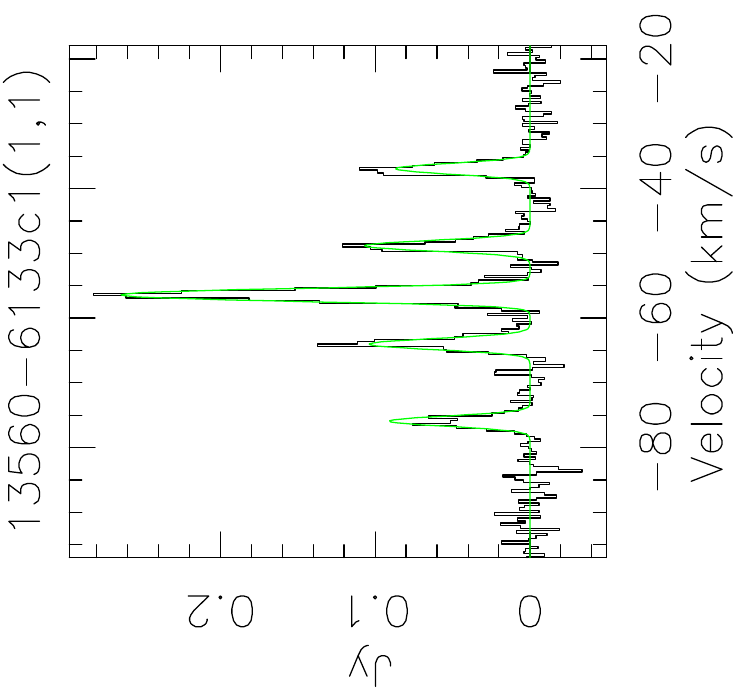}
 \includegraphics[angle=-90,width=0.24\textwidth]{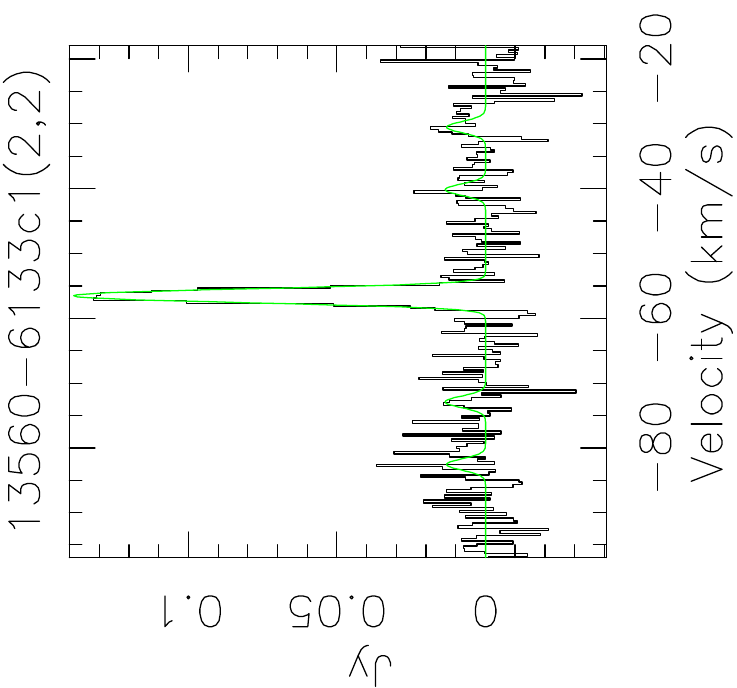}

 \includegraphics[angle=-90,width=0.24\textwidth]{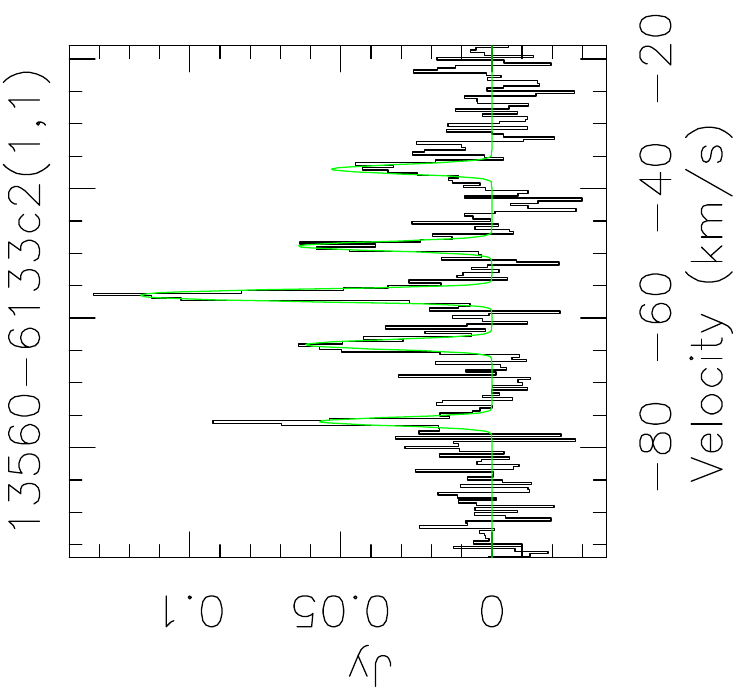}
 \includegraphics[angle=-90,width=0.24\textwidth]{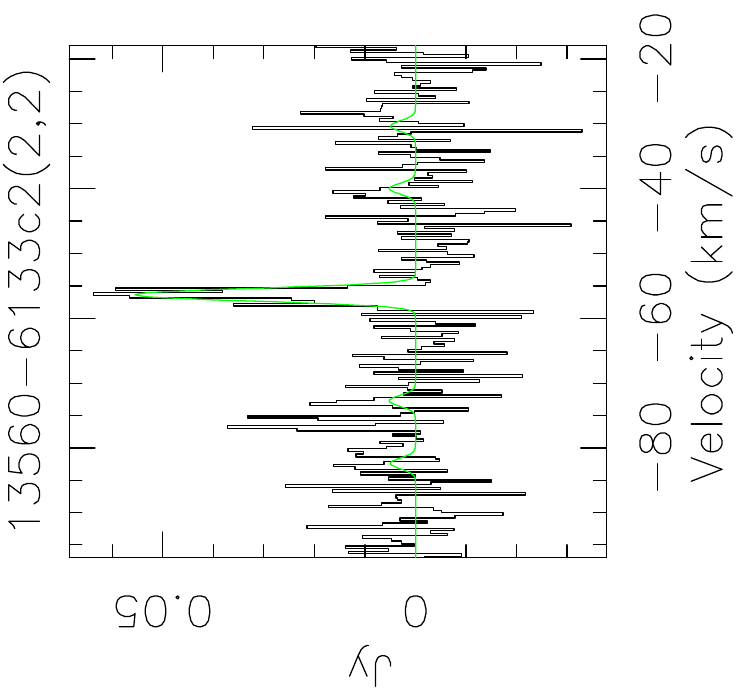}
 \includegraphics[angle=-90,width=0.24\textwidth]{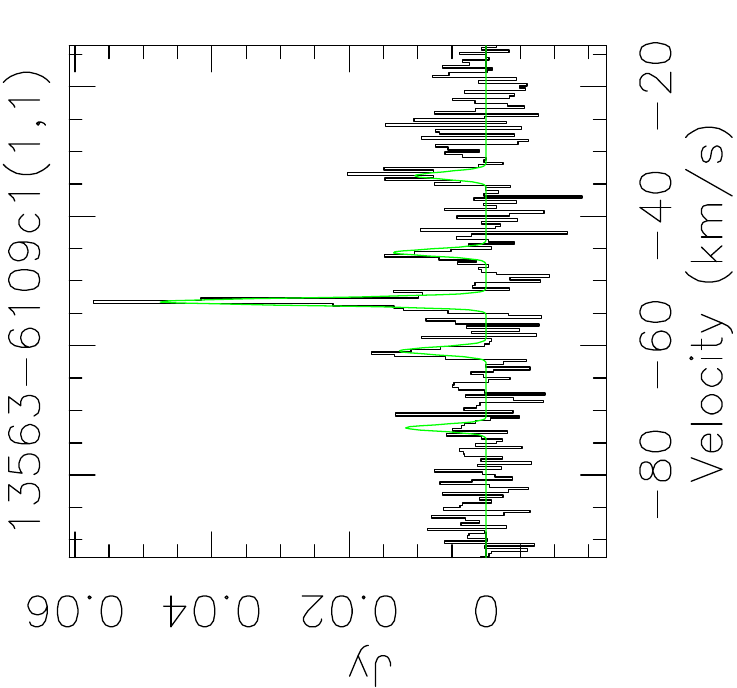}
 \includegraphics[angle=-90,width=0.24\textwidth]{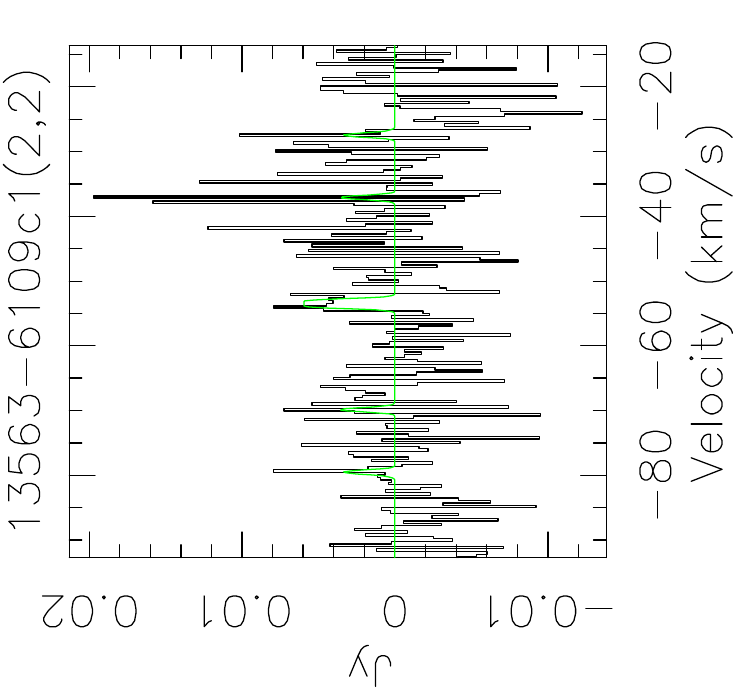}

 \includegraphics[angle=-90,width=0.24\textwidth]{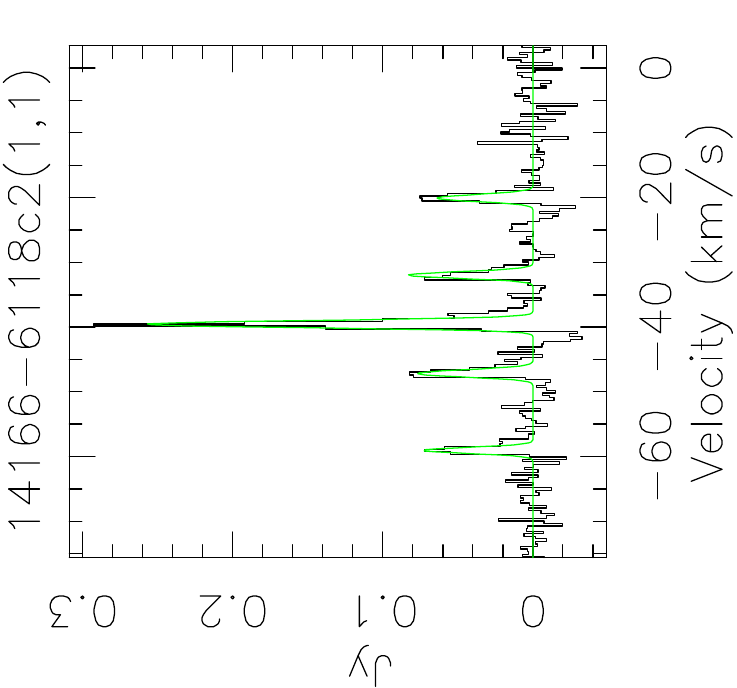}
 \includegraphics[angle=-90,width=0.24\textwidth]{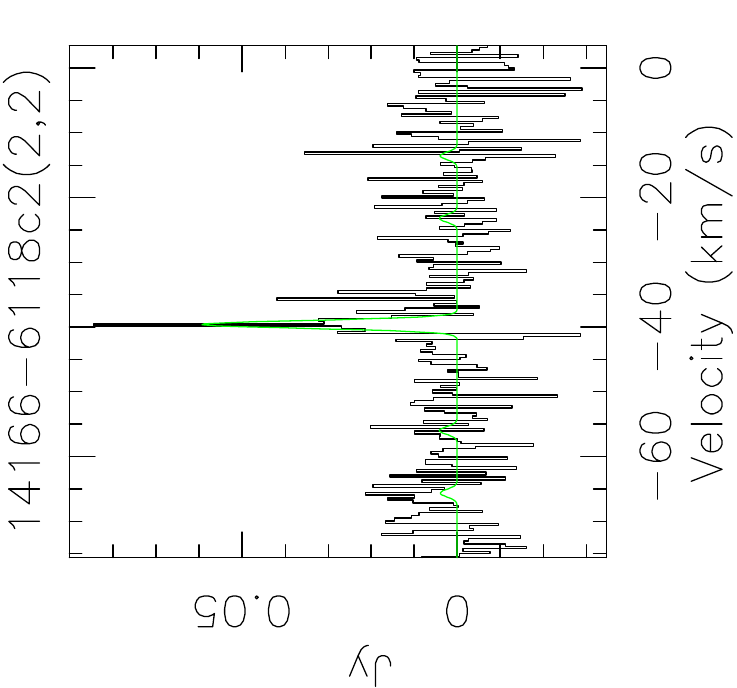}
 \includegraphics[angle=-90,width=0.24\textwidth]{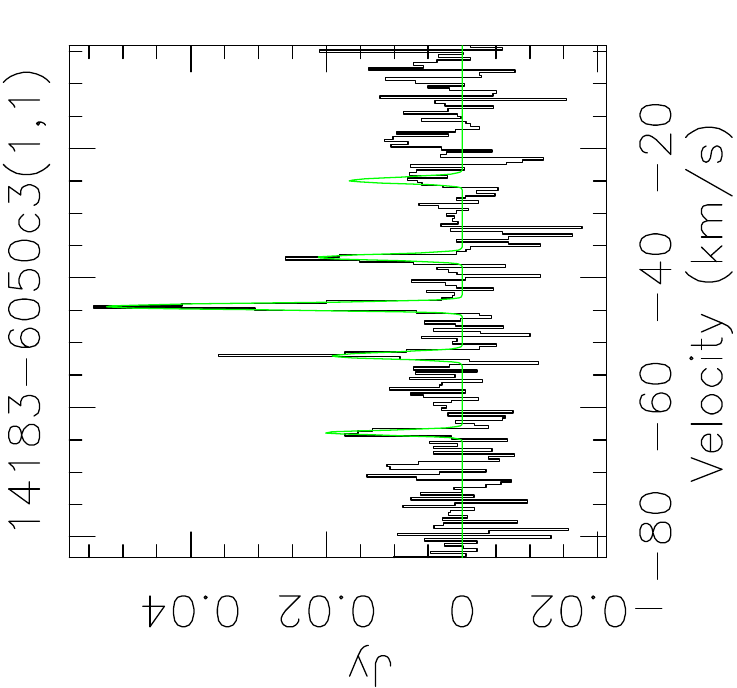}
 \includegraphics[angle=-90,width=0.24\textwidth]{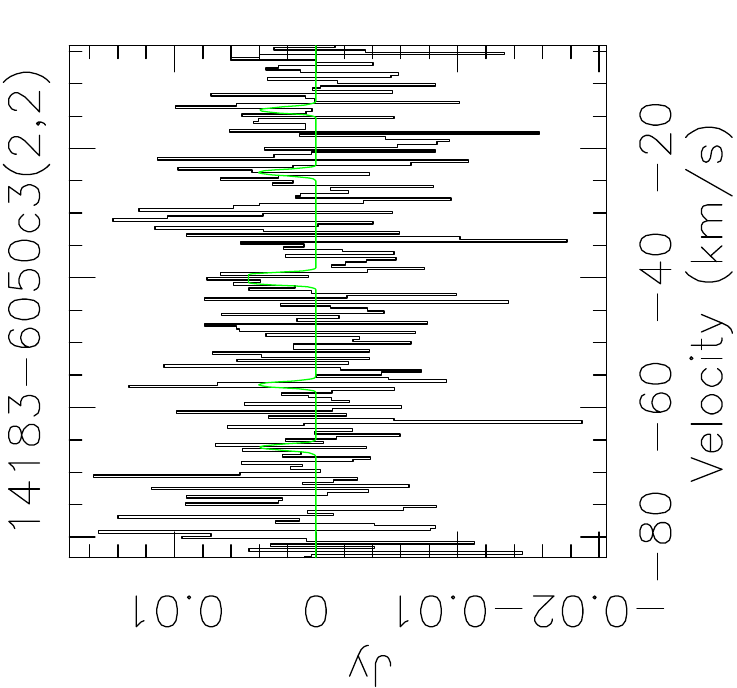}

 \caption{NH$_3$(1,1) and (2,2) spectra extracted at the peak of NH3(1,1) emission for all clumps. The clump name and transition are indicated above each panel.}
 \label{fig:nh3_spec}
\end{figure*}

\begin{figure*}[tbp]
 \ContinuedFloat
 \centering
 \includegraphics[angle=-90,width=0.24\textwidth]{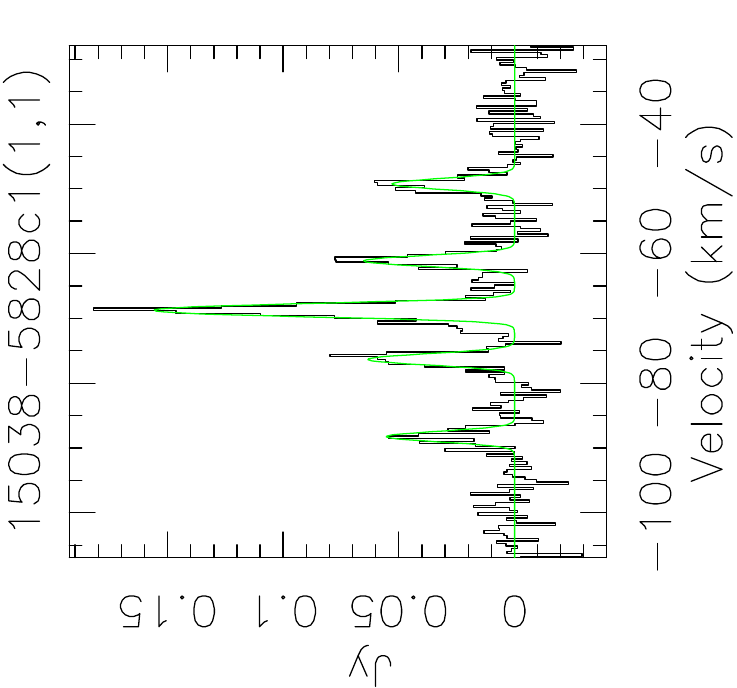}
 \includegraphics[angle=-90,width=0.24\textwidth]{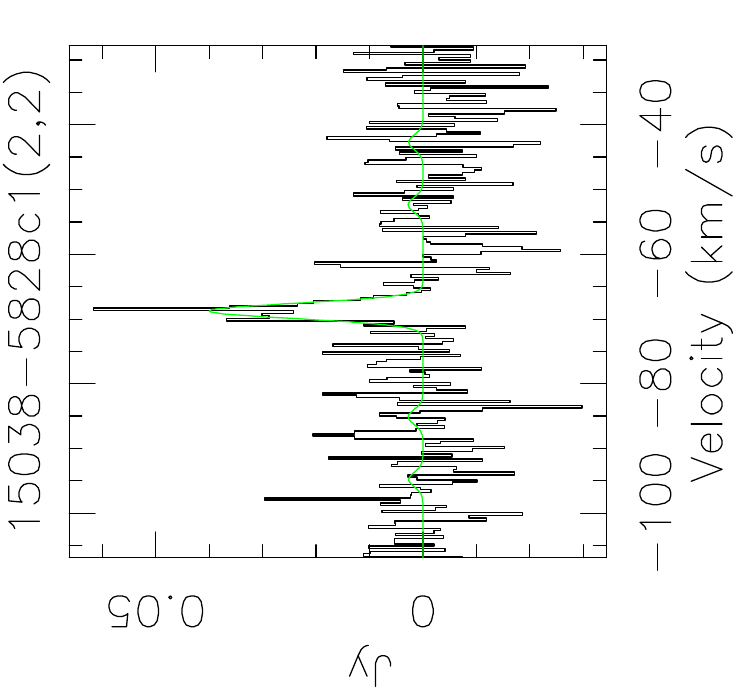}
 \includegraphics[angle=-90,width=0.24\textwidth]{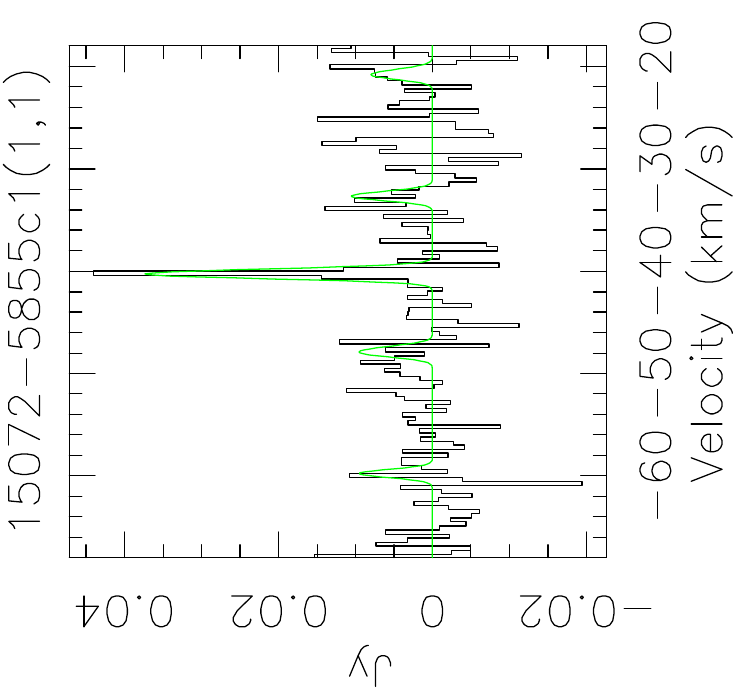}
 \includegraphics[angle=-90,width=0.24\textwidth]{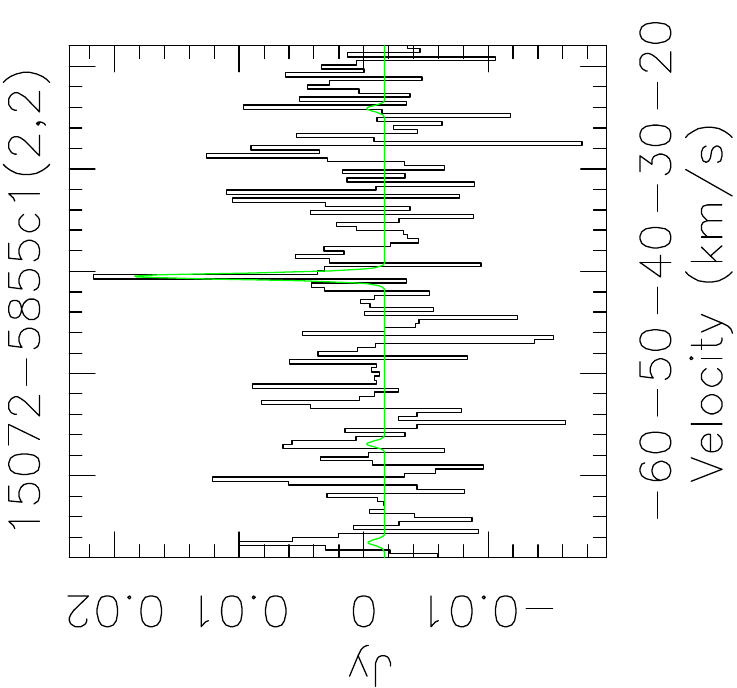}

 \includegraphics[angle=-90,width=0.24\textwidth]{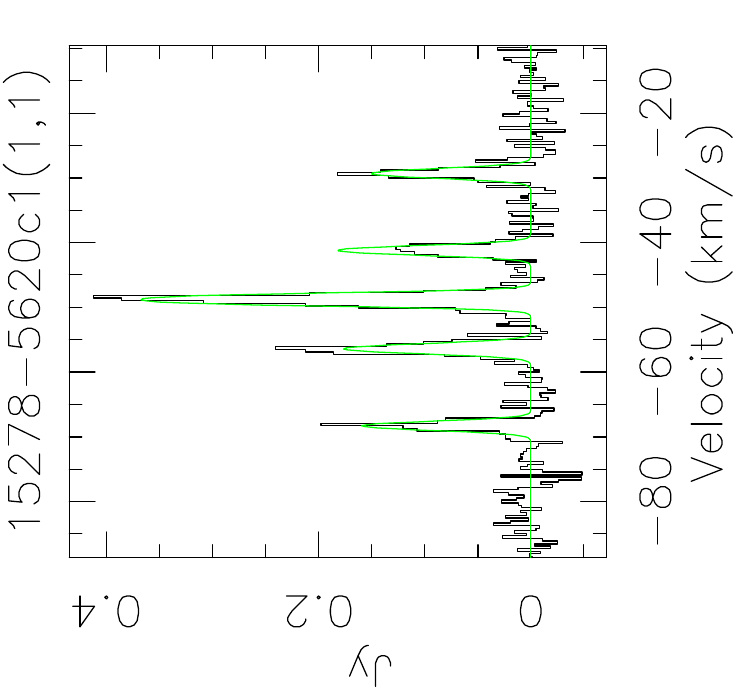}
 \includegraphics[angle=-90,width=0.24\textwidth]{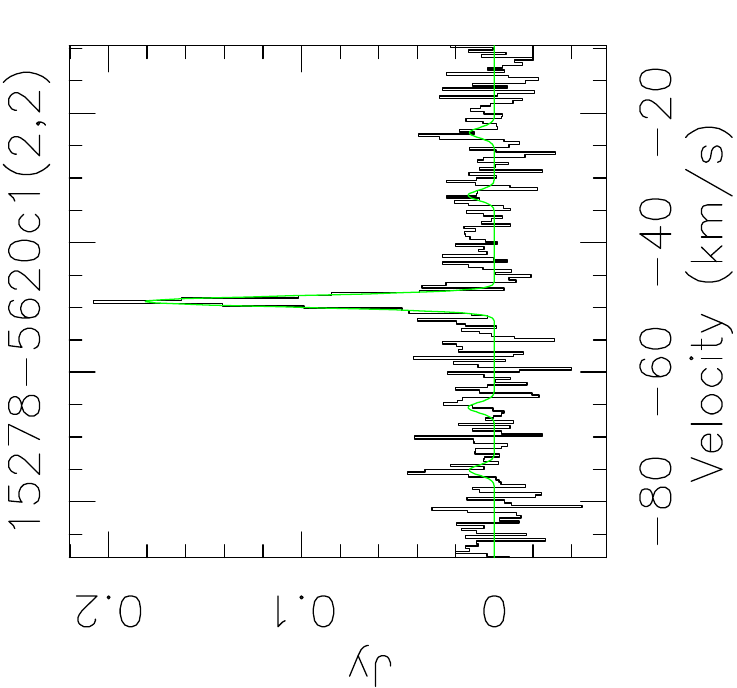}
 \includegraphics[angle=-90,width=0.24\textwidth]{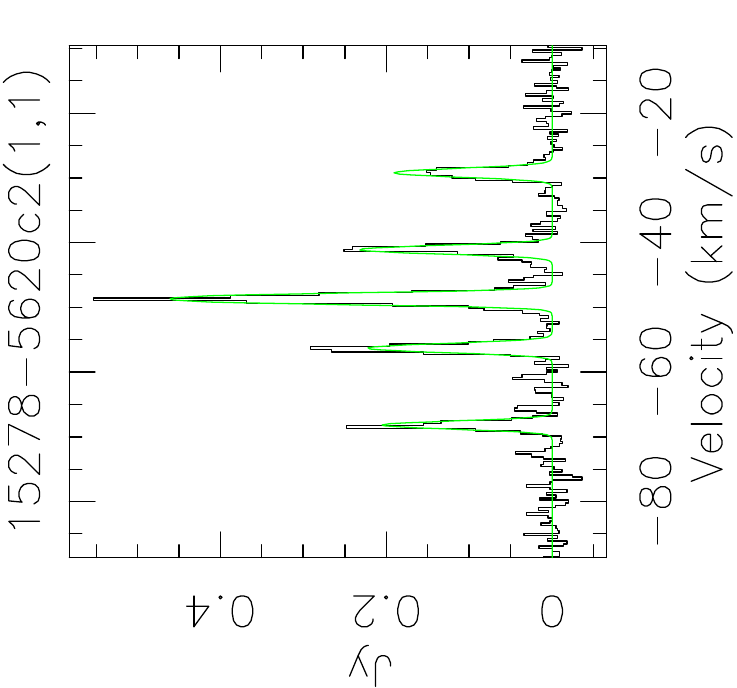}
 \includegraphics[angle=-90,width=0.24\textwidth]{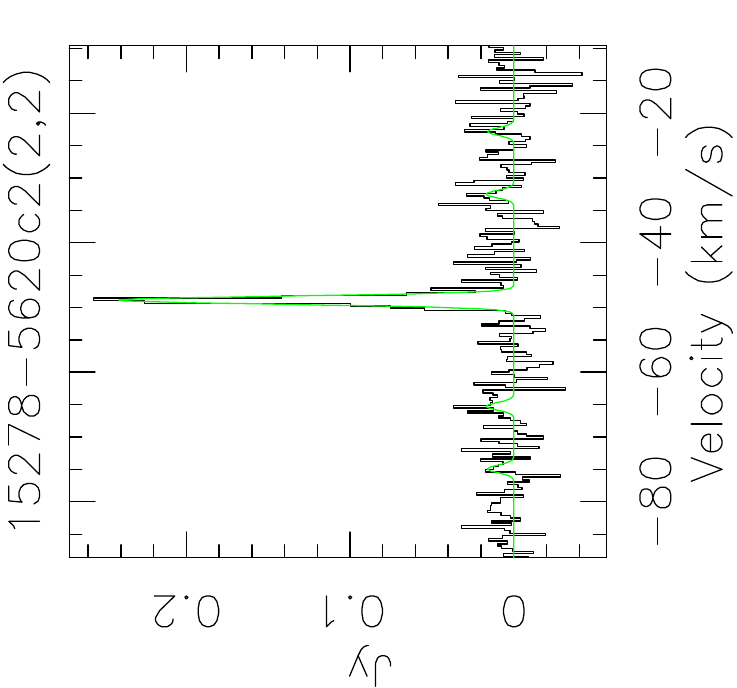}

 \includegraphics[angle=-90,width=0.24\textwidth]{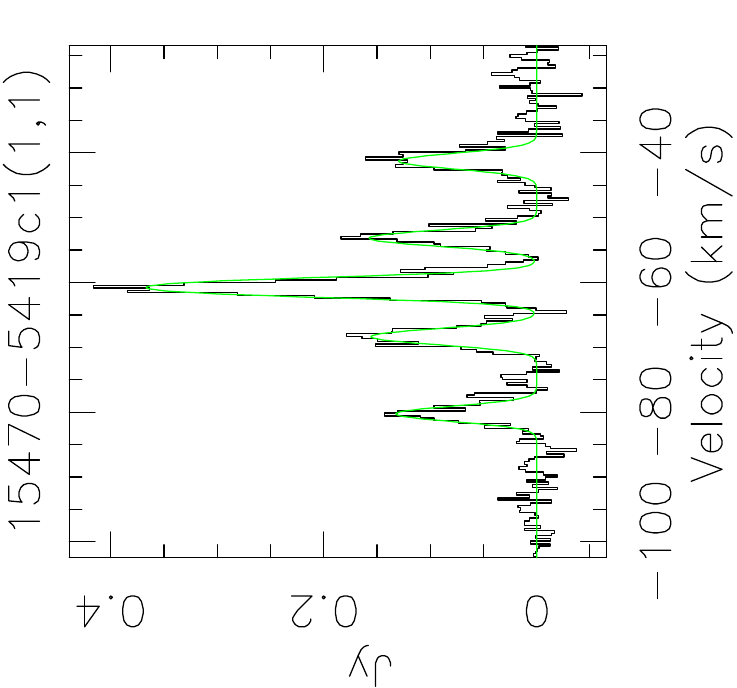}
 \includegraphics[angle=-90,width=0.24\textwidth]{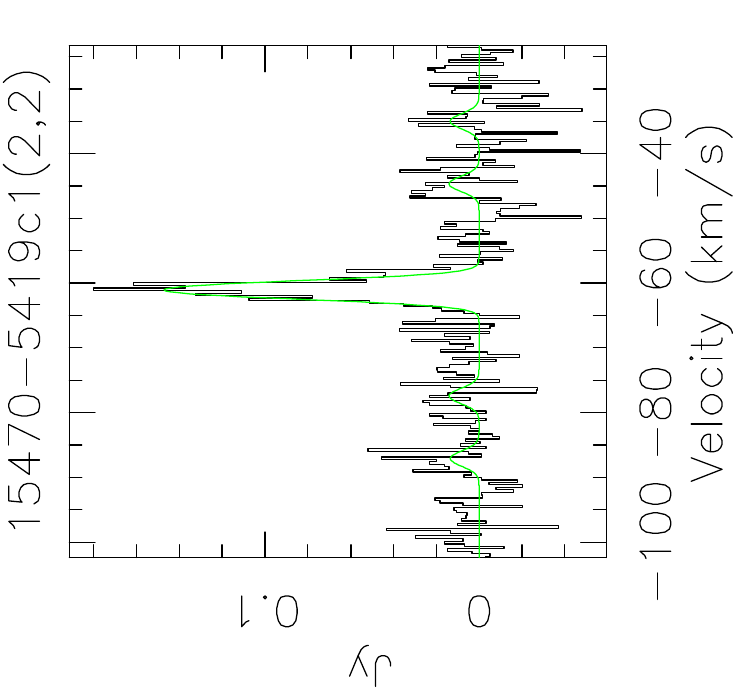}
 \includegraphics[angle=-90,width=0.24\textwidth]{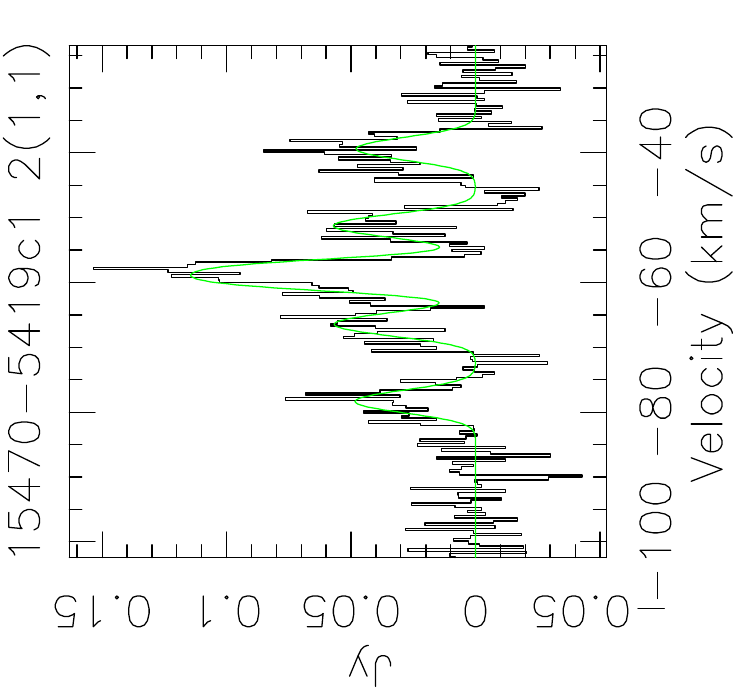}
 \includegraphics[angle=-90,width=0.24\textwidth]{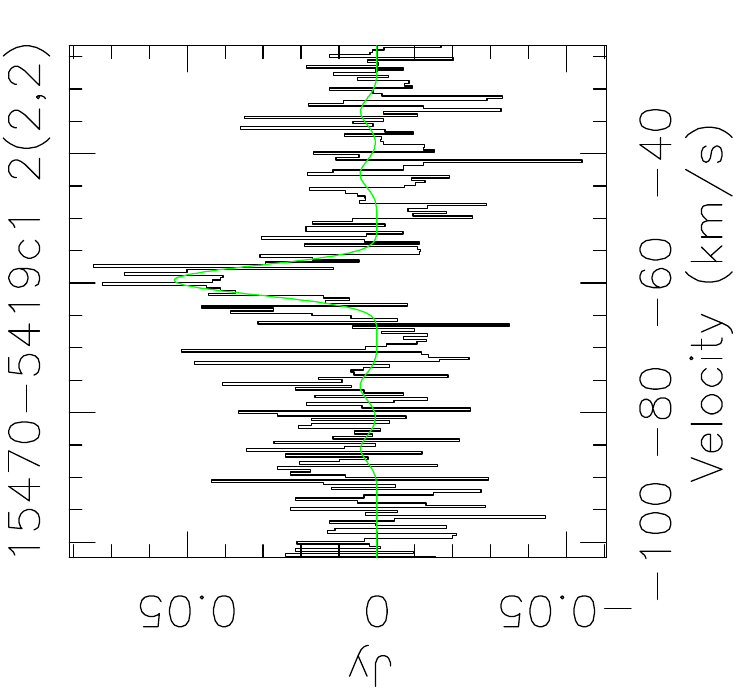}

 \includegraphics[angle=-90,width=0.24\textwidth]{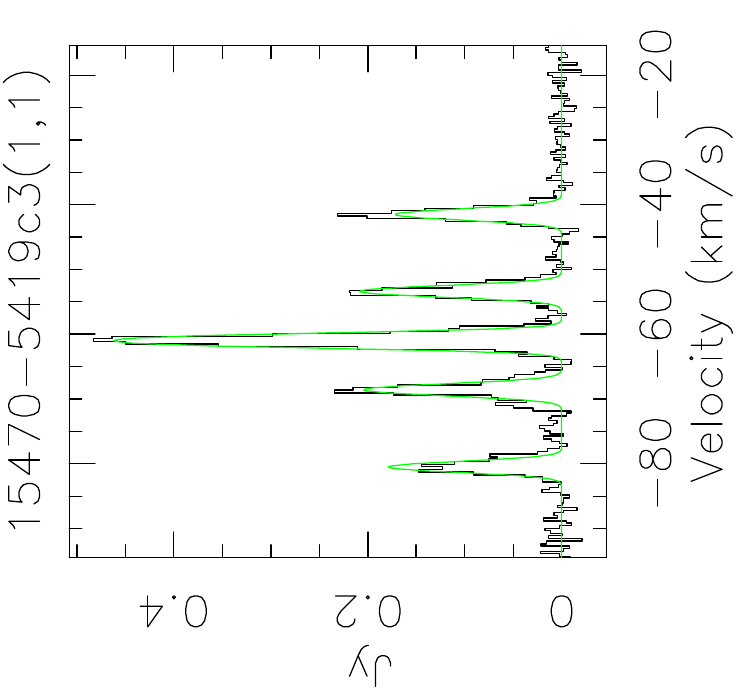}
 \includegraphics[angle=-90,width=0.24\textwidth]{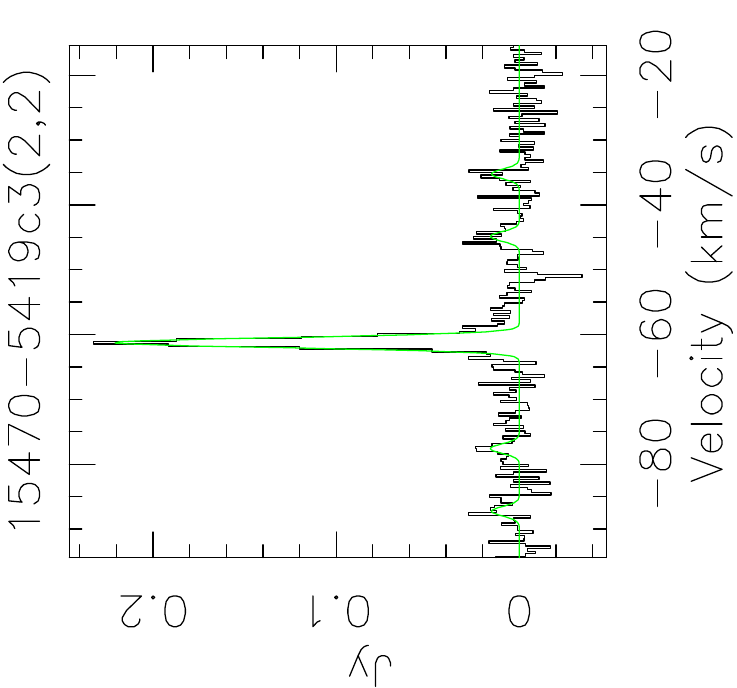}
 \includegraphics[angle=-90,width=0.24\textwidth]{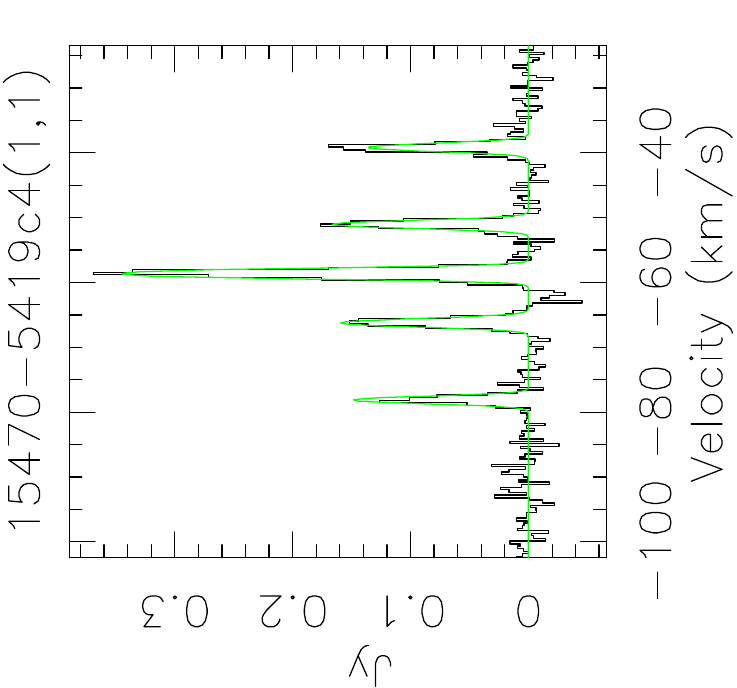}
 \includegraphics[angle=-90,width=0.24\textwidth]{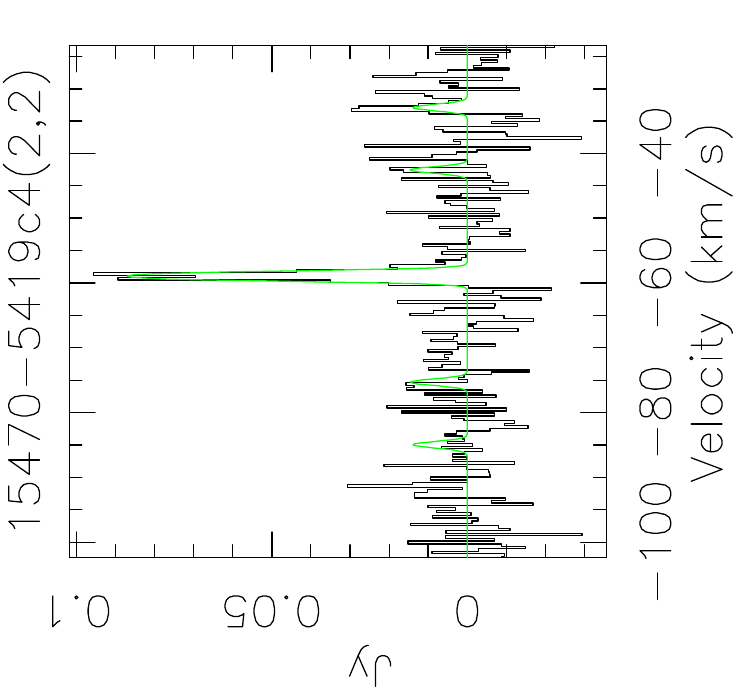}

 \includegraphics[angle=-90,width=0.24\textwidth]{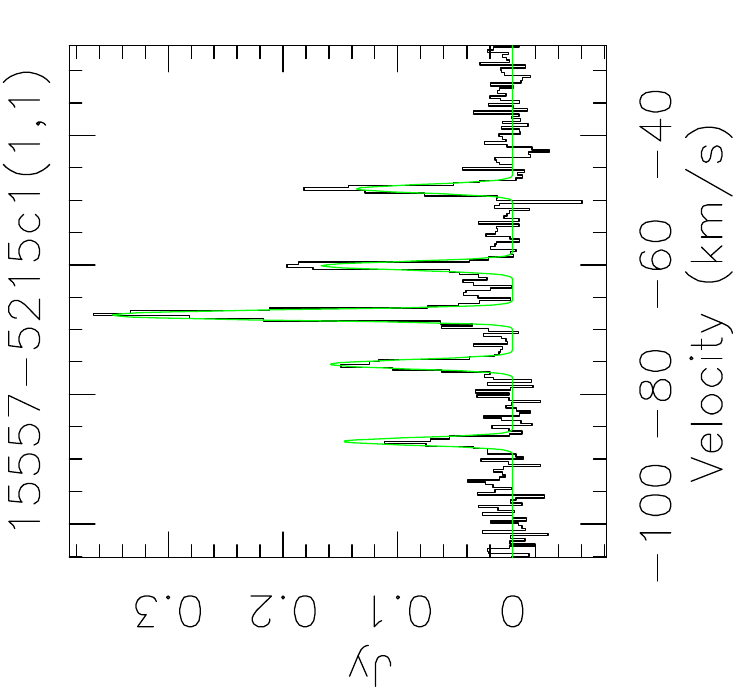}
 \includegraphics[angle=-90,width=0.24\textwidth]{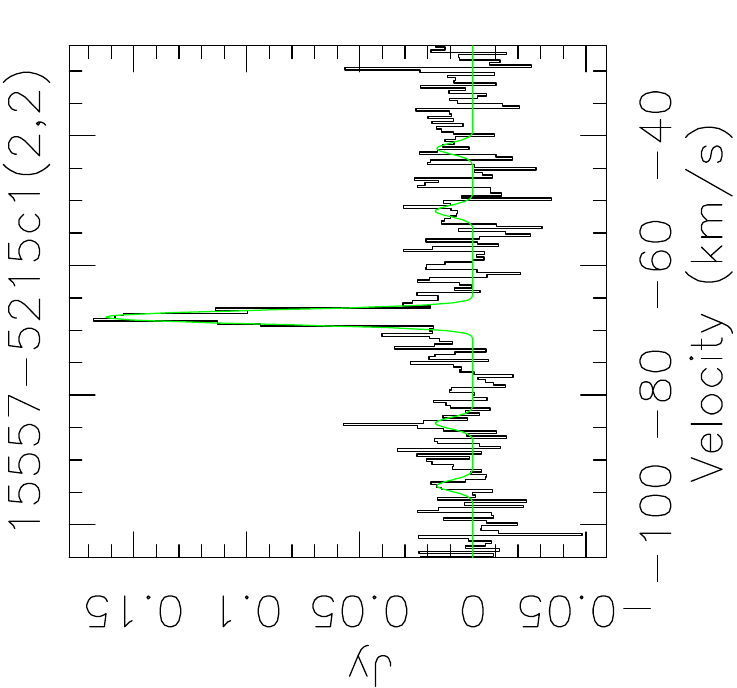}
 \includegraphics[angle=-90,width=0.24\textwidth]{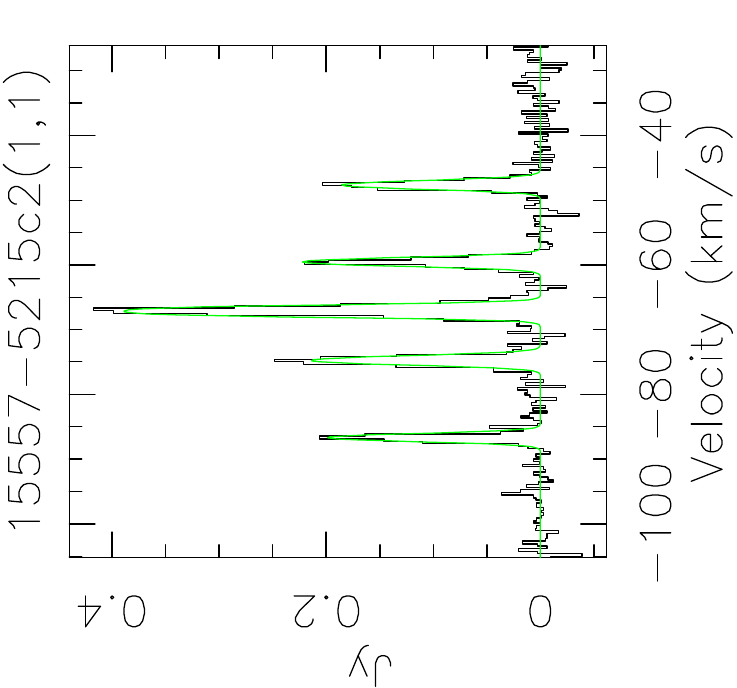}
 \includegraphics[angle=-90,width=0.24\textwidth]{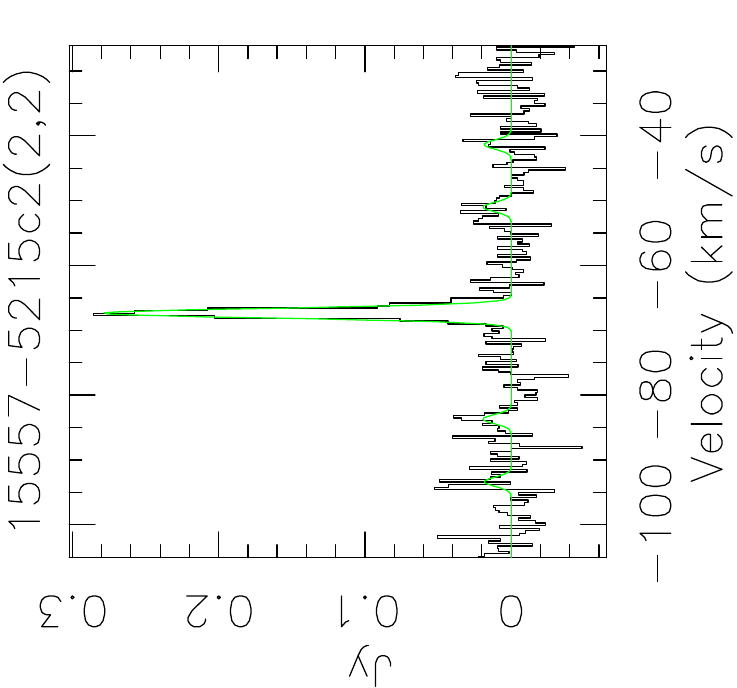}

 \caption{Continued.}
\end{figure*}

\begin{figure*}[tbp]
 \ContinuedFloat
 \centering
 \includegraphics[angle=-90,width=0.24\textwidth]{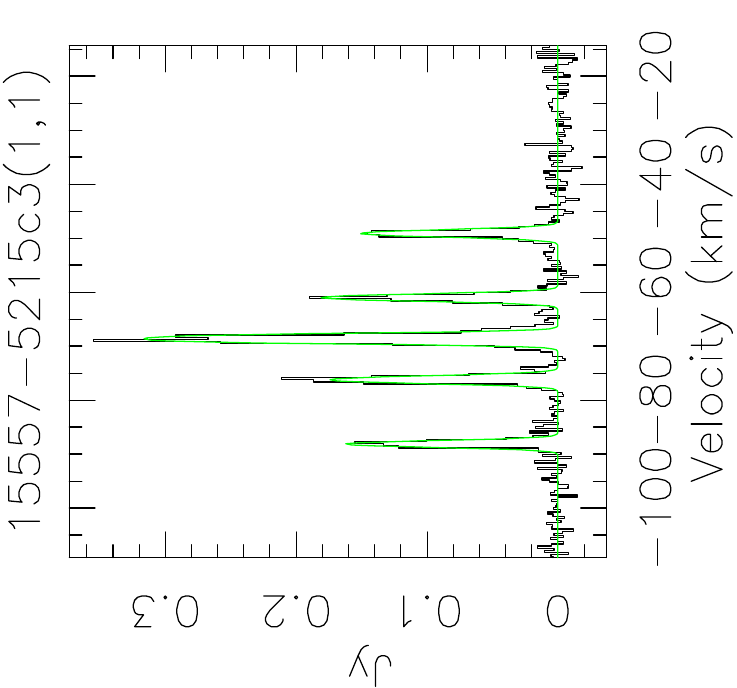}
 \includegraphics[angle=-90,width=0.24\textwidth]{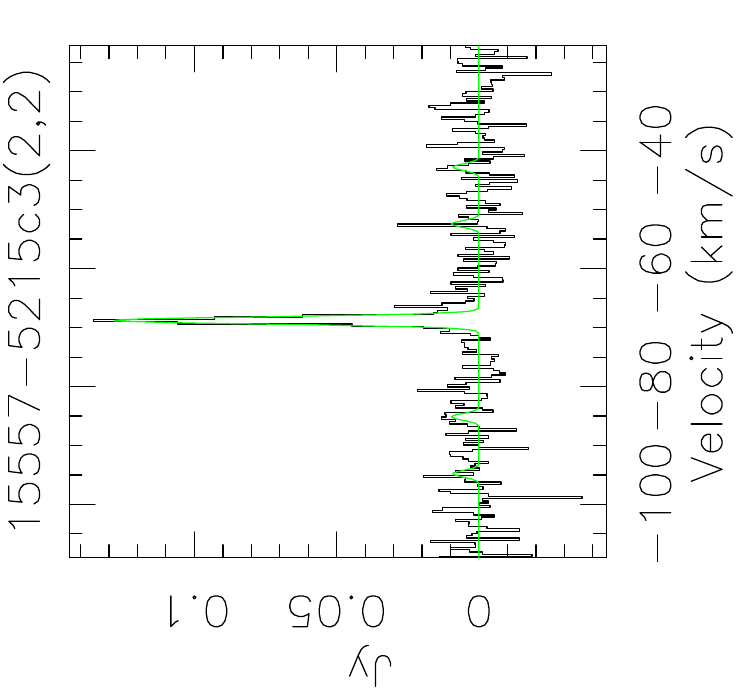}
 \includegraphics[angle=-90,width=0.24\textwidth]{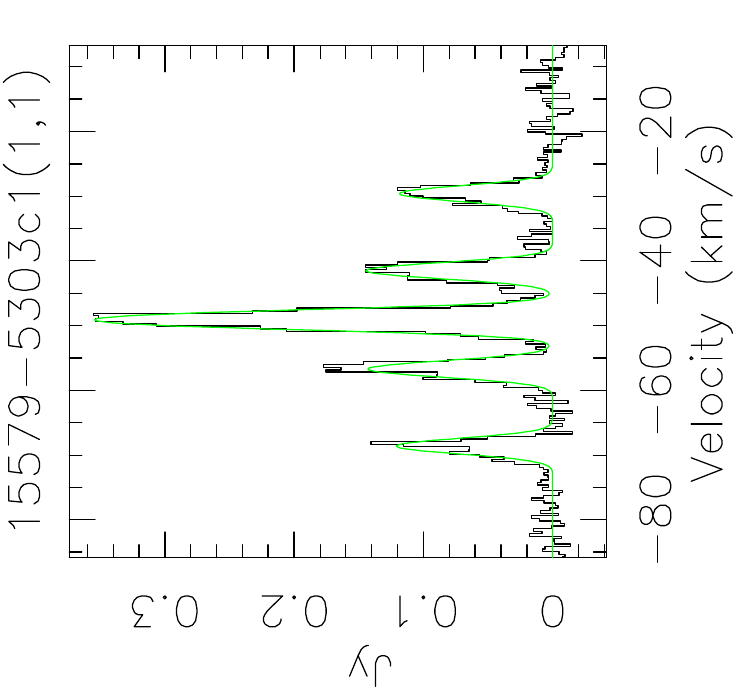}
 \includegraphics[angle=-90,width=0.24\textwidth]{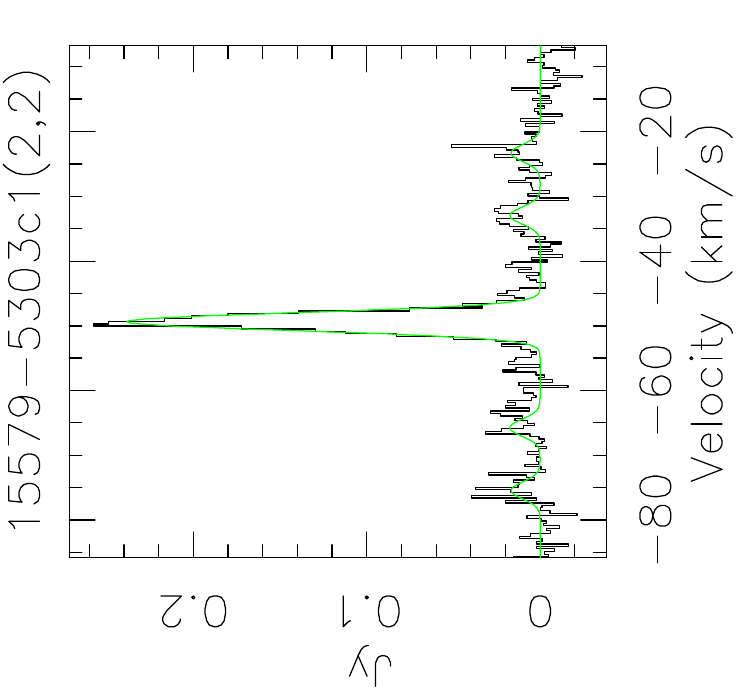}

 \includegraphics[angle=-90,width=0.24\textwidth]{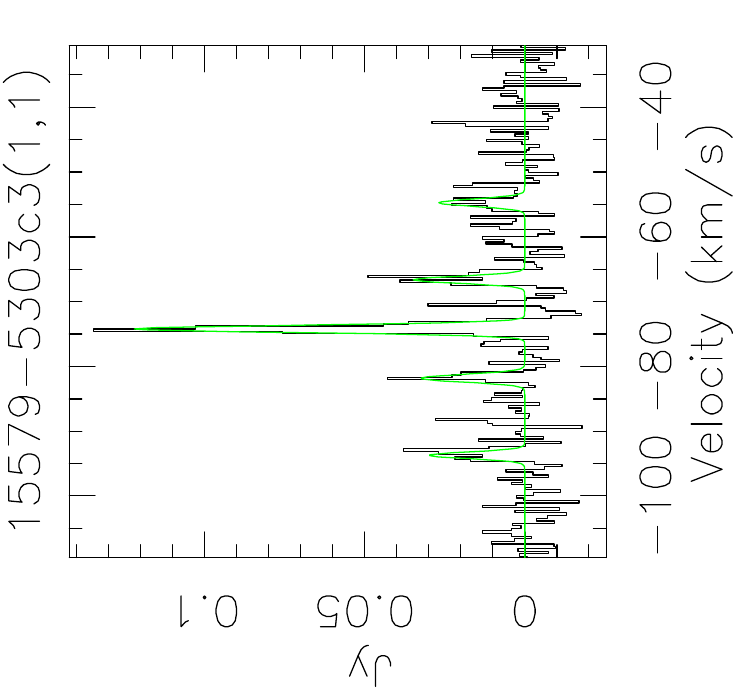}
 \includegraphics[angle=-90,width=0.24\textwidth]{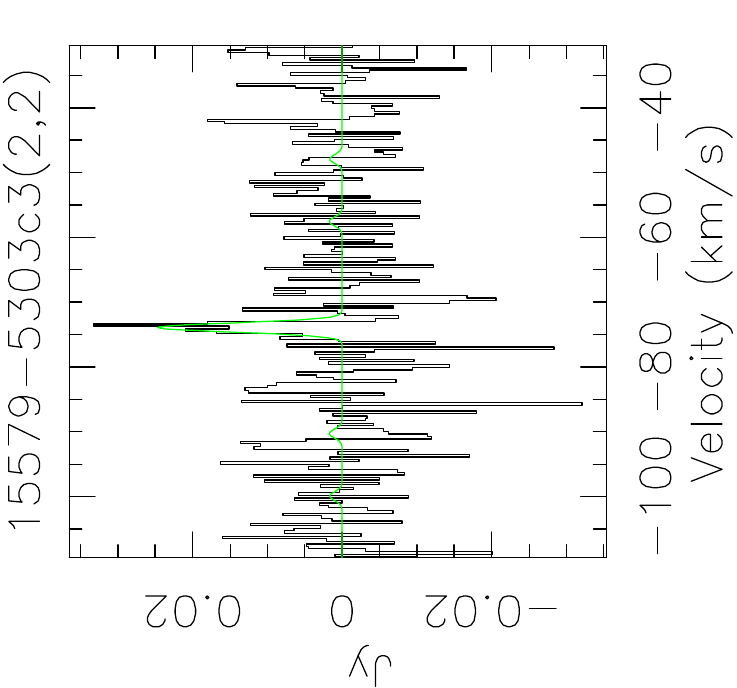}
 \includegraphics[angle=-90,width=0.24\textwidth]{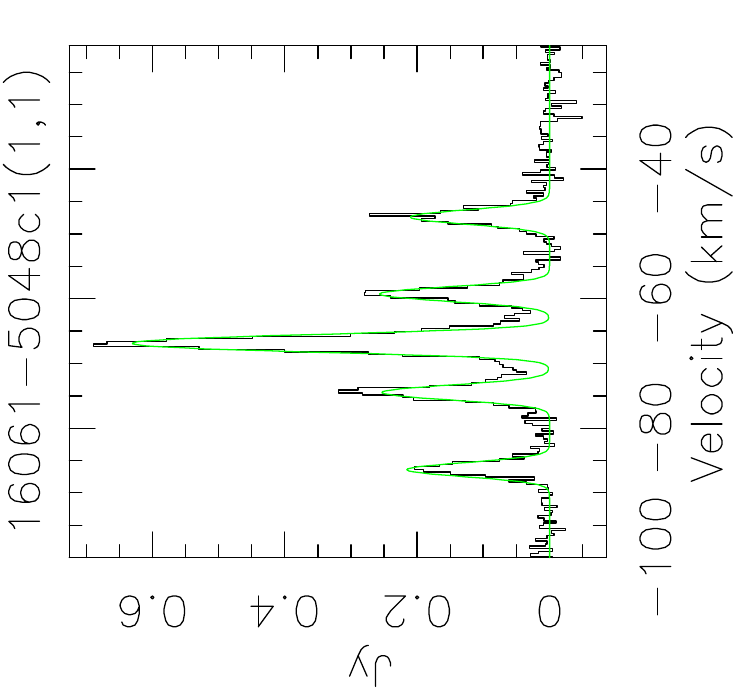}
 \includegraphics[angle=-90,width=0.24\textwidth]{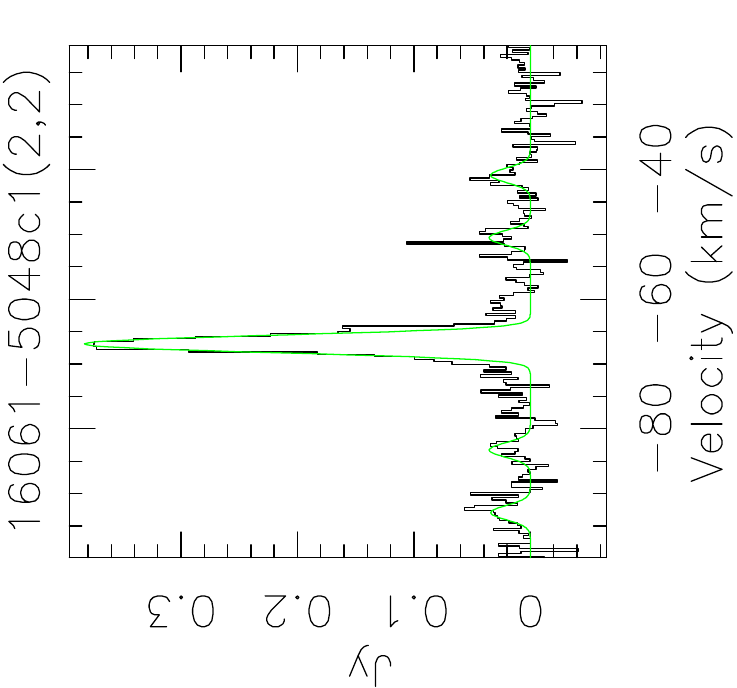}

 \includegraphics[angle=-90,width=0.24\textwidth]{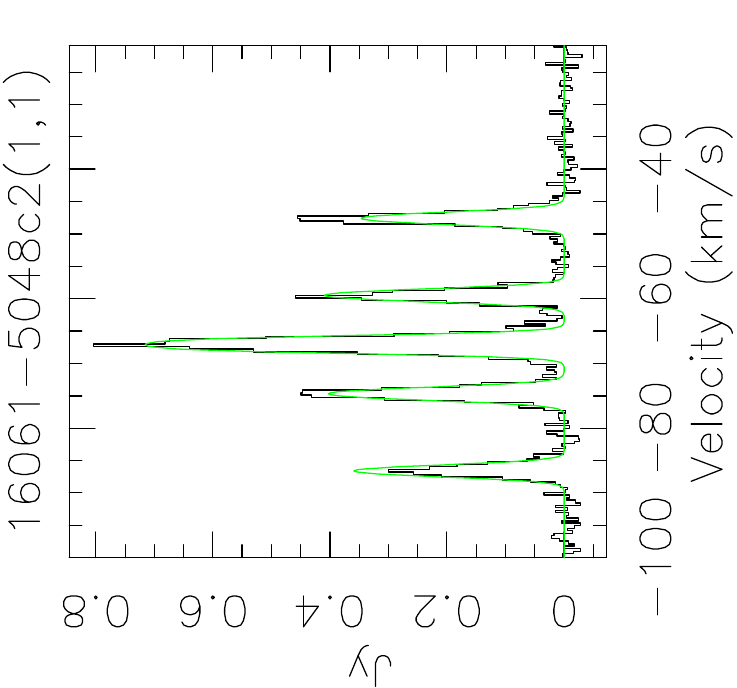} 
 \includegraphics[angle=-90,width=0.24\textwidth]{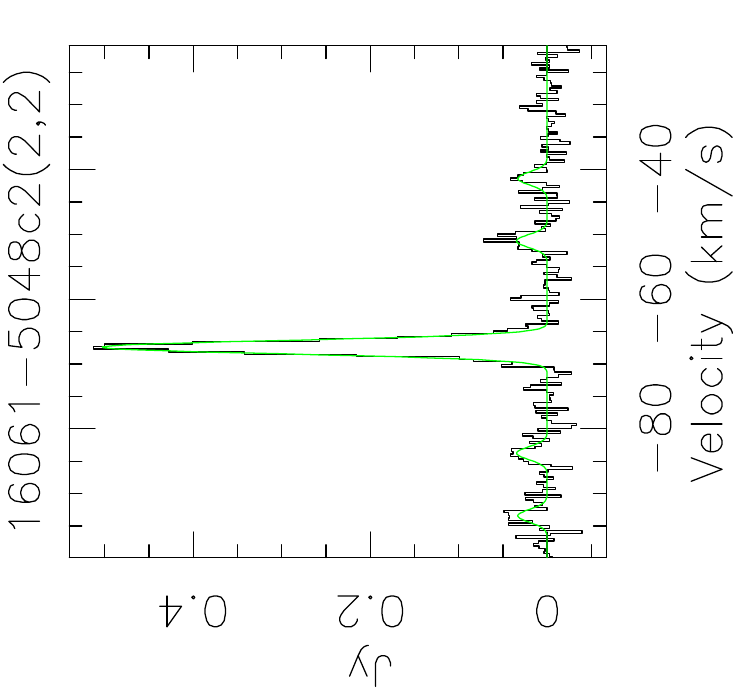} 
 \includegraphics[angle=-90,width=0.24\textwidth]{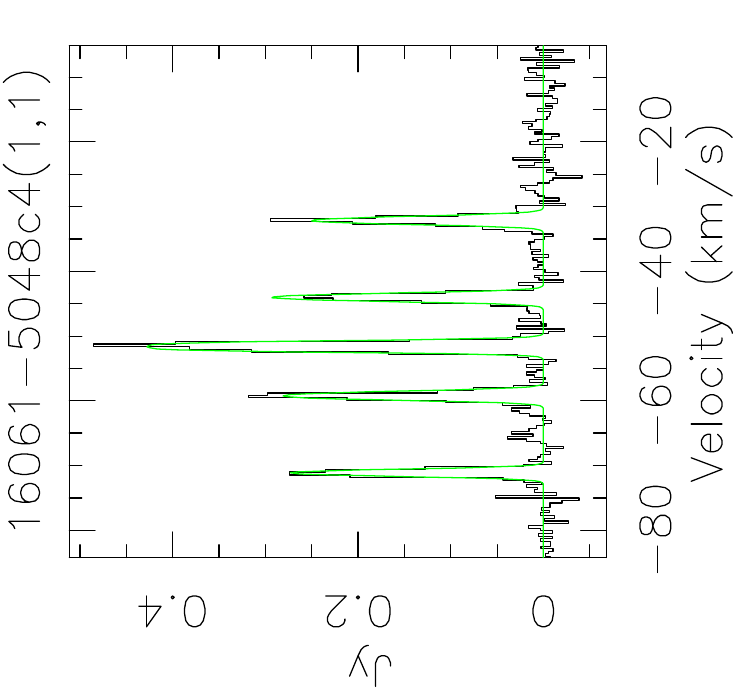}
 \includegraphics[angle=-90,width=0.24\textwidth]{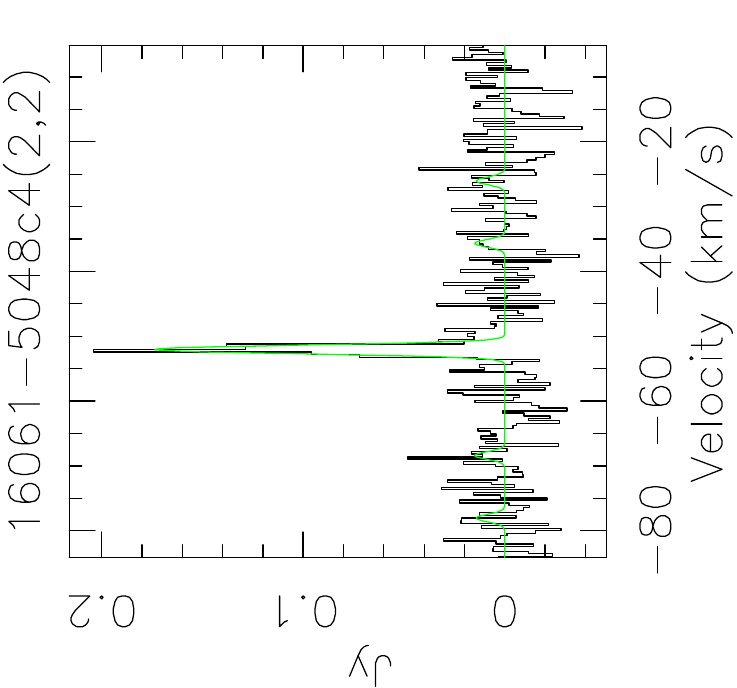}

 \includegraphics[angle=-90,width=0.24\textwidth]{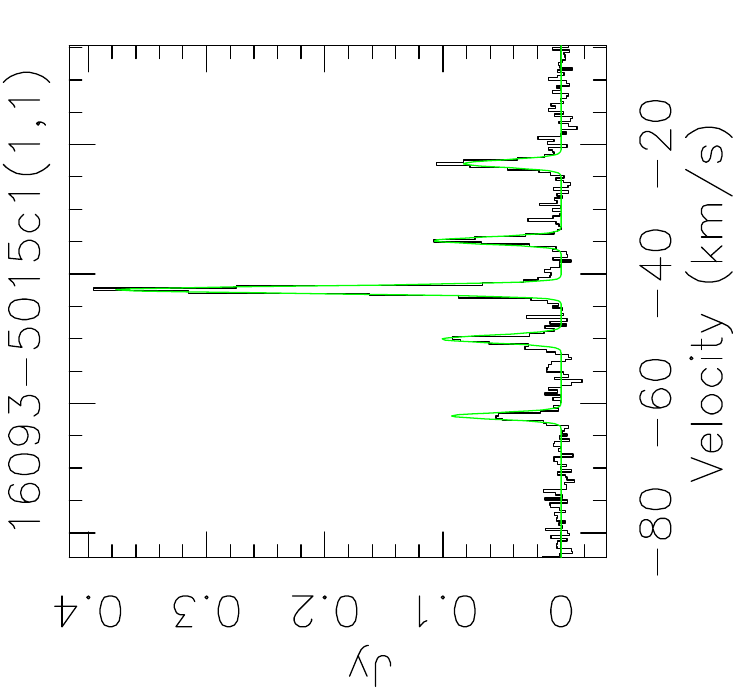}
 \includegraphics[angle=-90,width=0.24\textwidth]{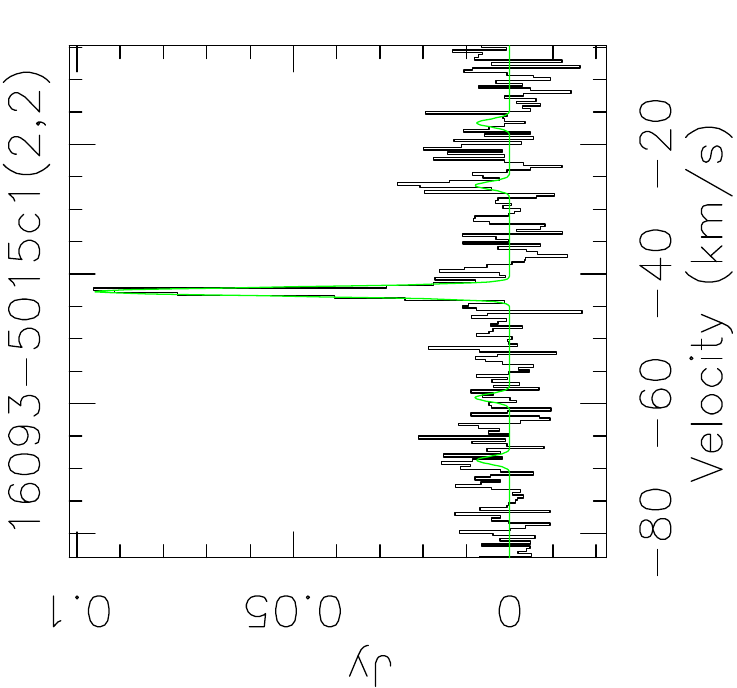}
 \includegraphics[angle=-90,width=0.24\textwidth]{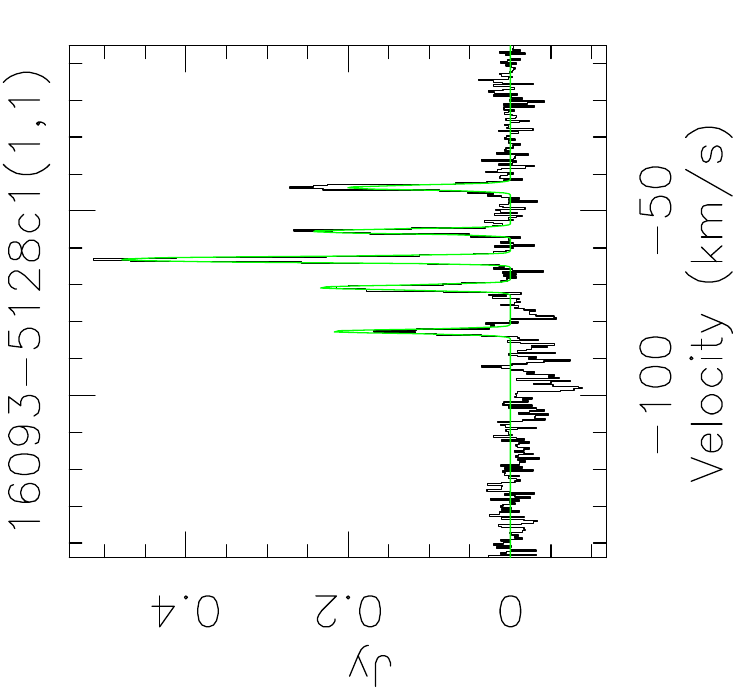}
 \includegraphics[angle=-90,width=0.24\textwidth]{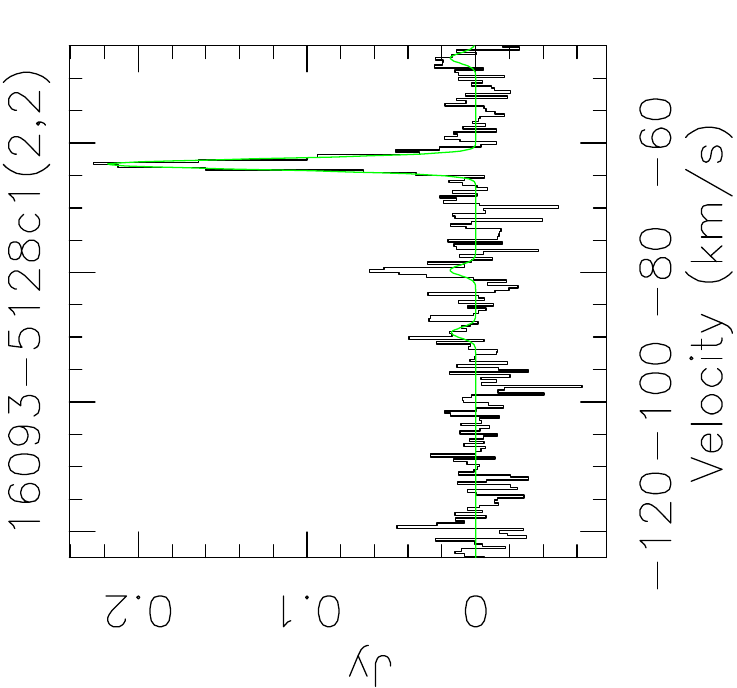}

 \includegraphics[angle=-90,width=0.24\textwidth]{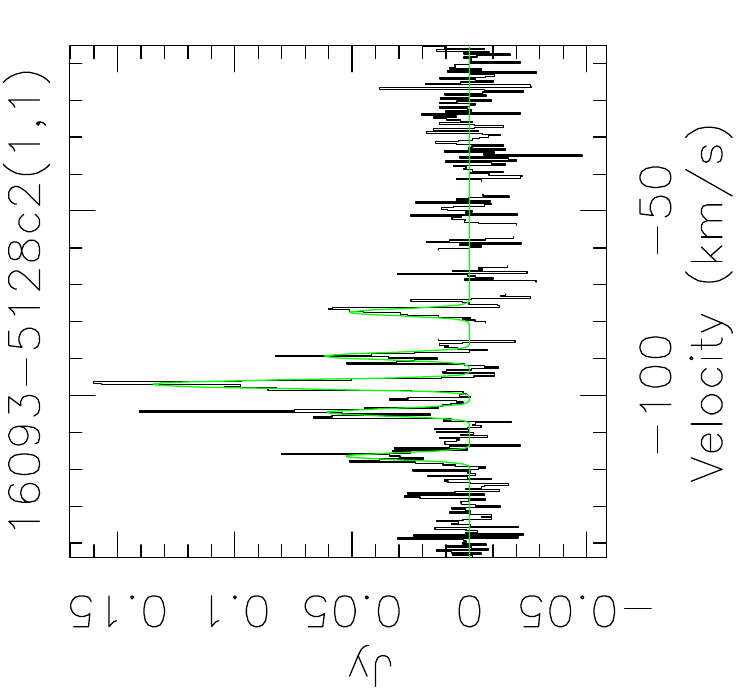}
 \includegraphics[angle=-90,width=0.24\textwidth]{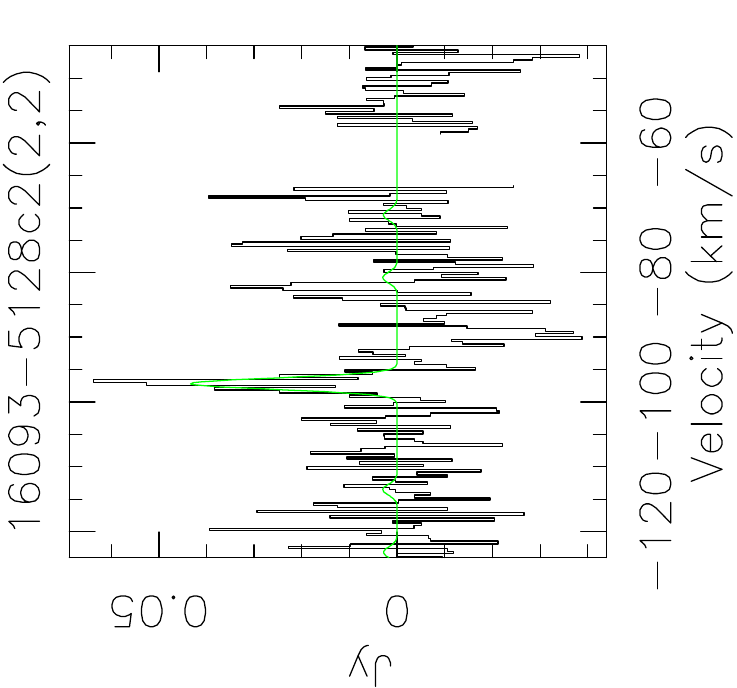}
 \includegraphics[angle=-90,width=0.24\textwidth]{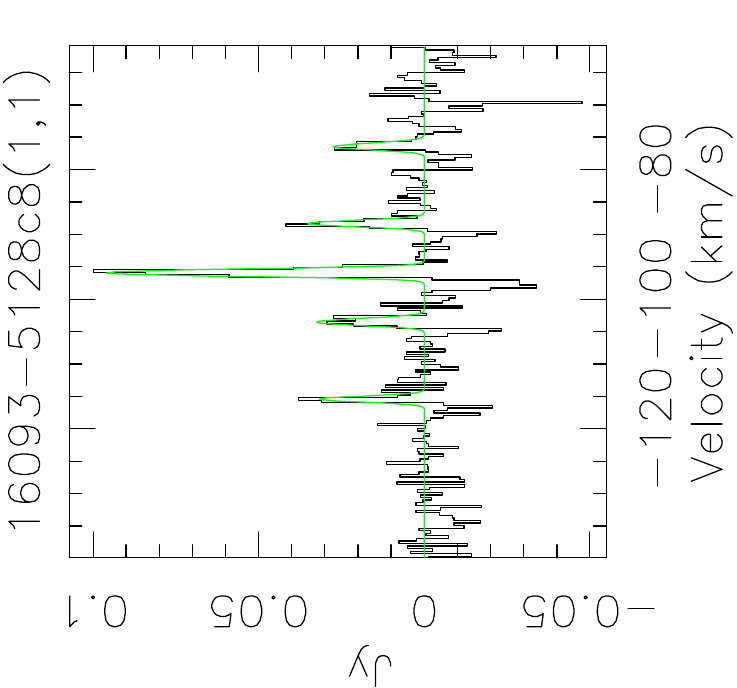}
 \includegraphics[angle=-90,width=0.24\textwidth]{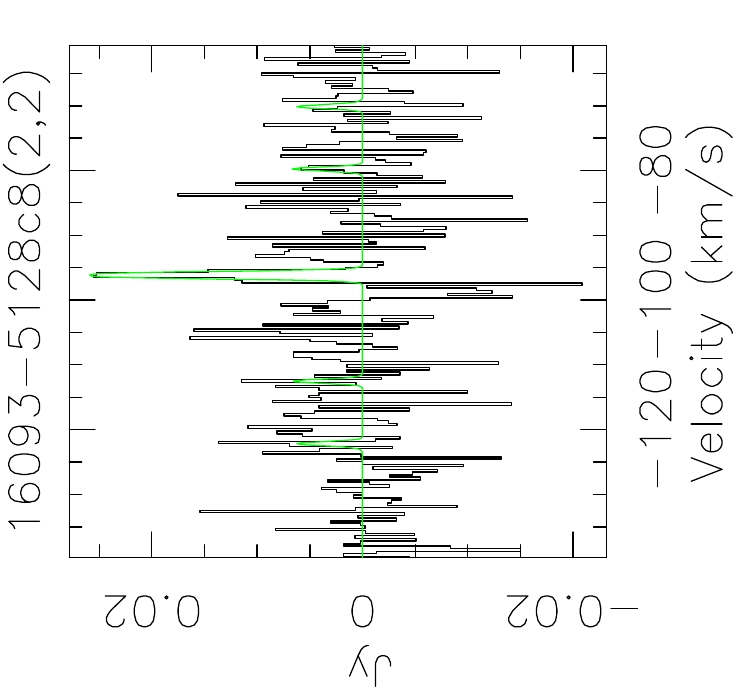}

 \caption{Continued.}
\end{figure*}

\begin{figure*}[tbp]
 \ContinuedFloat
 \centering
 \includegraphics[angle=-90,width=0.24\textwidth]{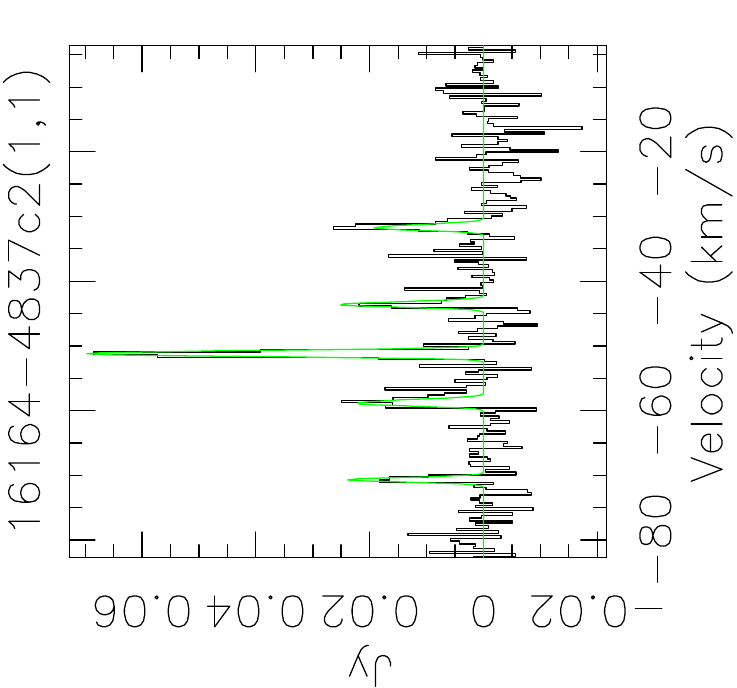}
 \includegraphics[angle=-90,width=0.24\textwidth]{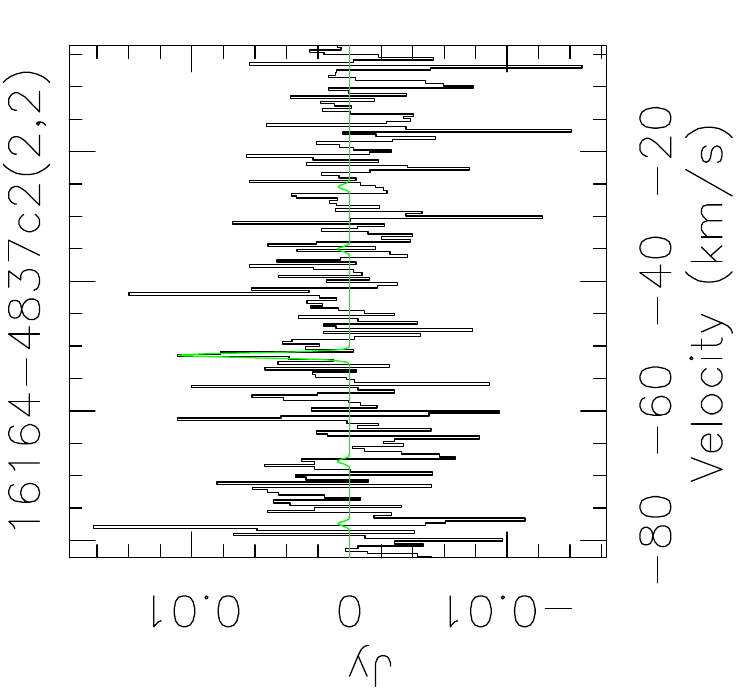}
 \includegraphics[angle=-90,width=0.24\textwidth]{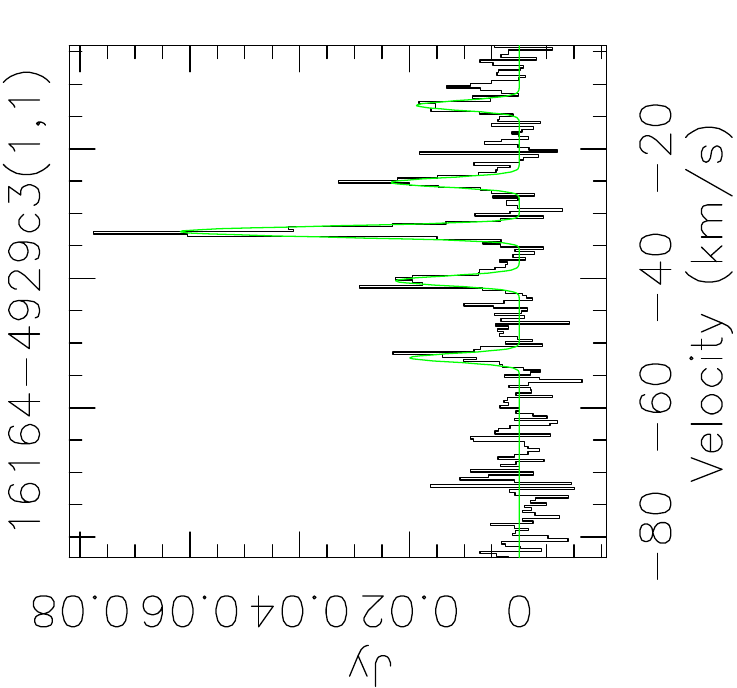}
 \includegraphics[angle=-90,width=0.24\textwidth]{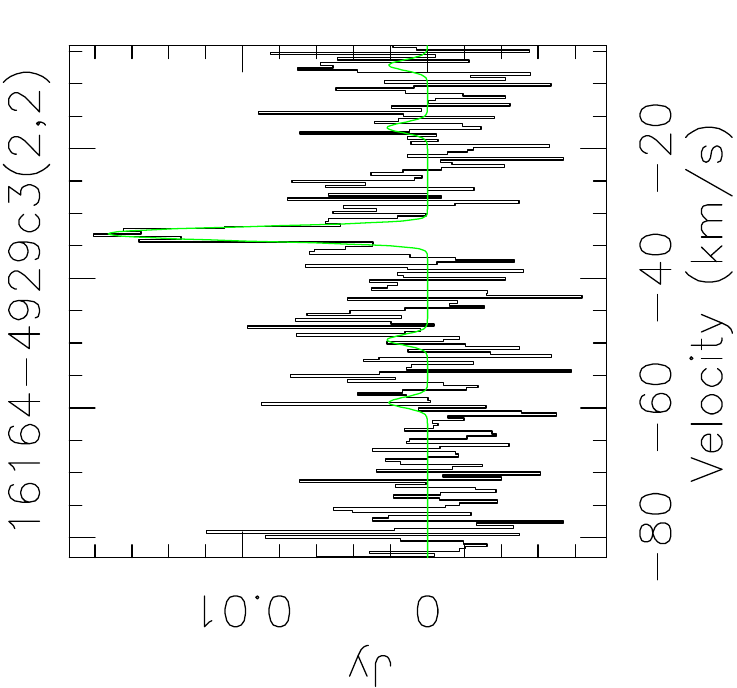}

 \includegraphics[angle=-90,width=0.24\textwidth]{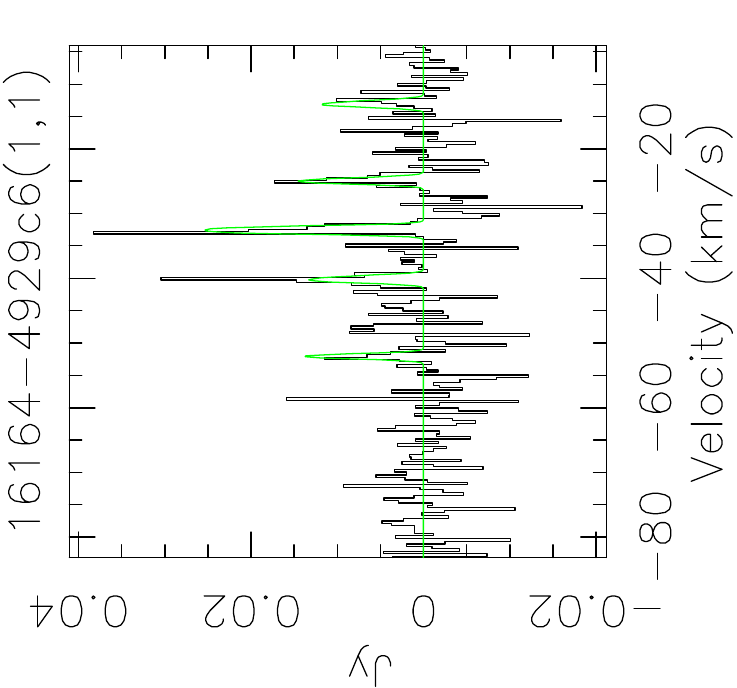}
 \includegraphics[angle=-90,width=0.24\textwidth]{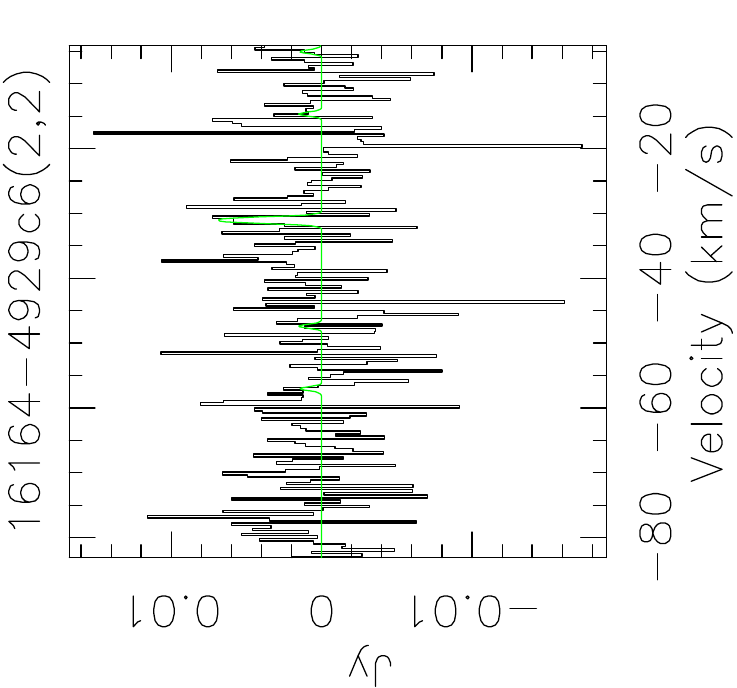}
 \includegraphics[angle=-90,width=0.24\textwidth]{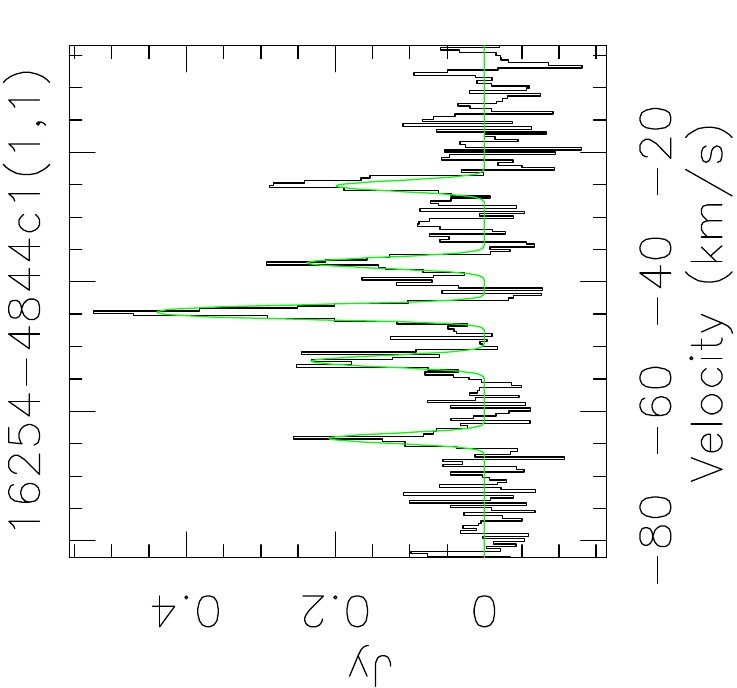}
 \includegraphics[angle=-90,width=0.24\textwidth]{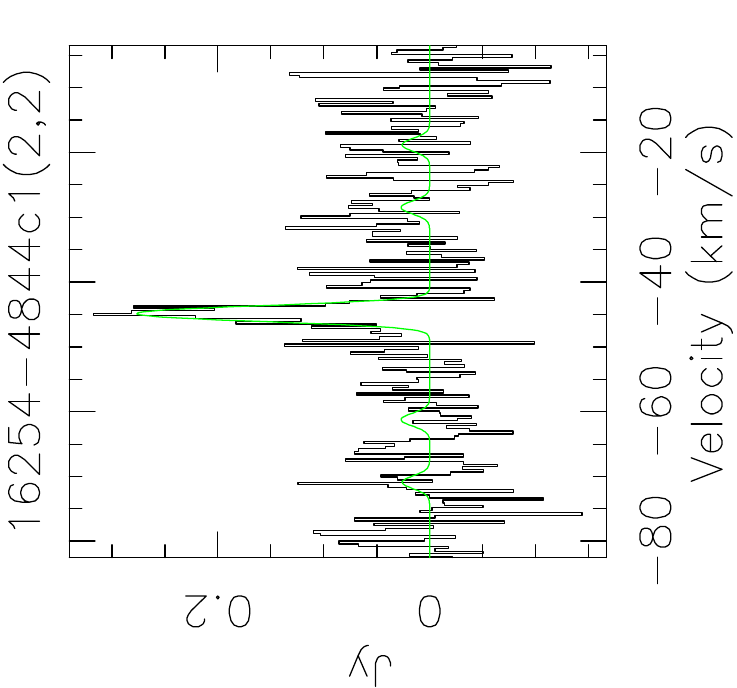}

 \includegraphics[angle=-90,width=0.24\textwidth]{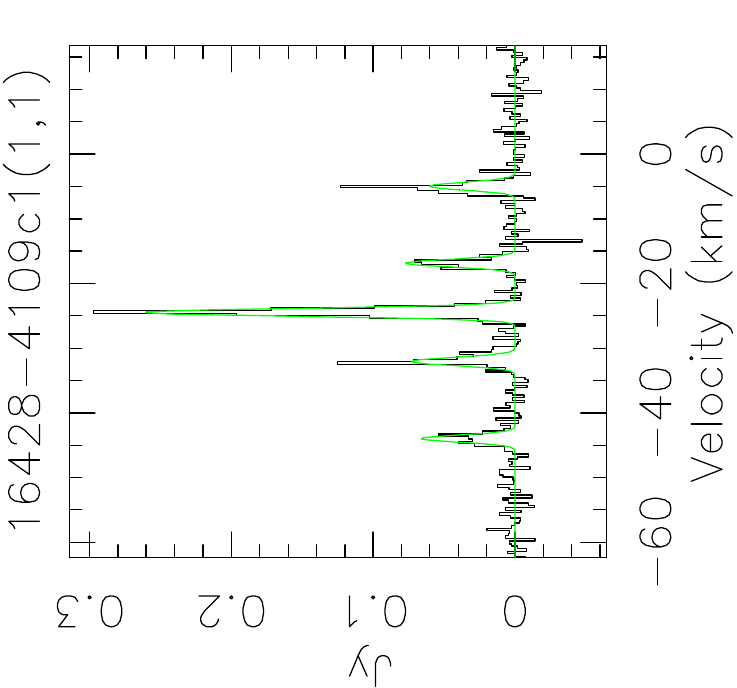}
 \includegraphics[angle=-90,width=0.24\textwidth]{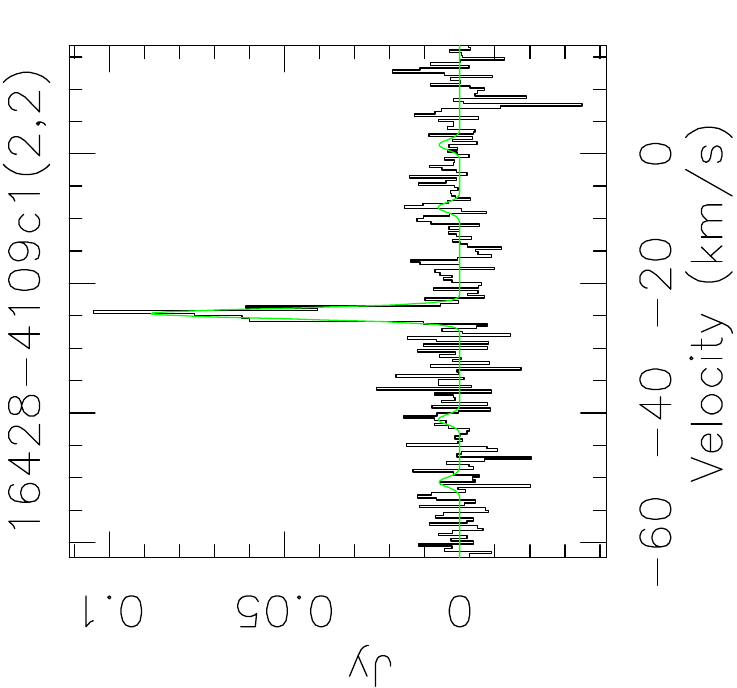}
 \includegraphics[angle=-90,width=0.24\textwidth]{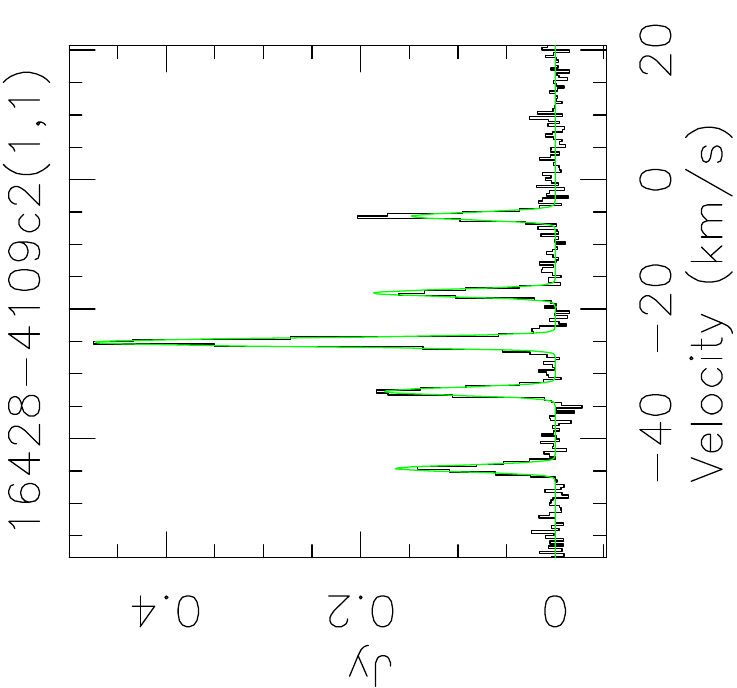}
 \includegraphics[angle=-90,width=0.24\textwidth]{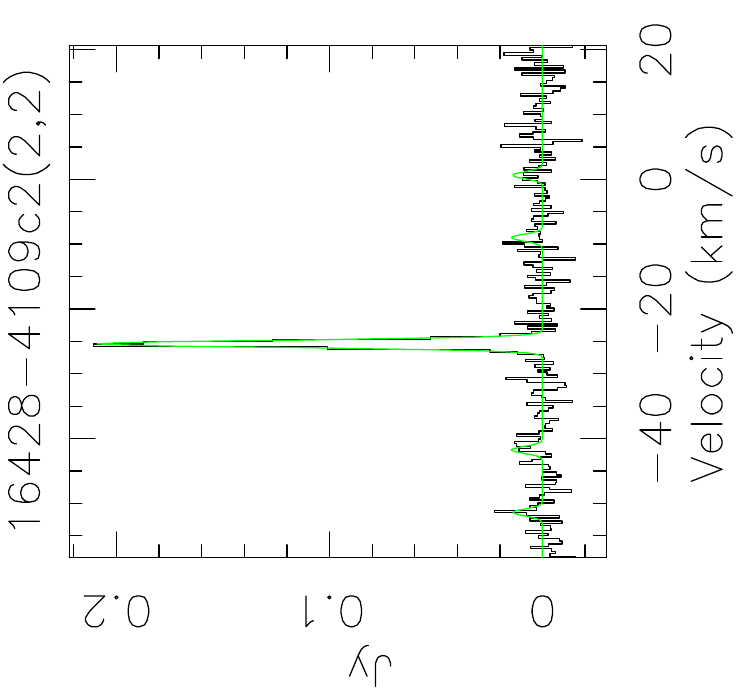}

 \includegraphics[angle=-90,width=0.24\textwidth]{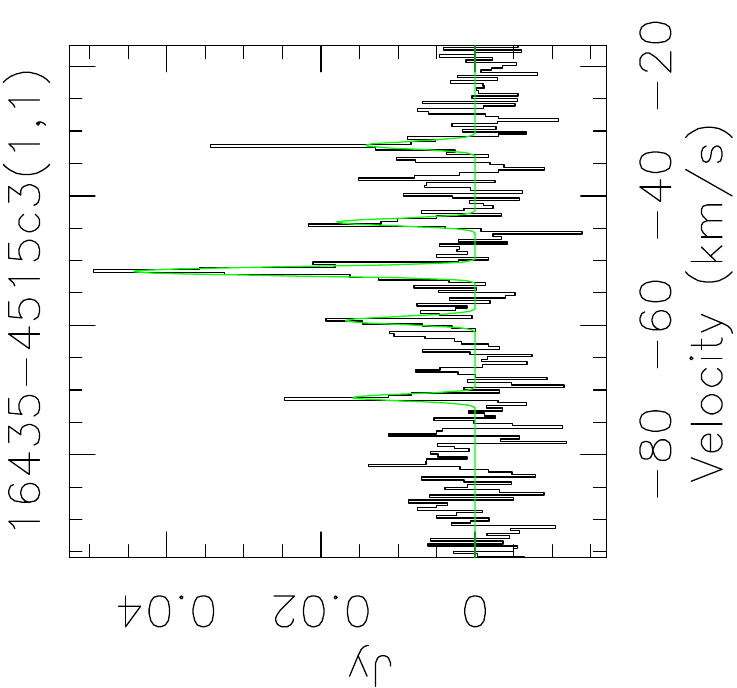}
 \includegraphics[angle=-90,width=0.24\textwidth]{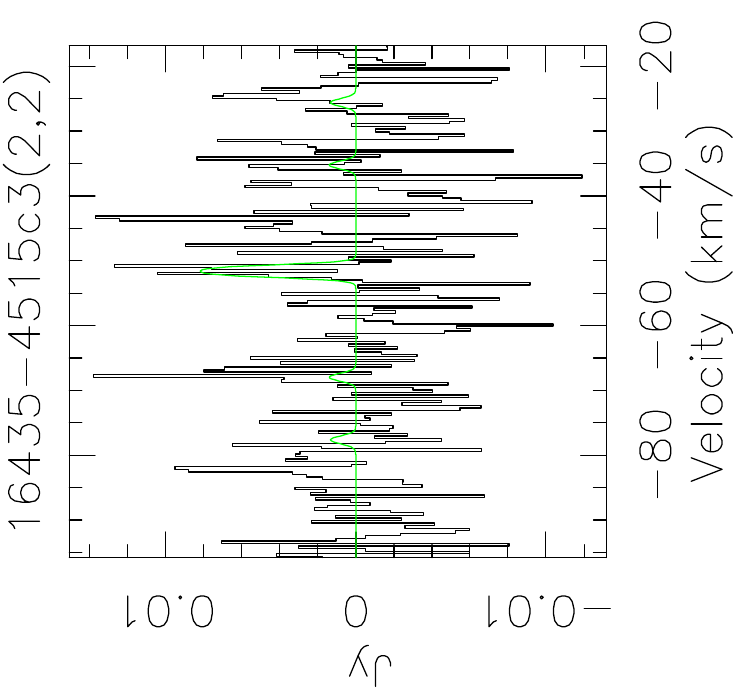}
 \includegraphics[angle=-90,width=0.24\textwidth]{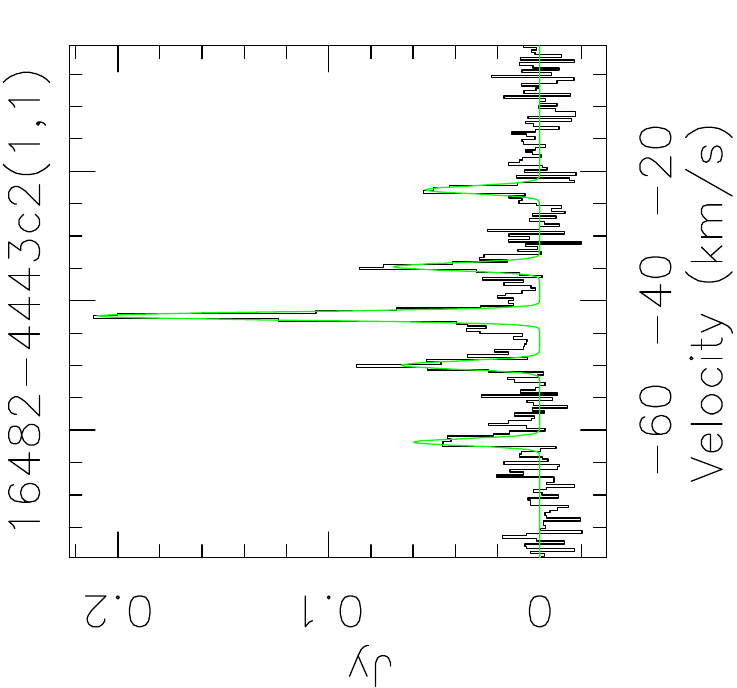}
 \includegraphics[angle=-90,width=0.24\textwidth]{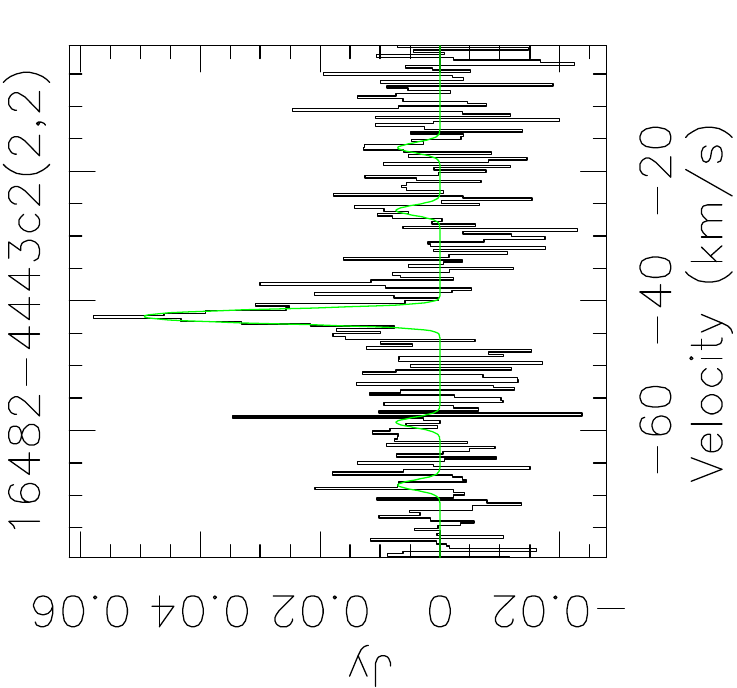}

 \includegraphics[angle=-90,width=0.24\textwidth]{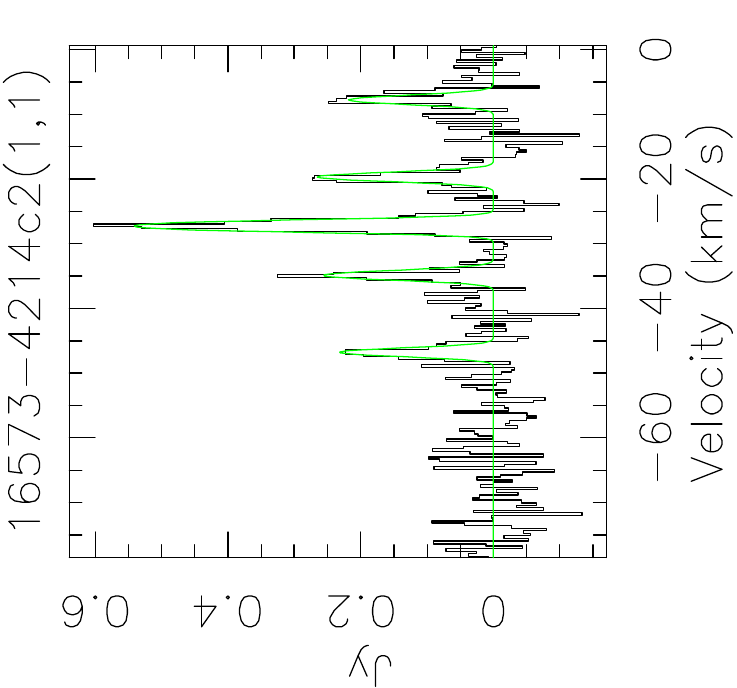}
 \includegraphics[angle=-90,width=0.24\textwidth]{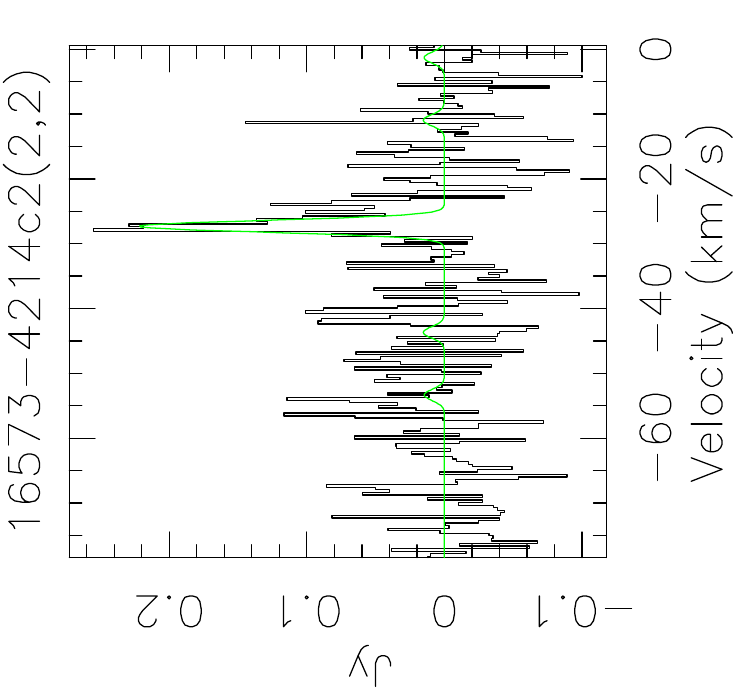}
 \includegraphics[angle=-90,width=0.24\textwidth]{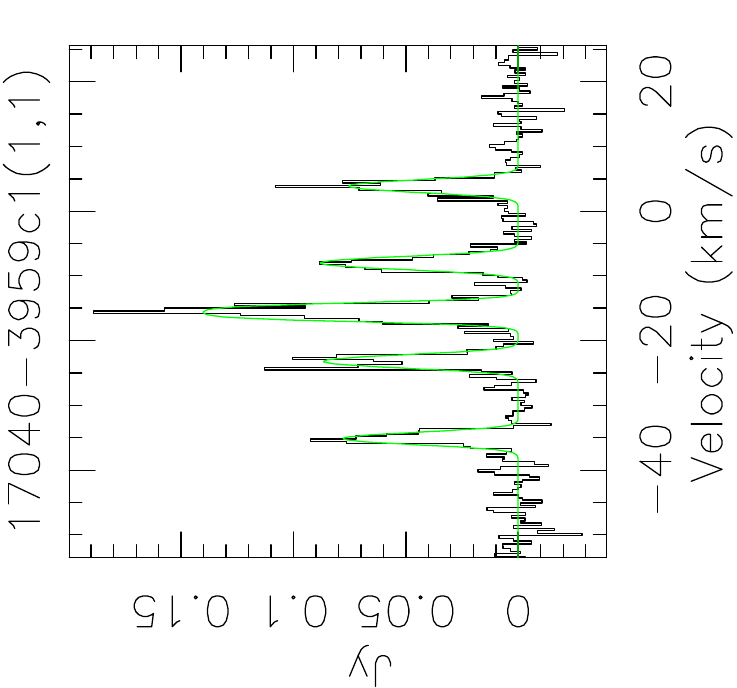}
 \includegraphics[angle=-90,width=0.24\textwidth]{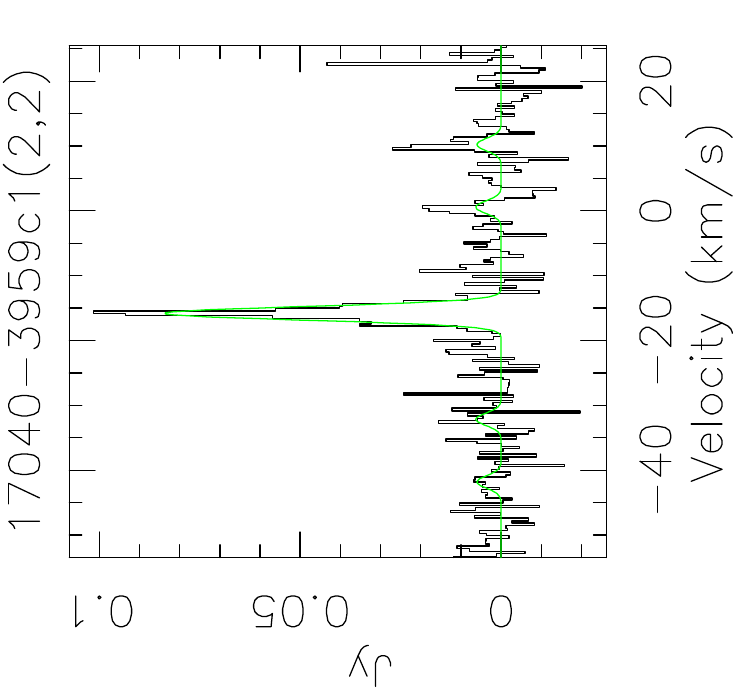}

 \caption{Continued.}
\end{figure*}

\begin{figure*}[tbp]
 \ContinuedFloat
 \centering
 \includegraphics[angle=-90,width=0.24\textwidth]{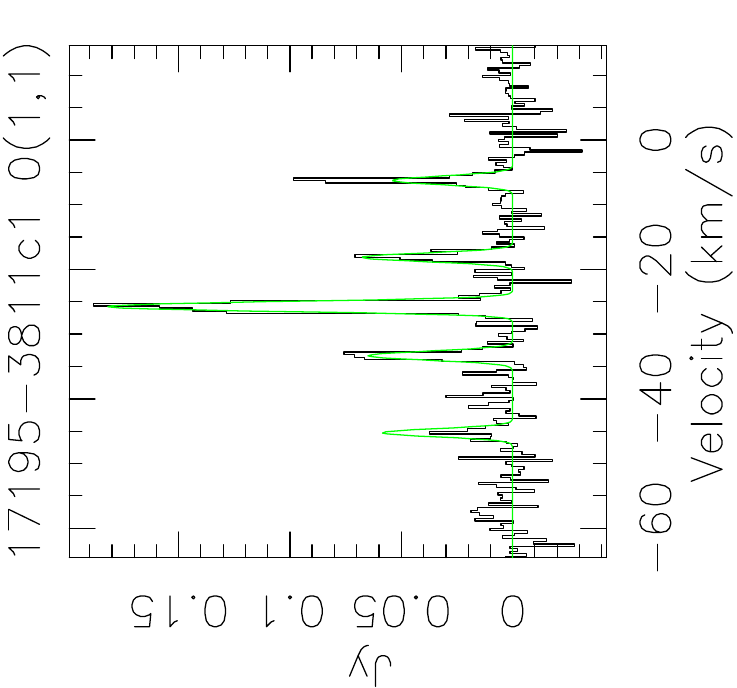}
 \includegraphics[angle=-90,width=0.24\textwidth]{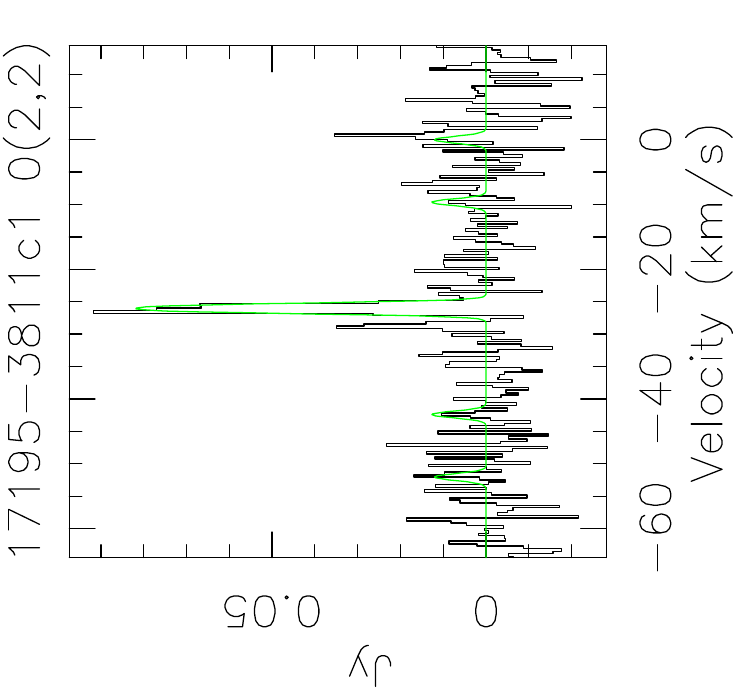}
 \includegraphics[angle=-90,width=0.24\textwidth]{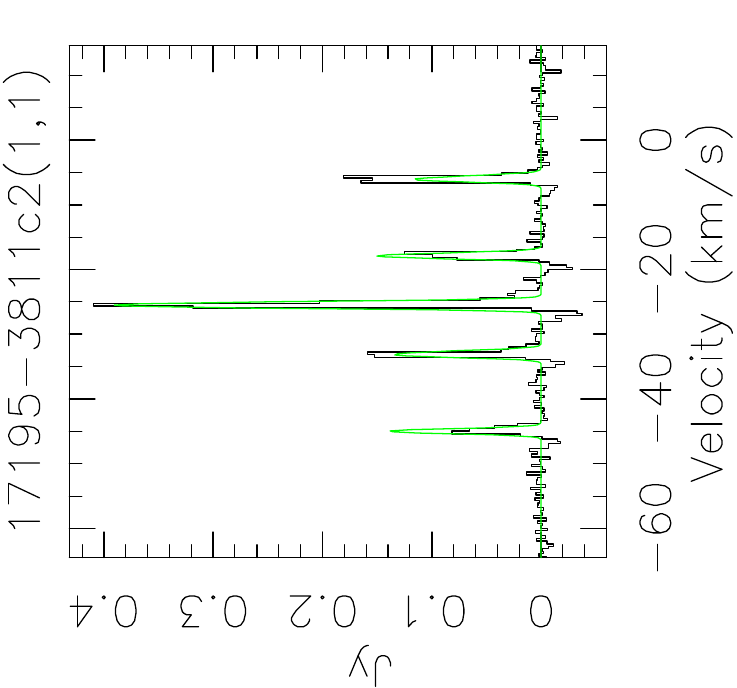}
 \includegraphics[angle=-90,width=0.24\textwidth]{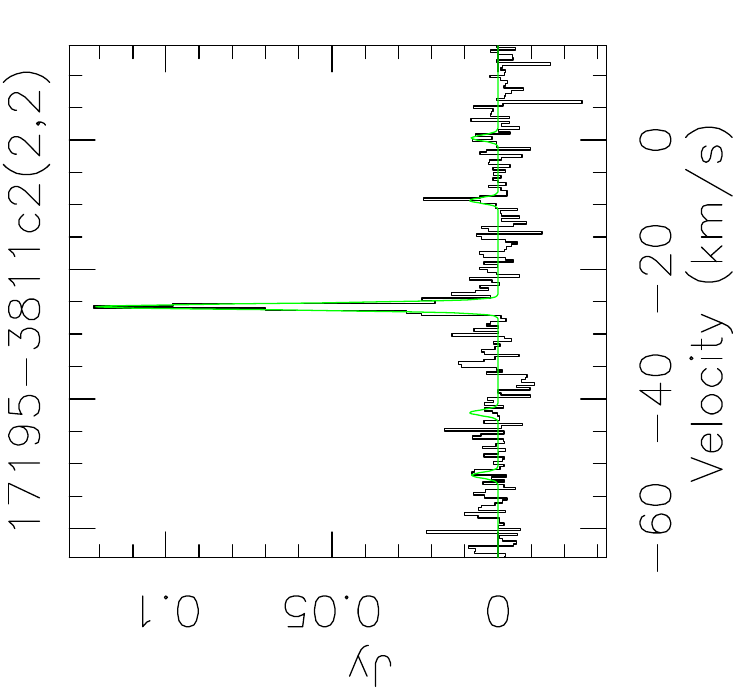}

 \includegraphics[angle=-90,width=0.24\textwidth]{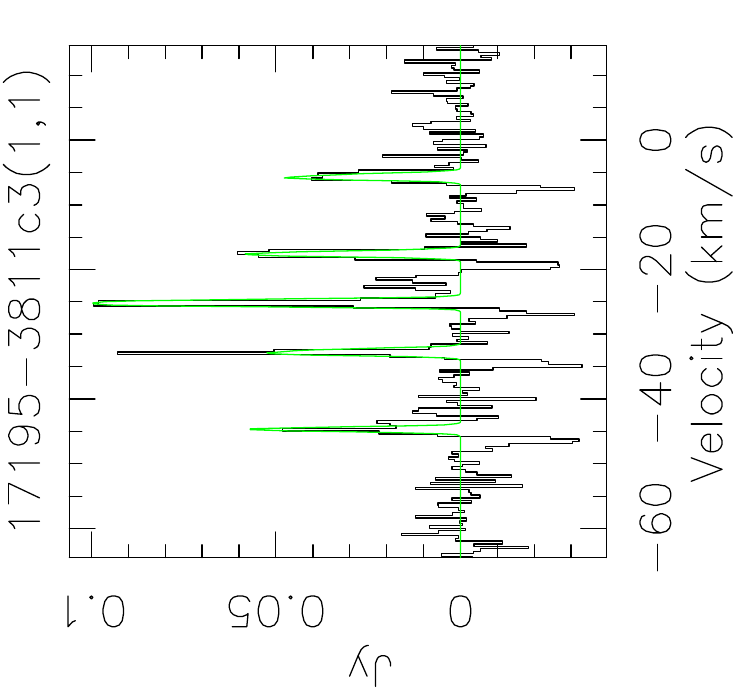}
 \includegraphics[angle=-90,width=0.24\textwidth]{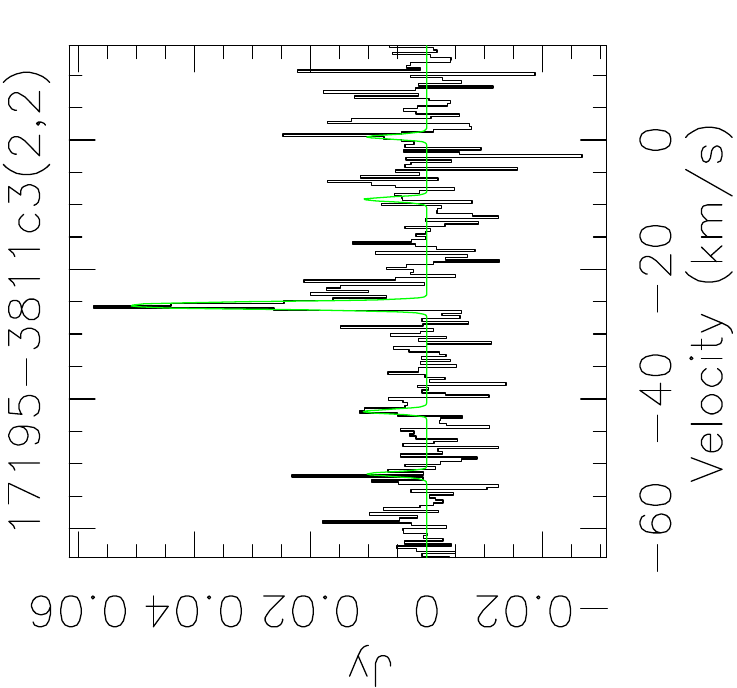}
 \includegraphics[angle=-90,width=0.24\textwidth]{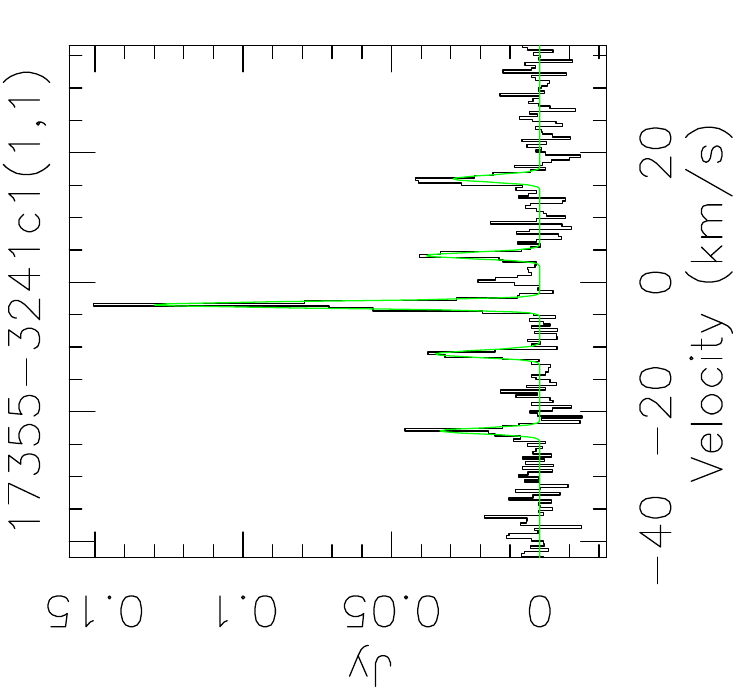}
 \includegraphics[angle=-90,width=0.24\textwidth]{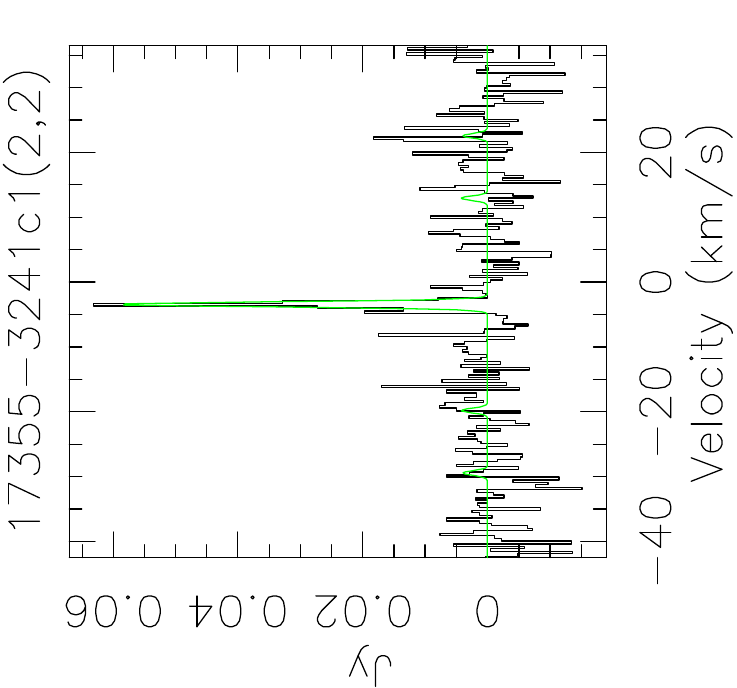}

 \caption{Continued.}
\end{figure*}

\begin{figure*}[tbp]
 \centering
 \includegraphics[angle=-90,width=0.24\textwidth]{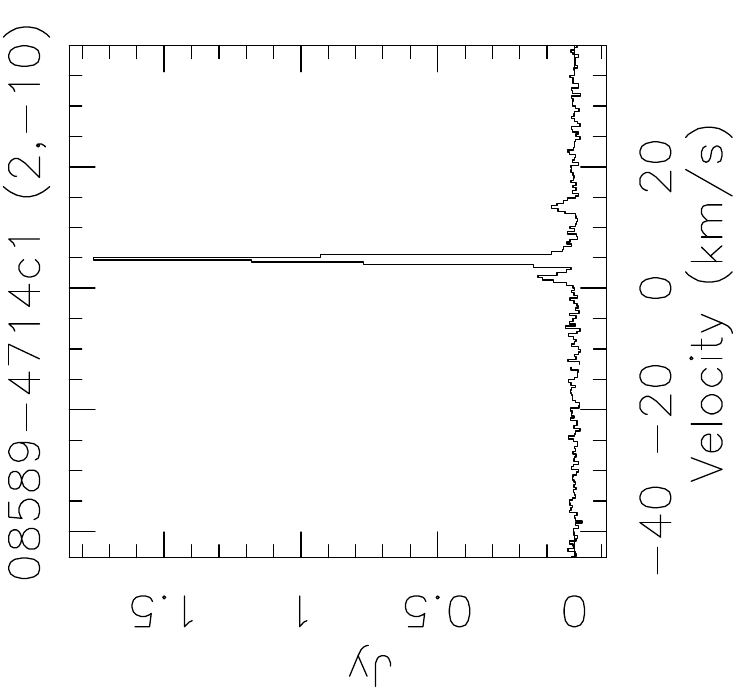}
 \includegraphics[angle=-90,width=0.24\textwidth]{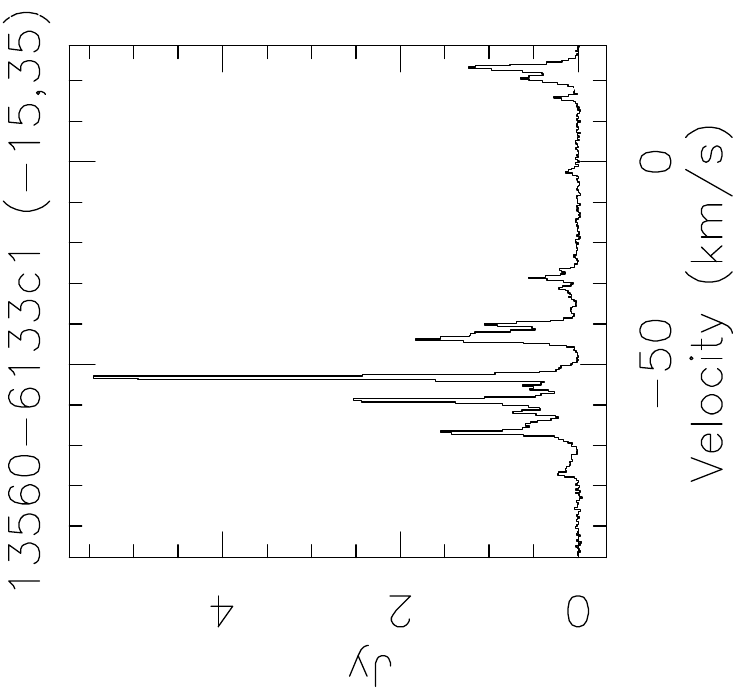}
 \includegraphics[angle=-90,width=0.24\textwidth]{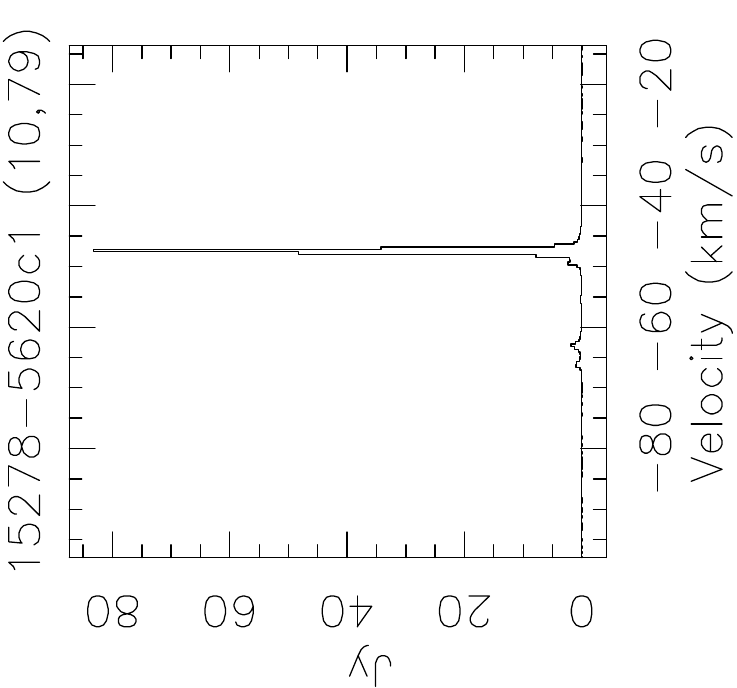}
 \includegraphics[angle=-90,width=0.24\textwidth]{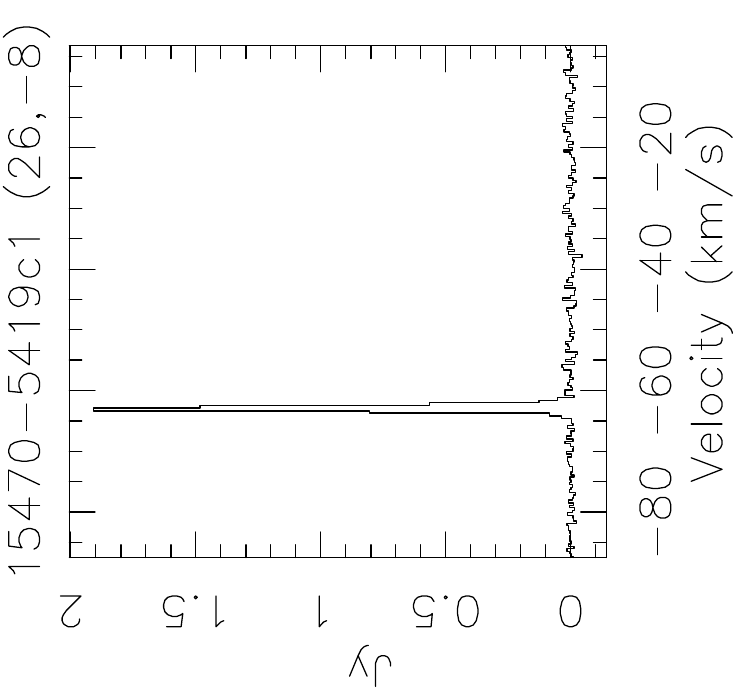}

 \includegraphics[angle=-90,width=0.24\textwidth]{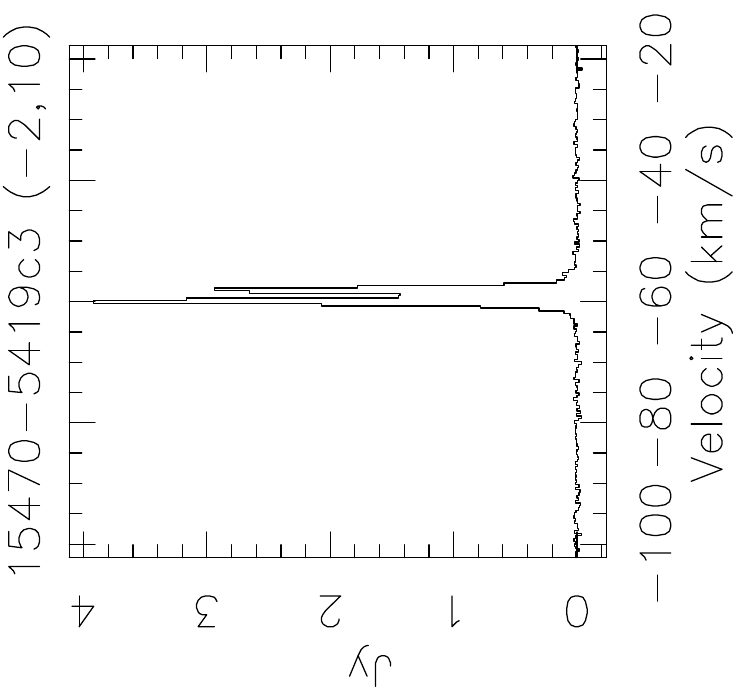}
 \includegraphics[angle=-90,width=0.24\textwidth]{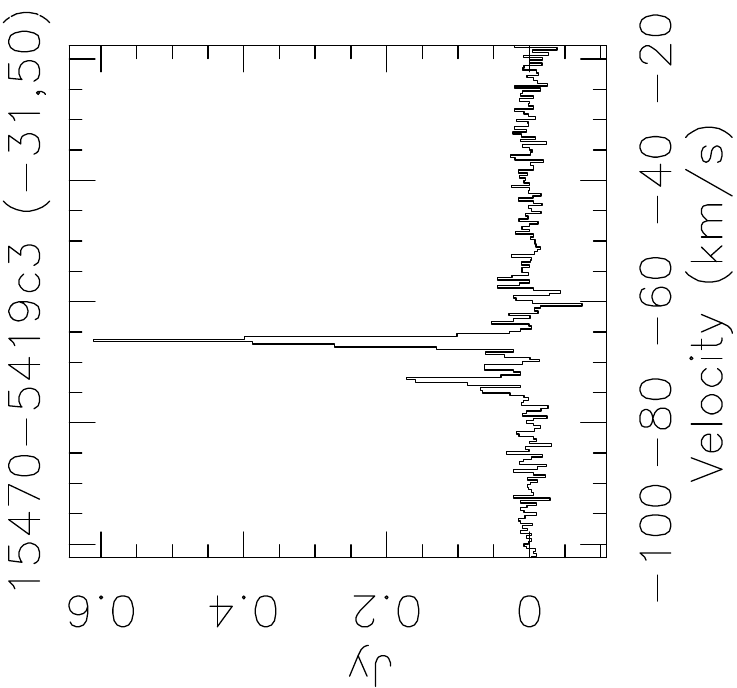}
 \includegraphics[angle=-90,width=0.24\textwidth]{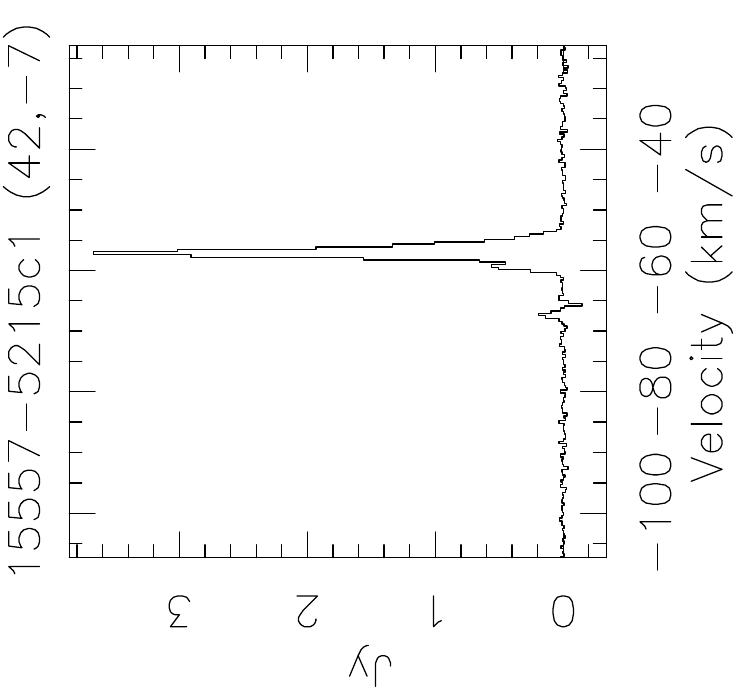}
 \includegraphics[angle=-90,width=0.24\textwidth]{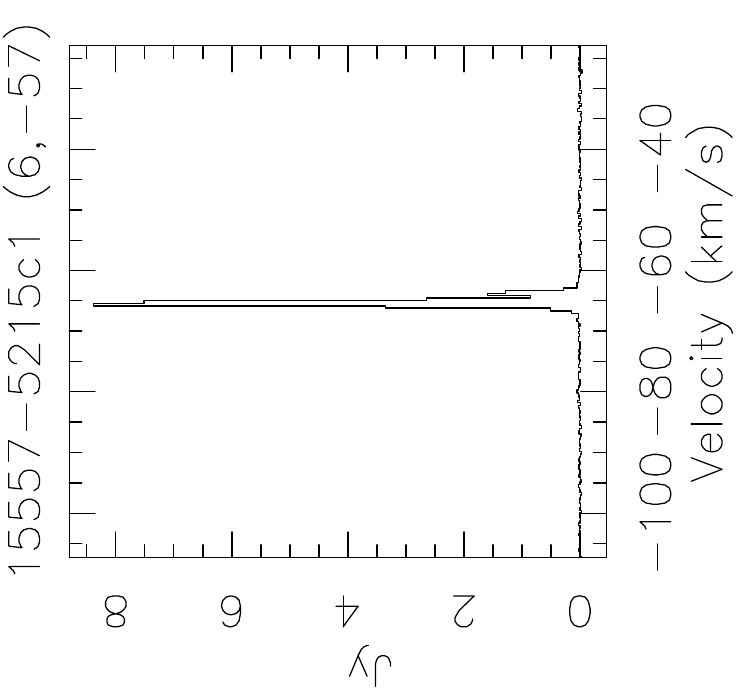}

 \includegraphics[angle=-90,width=0.24\textwidth]{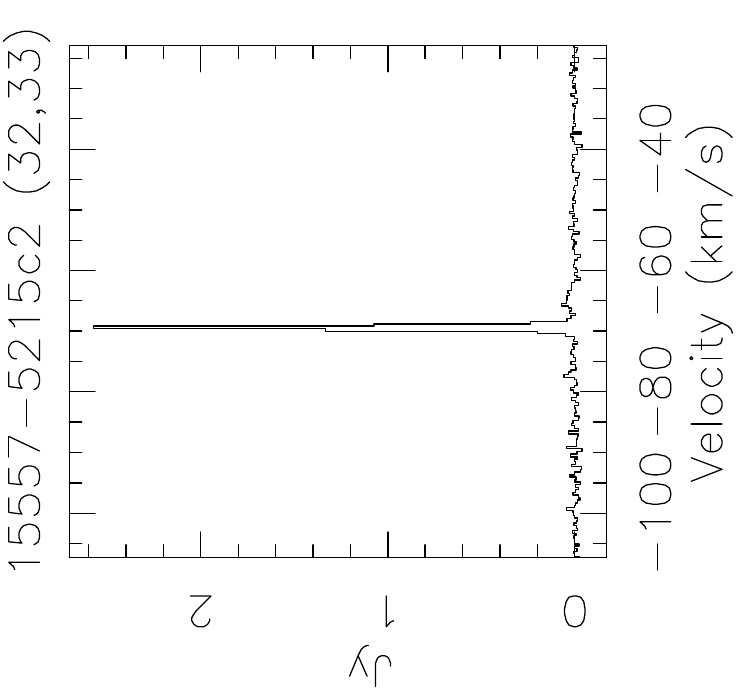}
 \includegraphics[angle=-90,width=0.24\textwidth]{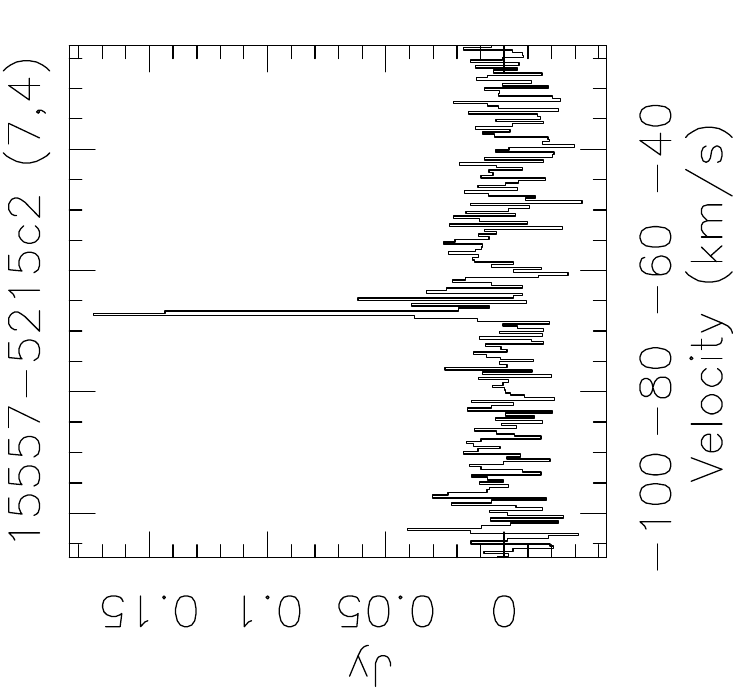}
 \includegraphics[angle=-90,width=0.24\textwidth]{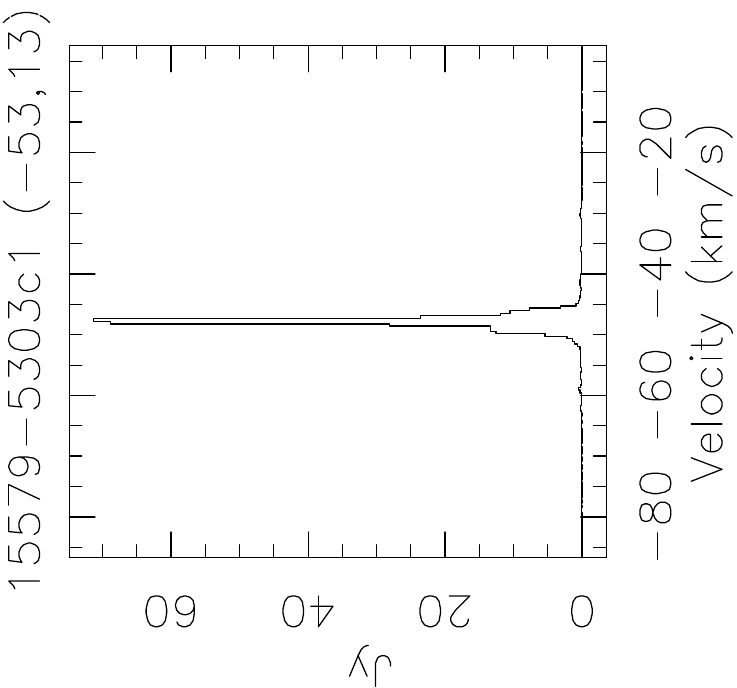}
 \includegraphics[angle=-90,width=0.24\textwidth]{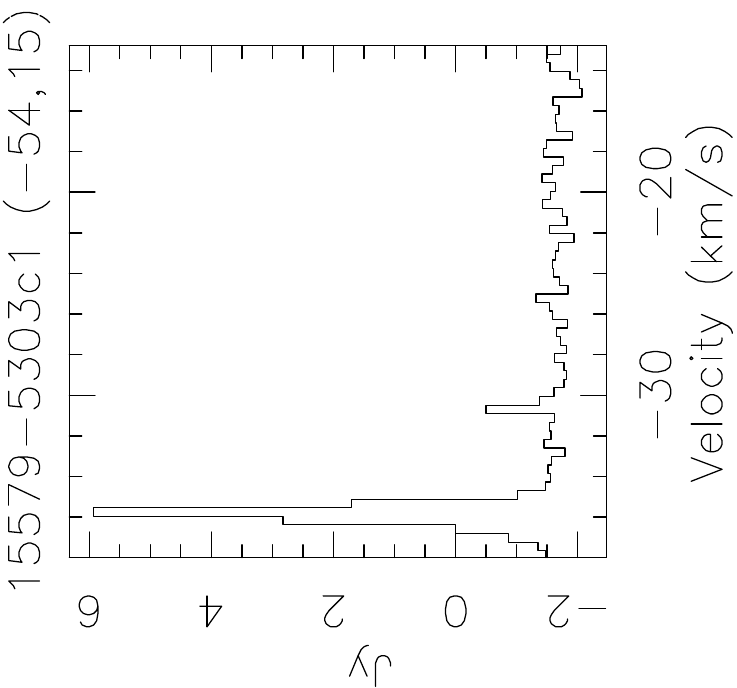}

 \includegraphics[angle=-90,width=0.24\textwidth]{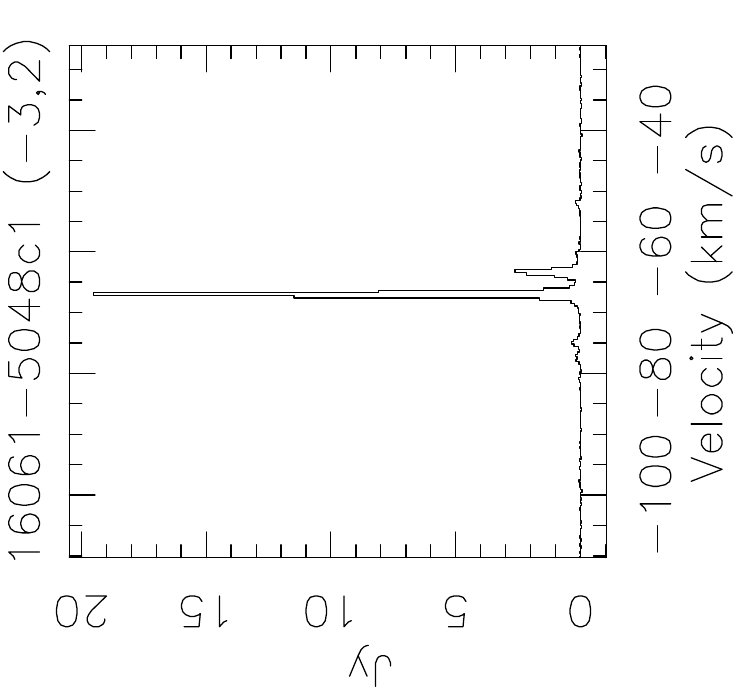}
 \includegraphics[angle=-90,width=0.24\textwidth]{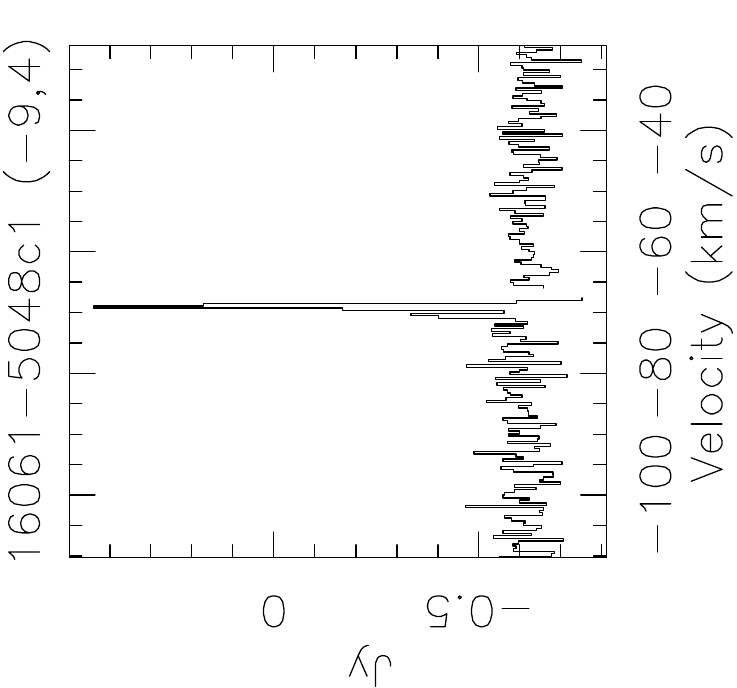}
 \includegraphics[angle=-90,width=0.24\textwidth]{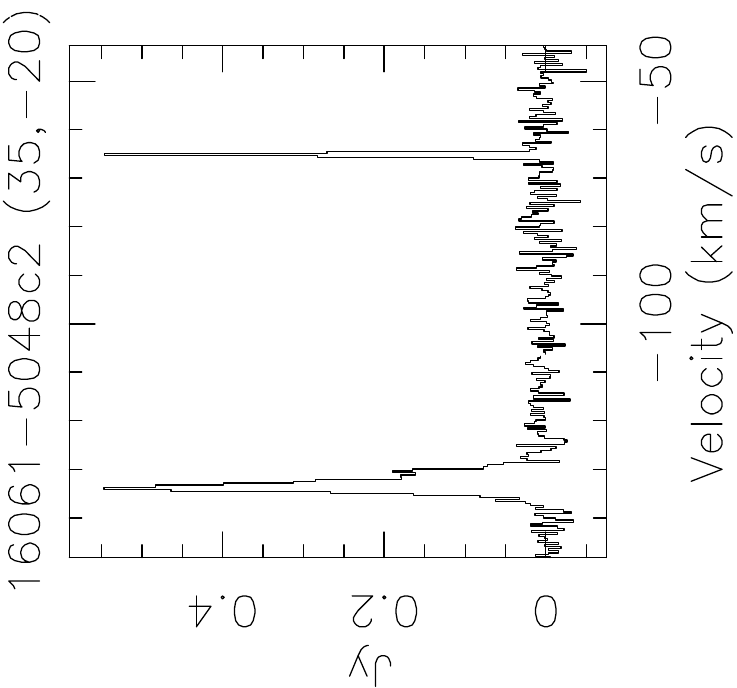}
 \includegraphics[angle=-90,width=0.24\textwidth]{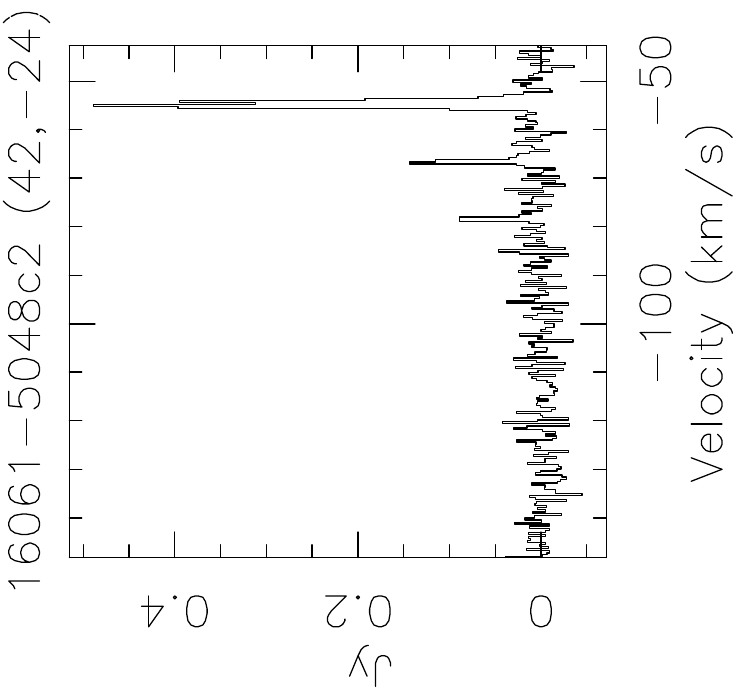}

 \includegraphics[angle=-90,width=0.24\textwidth]{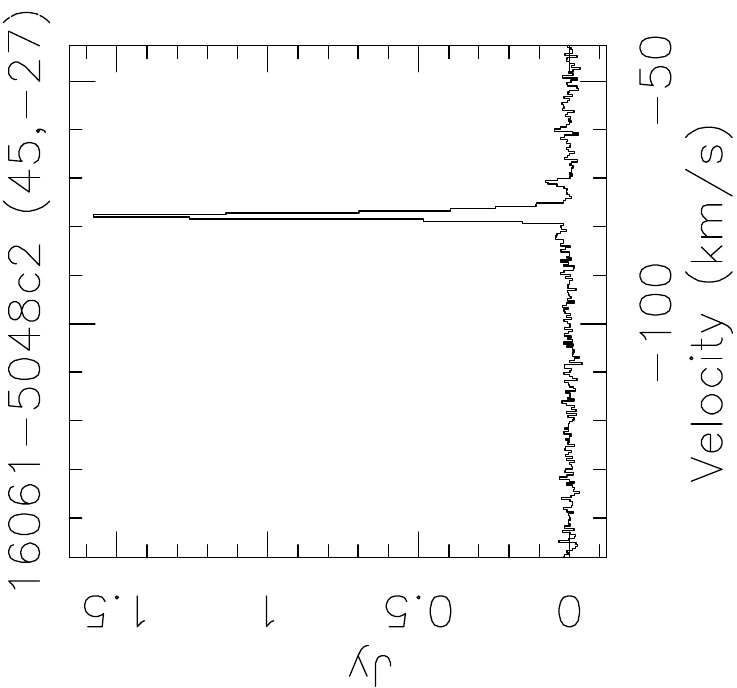}
 \includegraphics[angle=-90,width=0.24\textwidth]{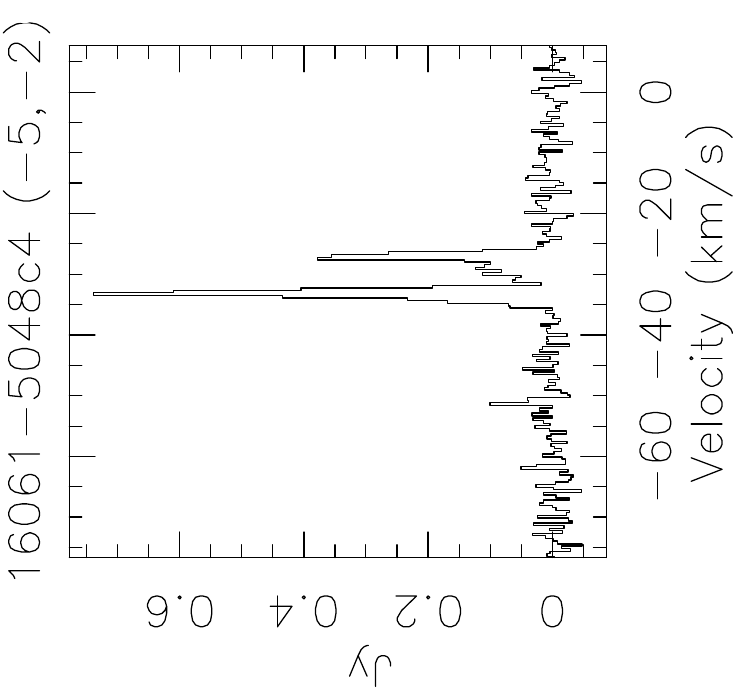}
 \includegraphics[angle=-90,width=0.24\textwidth]{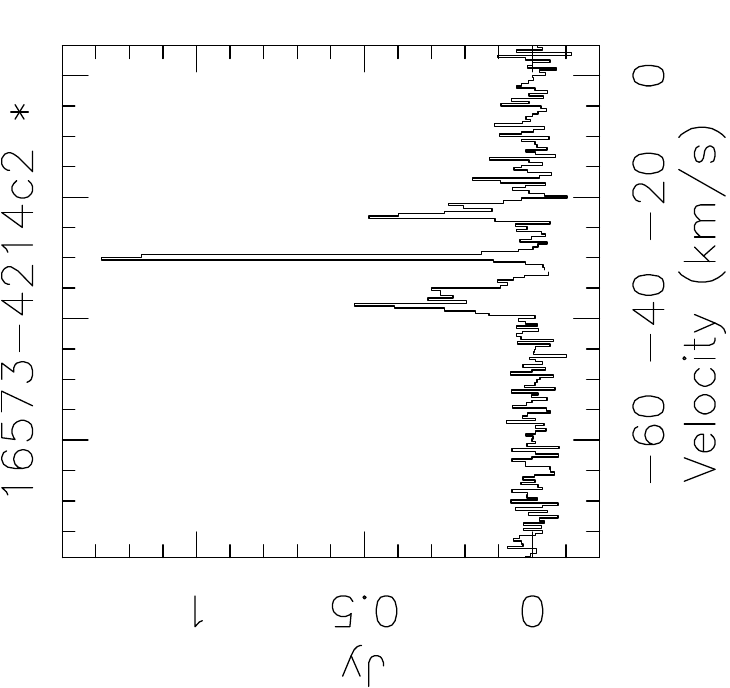}
 \includegraphics[angle=-90,width=0.24\textwidth]{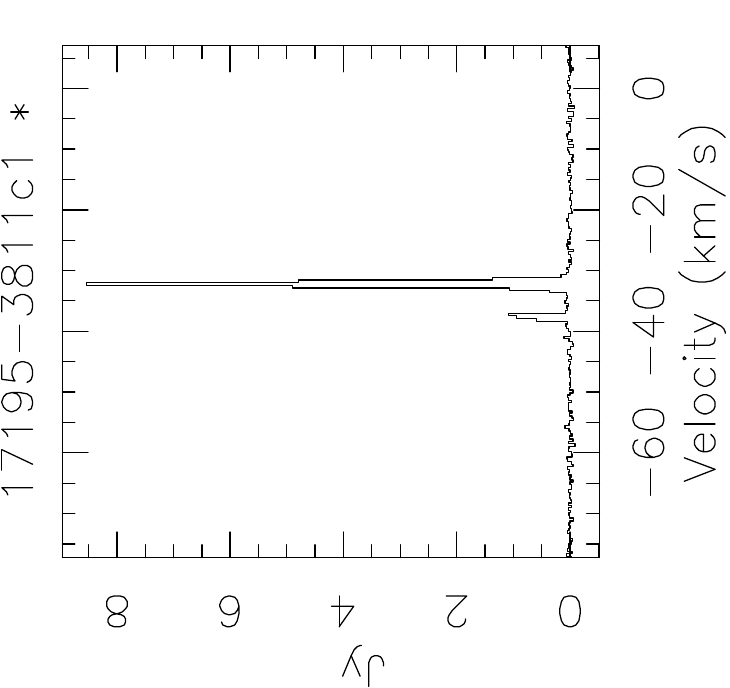}

 \caption{H$_2$O maser spectra. The clump name and the offset of the maser w.r.t. the phase centre are indicated above each panel.}
 \label{fig:maser_spec}
\end{figure*}

\begin{figure*}[tbp]
 \centering
 \includegraphics[width=0.3\textwidth]{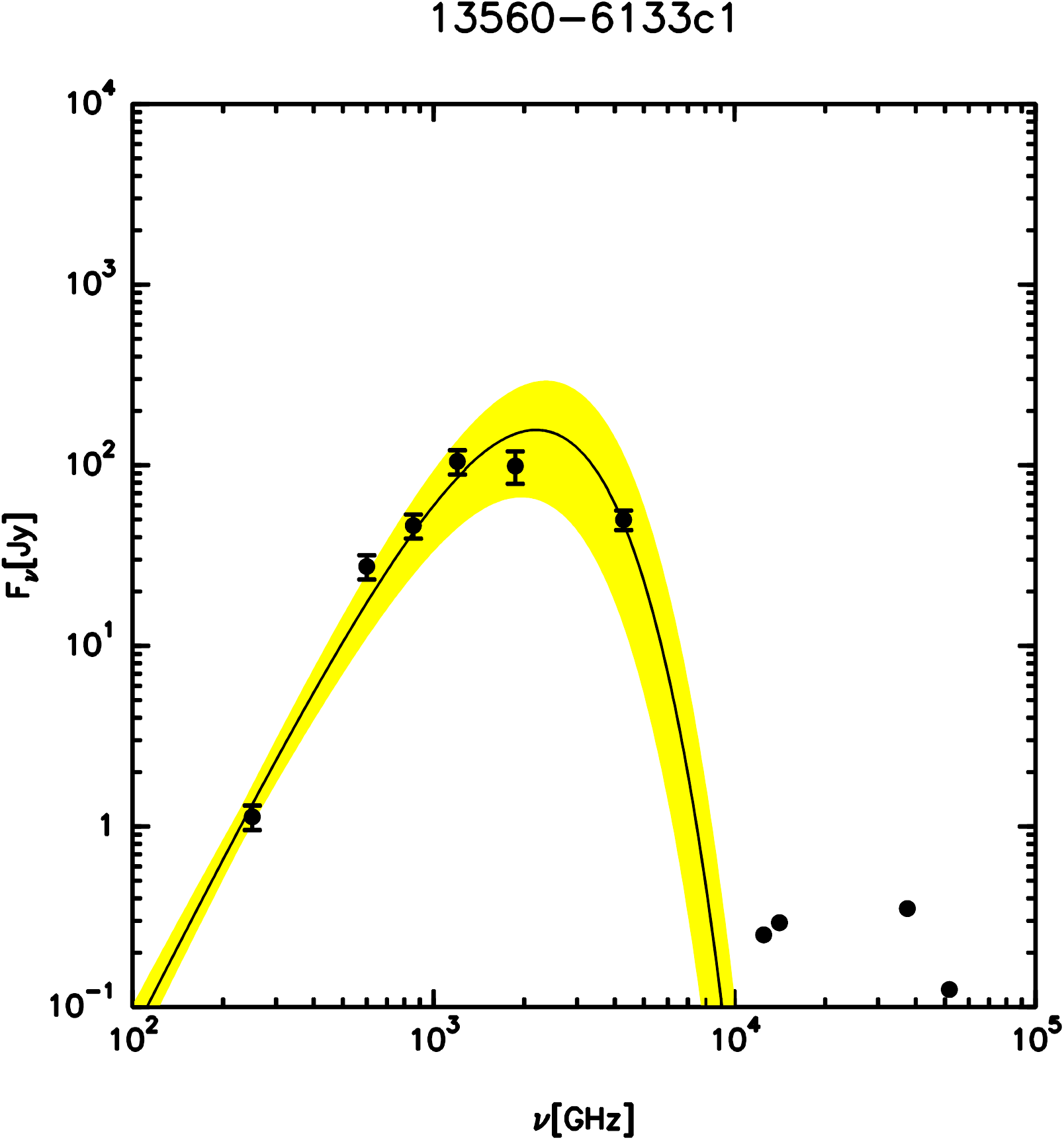}
 \includegraphics[width=0.3\textwidth]{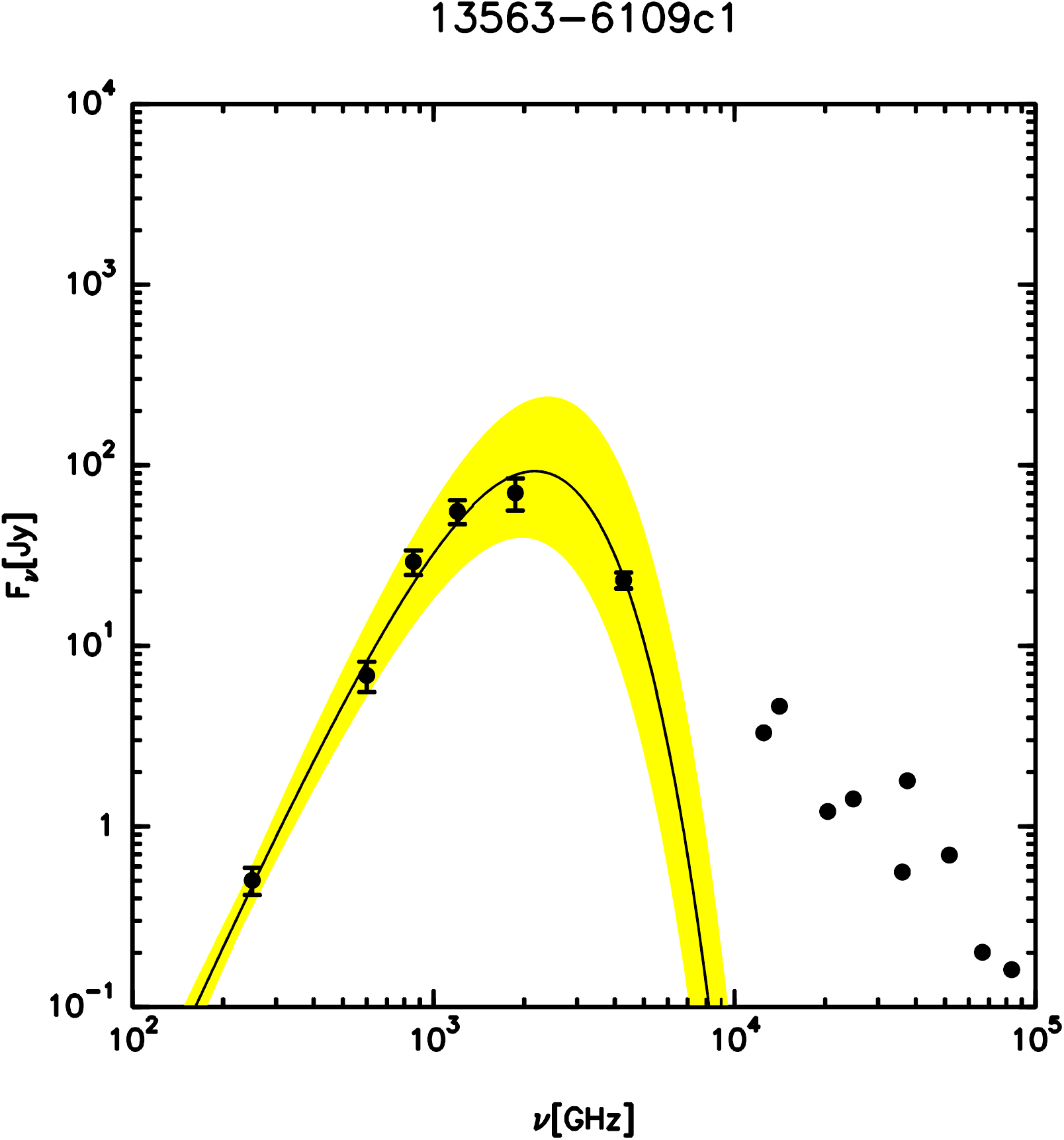}
 \includegraphics[width=0.3\textwidth]{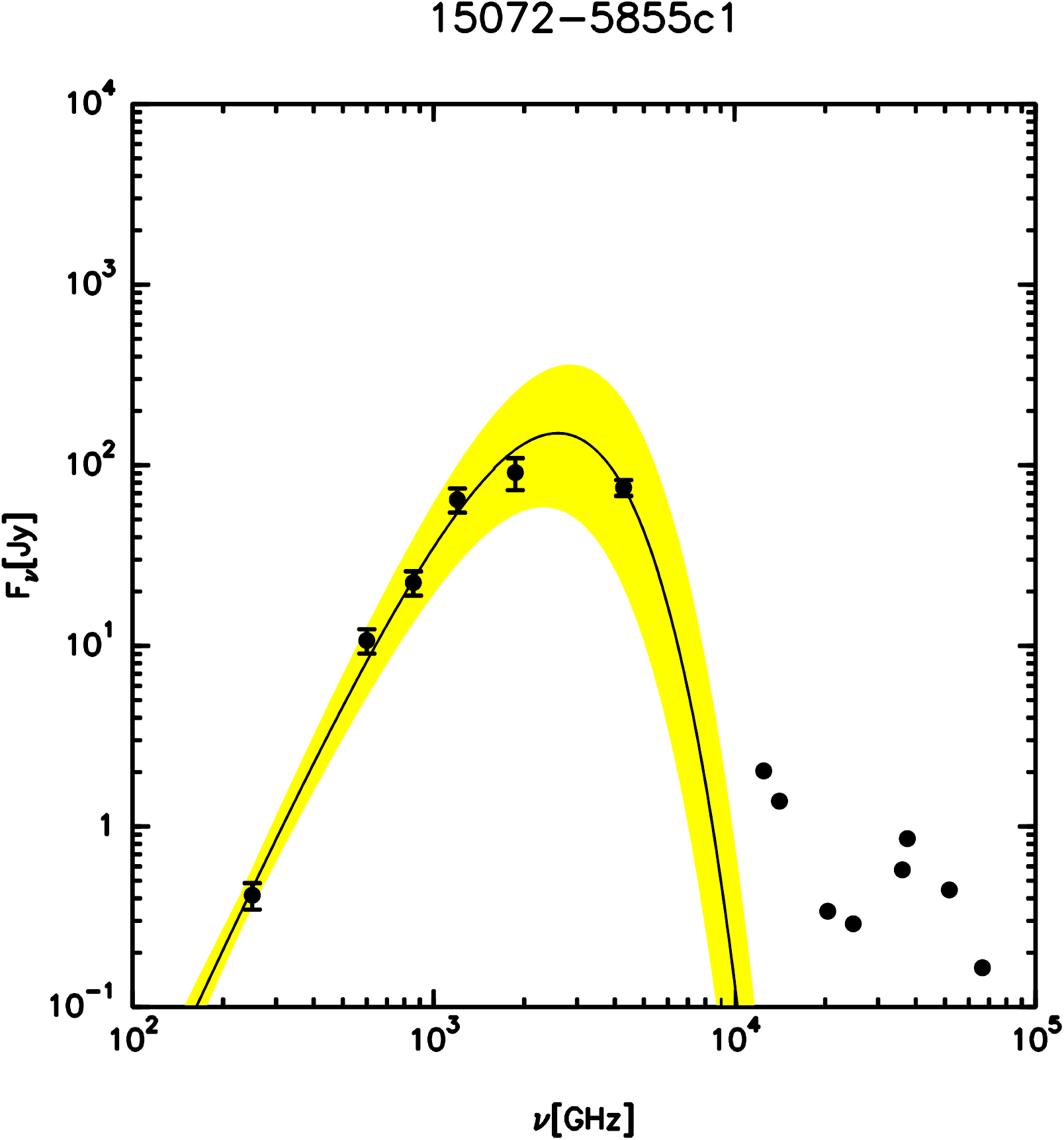}\\

 \includegraphics[width=0.3\textwidth]{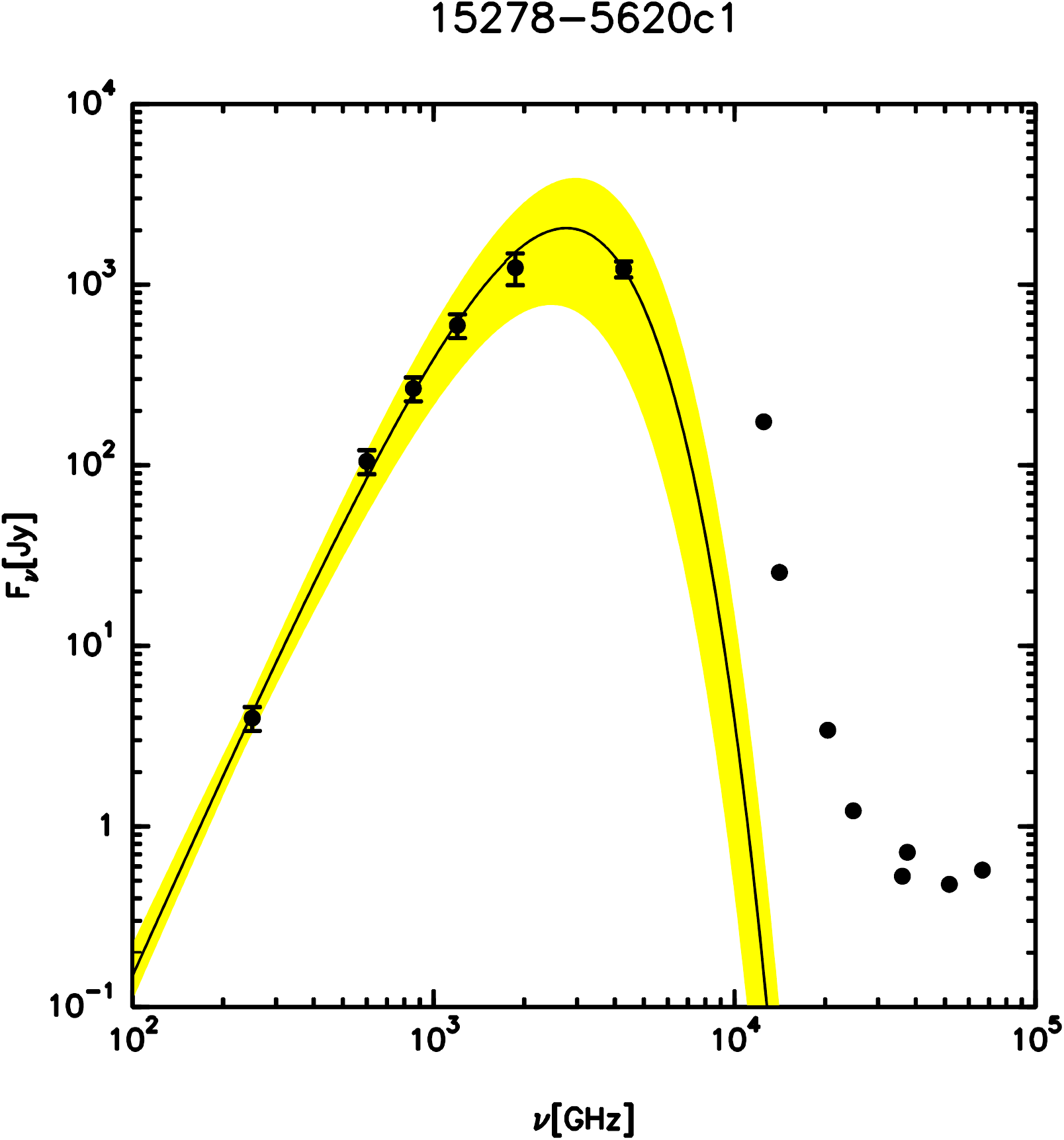}
 \includegraphics[width=0.3\textwidth]{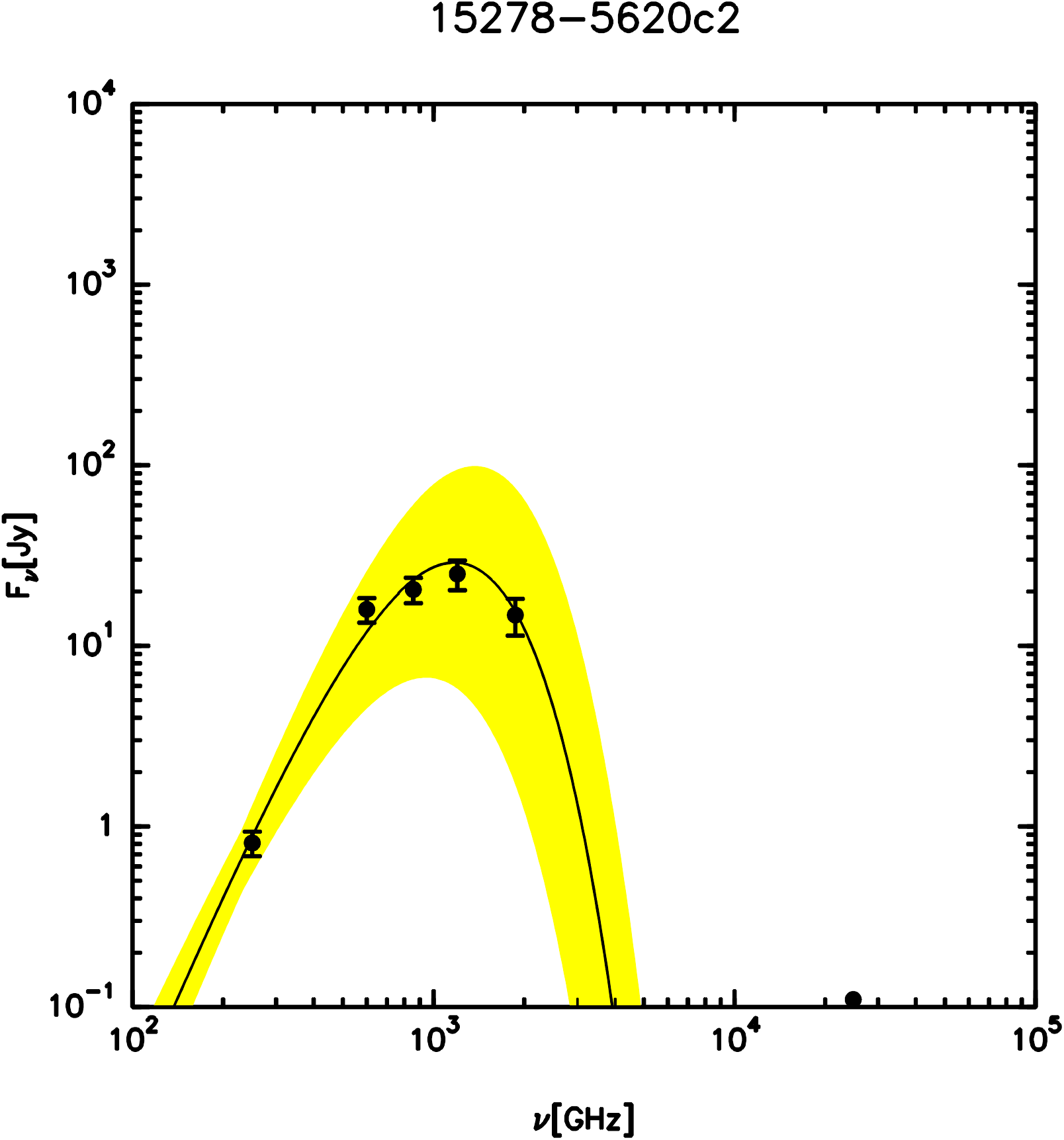}
 \includegraphics[width=0.3\textwidth]{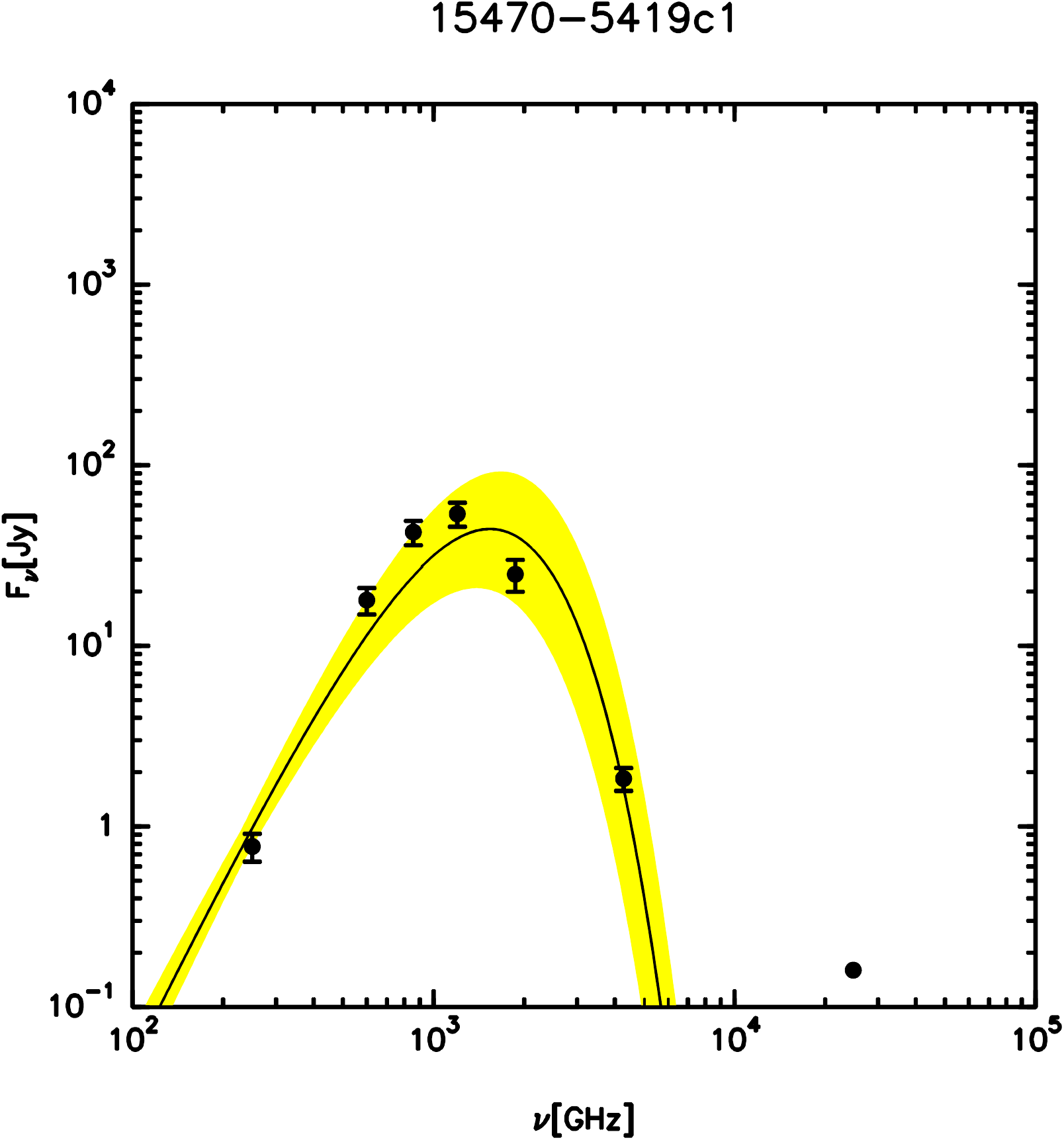}\\

 \includegraphics[width=0.3\textwidth]{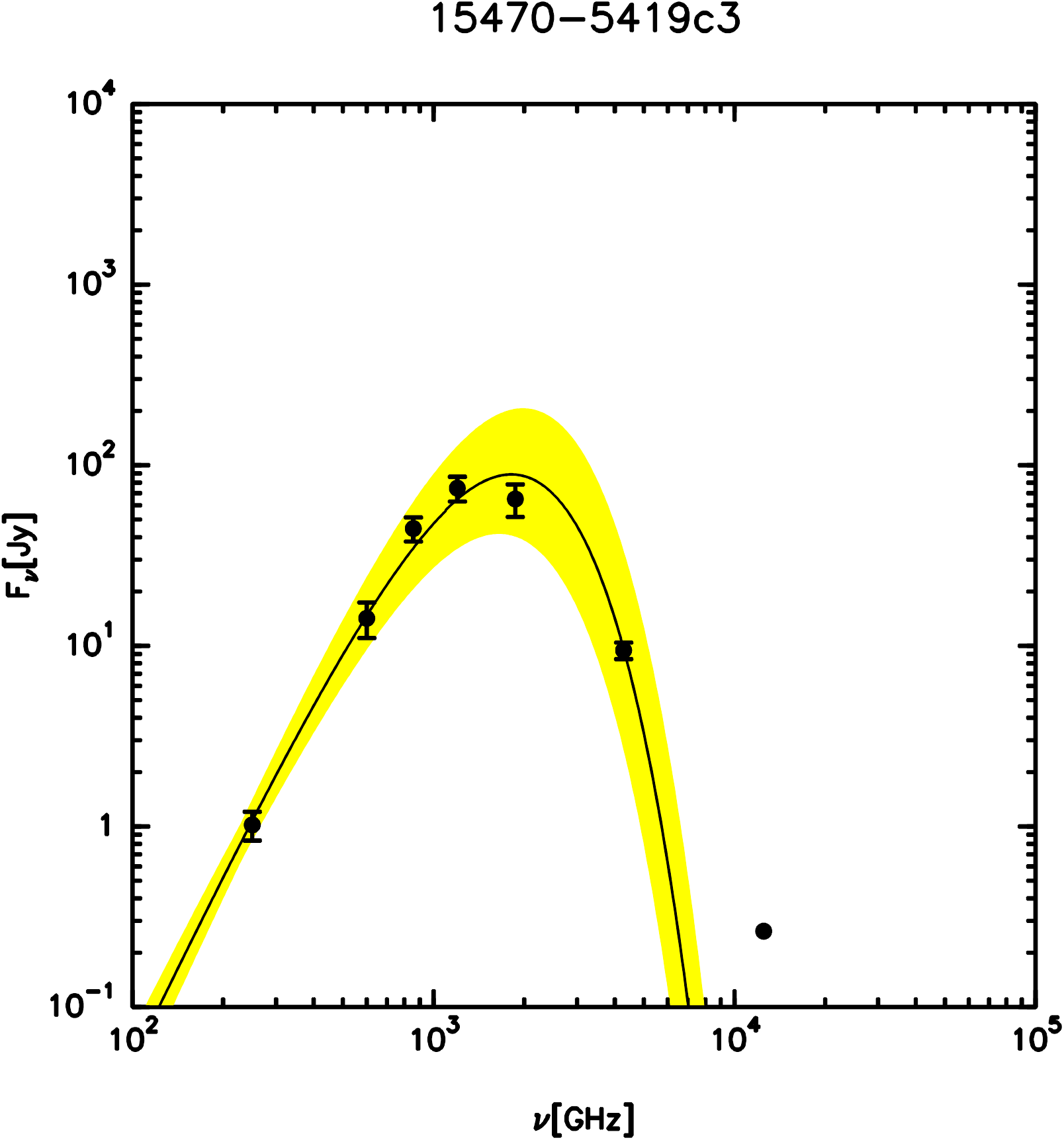}
 \includegraphics[width=0.3\textwidth]{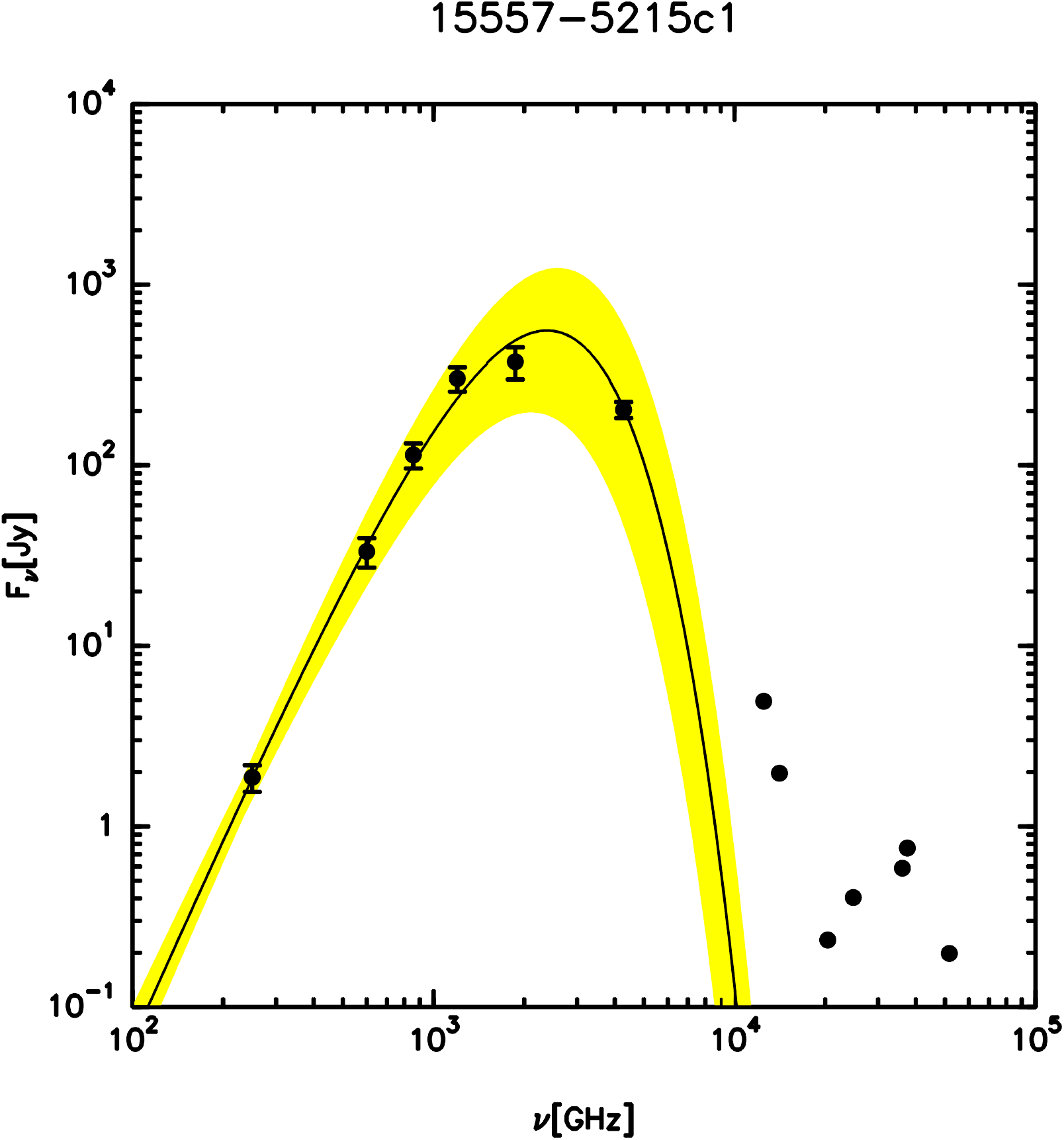}
 \includegraphics[width=0.3\textwidth]{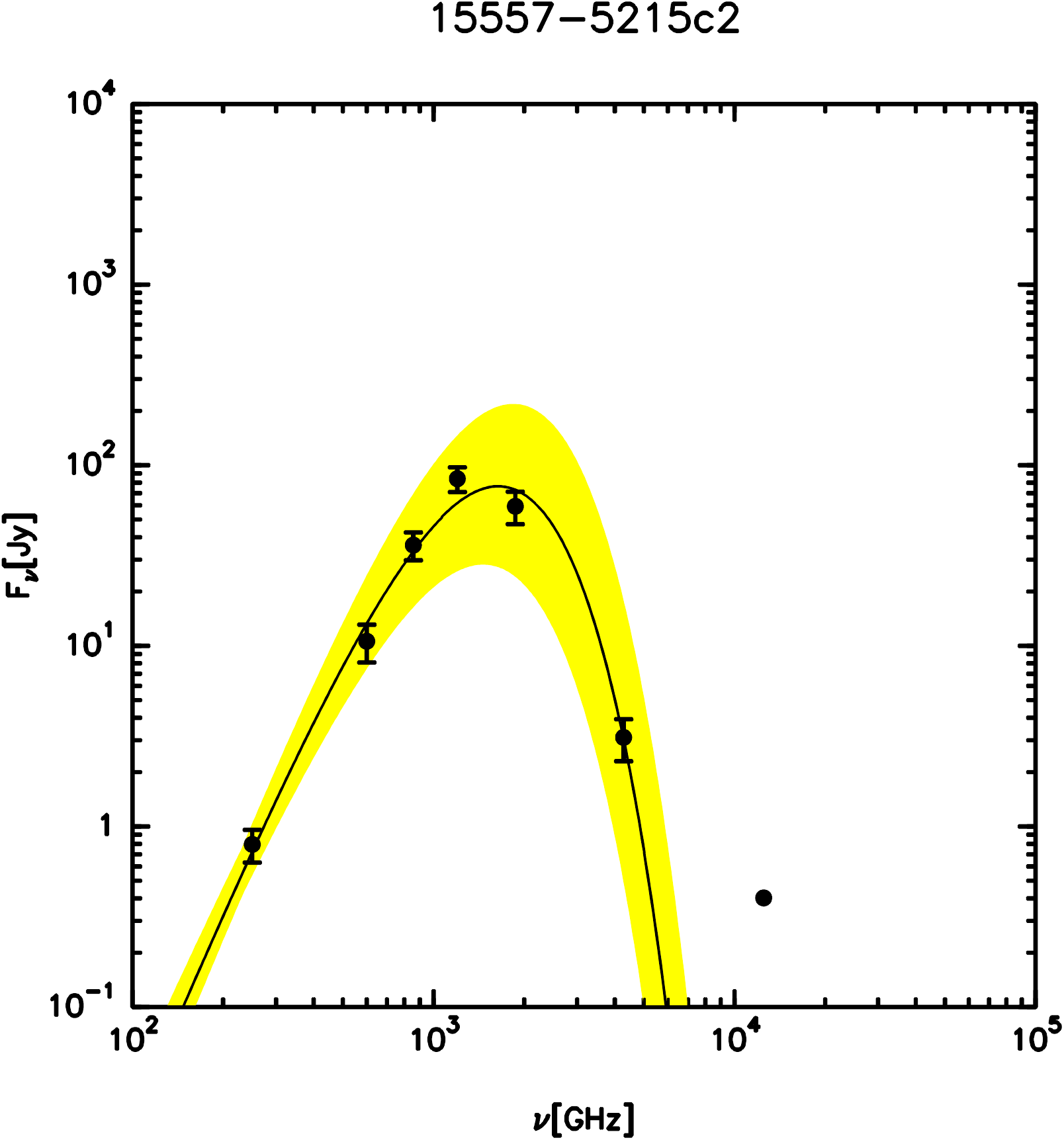}\\

 \includegraphics[width=0.3\textwidth]{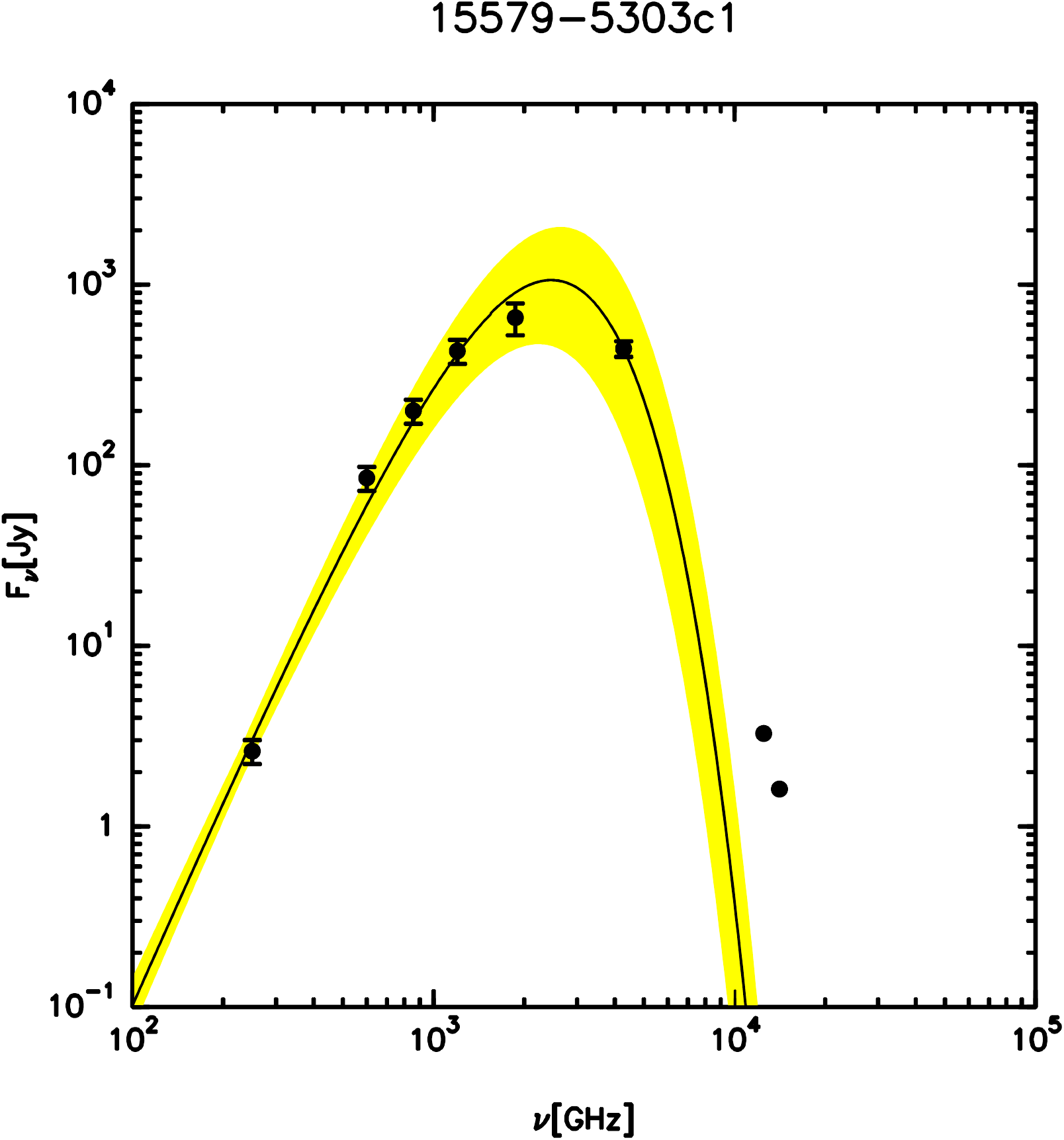}
 \includegraphics[width=0.3\textwidth]{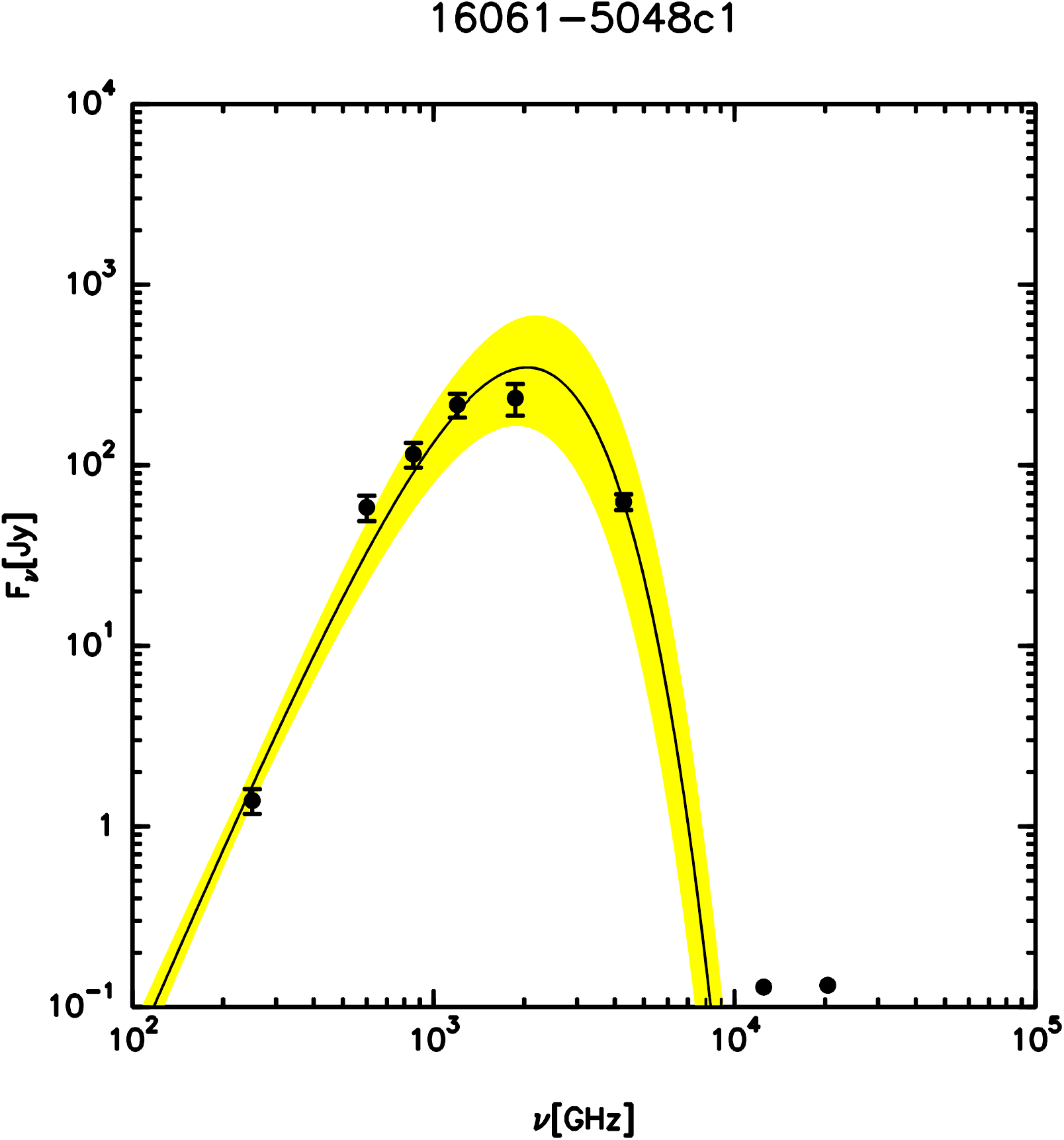}
 \includegraphics[width=0.3\textwidth]{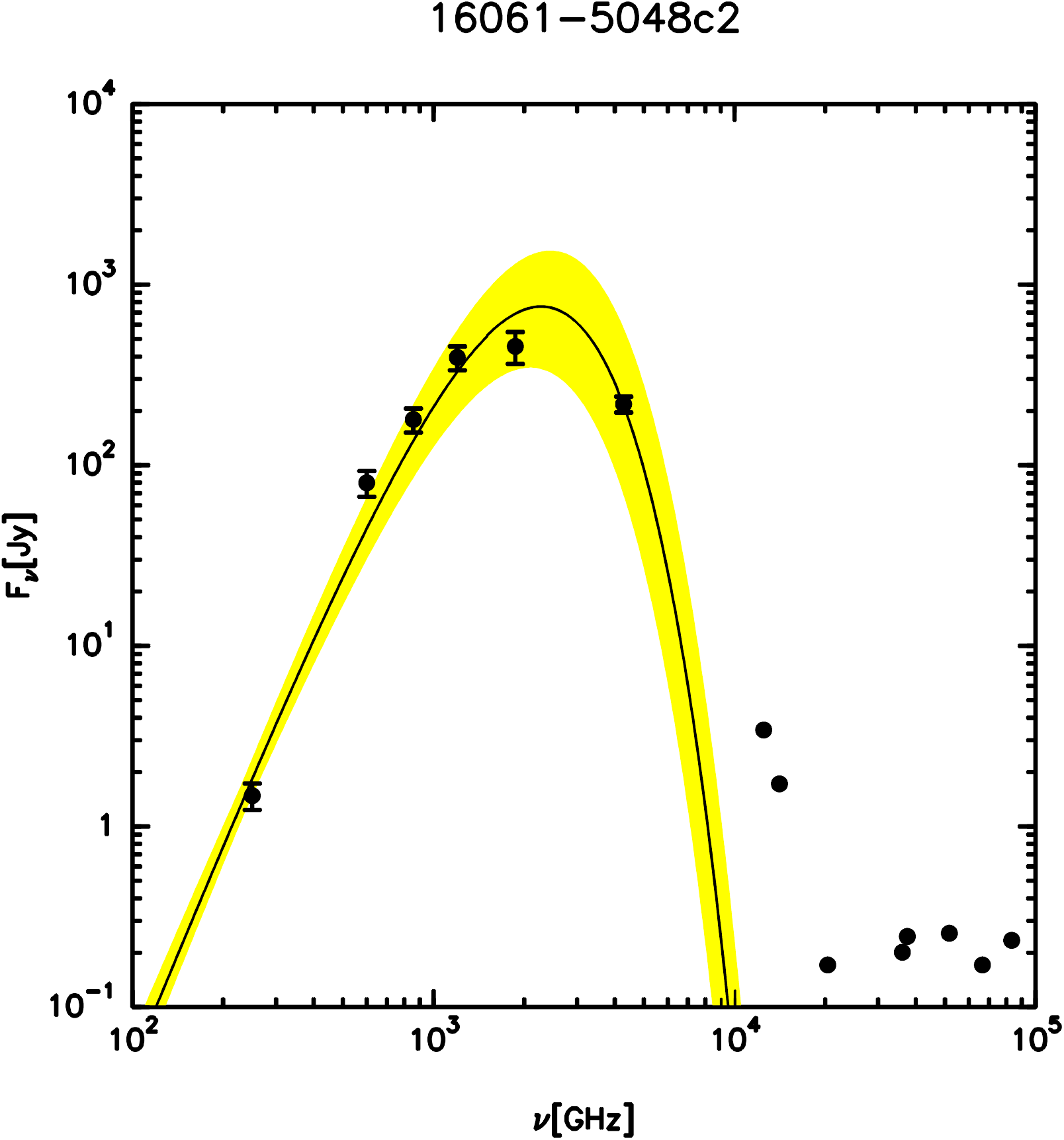}
 \caption{Spectral energy distributions for the SFS. Uncertainties in the fluxes are indicated. The yellow-shaded area shows the region encompassed by the extreme values of the fit-derived parameters for the modified black body.}
 \label{fig:sed_sfs}
\end{figure*}

\begin{figure*}[tbp]
 \ContinuedFloat
 \centering
 \includegraphics[width=0.3\textwidth]{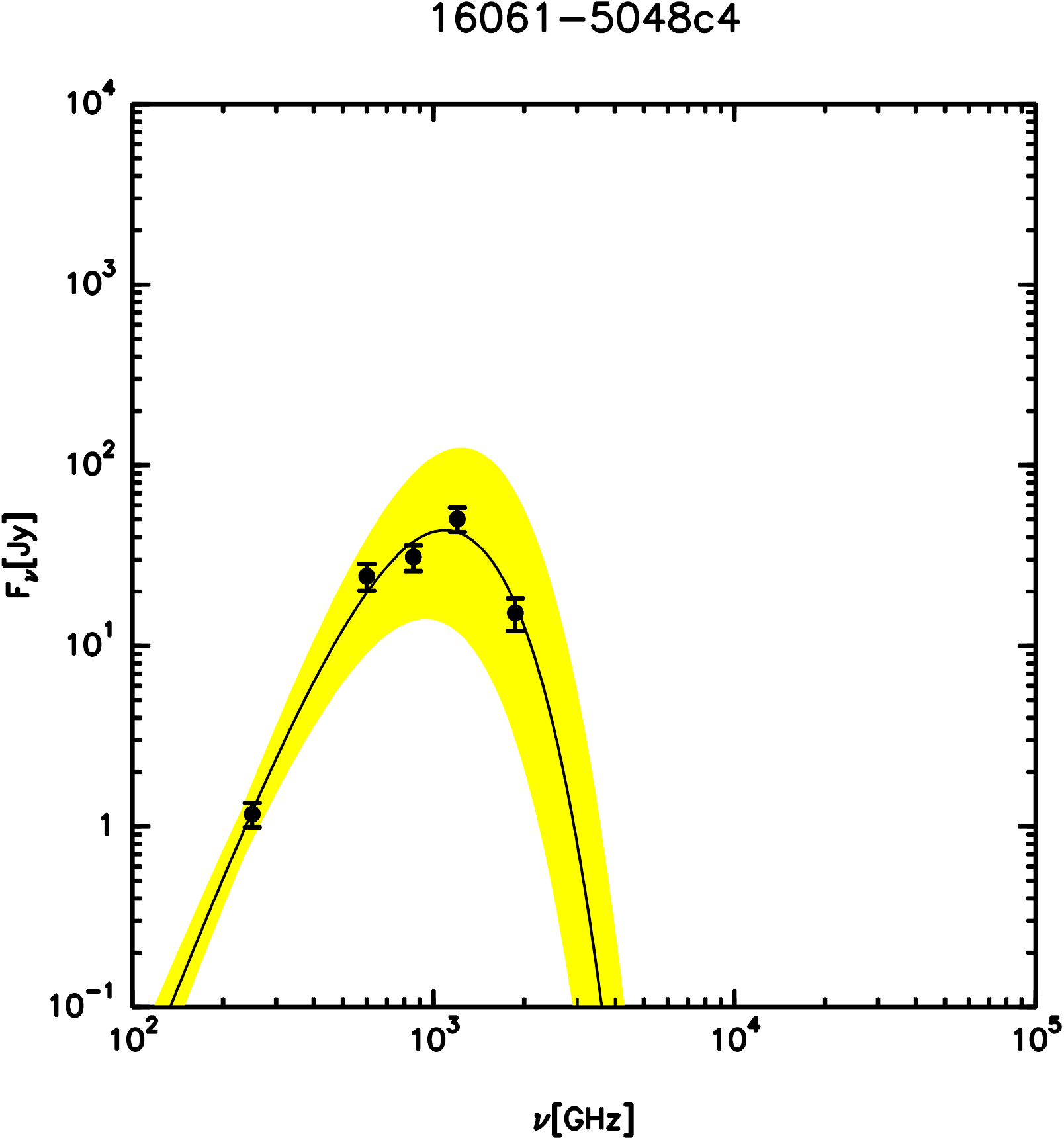}
 \includegraphics[width=0.3\textwidth]{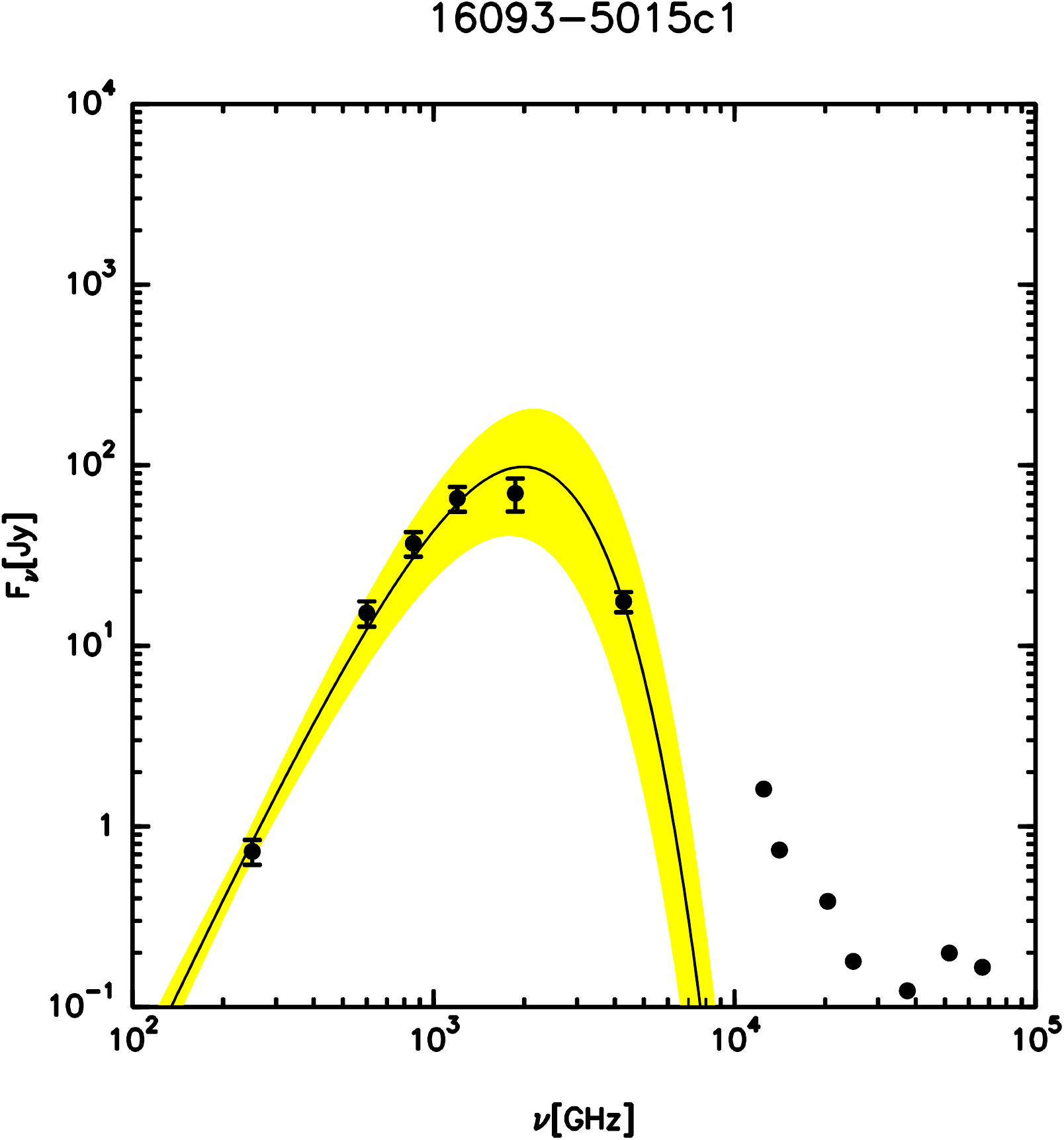}
 \includegraphics[width=0.3\textwidth]{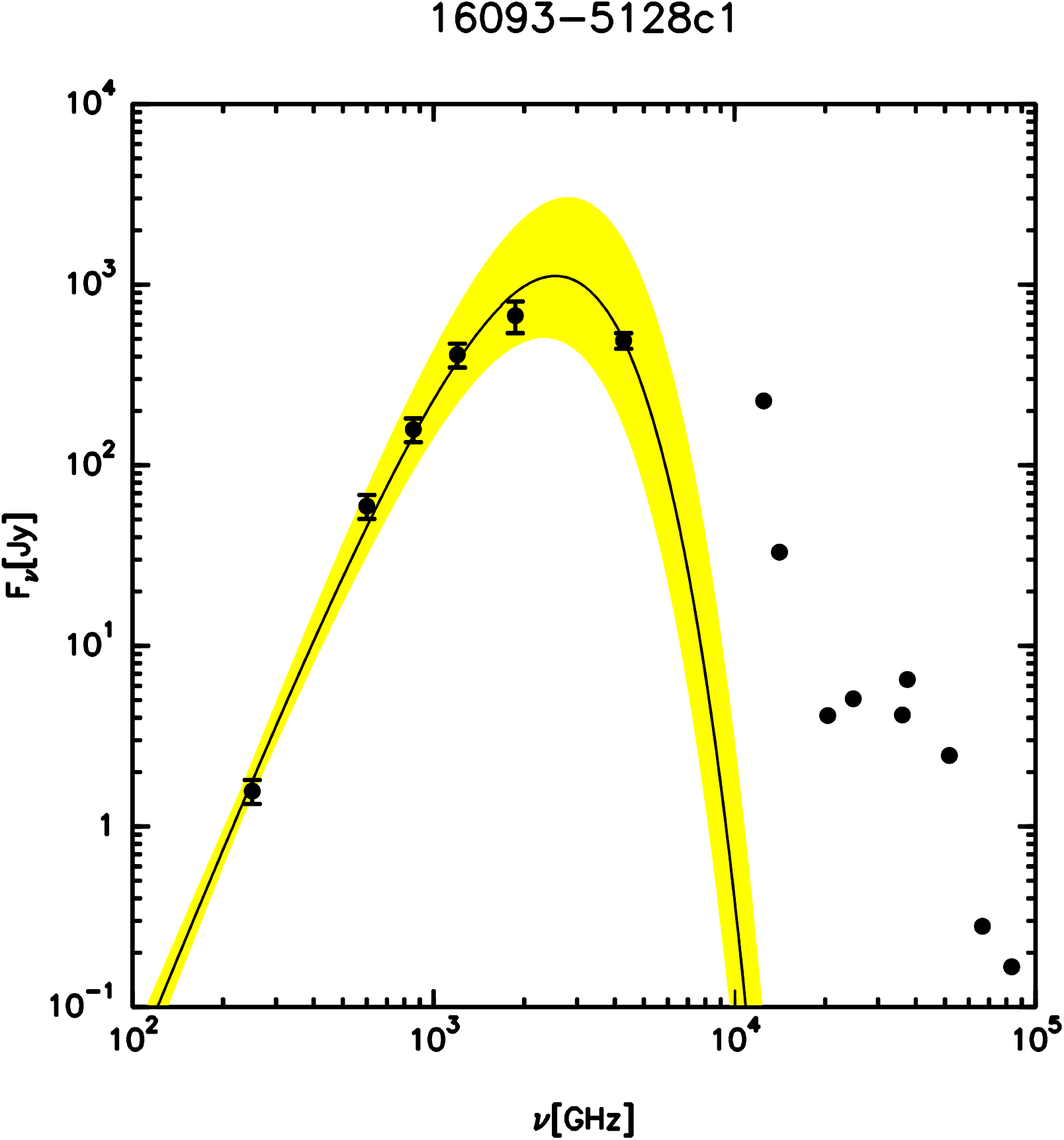}\\

 \includegraphics[width=0.3\textwidth]{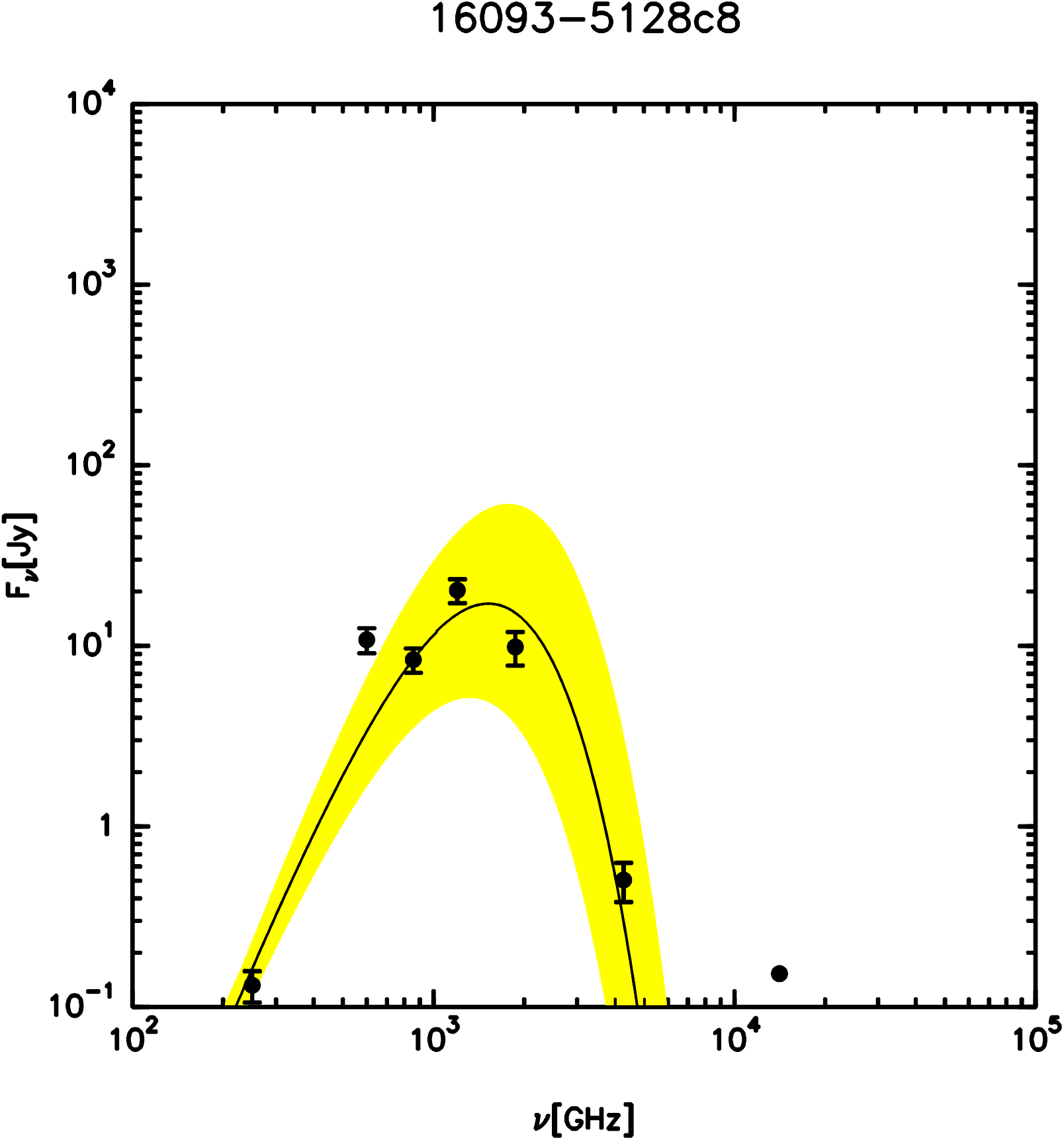}
 \includegraphics[width=0.3\textwidth]{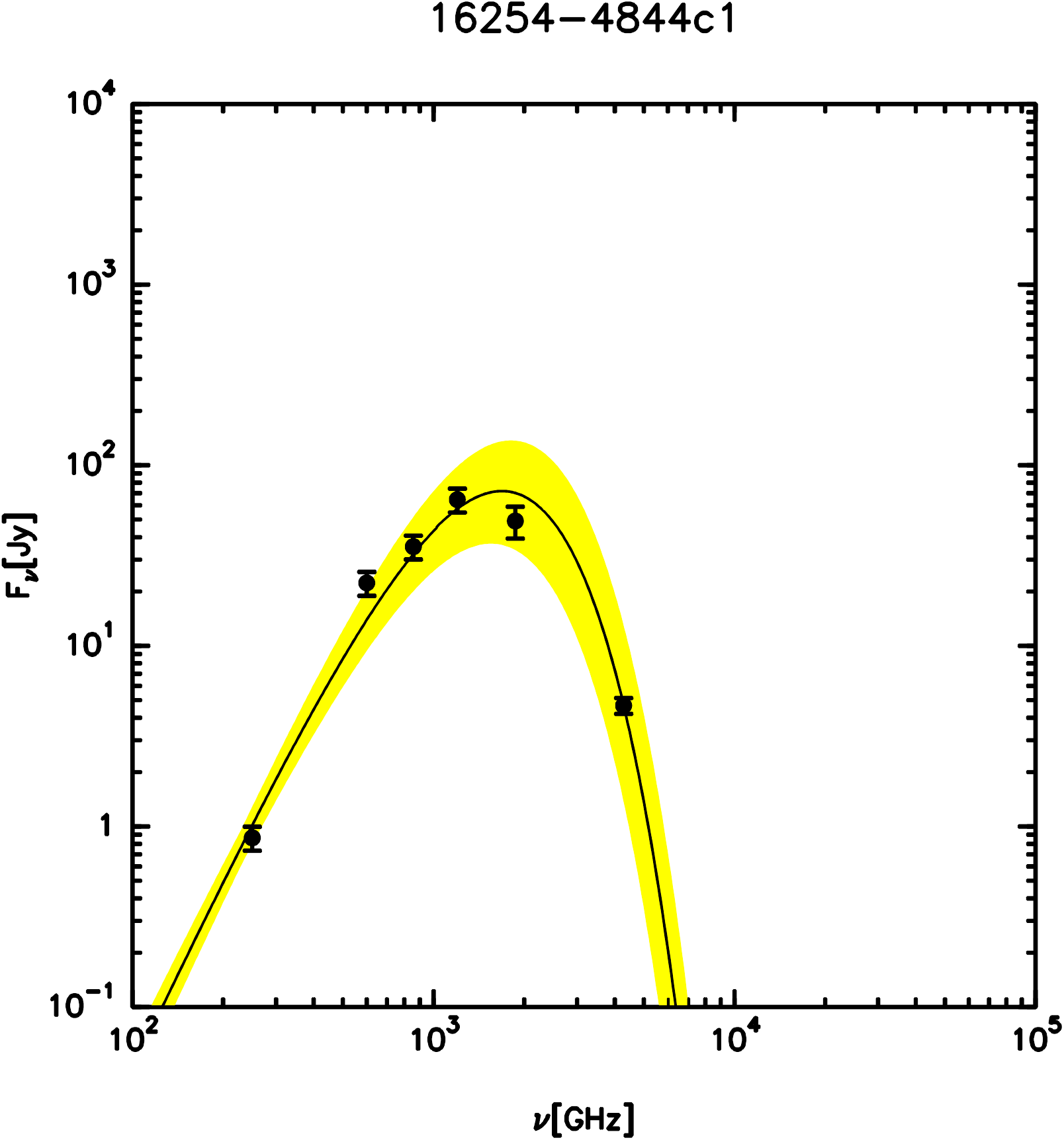}
 \includegraphics[width=0.3\textwidth]{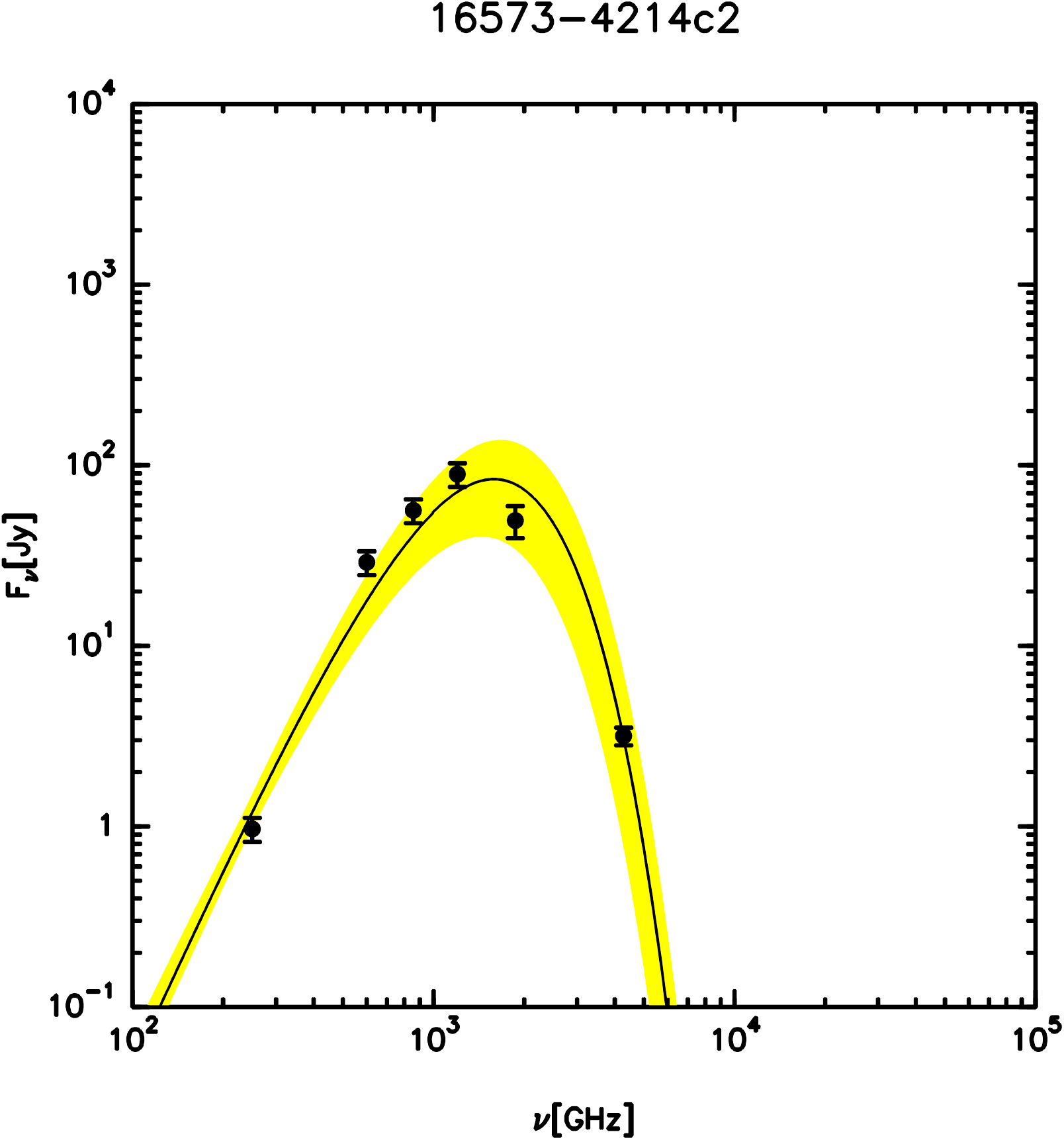}\\

 \includegraphics[width=0.3\textwidth]{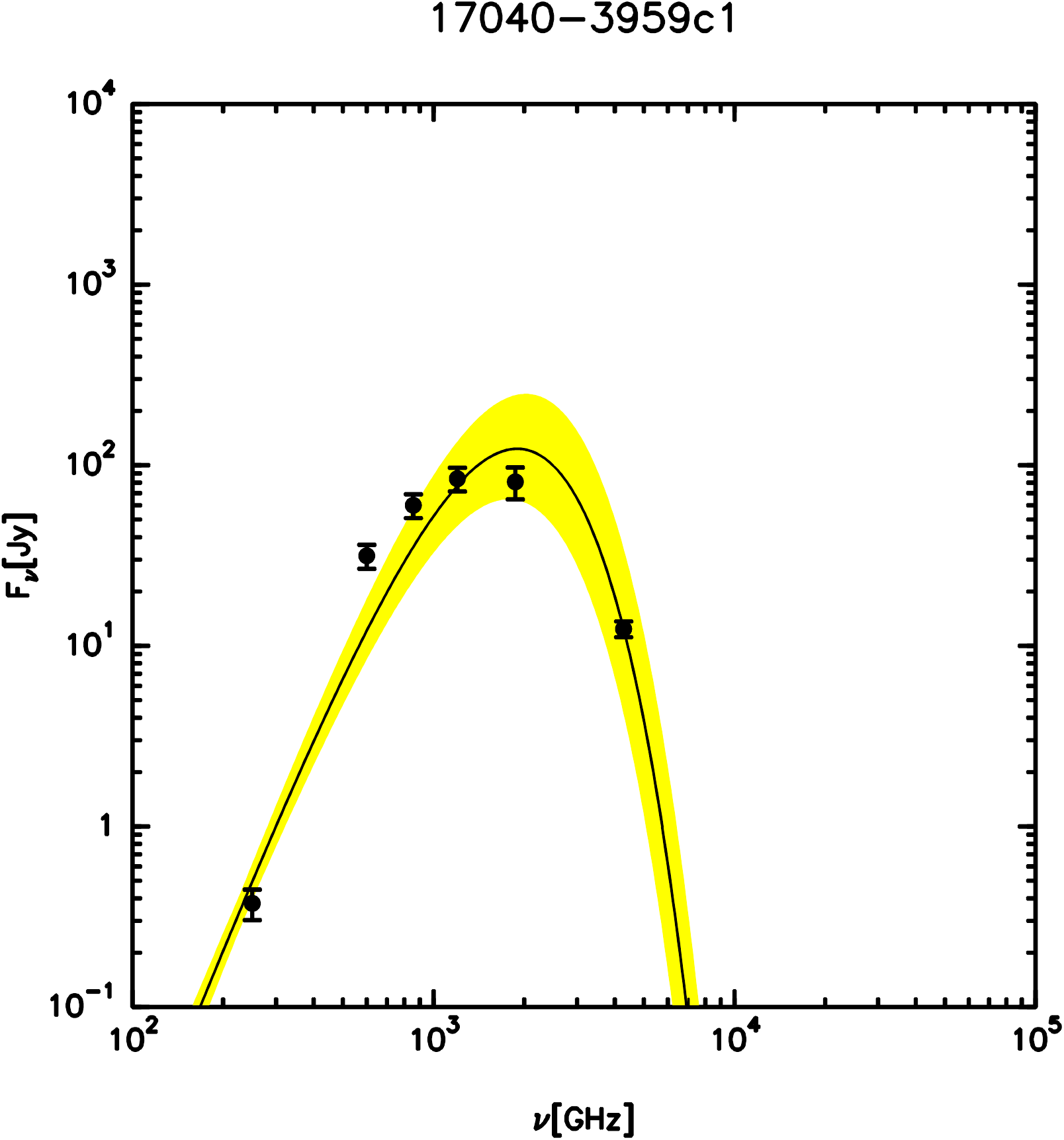}
 \includegraphics[width=0.3\textwidth]{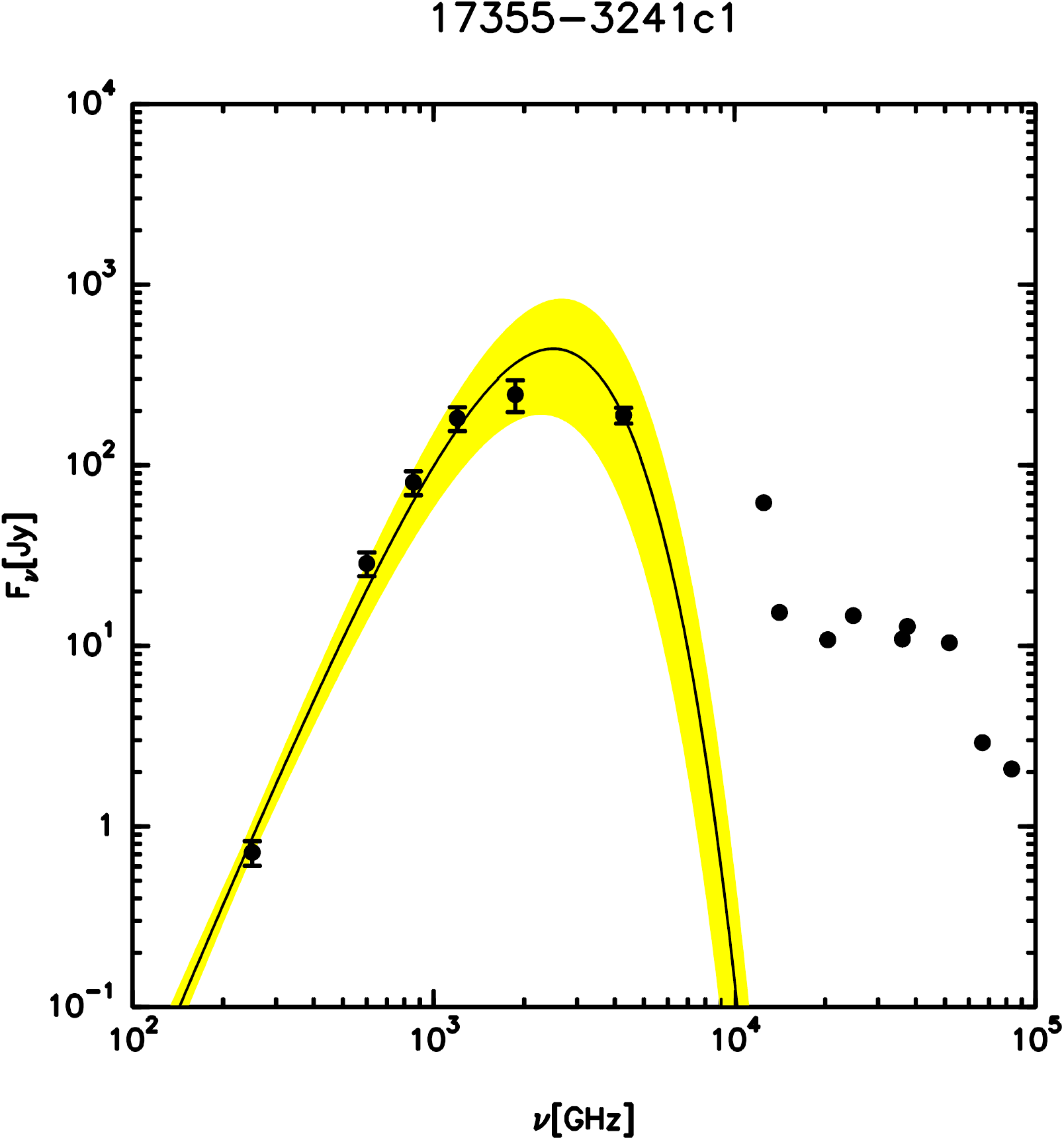}
  \caption{Continued.}
\end{figure*}

\begin{figure*}[tbp]
 \centering
 \includegraphics[width=0.3\textwidth]{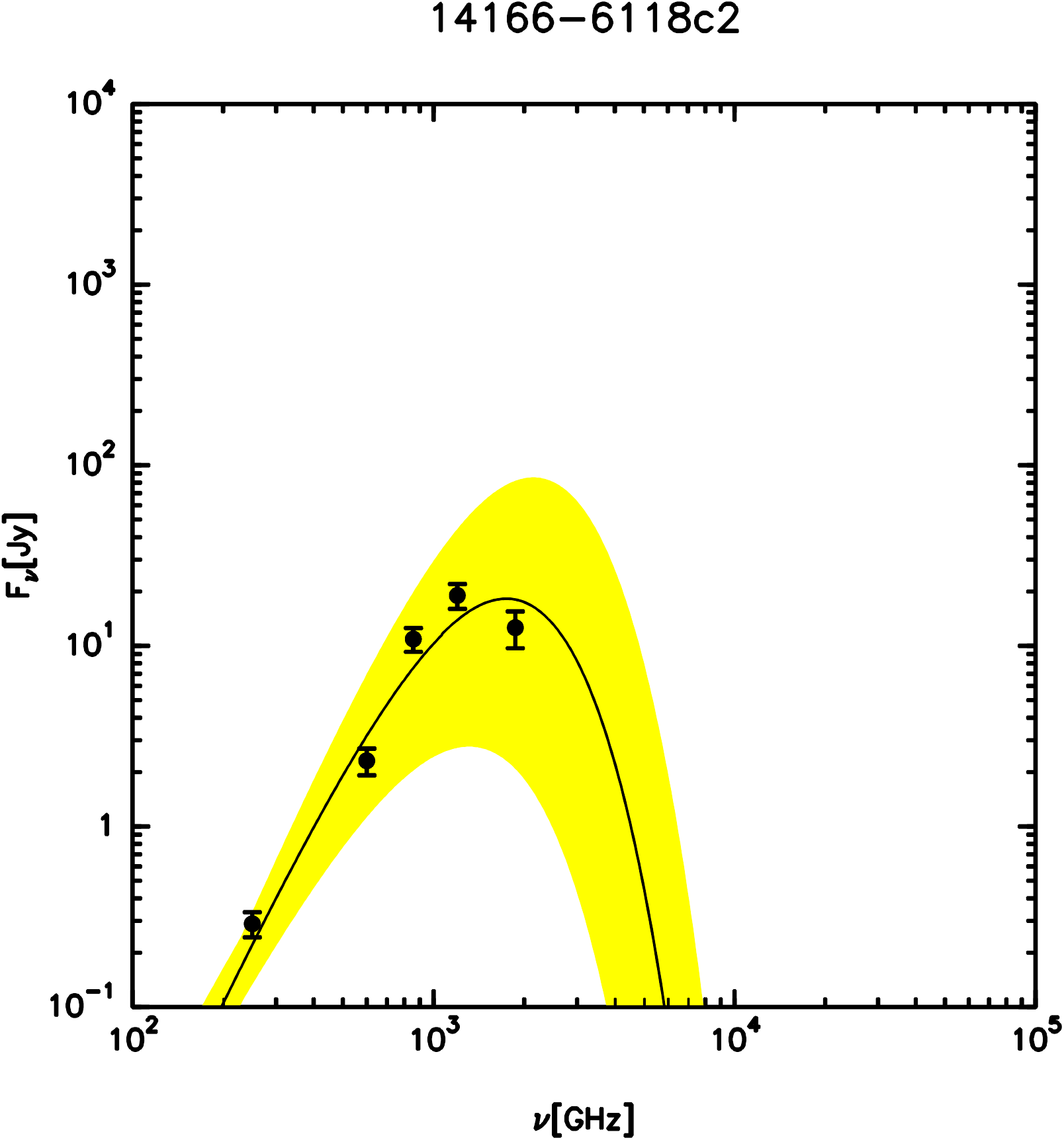}
 \includegraphics[width=0.3\textwidth]{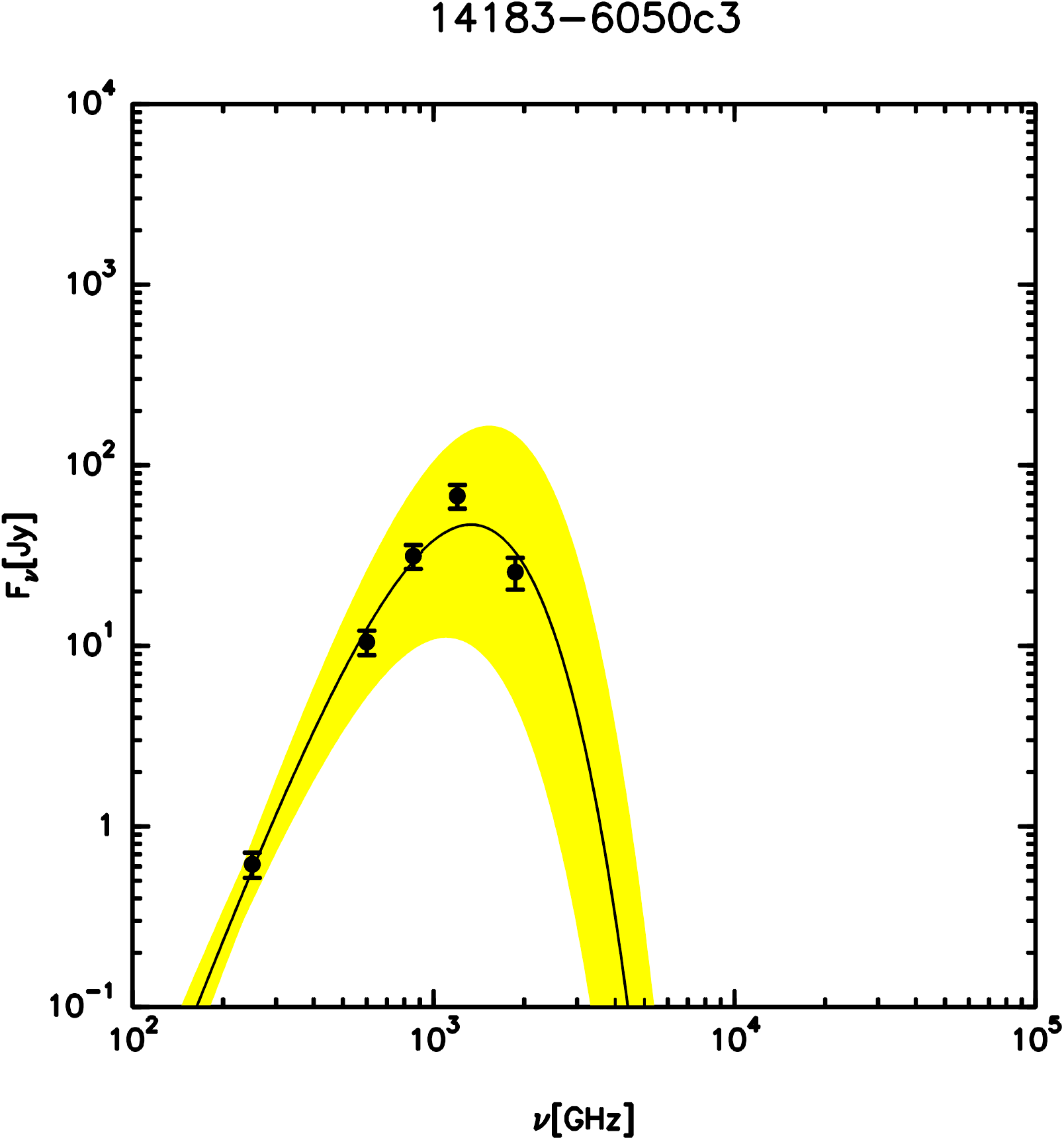}
 \includegraphics[width=0.3\textwidth]{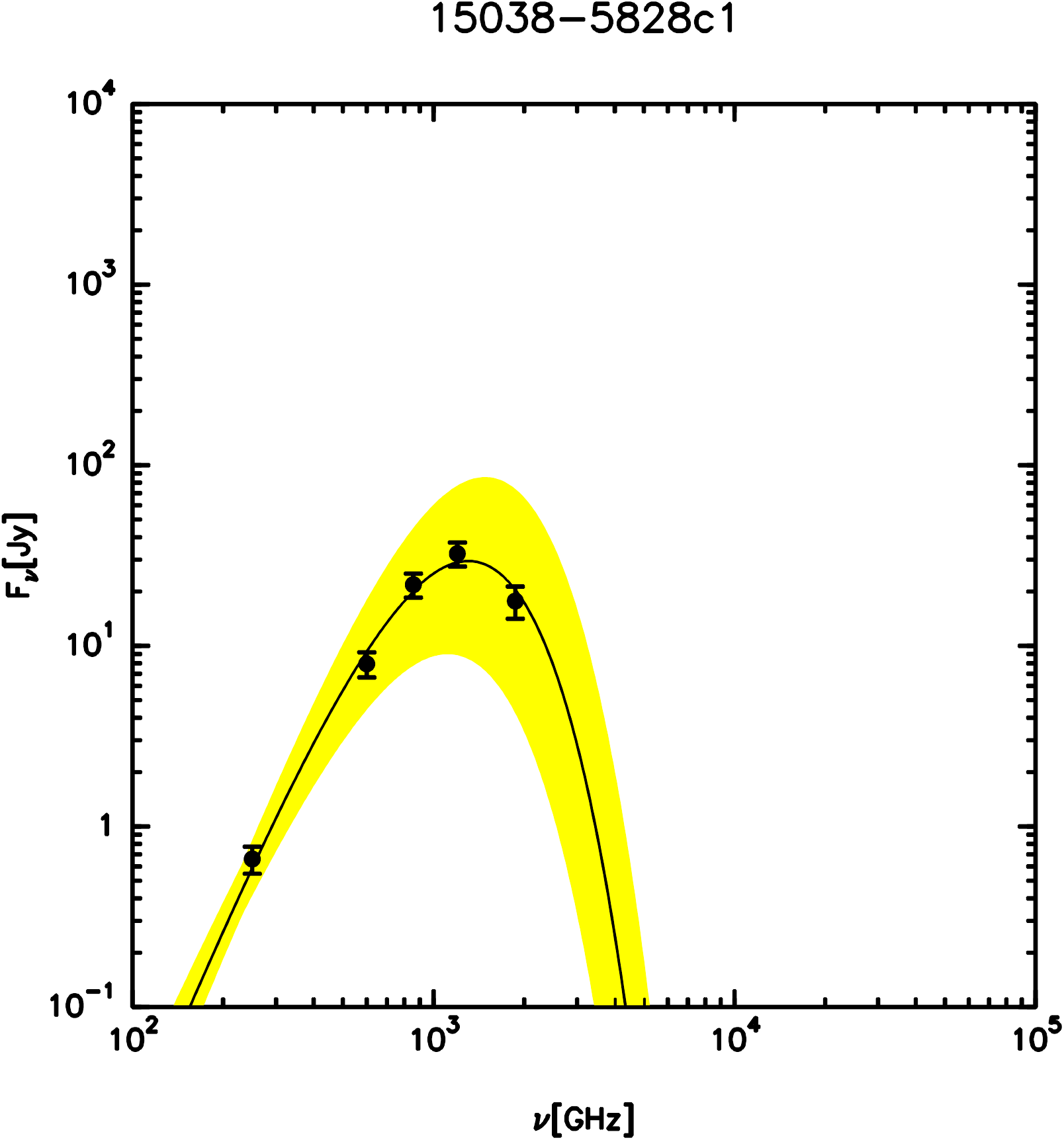}\\

 \includegraphics[width=0.3\textwidth]{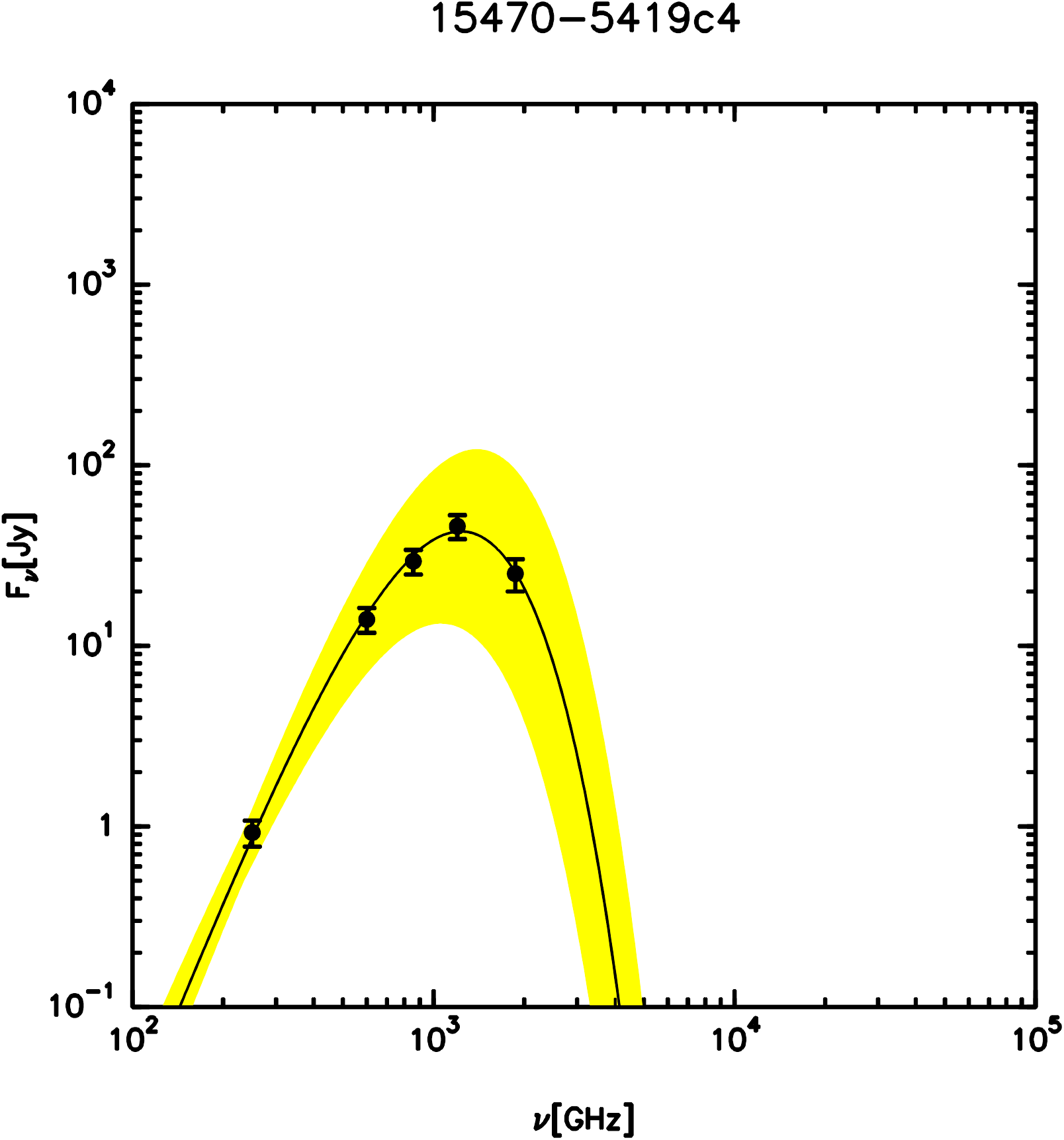}
 \includegraphics[width=0.3\textwidth]{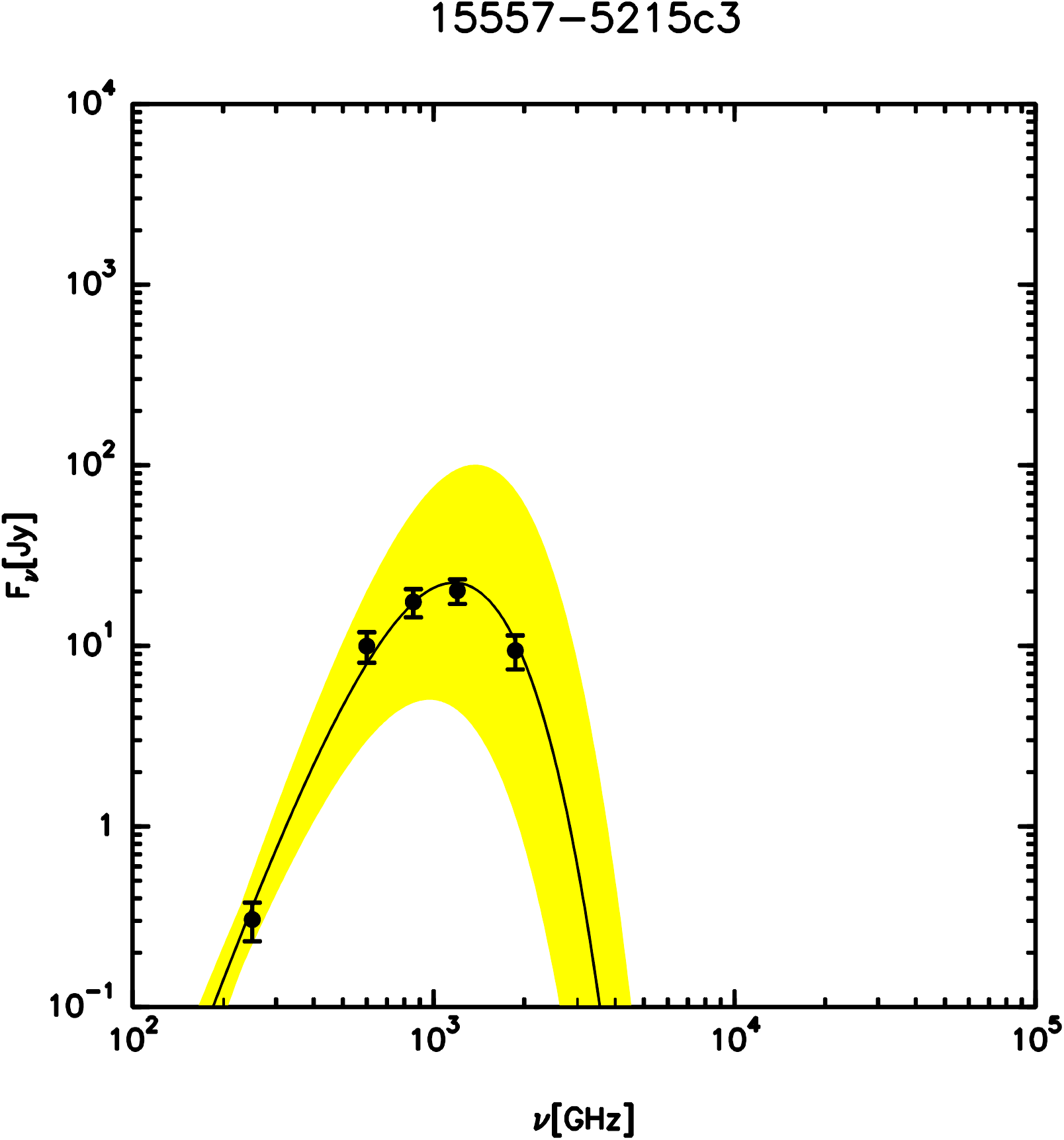}
 \includegraphics[width=0.3\textwidth]{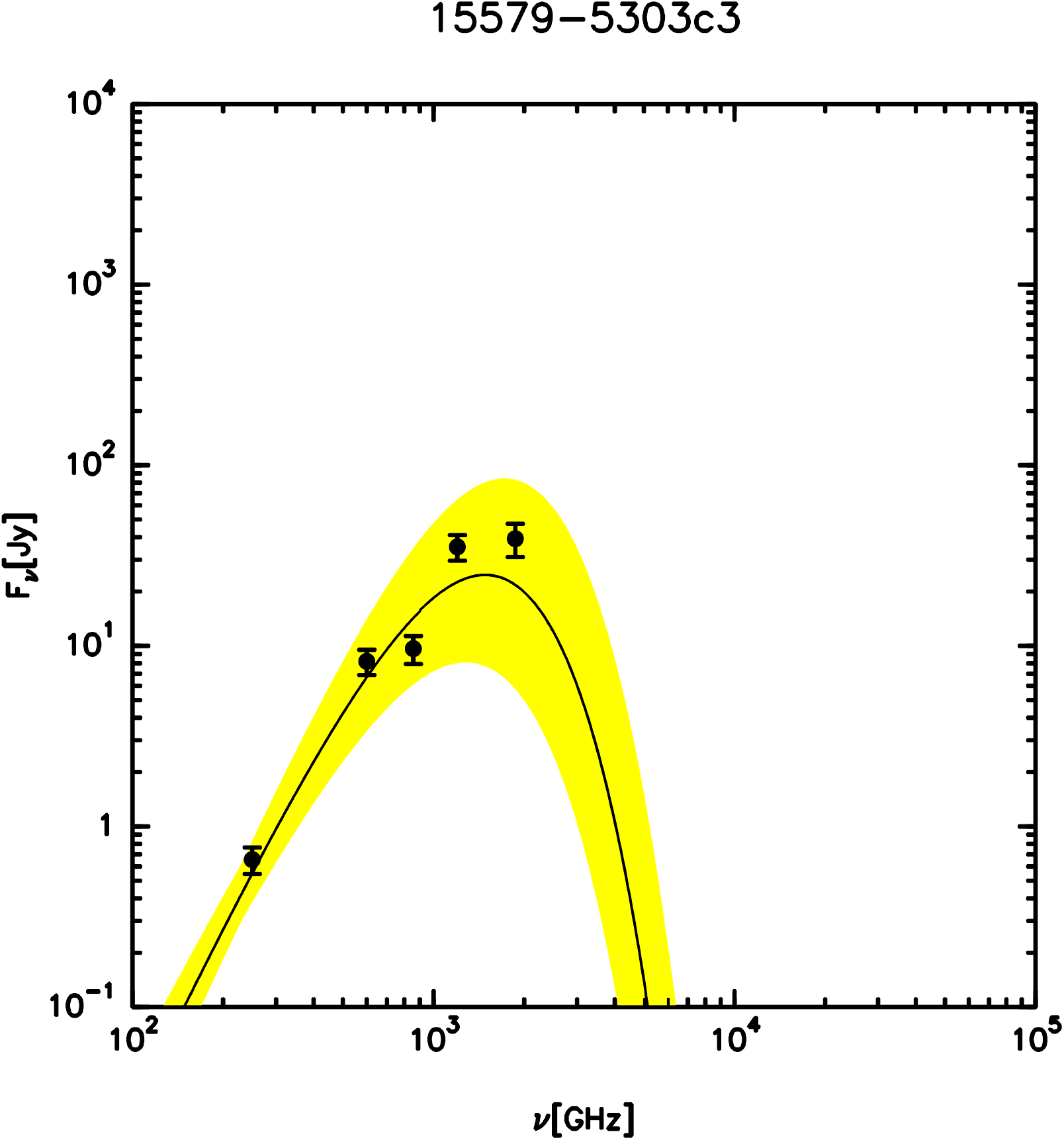}\\

 \includegraphics[width=0.3\textwidth]{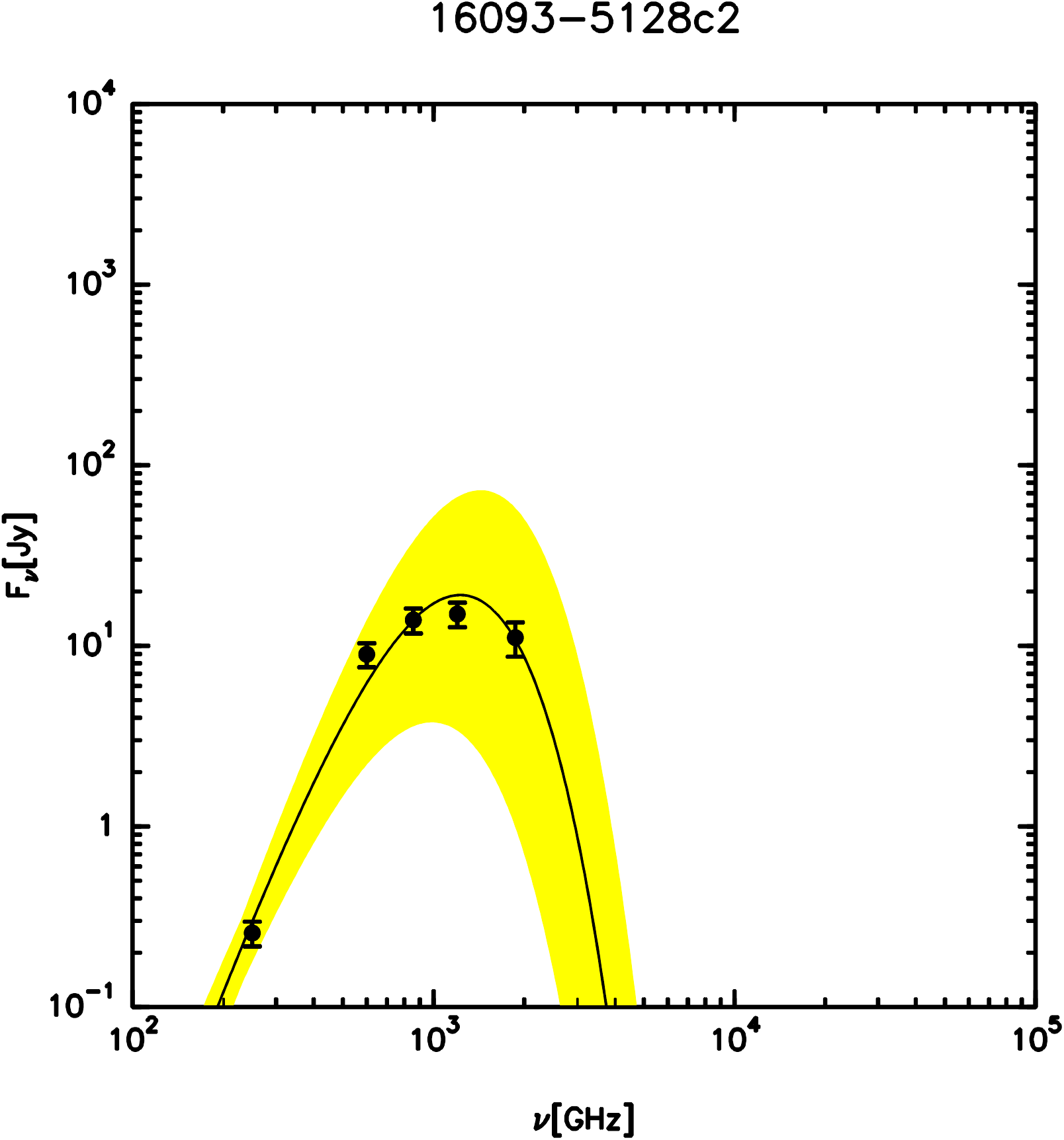}
 \includegraphics[width=0.3\textwidth]{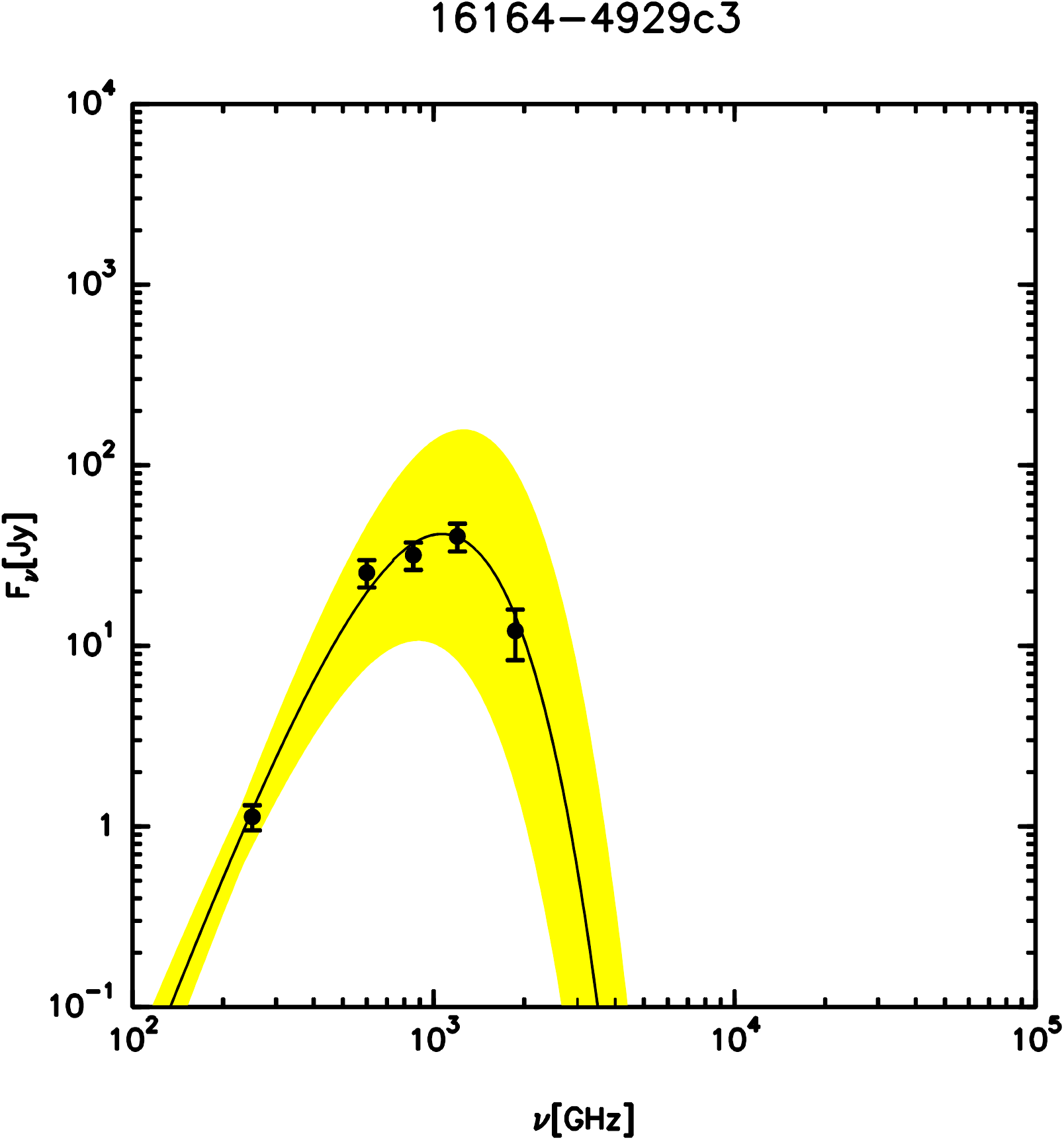}
 \includegraphics[width=0.3\textwidth]{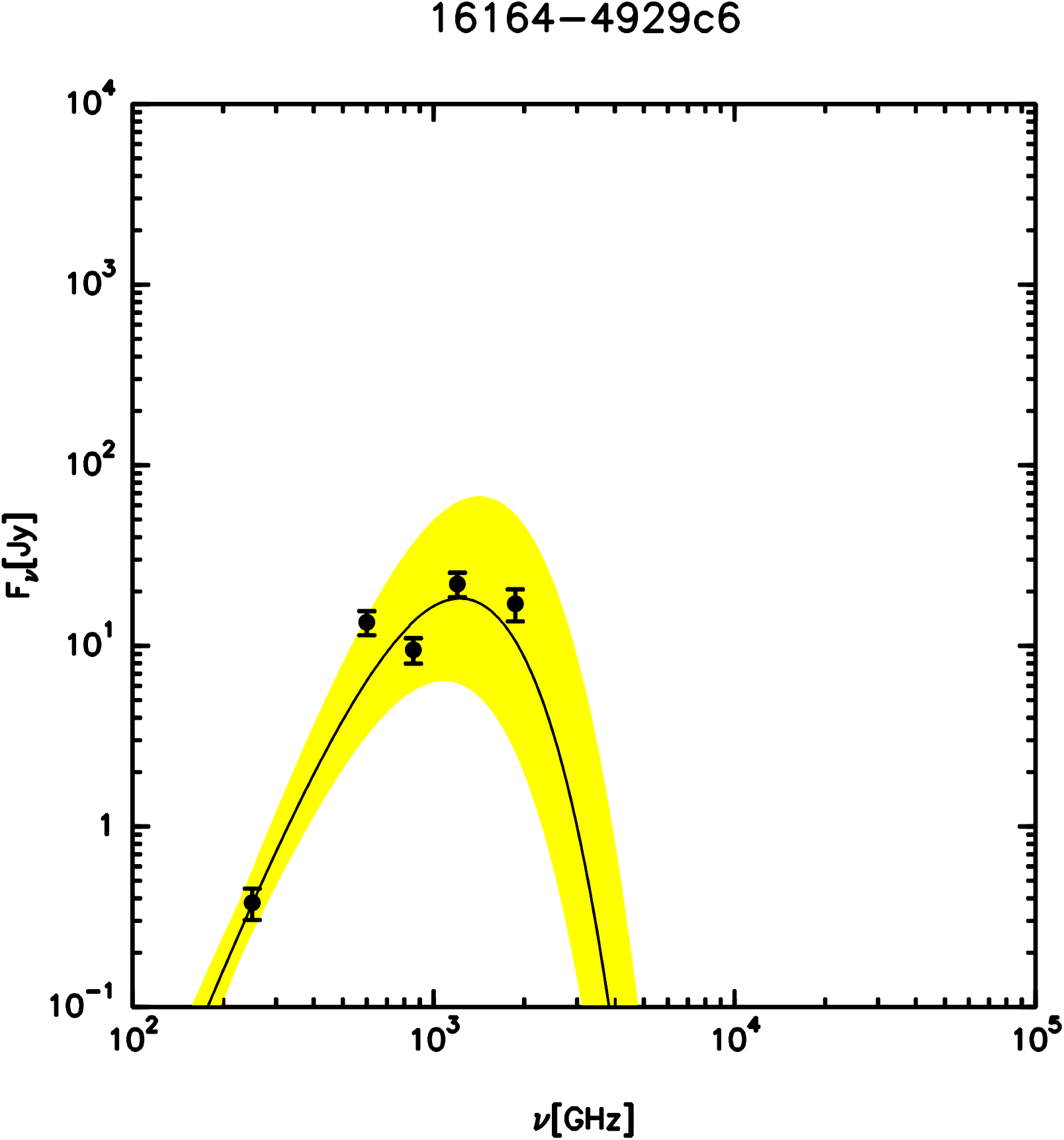}\\

 \includegraphics[width=0.3\textwidth]{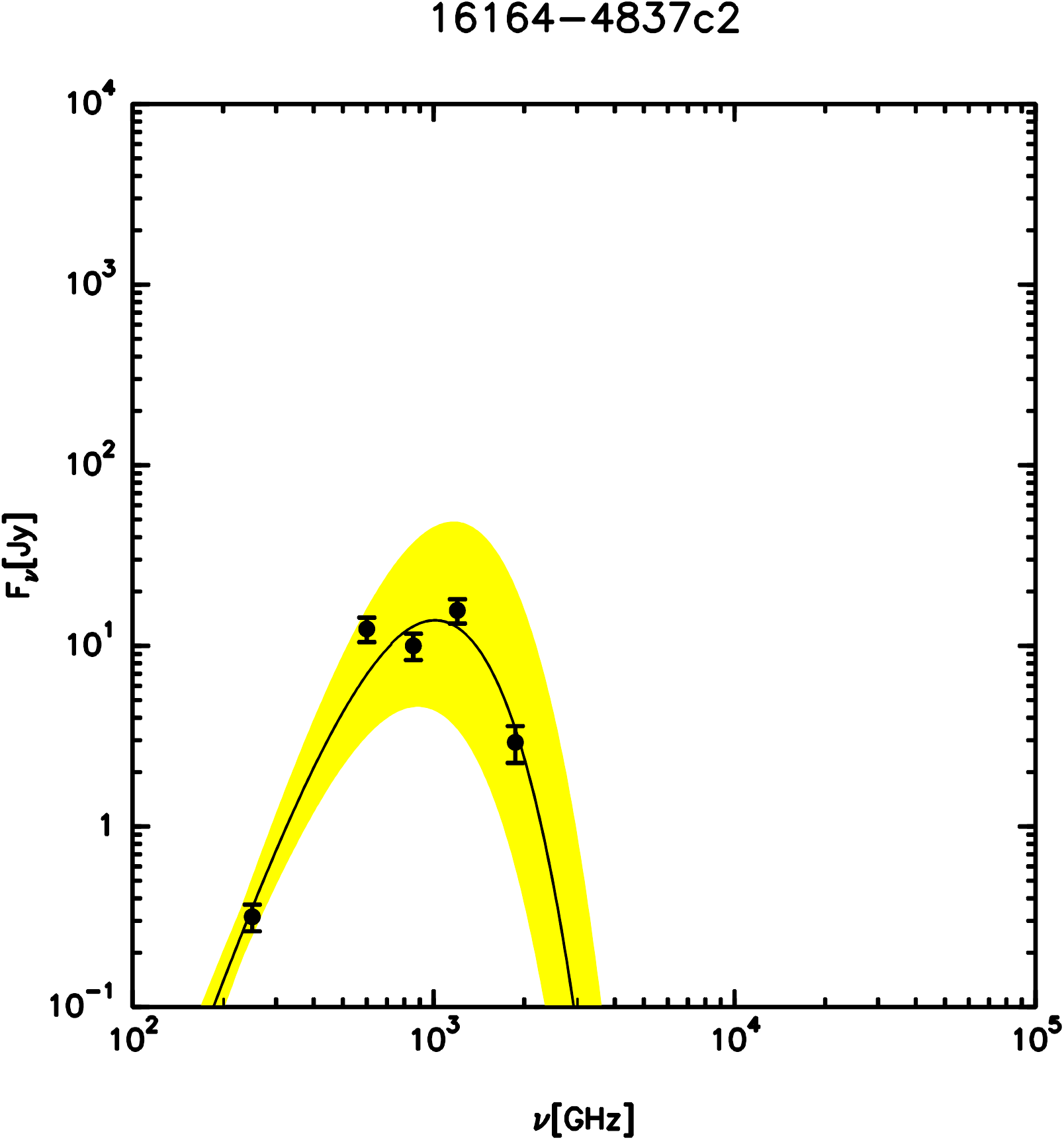}
 \includegraphics[width=0.3\textwidth]{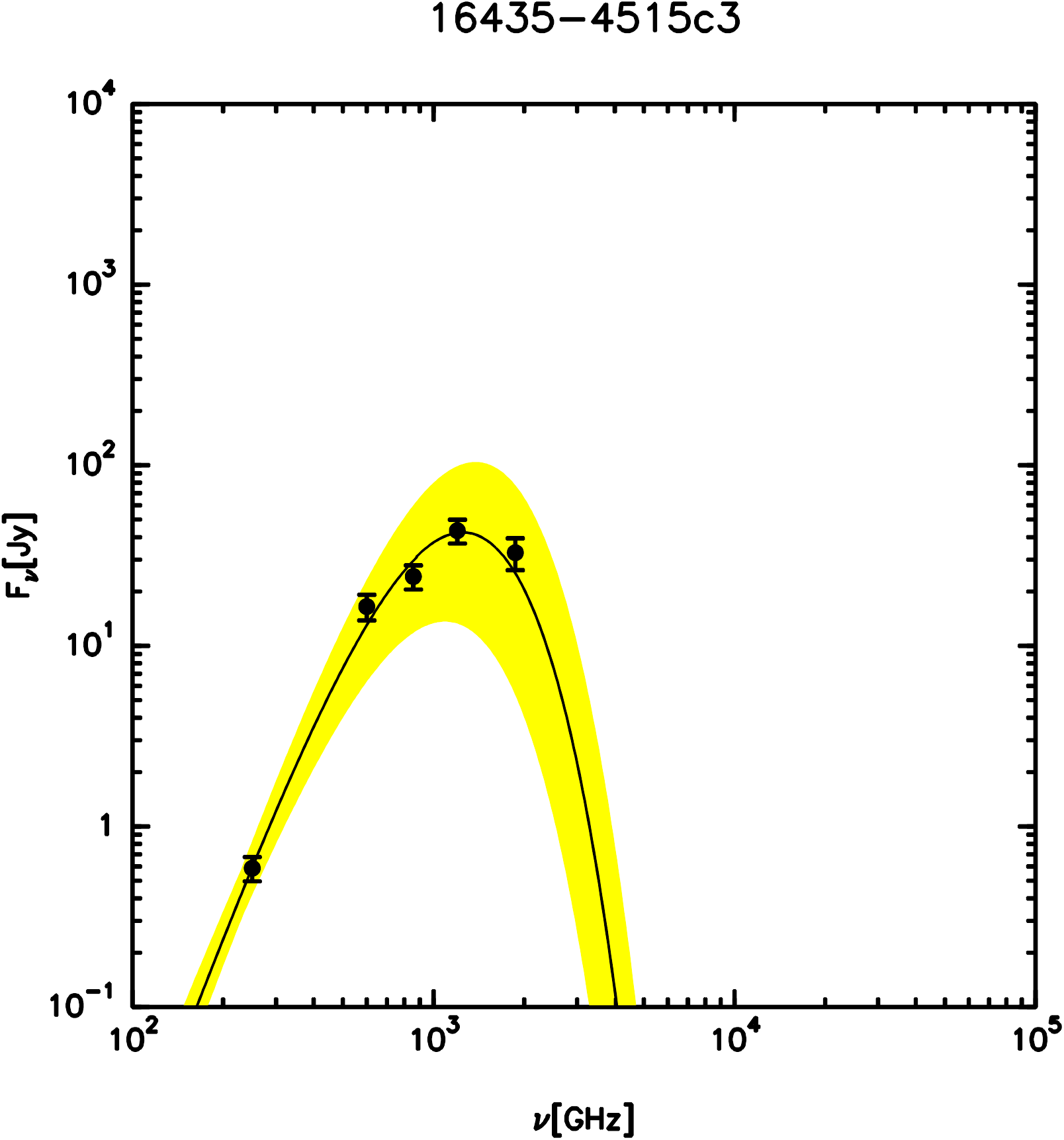}
 \includegraphics[width=0.3\textwidth]{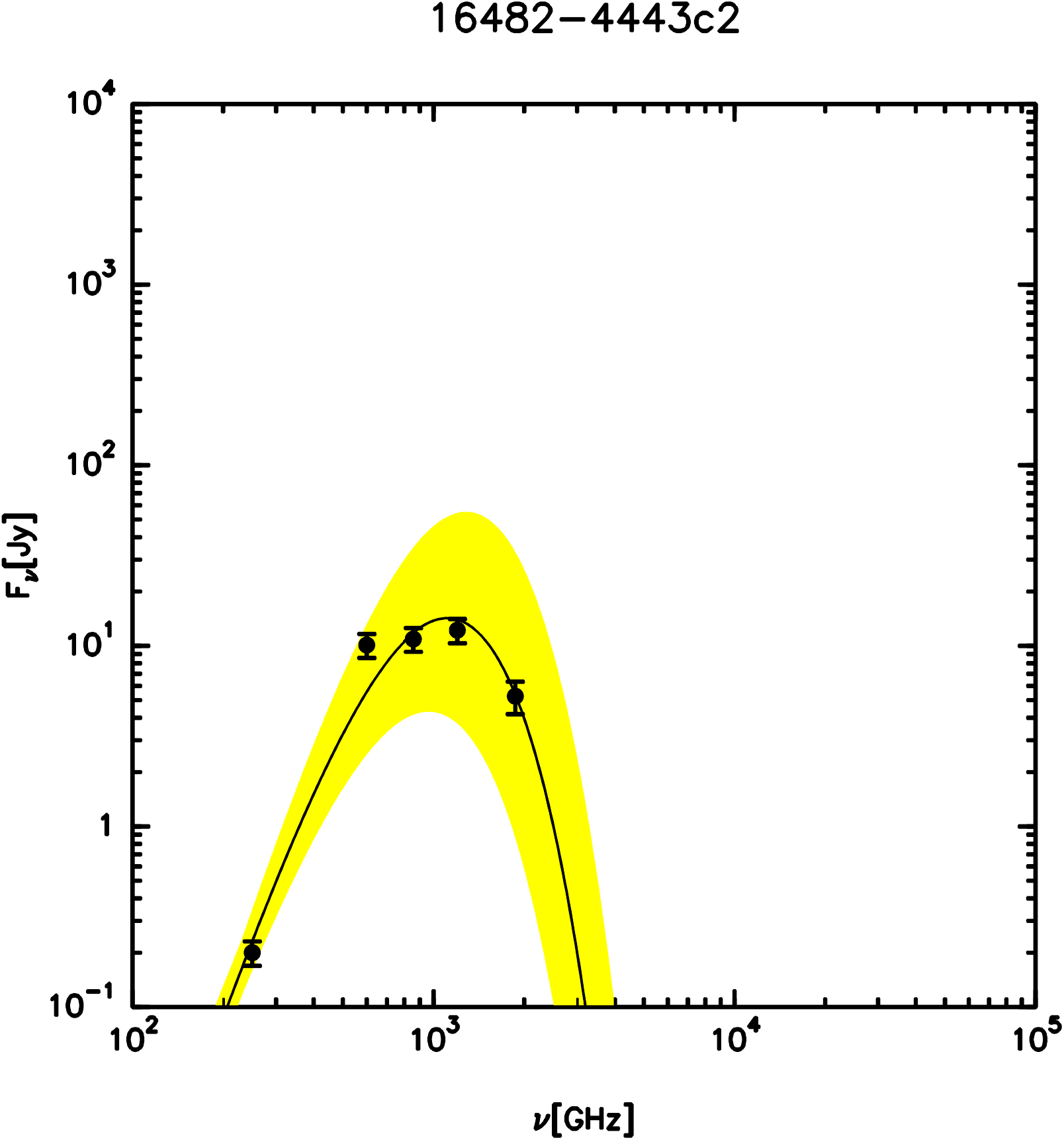}
 \caption{Same as Fig.~\ref{fig:sed_sfs}, but for the QS.}
 \label{fig:sed_qs}
\end{figure*}

\end{document}